\documentclass{article}
\usepackage[utf8]{inputenc}
\usepackage{graphicx}
\usepackage{geometry}
\usepackage{authblk}
\usepackage{amsmath}
\usepackage{amssymb}
\usepackage{xcolor}
\usepackage{hyperref}
\usepackage[sorting = none, backend = biber, style=numeric-comp]{biblatex}

\title{Mapping State Transition Susceptibility in Quantum Annealing}

\author[1]{Elijah Pelofske
\thanks{Email: epelofske@lanl.gov}}
\affil[1]{Los Alamos National Laboratory, CCS-3, Los Alamos, NM 87545, USA}

\date{\vspace{-6ex}}

\begin{document}

\maketitle

\begin{abstract}
Quantum annealing is a novel type of analog computation that aims to use quantum mechanical fluctuations to search for optimal solutions of Ising problems. Quantum annealing in the transverse field Ising model, implemented on D-Wave devices, works by applying a time dependent transverse field, which puts all qubits into a uniform state of superposition, and then applying a Hamiltonian over time which describes a user programmed Ising problem. 
We present a method which utilizes two control features of D-Wave quantum annealers, reverse annealing and an h-gain schedule, to quantify the susceptibility, or the distance, between two classical states of an Ising problem. The starting state is encoded using reverse annealing, and the second state is encoded on the linear terms of problem Hamiltonian. An h-gain schedule is specified which incrementally increases the strength of the linear terms, thus allowing a quantification of the h-gain strength required to transition the anneal into a specific state at the final measurement. By the nature of quantum annealing, the state tends towards global minima and therefore we restrict the second classical state to a minimum solution of the given Ising problem. This susceptibility mapping, when enumerated across all initial states, shows in detail the behavior of the quantum annealer during reverse annealing. The procedure is experimentally demonstrated on three small test Isings which were embedded in parallel on the D-Wave Advantage\_system4.1. Analysis of the state transition mapping shows detailed characteristics of the reverse annealing process including intermediate state transition paths, which are visually represented as state transition networks. 
\end{abstract}


\section{Introduction}
\label{section:introduction}
Quantum annealing in the transverse field Ising model was proposed as a novel analog computation model that utilizes quantum fluctuations in order to search for optimal solutions of combinatorial optimization problems \cite{doi:10.1126/science.284.5415.779, Kadowaki_1998, morita2008mathematical, das2008colloquium, finnila1994quantum, santoro2006optimization}. Adiabatic quantum computing is an ideal computation that is entirely isolated from the environment and slowly evolves the Hamiltonian \cite{grant2020adiabatic}. Adiabatic quantum computing (AQC) is of interest because it is equivalent to circuit model quantum computing. Quantum annealing (QA) is a practical heuristic implementation of adiabatic quantum computing \cite{grant2020adiabatic} where the coherence times are small \cite{hauke2020perspectives}, the system is not perfectly isolated from the environment, the variable connectivity is limited, and the time evolution is not necessarily slow. Quantum annealing has been implemented in hardware in a variety of contexts \cite{johnson2011quantum, hauke2020perspectives, boixo2014evidence, boixo2013experimental}, and D-Wave systems quantum annealing hardware is currently available as a cloud computing resource. D-Wave quantum annealers are implemented via typically sparse hardware graphs of superconducting flux qubits. Quantum annealing has been used as an experimental physics simulation tool \cite{harris2018phase, king2021qubit, zhou2021experimental, king2021scaling, munoz2019double} and as a computer to sample a wide variety of optimization problems \cite{10.1145/2482767.2482797, vert2021benchmarking, yarkoni2021quantum, boyda2017deploying} including, to name specific problem types, the graph coloring problem \cite{titiloye2011quantum, https://doi.org/10.48550/arxiv.2012.04470}, semiprime factorization \cite{jiang2018quantum, dridi2017prime, peng2019factoring, warren2019factoring, wronski2021practical, https://doi.org/10.48550/arxiv.2005.02268}, traveling salesperson problem \cite{silva2021mapping, papalitsas2019qubo}, air traffic management \cite{stollenwerk2019quantum}, maximum clique \cite{pelofske2022parallel, chapuis2019finding, pelofske2019solving}, graph partitioning \cite{ushijima2017graph, pelofske2021reducing}, boolean tensor networks \cite{o2020tucker, pelofske2021boolean, pelofske2022quantum}, community detection \cite{negre2020detecting}, spanning trees \cite{novotny2016spanning}, fault detection \cite{perdomo2015quantum}, and maximum cut \cite{pelofske2020advanced, barbosa2020optimizing}. Following with the theme of sampling optimization problems that are of interest for many possible applications, there have been numerous studies developing methods to improve the capabilities of modern quantum annealers using different parameter tuning techniques and algorithms \cite{berwald2021understanding, 9485068, chancellor2019domain, 9485068, PhysRevApplied.17.044005, barbosa2020optimizing, mishra2016performance, pudenz2015quantum, vinci2015quantum}. Overall, quantum annealing, and in particular the the cloud based D-Wave systems quantum annealers, are an active topic of study because of the potential heuristic capability of quantum annealing to sample optimization problems of interest in the NISQ-era \cite{preskill2018quantum} where hardware error rates and limited hardware connectivity \cite{PRXQuantum.2.040322, boothby2016fast, boothby2020next} prohibit more exact computation. For D-Wave quantum annealers, the Hamiltonian that is implemented in hardware can represented in Equation \ref{equation:QA_Hamiltonian} as a sum of the initial transverse field Hamiltonian and the user encoded Hamiltonian. 

\begin{equation}
    H_{ising} = - \frac{A(s)}{2} \Big( \sum_i \hat{\sigma}_{x}^{(i)} \Big) + \frac{B(s)} {2} \Big( \sum_i h_i \hat{\sigma_z}^{(i)} + \sum_{i>j} J_{i, j} \hat{\sigma_z}^{(i)} \hat{\sigma_z}^{(j)} \Big)
    \label{equation:QA_Hamiltonian}
\end{equation}

Where $\hat{\sigma}_{x,z}$ are Pauli matrices operating on qubit $i$. $h_i \in \mathbb{R}$ are the qubit biases and $J_{i,j} \in \mathbb{R}$ are the coupling strengths. D-Wave quantum annealers also allow users to change many parameters of the anneal beyond the problem Ising that is mapped to the device. For example, the users can modify the annealing schedule that defines where the anneal fraction $s$ is at each time step in the anneal; in normal forward annealing the points of the anneal schedule are defined as a linear interpolation from $s=0$ at the start of the anneal and $s=1$ at the end of the anneal \cite{https://doi.org/10.48550/arxiv.quant-ph/0001106}. Another relevant parameter that the user of cloud based D-Wave quantum annealers can specify is the \emph{annealing time} which specifies the anneal duration in microseconds (for the D-Wave \texttt{Advantage\_system4.1} for example, the annealing times can be in range from $0.5$ to $2000$). Equation \ref{equation:var_assignment_Hamiltonian} defines the problem Hamiltonian that is minimized during quantum annealing as a variable assignment problem.

\begin{equation}
    H(x_1,\ldots,x_n) = \sum_{i}^n h_i x_i + \sum_{i>j} J_{i, j} x_i x_j
    \label{equation:var_assignment_Hamiltonian}
\end{equation}

The linear weights $h_i \in \mathbb{R}$ and the quadratic couplers $J_{ij} \in \mathbb{R}$ define the discrete optimization problem, where the goal is to find the assignment of unknown binary variables $x_i$, $i \in \{1,\ldots,n\}$ that minimizes Equation \ref{equation:var_assignment_Hamiltonian}. Equation \ref{equation:var_assignment_Hamiltonian} is called a Quadratic Unconstrained Binary Optimization problem (\textit{QUBO}) problem if $x_i \in \{0,1\}$, and an \textit{Ising} problem if $x_i \in \{-1,+1\}$, where $i \in \{1,\ldots,n\}$. The reason that quantum annealing can be applied to so many different problem types, including NP-Hard problems, is because they can be formulated as Isings \cite{lucas2014ising}, or equivalently QUBOs \cite{https://doi.org/10.48550/arxiv.1811.11538, zaman2021pyqubo}, which can be mapped to a problem Hamiltonian of the form of Equation \ref{equation:var_assignment_Hamiltonian}. 

Alongside expanding its possible application domains, understanding the dynamics of how quantum annealers sample problems has been a subject of research \cite{Abel_2021, abel2021quantum, PhysRevA.105.022410, pelofske2020inferring, pelofske2019peering}. For example, one of the examples of a notable property of quantum annealing in the transverse field Ising model is that it does not sample ground states fairly for optimization problems which have multiple optimal solutions \cite{matsuda2009ground, pelofske2021sampling, konz2019uncertain, kumar2020achieving, mandra2017exponentially, yamamoto2020fair, PhysRevE.99.063314, PhysRevE.99.043306}. Fairly sampling optimal solutions, while not always necessary when solving optimization problems, is important for a variety of applications \cite{eslami2014shape, hinton2002training, gomes2021model, jerrum1986random, azinovic2017assessment, douglass2015constructing, weaver2012satisfiability, schaefer1978complexity, https://doi.org/10.48550/arxiv.2104.01941}. 

In this article we present a state transition susceptibility mapping methodology where a quantification can be made of the susceptibility of transitioning between two classical states in a quantum annealer, which can then be enumerated across all possible initial states for small Isings. This state mapping is accomplished by specifying an initial classical state as the starting point of the anneal (this is accomplished using \emph{reverse annealing}) and h-gain state encoding guides the anneal towards a specific intended classical ground state. This measure can be regarded as a type of susceptibility because it is measuring the response to an applied magnetic field during the quantum annealing process. This data is then analyzed using several different metrics which show correlations between initial state energy and hamming distance proportions. The data is then further characterized by creating state transition networks to show intermediate states. The results show that there are some initial states which are significantly more susceptible to being moved into an intended ground state, but are not necessarily near the ground state in terms of energy. The results are also analyzed to determine if they have a biased state transition susceptibility towards some ground states, and if that correlates to unfair sampling in forward annealing. 

All raw data from the experiments are publicly available datasets \cite{elijah_pelofske_2023_7676291, elijah_pelofske_2023_7676350, elijah_pelofske_2023_7676389, elijah_pelofske_2023_7677424}.


\section{Methods}
\label{section:methods}
In this section we outline the experimental methods used to map the susceptibility from input states to ground states of small test Isings. First, in section \ref{section:methods_problem_isings} we outline the three Ising problems that will be investigated. In section \ref{section:methods_state_encoding_methods} the settings and methods used to encode the initial state and the intended ground state of the Isings. In section \ref{section:methods_metrics}, the different metrics that the results will be analyzed with are defined. All figures generated in this article use the python packages Matplotlib \cite{matplotlib, thomas_a_caswell_2022_6513224} and Networkx \cite{hagberg2008exploring}. 

\begin{figure}[t]
    \centering
    \includegraphics[width=0.32\textwidth]{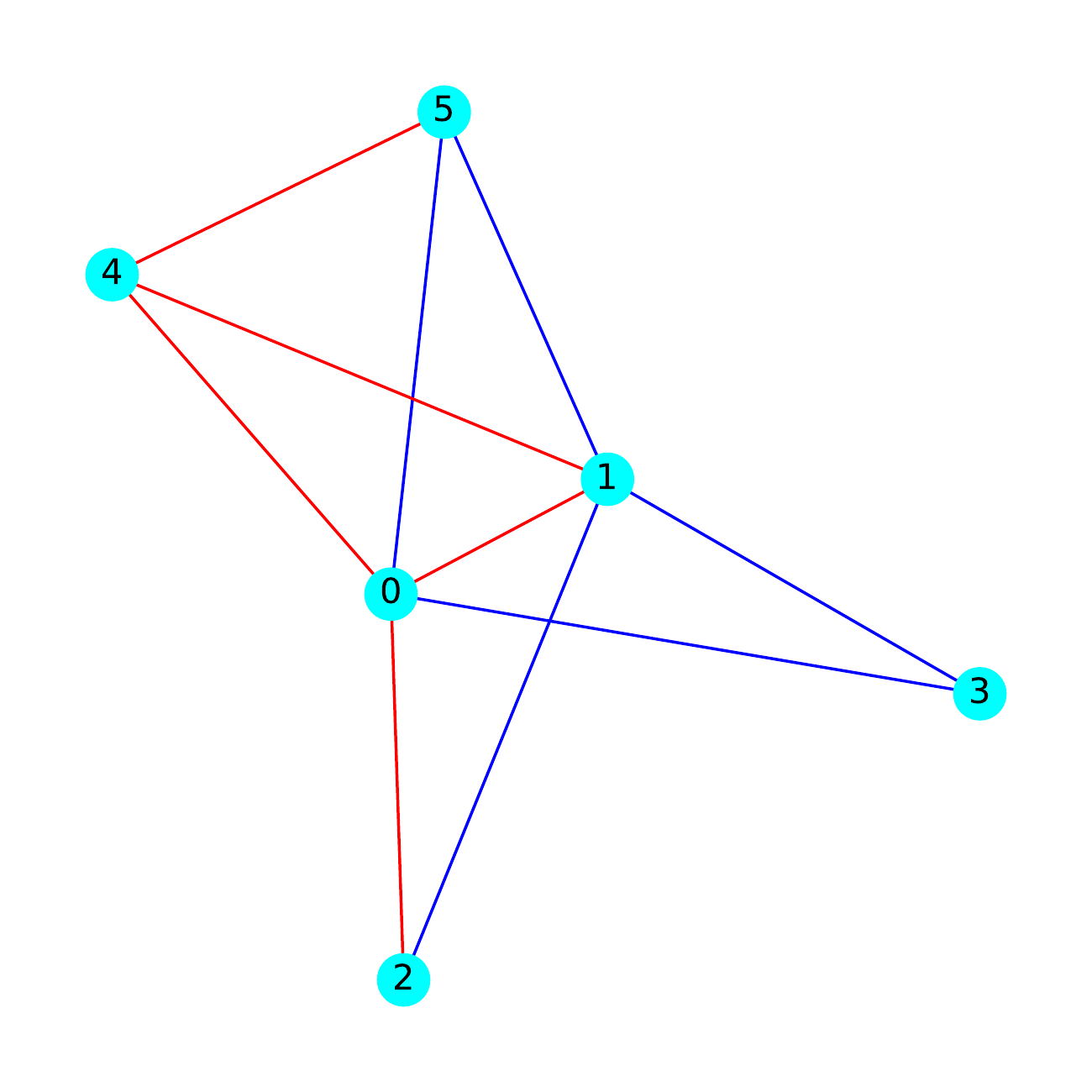}
    \includegraphics[width=0.32\textwidth]{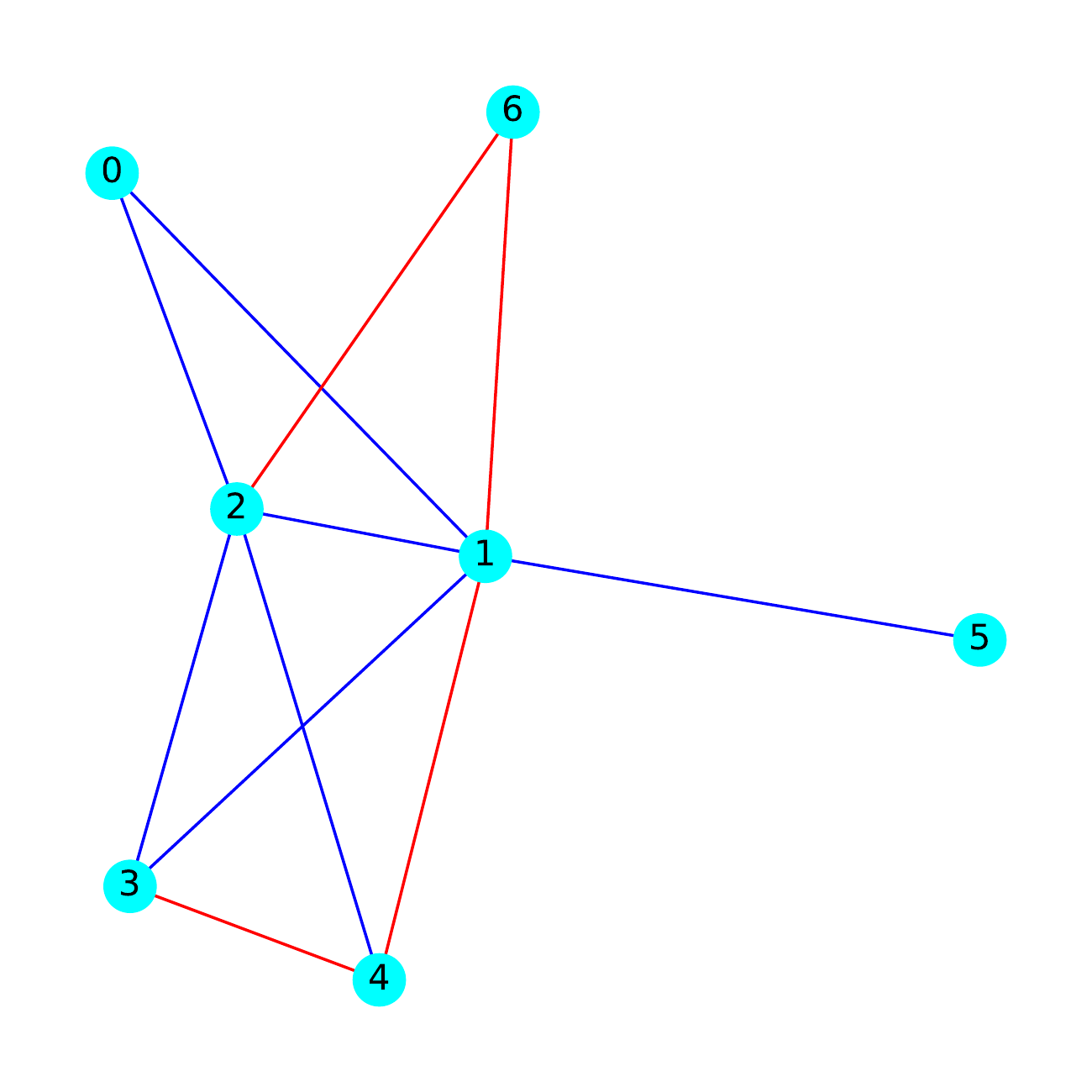}
    \includegraphics[width=0.32\textwidth]{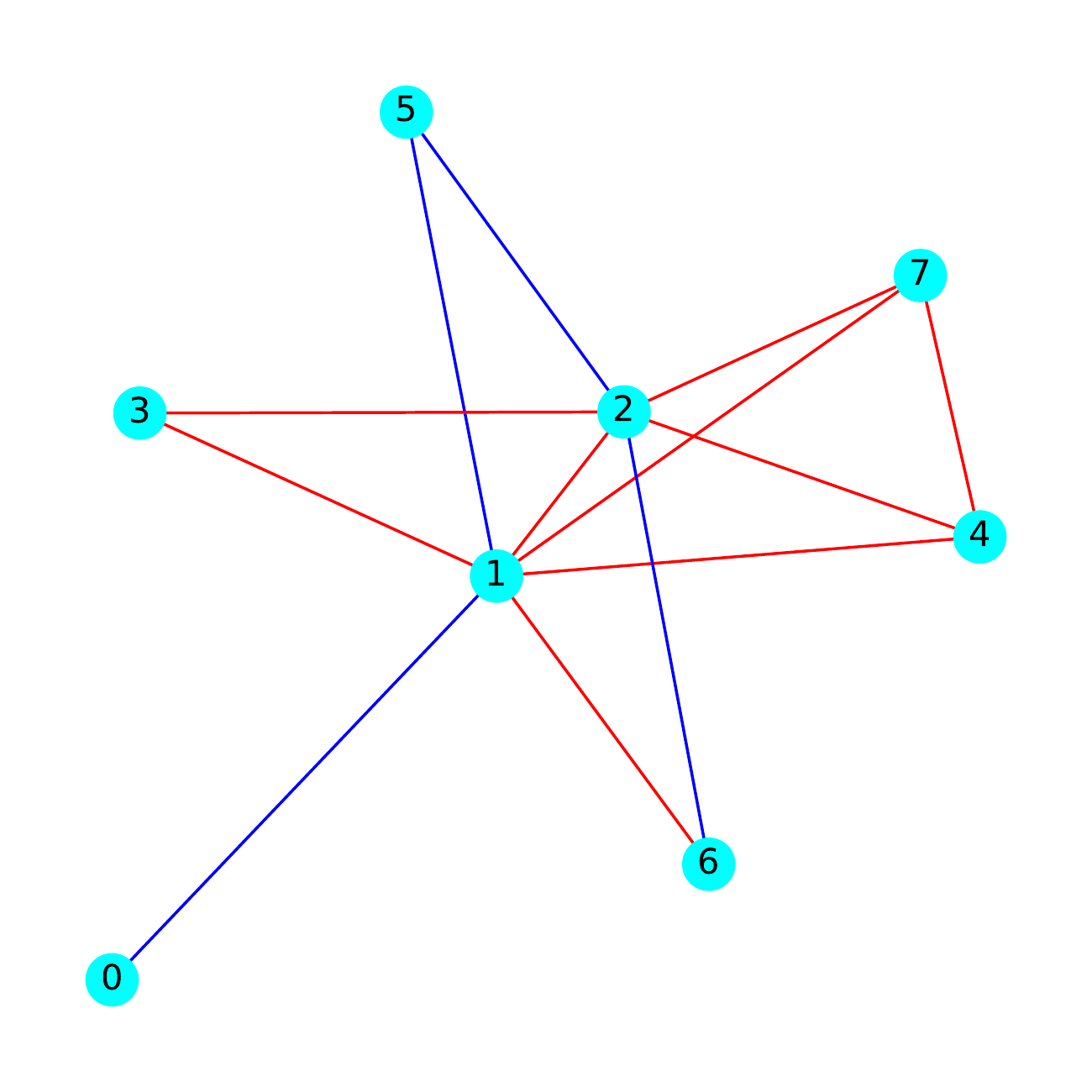}
    \caption{The logical $N_6$ Ising, $N_7$ and $N_8$ in order from left to right. Quadratic weights of $+1$ are encoded as blue edges in these problem graphs, and quadratic weights of $-1$ are encoded as red edges in these problem graphs. The $N_6$ Ising has 4 ground states (optimal solutions), $N_7$ has 2 ground states and $N_8$ has 8 ground states. All linear terms are set to $0$. The variable indices are drawn on the node labels. Each of these Isings are subgraphs of the Pegasus graph topology. }
    \label{fig:logical_isings}
\end{figure}

\subsection{Problem Isings and hardware embedding}
\label{section:methods_problem_isings}

\begin{figure}[t]
    \centering
    \includegraphics[width=0.32\textwidth]{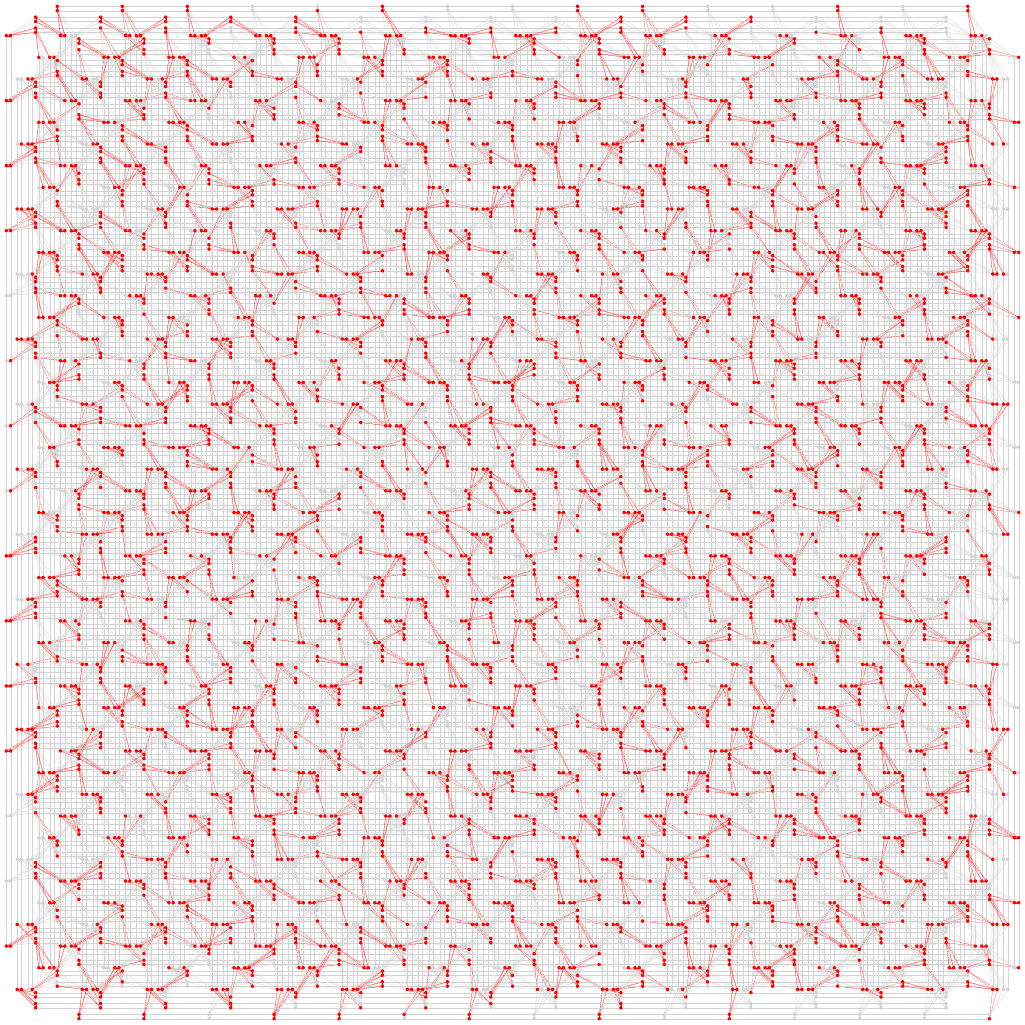}
    \includegraphics[width=0.32\textwidth]{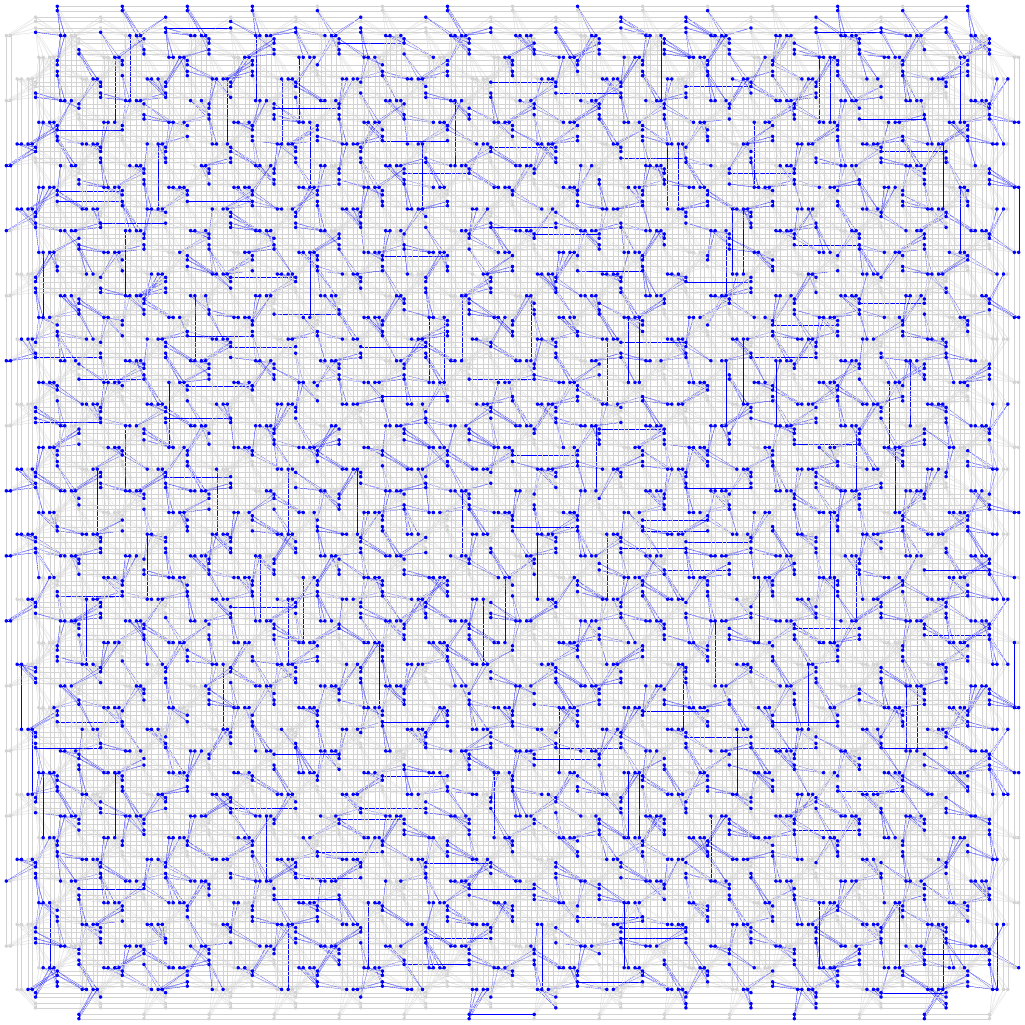}
    \includegraphics[width=0.32\textwidth]{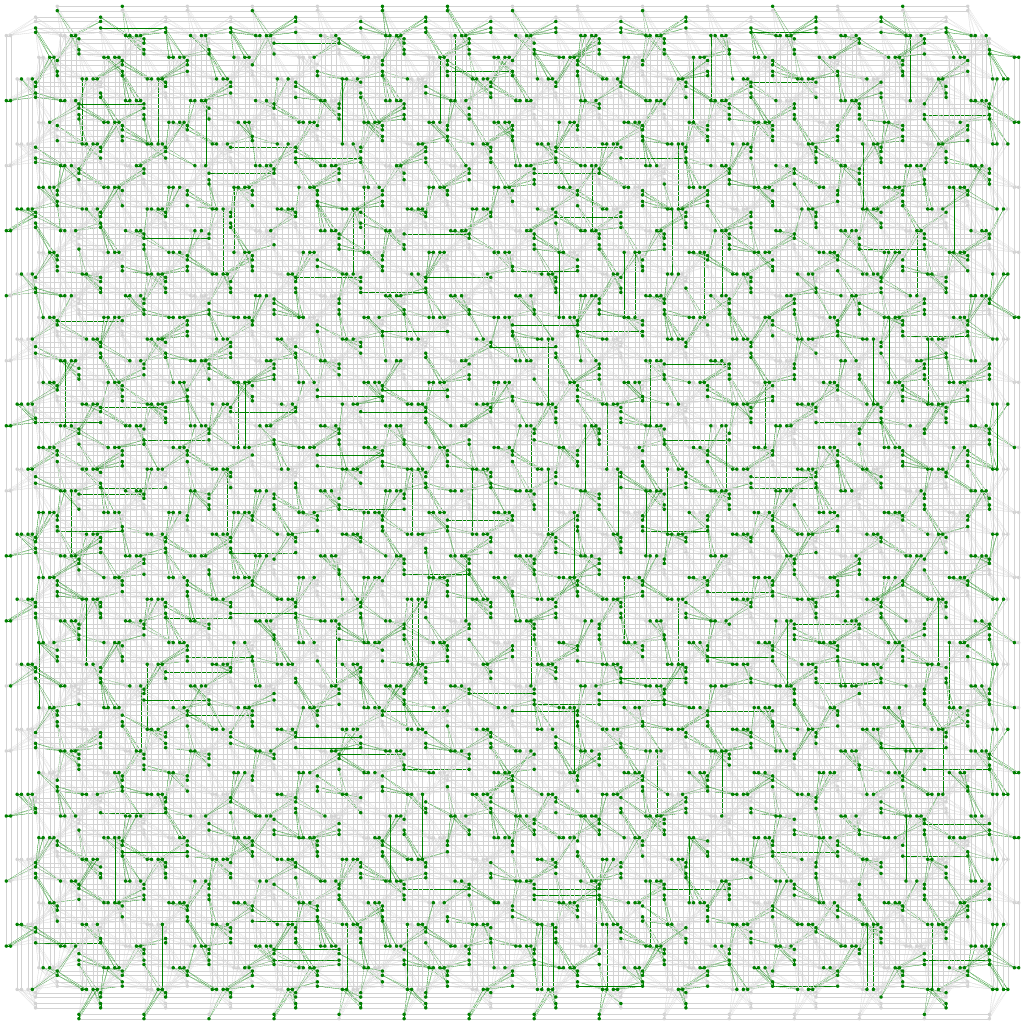}
    \caption{Test Isings embedded in parallel onto the hardware connectivity graph of Advantage\_system4.1. The $N_6$ Ising is embedded 722 times (left). The $N_7$ Ising is embedded 625 times (middle). The $N_8$ Ising is embedded 525 times (right). The nodes and edges used in the hardware graph embeddings are colored red (left), blue (middle), and green (right) and unused qubits and couplers are colored with high transparency grey. The Advantage\_system4.1 device has a Pegasus $P_{16}$ topology, however it also has some missing hardware components (qubits and couplers) which are unintentional manufacturing defects, therefore these parallel (or \emph{tiled}) embeddings must account for these hardware defects. }
    \label{fig:embeddings}
\end{figure}

Figure \ref{fig:logical_isings} defines the three problem Isings we will investigate. Importantly, these three Isings are natively embeddable onto the D-Wave Pegasus Quantum Processing Unit (QPU) topology \cite{dattani2019pegasus, zbinden2020embedding, boothby2020next}. This allows for these small Isings to be repeatedly embedded onto the chip connectivity which allows us to execute a large sample size of these small Isings \emph{in parallel} during the same annealing cycle - i.e. \emph{parallel quantum annealing} \cite{pelofske2022quantum, pelofske2022parallel}. These parallel embeddings are also referred to as \emph{tiling} \footnote{\url{https://dwave-systemdocs.readthedocs.io/en/samplers/reference/composites/tiling.html}}. Embedding many problem instances onto the hardware ensures not only that robust statistics of the problem will be gathered, but it also aims to average out any specific biases that may exist for specific groups of qubits. Not using minor-embedding also mitigates the additional potential problems that can come with minor embedding \cite{Marshall_2022}, including resolving chain breaks \cite{grant2022benchmarking, pelofske2019solving} and the ferromagnetic coupling chain strength dominating the programmed energy scale on the chip \cite{osti_1498001}. All experiments use the D-Wave \texttt{Advantage\_system4.1}; Figure \ref{fig:embeddings} shows these three Isings embedded onto the \texttt{Advantage\_system4.1} connectivity at least several hundred times using the minorminer heuristic minor embedding tool \cite{https://doi.org/10.48550/arxiv.1406.2741}. These test Isings were selected for several reasons. 

\begin{enumerate}
    \item They match the native connectivity of the Pegasus graph which means that minor embedding will not be required in order to execute these Isings on D-Wave Advantage\_system4.1. 
    \item They have multiple optimal solutions which allows us to examine any differences, or similarities, in the behavior of the h-gain response curves for the different ground states. 
    \item They are simple random spin glasses, which are typical for test problems on quantum annealers \cite{PhysRevX.5.031040, konz2019uncertain, https://doi.org/10.48550/arxiv.2202.03044, PhysRevB.95.184416}, and in general this class of problems is NP-Hard \cite{barahona1982computational}.
    \item They are small enough that we can easily compute their optimal solutions and we can also enumerate over all possible initial states with the reverse annealing and h-gain state encoding methods. 
    \item Linear terms are intentionally not included on the problem Ising so that the \emph{h-gain state encoding} can be utilized. 
\end{enumerate}

Because these Isings are sufficiently small it is possible to enumerate over all possible solutions of the variable assignment problem $\{+1, -1\}$ and determine which states are the optimal solutions. These types of Ising problems (meaning edge coefficients with weight of either $1$ or $-1$ and there no linear terms) naturally will have complementary ground states, meaning that there are always at least two optimal solutions and they will be complements of each other. Here we provide these Isings optimal solutions:

\begin{enumerate}
    \item The $N_6$ Ising has exactly $4$ optimal solutions, each with an energy of $-6$. These optimal variable assignments are $[-1, -1, -1, +1, -1, +1]$, $[-1, -1, +1, +1, -1, +1]$, $[+1, +1, -1, -1, +1, -1]$, $[+1, +1, +1, -1, +1, -1]$
    \item The $N_7$ Ising has exactly $2$ optimal solutions, each with an energy of $-7$. These optimal variable assignments are $[-1, +1, +1, -1, -1, -1, +1]$, $[+1, -1, -1, +1, +1, +1, -1]$
    \item The $N_8$ Ising has exactly $8$ optimal solutions, each with an energy of $-11$. These optimal variable assignments are $[-1, +1, -1, +1, +1, -1, -1, +1]$, $[-1, +1, -1, +1, +1, -1, +1, +1]$, $[-1, +1, +1, -1, -1, -1, +1, +1]$, $[-1, +1, +1, -1, +1, -1, +1, +1]$, $[+1, -1, -1, +1, -1, +1, -1, -1]$, $[+1, -1, -1, +1, +1, +1, -1, -1]$,
    \newline $[+1, -1, +1, -1, -1, +1, -1, -1]$, $[+1, -1, +1, -1, -1, +1, +1, -1]$
\end{enumerate}

\subsection{State encoding methods}
\label{section:methods_state_encoding_methods}

The h\_gain\_schedule parameter on D-Wave quantum annealers works by introducing a time dependent function $g(t)$ to the linear terms of the problem Hamiltonian, resulting in a modified version of Equation \ref{equation:QA_Hamiltonian}. 

\begin{equation}
    H_{ising} = - \frac{A(s)}{2} \Big( \sum_i \hat{\sigma}_{x}^{(i)} \Big) + \frac{B(s)} {2} \Big( g(t) \sum_i h_i \hat{\sigma_z}^{(i)} + \sum_{i>j} J_{i, j} \hat{\sigma_z}^{(i)} \hat{\sigma_z}^{(j)} \Big)
    \label{equation:QA_Hamiltonian_h_gain}
\end{equation}

The h-gain feature of D-Wave quantum annealers has been used before in a variety of contexts \cite{harris2018phase, PRXQuantum.2.030317, Abel_2021, pelofske2020advanced, https://doi.org/10.48550/arxiv.2208.09068}. The h-gain feature could also be used for other interesting simulation methods such as implementing adiabatic reverse annealing \cite{PhysRevA.105.032431, PhysRevA.100.052321}. The h-gain field allows the user to program a longitudinal magnetic field into the system according to a specific schedule, meaning that it can vary over time. The h-gain encoding of the ground state solutions involves specifying the linear weights of the Ising as the complements (multiply all variables by $-1$) of the ground state. Then we also specify the h\_gain\_schedule (which starts out at $0$ strength). More details of this implementation can be found in ref. \cite{pelofske2020advanced}. For simplicity the Isings we consider in this study do not have linear terms, however it is important to note that by introducing an additional slack variable (which may require minor-embedding depending on the problem) one can still utilize this h-gain encoding method for other problems. For consistency, in the article when referring to this specific encoding method the term \emph{h-gain state encoding} will be used. 

The h\_gain\_schedule is defined to be \texttt{[[0.0, 0], [0.05, 0], [0.1, h], [99.1, h], [99.15, 0], [100, 0]]} where the first coordinate is time in microseconds, and the second coordinate is the applied factor by which the linear terms are multiplied (i.e. this schedule defines the function $g(t)$). The parameter $h$ is varied from $0.1$ to $3$ in steps of $0.1$; $3$ is the maximum h-gain strength allowed on the \texttt{Advantage\_system4.1} hardware. The intent of this schedule is to begin applying the h-gain field very quickly once the anneal starts using a high slope ramp up, and then to ramp down just before the anneal ends. However, the applied h\_gain field is turned off both at the very start and very end of the anneal, with the field being turned off nearly $1$ microsecond before the end of the anneal. The intention of beginning and ending the schedule in this manner is to allow the state transition measurement at the end of the anneal be entirely based on the state of the system having been allowed to evolve a small amount before measurement without the applied h-gain field. The exact choice if timing here is based only on empirical results where the applied field was strong enough to get many states into the intended ground state, but not so strong as to overwhelm the anneal immediately into the intended ground state. Other choices of h\_gain\_schedule also have these properties and could be utilized in the future as well; for example smaller pulses of h-gain fields could be used to induce smaller ``pushes" towards an intended state. This choice of h-gain schedule satisfies another constraint of the Advantage\_system4.1, which is that the maximum number of points we can use to define the $g(t)$ function is $20$. The maximum slope present in the h\_gain\_schedule is also constrained, which is the reason for the jumps of $0.05$ microseconds. Although not used in our methodology, the h\_gain\_schedule is also allowed to have negative values. When instantiated in the hardware, the user programmed points are linearly interpolated between and a continuous function over time is created which is then implemented in the hardware. 

The initial variable state is specified using the \emph{reverse annealing} control feature of the D-Wave quantum annealer. Reverse annealing is a variant of quantum annealing which begins and ends the anneal in classical states, where the beginning state is programmed by the user, and over the duration of the anneal the system can be set to different anneal fractions $s$ over time with the goal of improving upon the initial state the anneal began with \cite{ohkuwa2018reverse, golden2021reverse, venturelli2019reverse, kumar2020achieving, pelofske2020advanced, perdomo2011study, chancellor2017modernizing, marshall2019power}. Programming reverse annealing consists of specifying three different parameters. First, the boolean state \texttt{reinitialize\_state} must be specified, which dictates whether the state is re-initialized after each read out or not. For all experiments we set \texttt{reinitialize\_state} to True. Next we set the anneal schedule is chosen, which needs to begin and stop in the classical state (meaning the anneal fraction $s=1$). The reverse annealing schedule is fixed to \texttt{[[0.0, 1.0], [20, 0.65], [80, 0.65], [100, 1.0]]} where the first coordinate is time in microseconds and the second coordinate is the anneal fraction $s$. The reverse annealing schedule was chosen to have a standard ramp up and ramp down that is symmetric, both with a duration of 20 microseconds. The anneal fraction to pause at was chosen empirically so as to not overwhelm the samples with either the initial state or the intended state - in other words these parameters were chosen such that the response curves were not flat. A smaller anneal fraction during the pause (for example $0.5$ or $0.4$) would make the state transitions overall more susceptible, and a larger anneal fraction (for example $0.8$) would make the state transitions less susceptible. Lastly we need to specify the initial classical state for all active qubits. This is done using the \texttt{initial\_state} parameter, which encodes the variable states as spins of either $1$ or $-1$. As with the h\_gain\_schedule, the programmed points that define the anneal schedule are linearly interpolated between in order to create a continuous anneal schedule over the anneal time which is implemented in the hardware. 

Each device call uses exactly $1000$ anneals. Every other parameter is set to default; for example \texttt{auto\_scale} is set to True. The \texttt{annealing\_time} is always $100$ microseconds. Both the \texttt{programming\_thermalization} and \texttt{readout\_thermalization} are set to $0$ microseconds. 

Combining reverse annealing to specify an arbitrary initial state and the \emph{h-gain state encoding} to specify a final state results in the unified \emph{state transition susceptibility mapping} technique. For simplicity and because the annealing process naturally tends towards ground states by design, we restrict the intended final states to be those of optimal solutions, or ground states, of the problem Isings. Data which shows the success proportion of how many final states were the intended ground state as a function of increasing the parameter $h$ for the h-gain schedule will be referred to \emph{h-gain response curves}. In order to denote this mapping procedure we will use $A \rightarrow B$, where $A$ is any classical initial state and $B$ is any classical ground state of the problem Ising.

\subsection{Metrics and algorithms}
\label{section:methods_metrics}

The primary quantity of interest in these experiments is the proportion of samples, out of all measured samples across the $1000$ anneals and hundreds of embedded instances which we will denote as $n\_samples$, that are in the ground state that was encoded in the linear terms of the problem. Given we find that $n\_{GS}$ samples are in this specific groundstate, we will define this proportion of ground state samples:

\begin{equation}
    \label{eq:proportion_of_groundstates}
    P_{GS} = \frac{n\_{GS}}{n\_samples}
\end{equation}

Because each D-Wave backend call requests $1000$ samples and the Ising is embedded multiple times on the hardware, the total number of samples for each device call is $n\_samples = 722000$ for the $N_6$ Ising, $n\_samples = 625000$ for the $N_7$ Ising, and $n\_samples = 525000$ for the $N_8$ Ising. This experimental setup allows for robust sample sizes to be collected and to get high accuracy simulation results. 

Having enumerated across all $h$ strengths $\in [0, 3]$ in steps of $0.1$ indexed by the variable $j$, we get a $P_{GS}$ measure for each of those different $h$ strengths. From these measures we create a \emph{susceptibility} metric defined in Equation \ref{eq:susceptibility} which describes the amount of h-gain strength that was required to move the anneal from the initial state (encoded using reverse annealing) to the specified ground state (encoded using the h-gain state encoding method). 

\begin{equation}
    \label{eq:susceptibility}
    \chi = \frac{\sum_{j=0}^{30} P_{{GS}_j}}{30}
\end{equation}

The susceptibility metric captures across all measured h-gain strengths how successful this mechanism was at causing the system to end in the intended ground state. If $\chi=1$ (e.g., high susceptibility), at every amount of applied $h$ the system $A \rightarrow B$ was \emph{very susceptible} to transitioning to the target ground state. If $\chi=0$ (e.g., low susceptibility), there were no samples from the quantum annealer at any value of $h$ that were measured in the intended ground state - indicating that the mapping system $A \rightarrow B$ was \emph{not susceptible} to transitioning to the target ground state. In practice, no states will have exactly $\chi=0$ or $\chi=1$ due to the high measurement count and the noise present on the device. Note that this metric is similar to measuring magnetic susceptibility in other contexts of quantum annealing (see \cite{katzgraber2014glassy, PRXQuantum.2.030317, doi:10.1126/science.284.5415.779}), however it is slightly different in that it is a sum over several h-gain strengths, not a single pulse, and the system was in a very specific initial state due to reverse annealing. 

\begin{figure}[t!]
    \centering
    \includegraphics[width=0.19\textwidth]{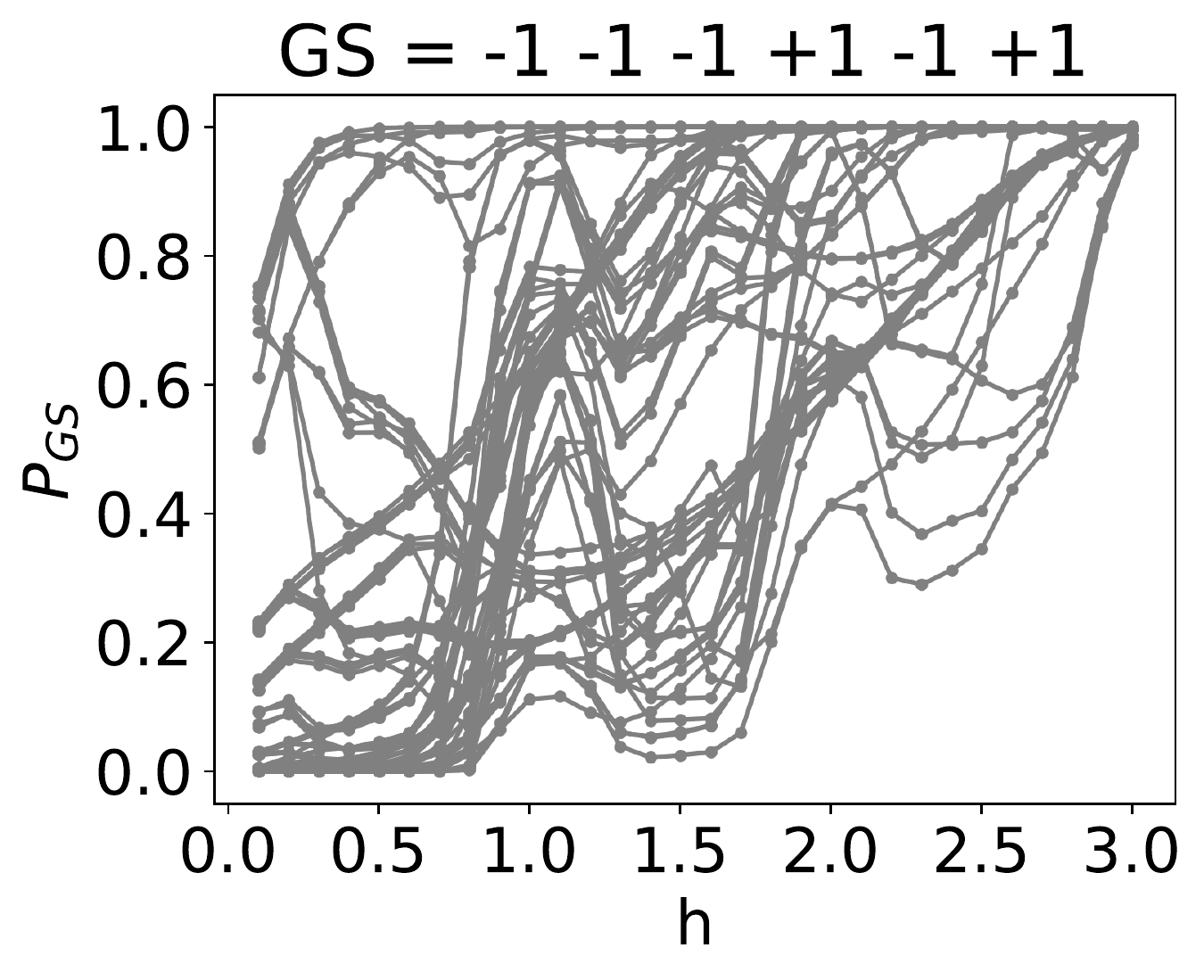}
    \includegraphics[width=0.19\textwidth]{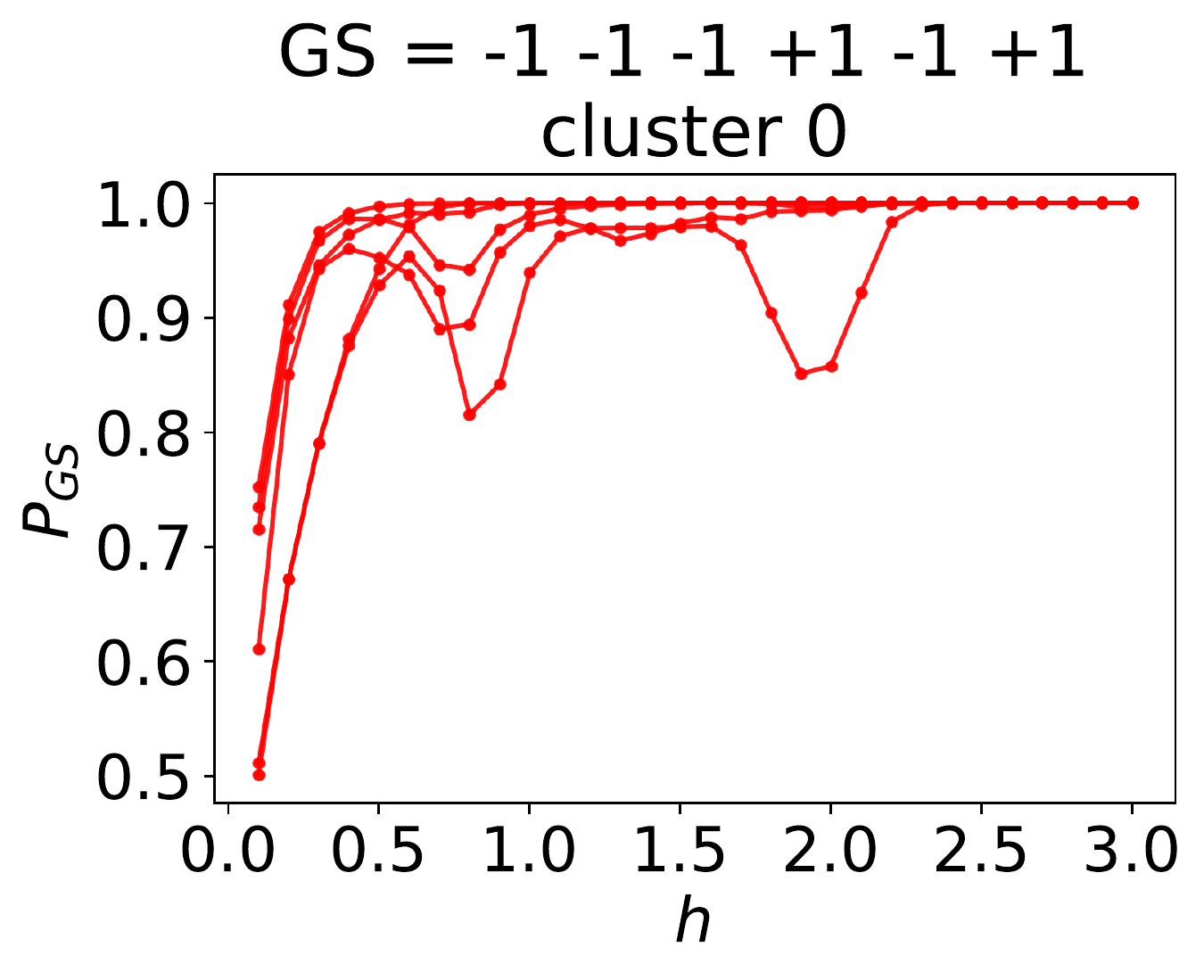}
    \includegraphics[width=0.19\textwidth]{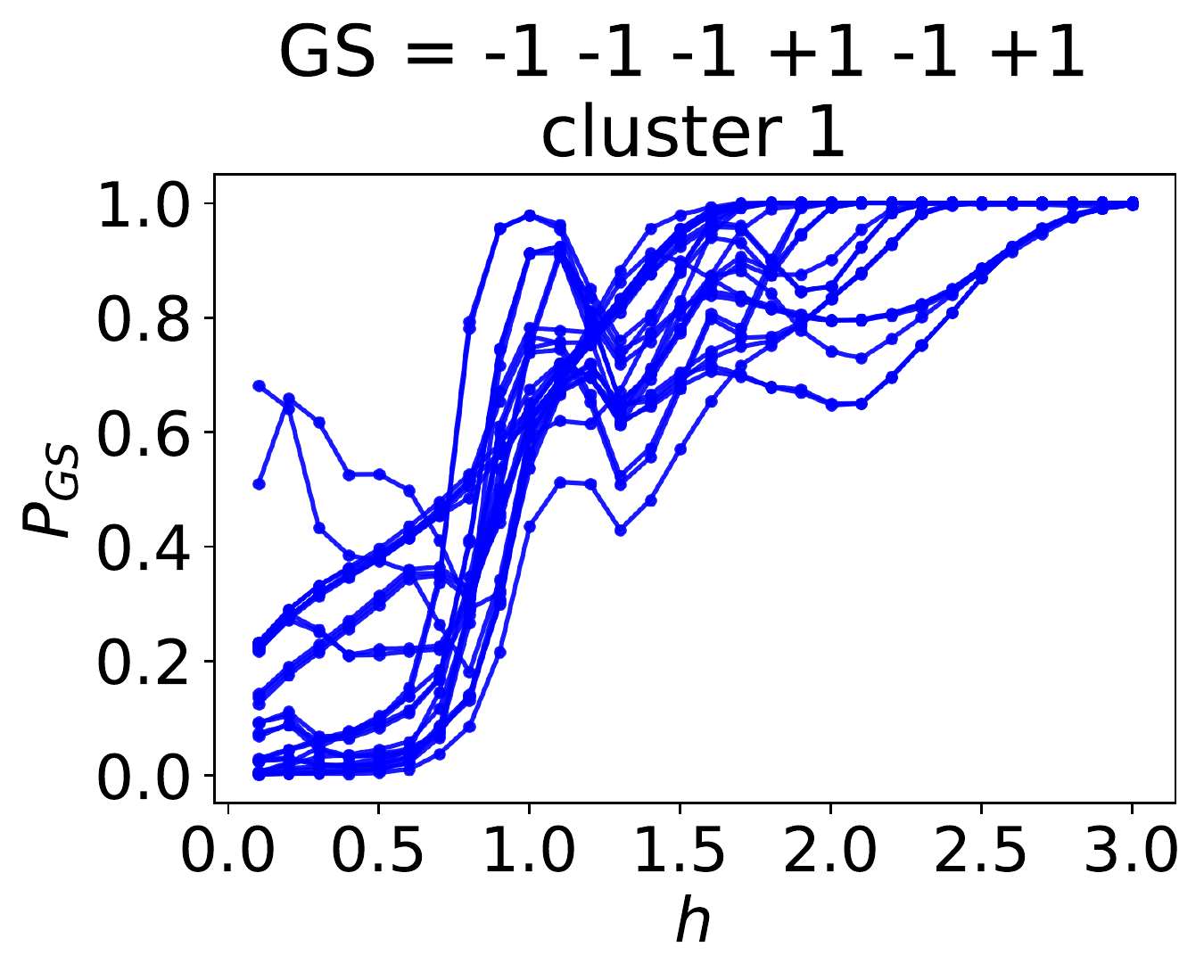}
    \includegraphics[width=0.19\textwidth]{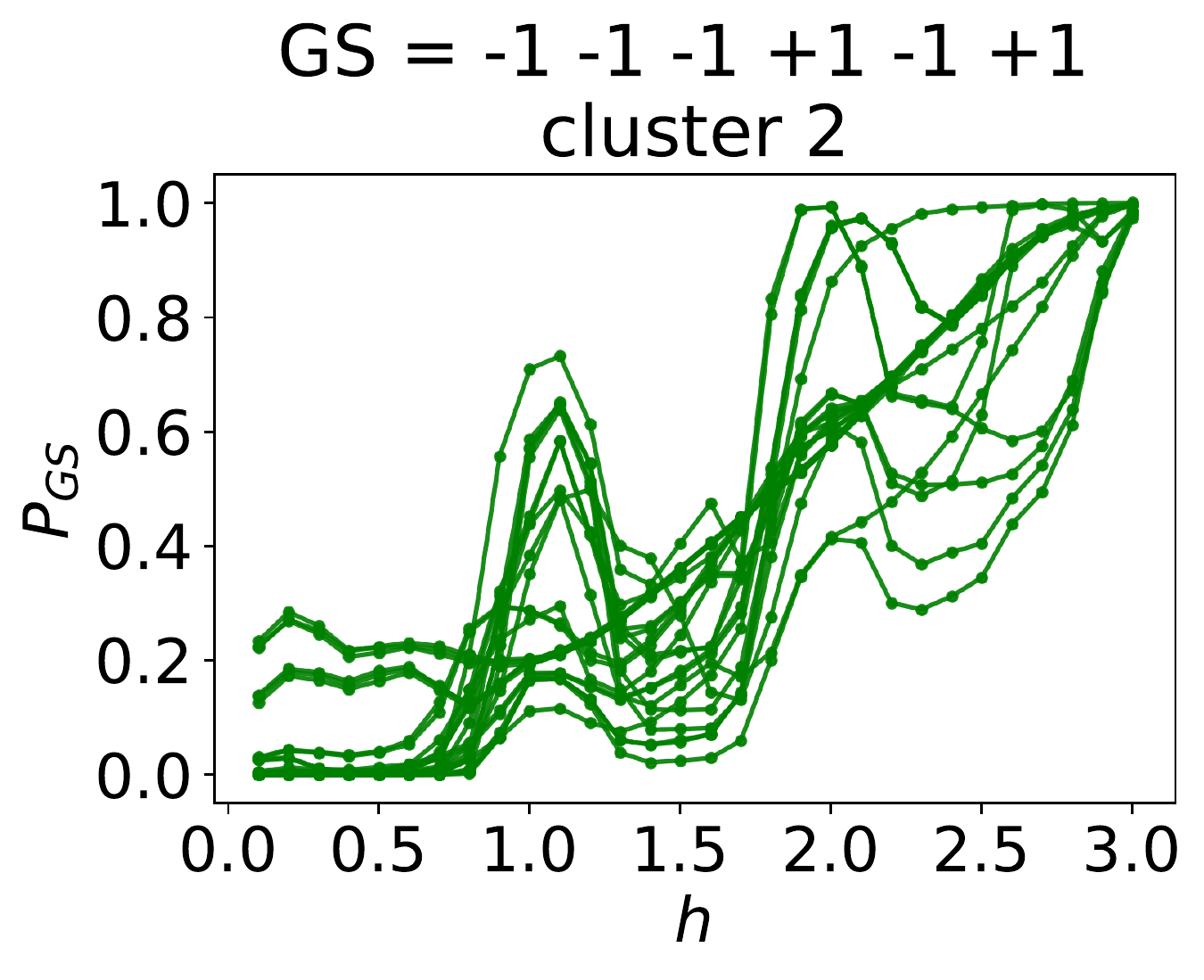}
    \includegraphics[width=0.19\textwidth]{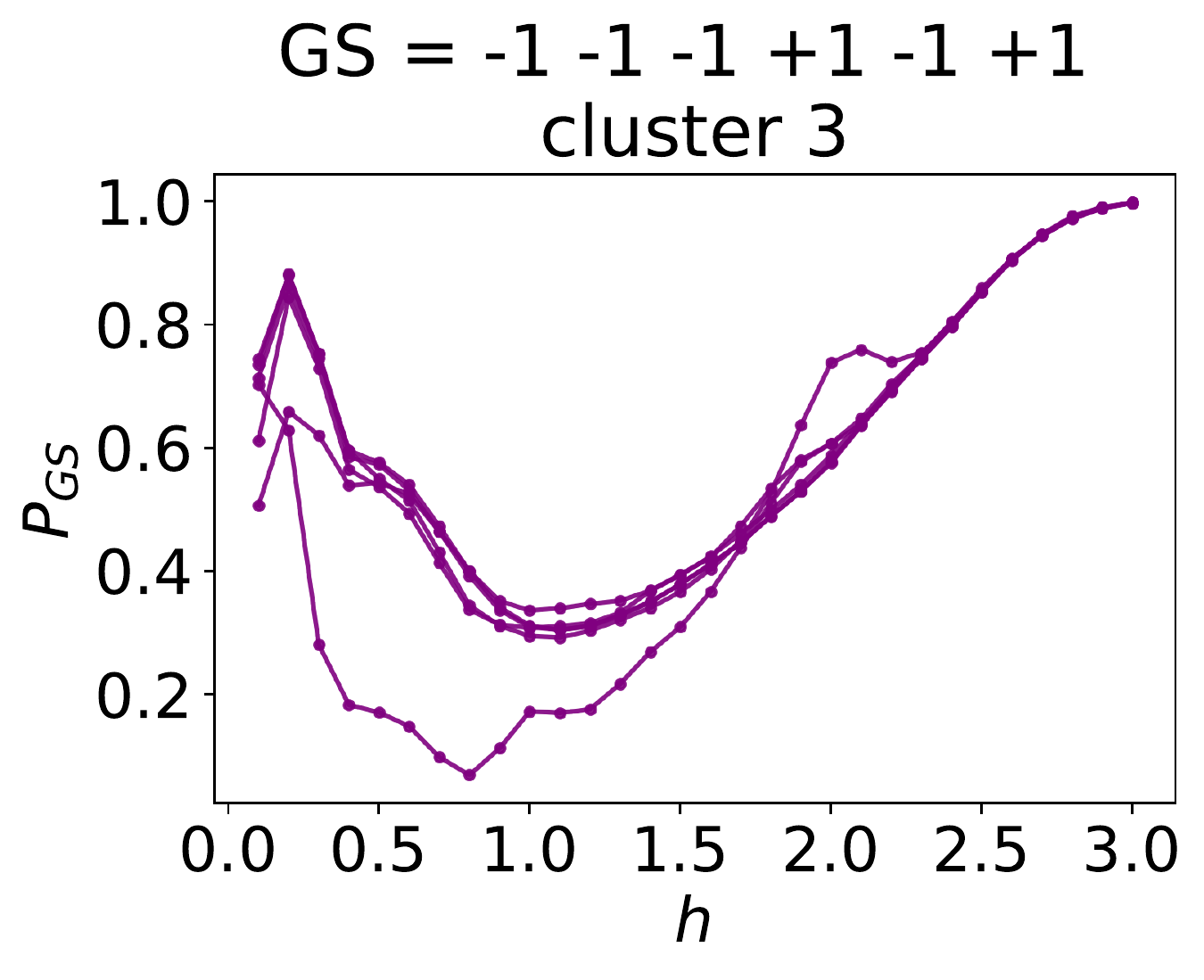}
    \caption{The left hand gray colored plot shows the distribution of $P_{GS}$ as $h$ increases for transitioning all $2^6$ input states into a single ground state (the exact ground state is shown in the title of each of the sub-figures). The four right-hand most plots split up this data into $4$ distinct clusters using unsupervised spectral clustering of the vectors of $P_{GS}$ values across the increasing $h$ strengths. Each cluster of vectors is colored uniquely, and the coloring arbitrary and randomly chosen. }
    \label{fig:h_strength_vs_GSP_n6_clustered_GS0}
\end{figure}

\begin{figure}[t!]
    \centering
    \includegraphics[width=0.19\textwidth]{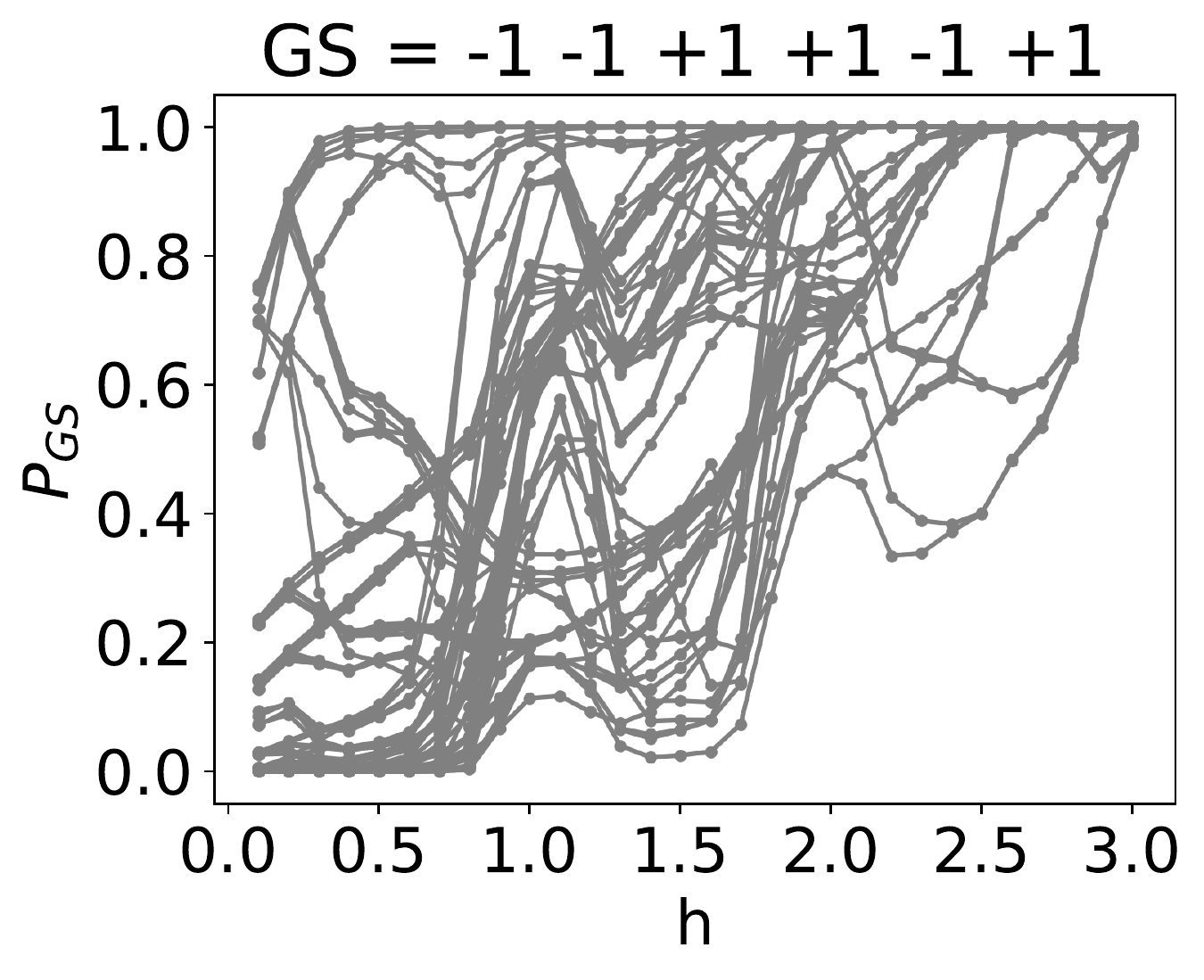}
    \includegraphics[width=0.19\textwidth]{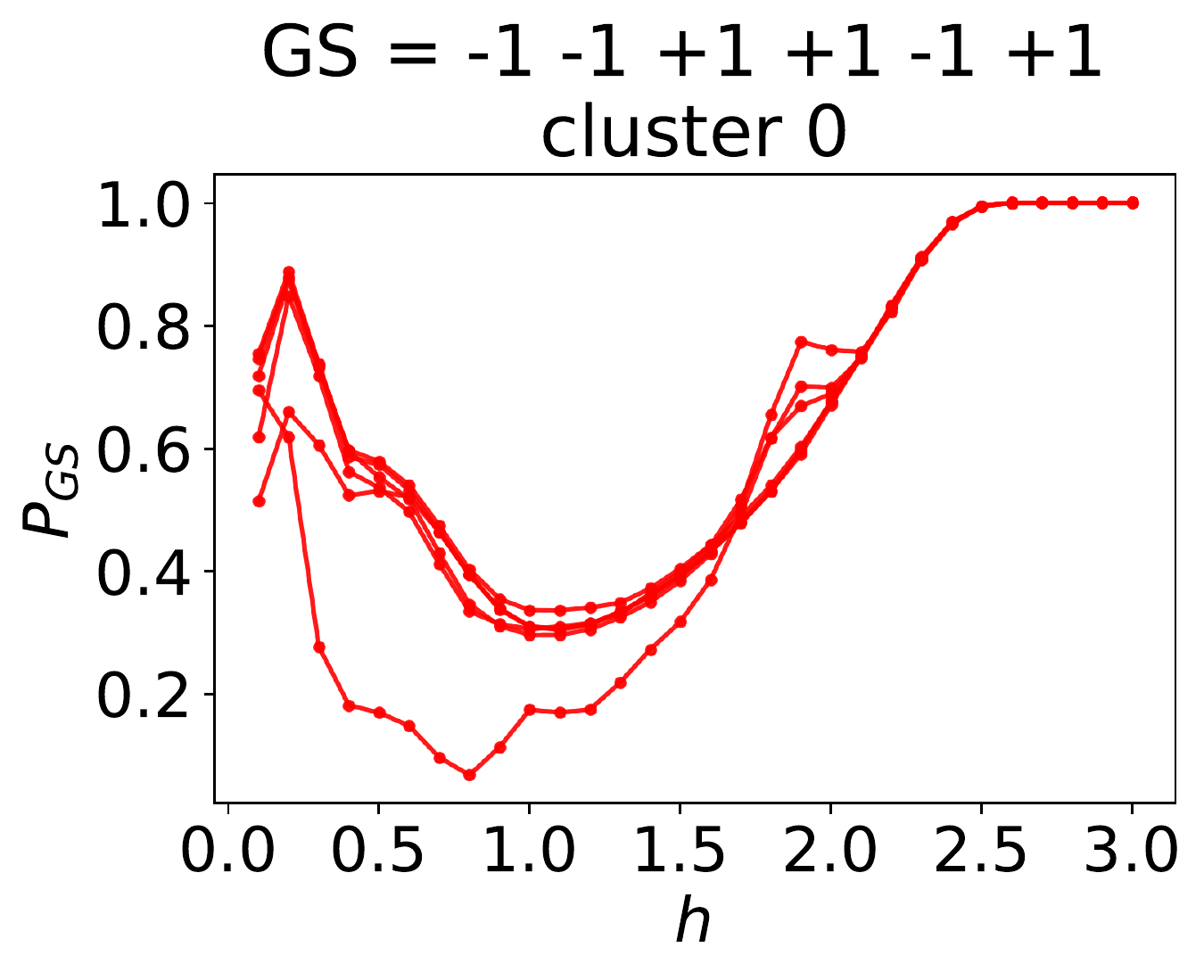}
    \includegraphics[width=0.19\textwidth]{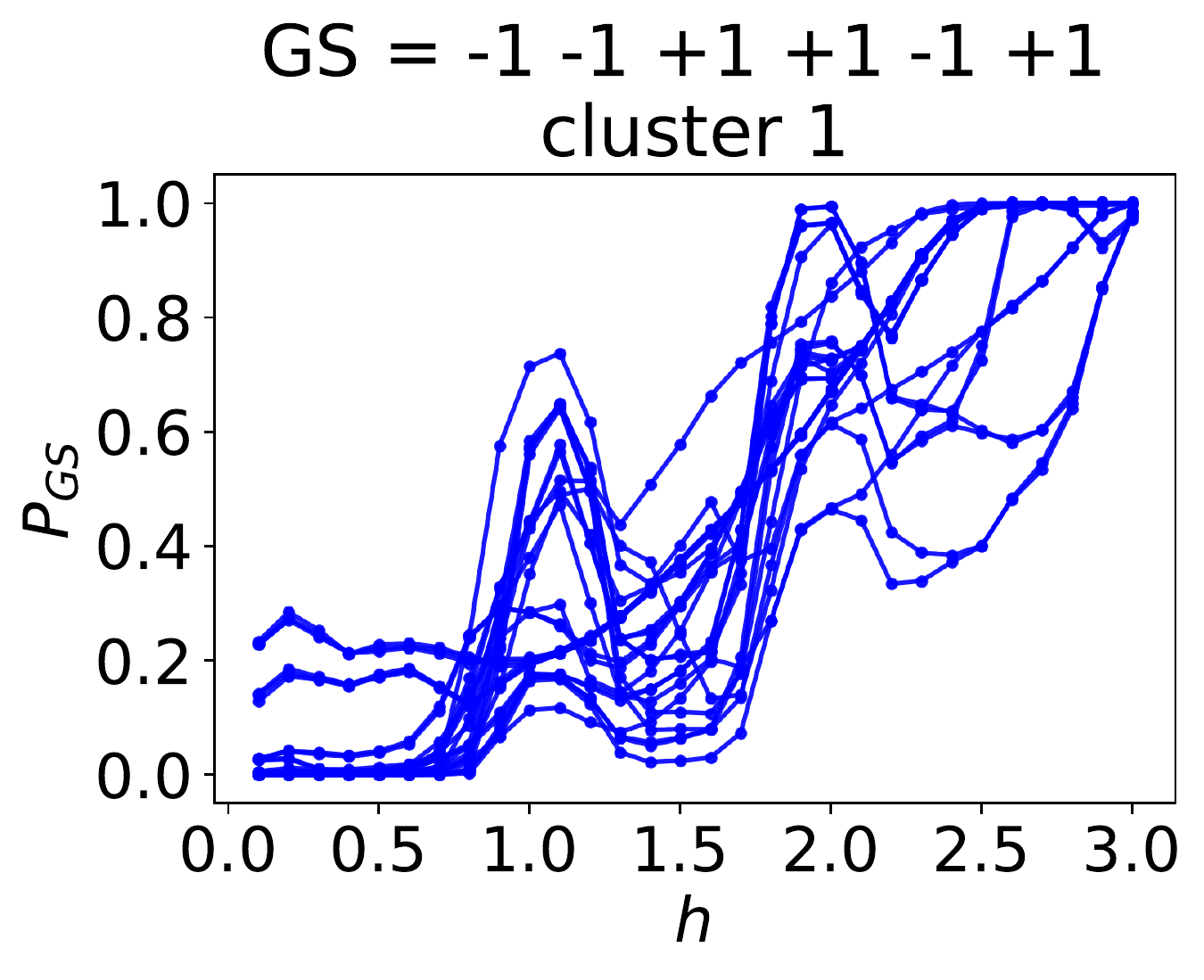}
    \includegraphics[width=0.19\textwidth]{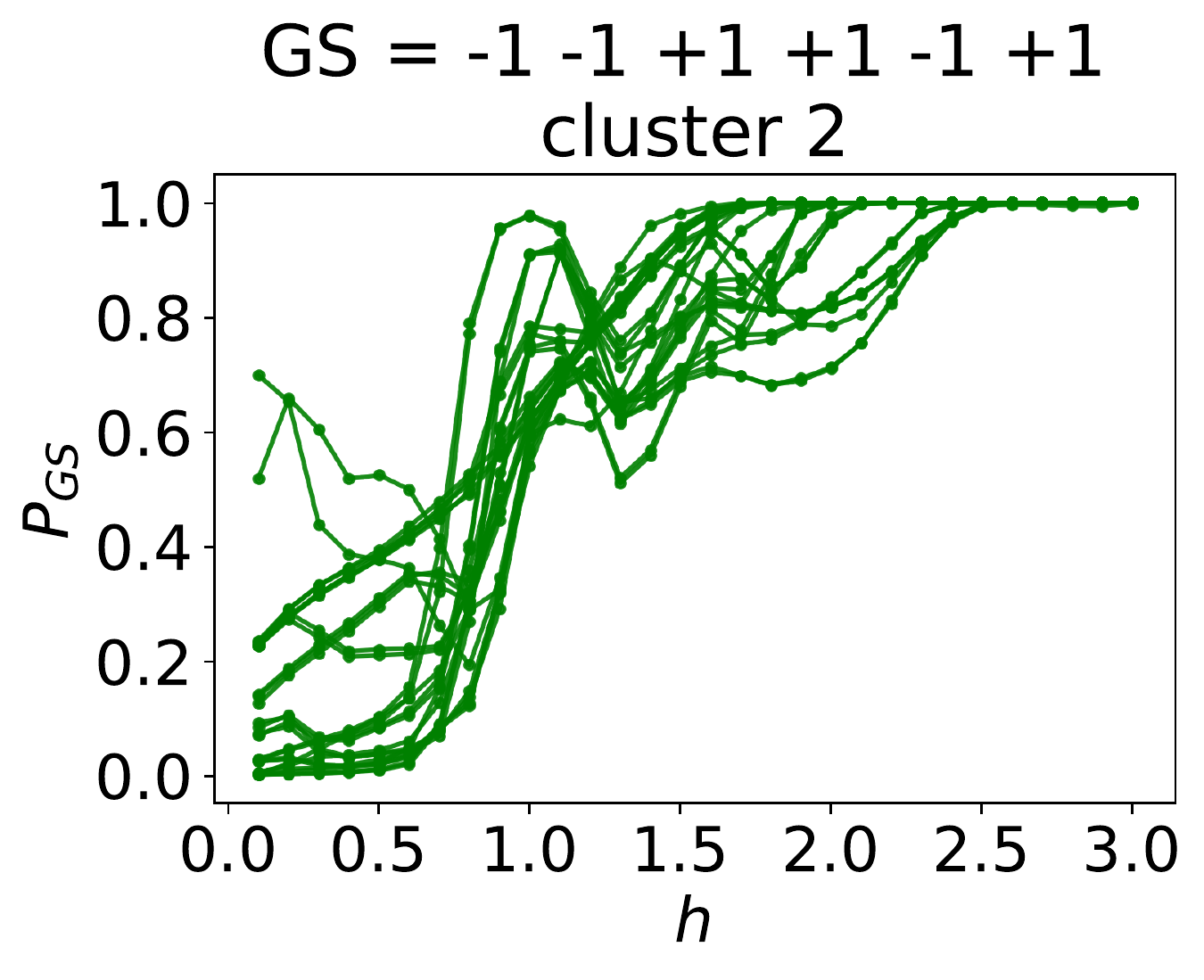}
    \includegraphics[width=0.19\textwidth]{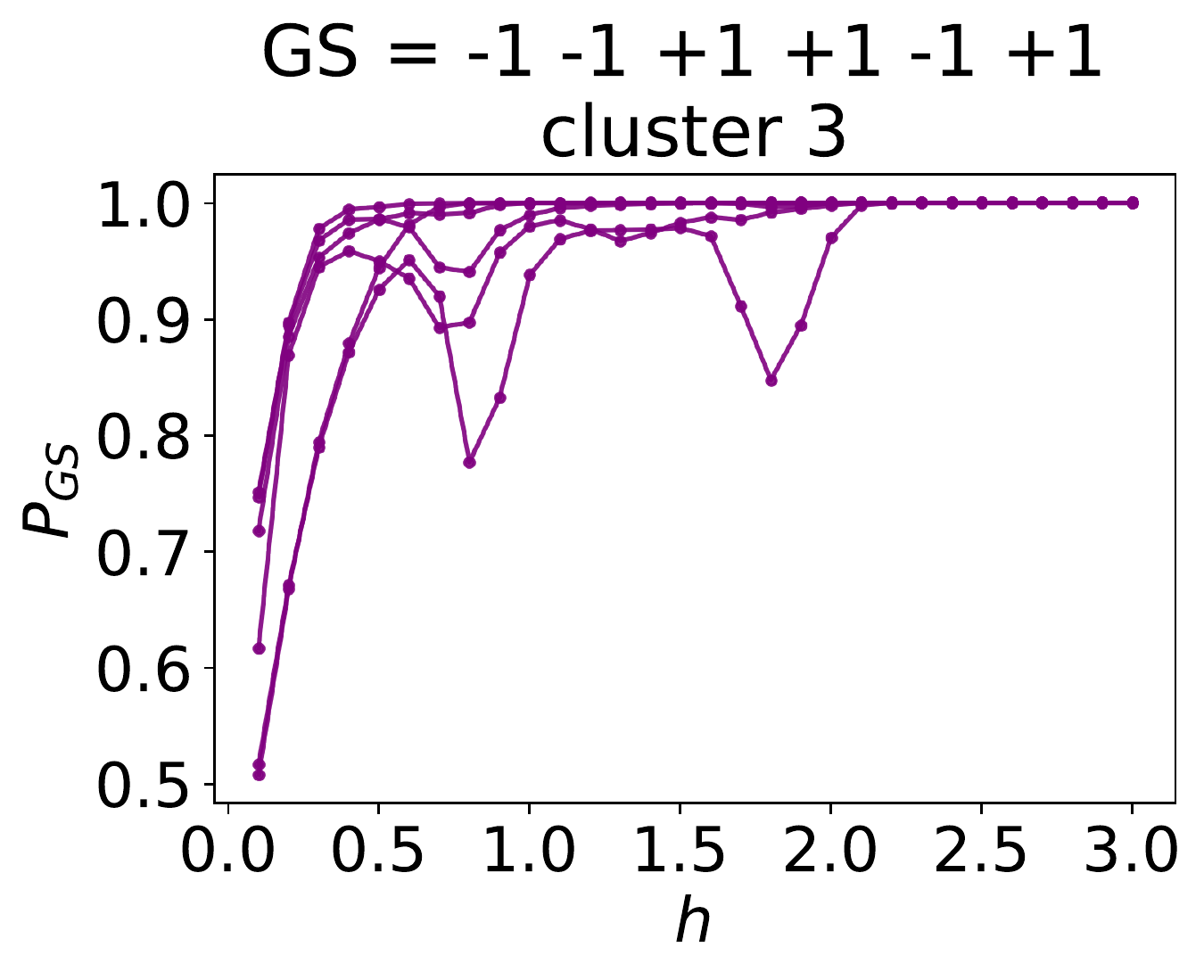}
    \caption{The left hand gray colored plot shows the distribution of $P_{GS}$ as $h$ increases for transitioning all $2^6$ input states into the single ground state (the exact ground state is shown in the title of each of the sub-figures). The four right-hand most plots split up this data into $4$ distinct clusters using unsupervised spectral clustering of the vectors of $P_{GS}$ values across the increases $h$ strengths. Note that this plot overall is very similar to Figure \ref{fig:h_strength_vs_GSP_n6_clustered_GS0}, but is showing data for a different ground state encoding (this ground state it turns out is hamming distance 1 away from the ground state in Figure \ref{fig:h_strength_vs_GSP_n6_clustered_GS0}) and the ordering of the clusters are different. }
    \label{fig:h_strength_vs_GSP_n6_clustered_GS1}
\end{figure}

\begin{figure}[t!]
    \centering
    \includegraphics[width=0.48\textwidth]{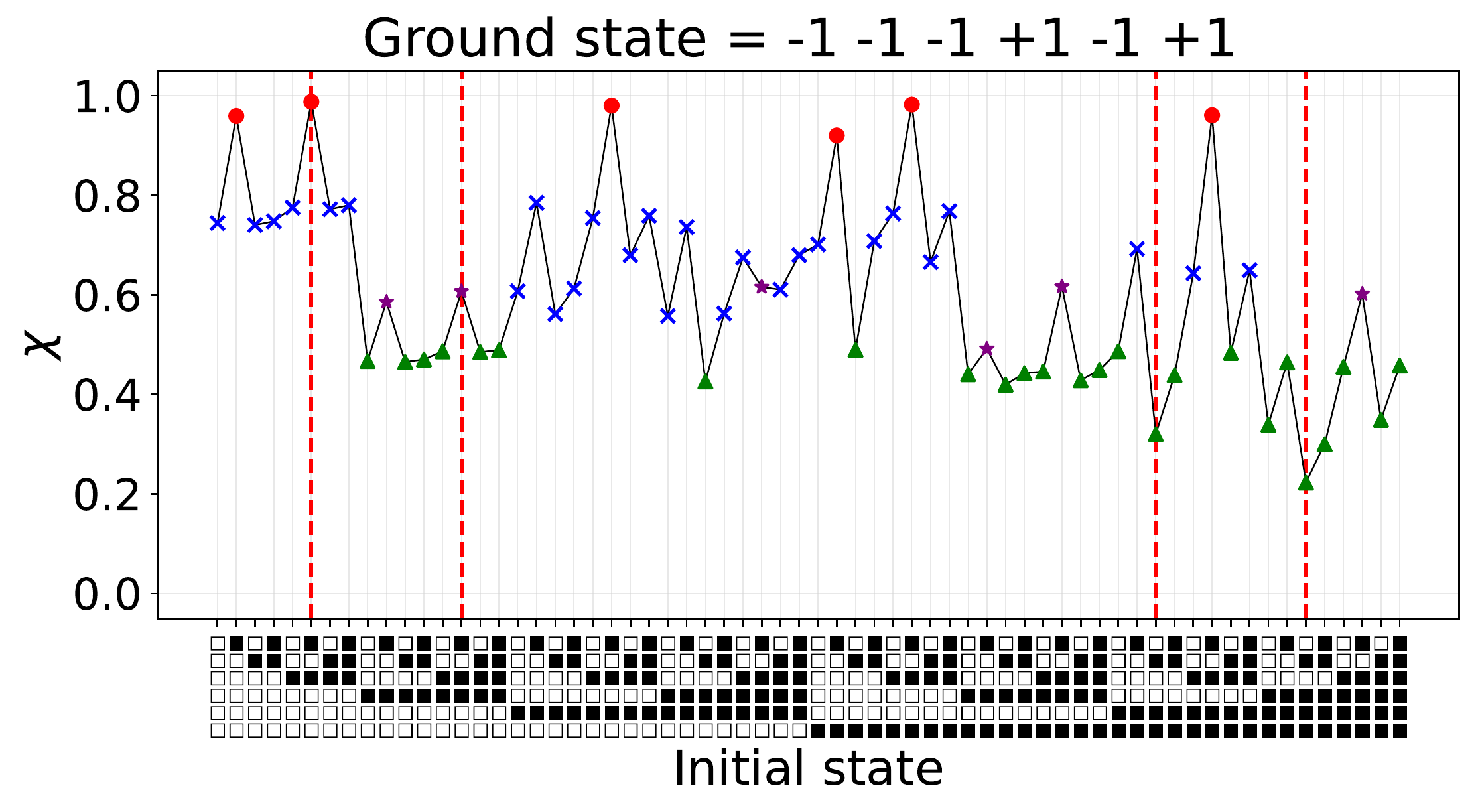}
    \includegraphics[width=0.48\textwidth]{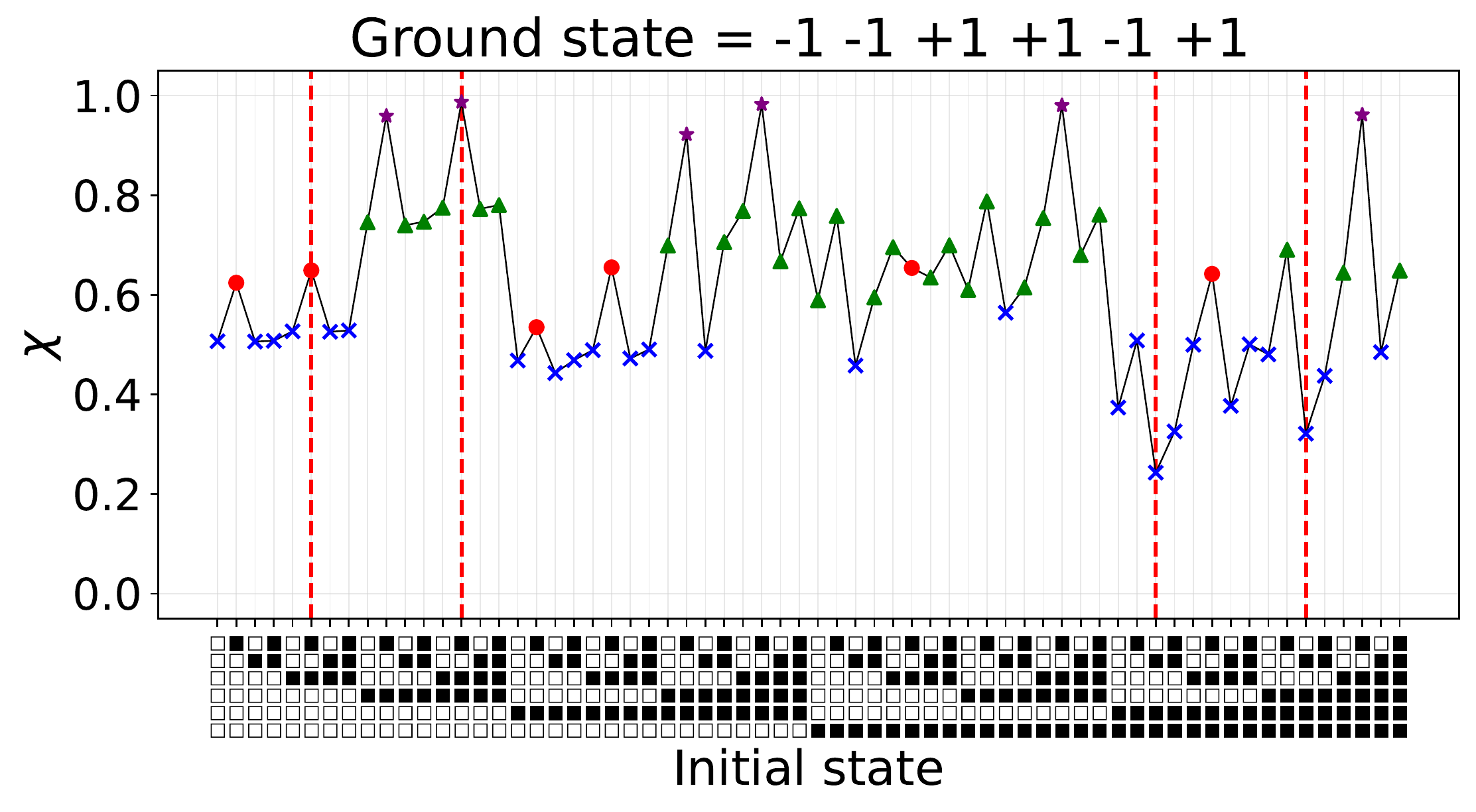}\\
    \includegraphics[width=0.48\textwidth]{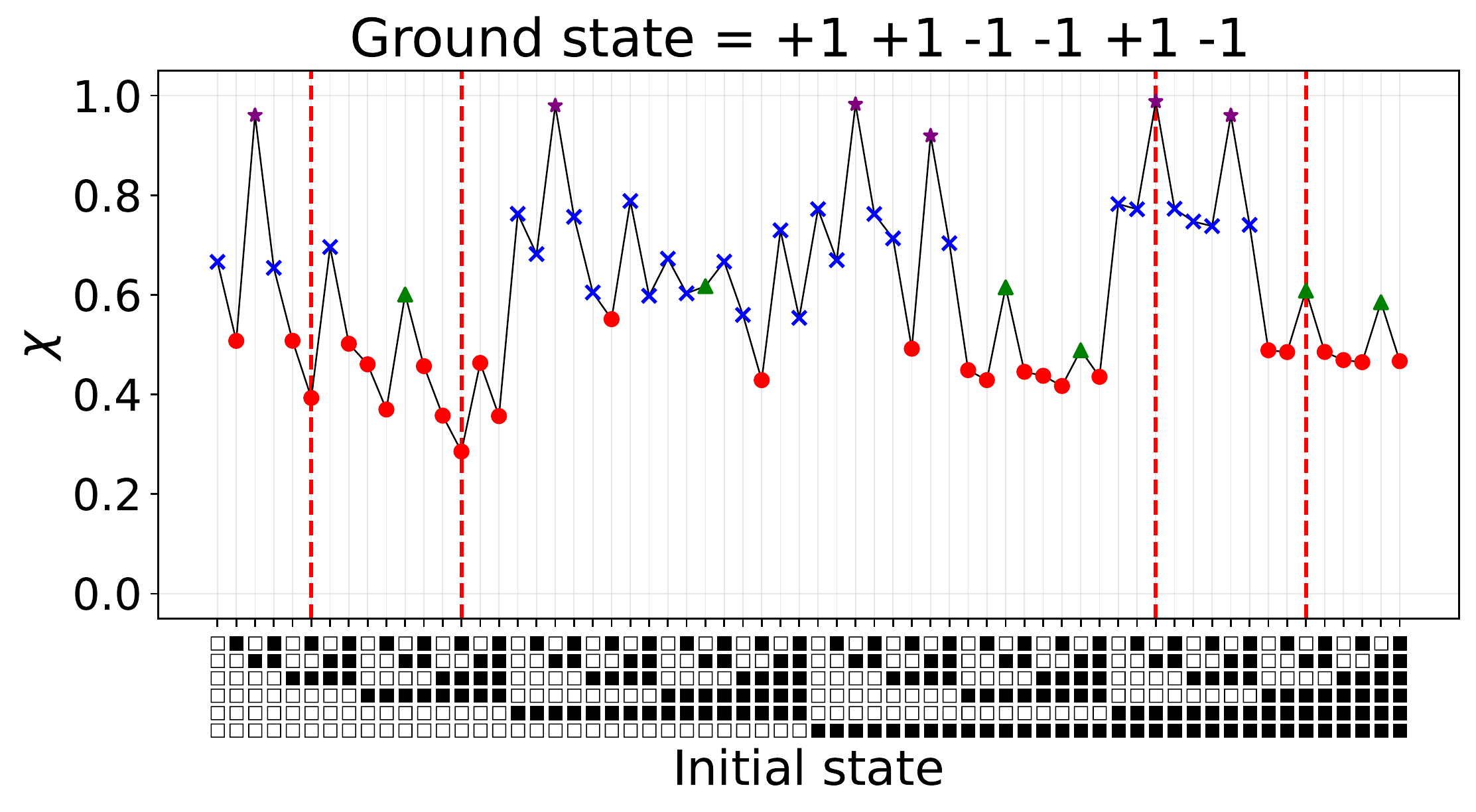}
    \includegraphics[width=0.48\textwidth]{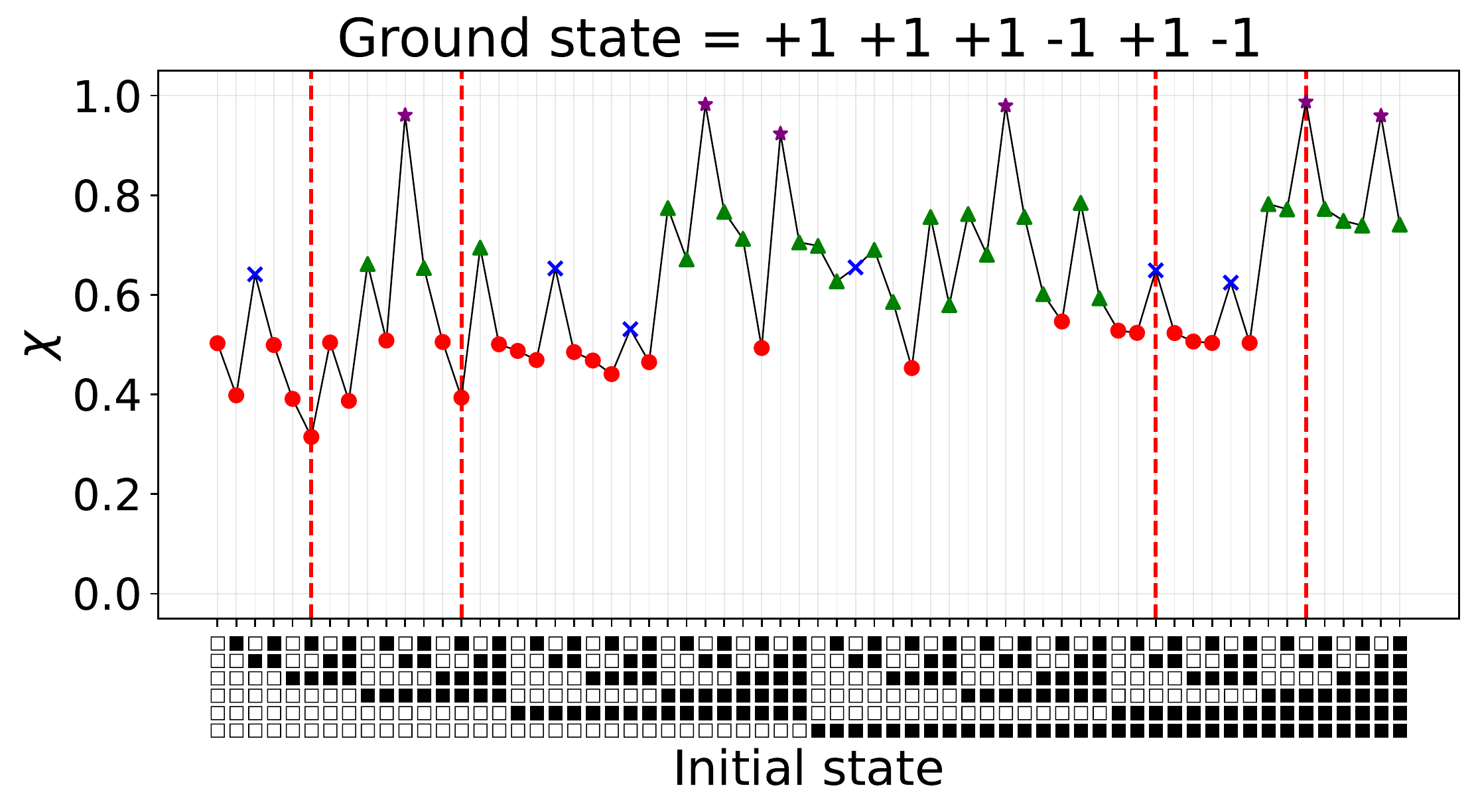}
    \caption{$N_6$ Ising susceptibility across all $2^6$ initial states when applying the h-gain schedule to force the system into the each of the four ground states. The x-axis encodes these initial states as vectors of vertical blocks where $\blacksquare$ denotes a variable state of $+1$ and $\square$ denotes a variable state of $-1$. The initial state vectors are read from bottom to top where the bottom is the first index which corresponds to variable $0$ in the problem Ising. All initial states which are also other ground states are marked with dashed red vertical lines. For each sub-figure, the reflexive ground state mapping (i.e. where the initial state and the intended state are the same ground state) case can be found visually as the state marked with a red vertical dashed line which has the maximum susceptibility measure among all of the initial states. }
    \label{fig:HGain_susceptibility_to_groundstate_n6}
\end{figure}

One of the interesting trends that is observed in the data, which will be explored in detail in Section \ref{section:results}, is a consistent lower overall susceptibility measure for some states that are very near the ground state in terms of hamming distance, but not necessarily close in terms of other relevant metrics such as the energy of the state. It turns out that these states that exhibit very small susceptibility towards the ground state are several bits flip away from the ground state; but the index of that bit flip is the highest degree node in the Ising graph. In order to determine whether this trend is consistent across other states (and across all 3 test Isings), we define a metric $\delta$ on states to be maximized where the state is exactly one highest degree variable away from a ground state, and minimized for the state that is the complement of the ground state. This metric is based on defining the subset $V$ of variables out of the variables in a given initial state $i$ who's state is opposite that of the ground state we are pushing the system into. Thus if $V$ contains 1 variable then our state $i$ is exactly one bit-flip away from the ground state and if $V$ contains the same number of variables as the Ising, then the given state $i$ is exactly the complement of the ground state. Assume the procedure $G\_deg$ computes the degree of the variable $v$ of the specific Ising graph $G$. 

\begin{equation}
    \label{eq:delta_metric}
    \delta = \frac{\sum_{v \in V} \text{G\_deg}(v) }{ \text{length}(V) } \cdot \frac{1}{ \text{length}(V) }
\end{equation}

Equation \ref{eq:delta_metric} is undefined when the initial state is the same as the ground state because in that case $V =\varnothing$; therefore when this case occurs we set $\delta = M+1$ where $M$ is the largest degree in the Ising graph. 

Two other metrics that will be utilized are:
\begin{enumerate}
    \item The energy of the initial state evaluated on the given test Ising (either $N_6$, $N_7$, or $N_8$). We denote the energy metric as $E(i)$ for the initial state vector $i$
    \item The hamming distance between the initial state and the intended ground state as a proportion out of the number of variables. The hamming distance proportion is denoted as $d(i, GS)$ for the initial state vector $i$ and the ground state vector $GS$. 
\end{enumerate}

\begin{figure}[t!]
    \centering
    \includegraphics[width=0.24\textwidth]{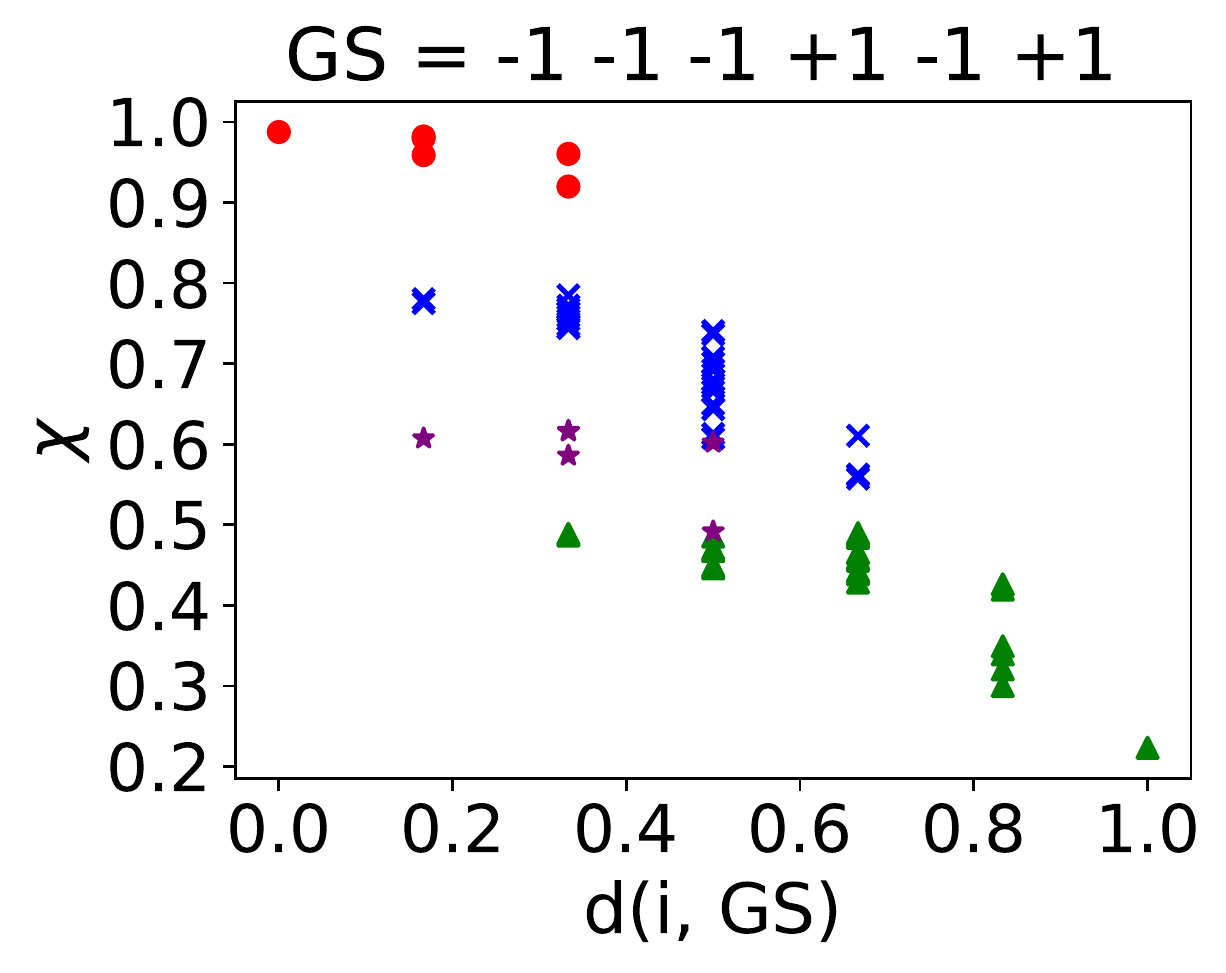}
    \includegraphics[width=0.24\textwidth]{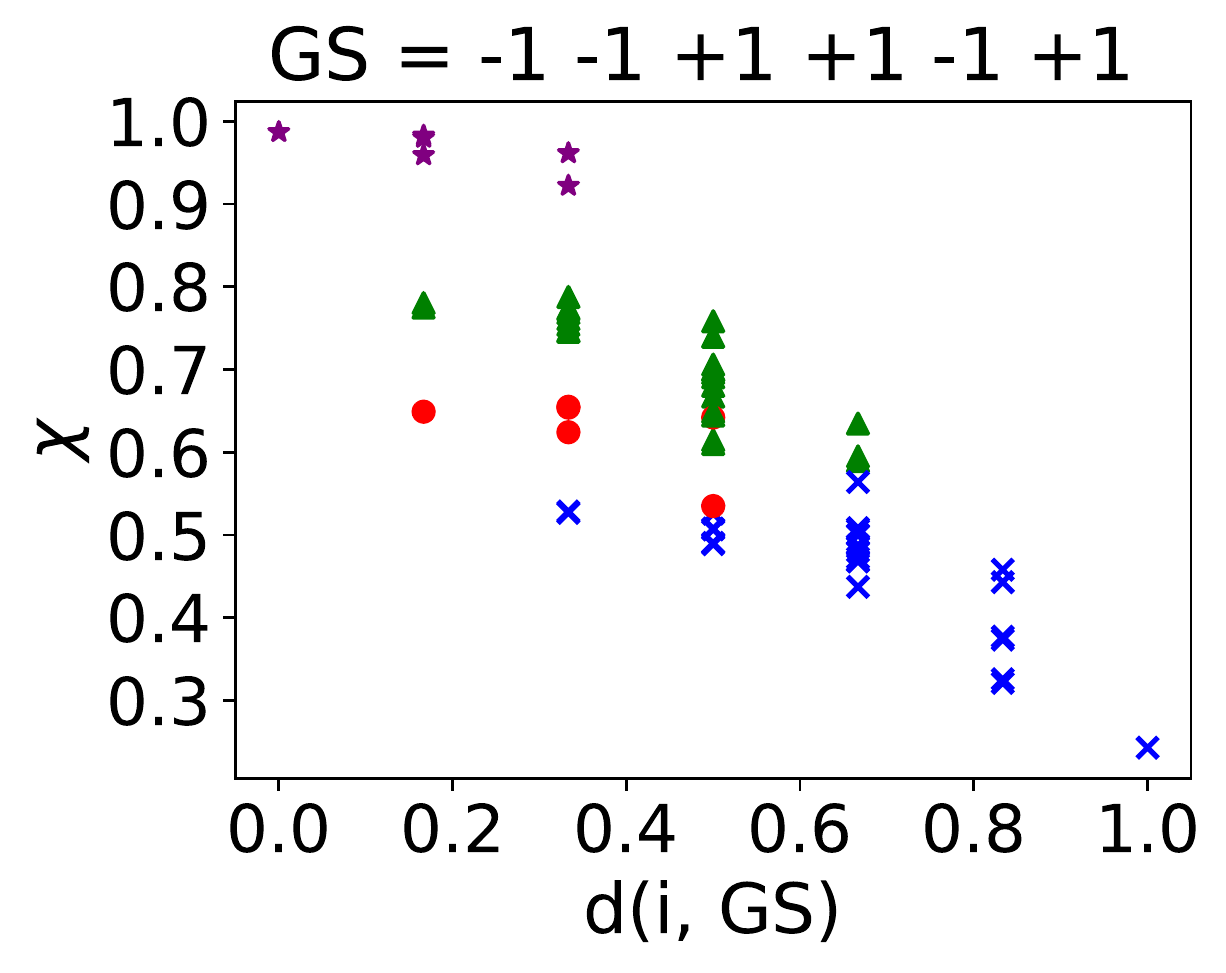}
    \includegraphics[width=0.24\textwidth]{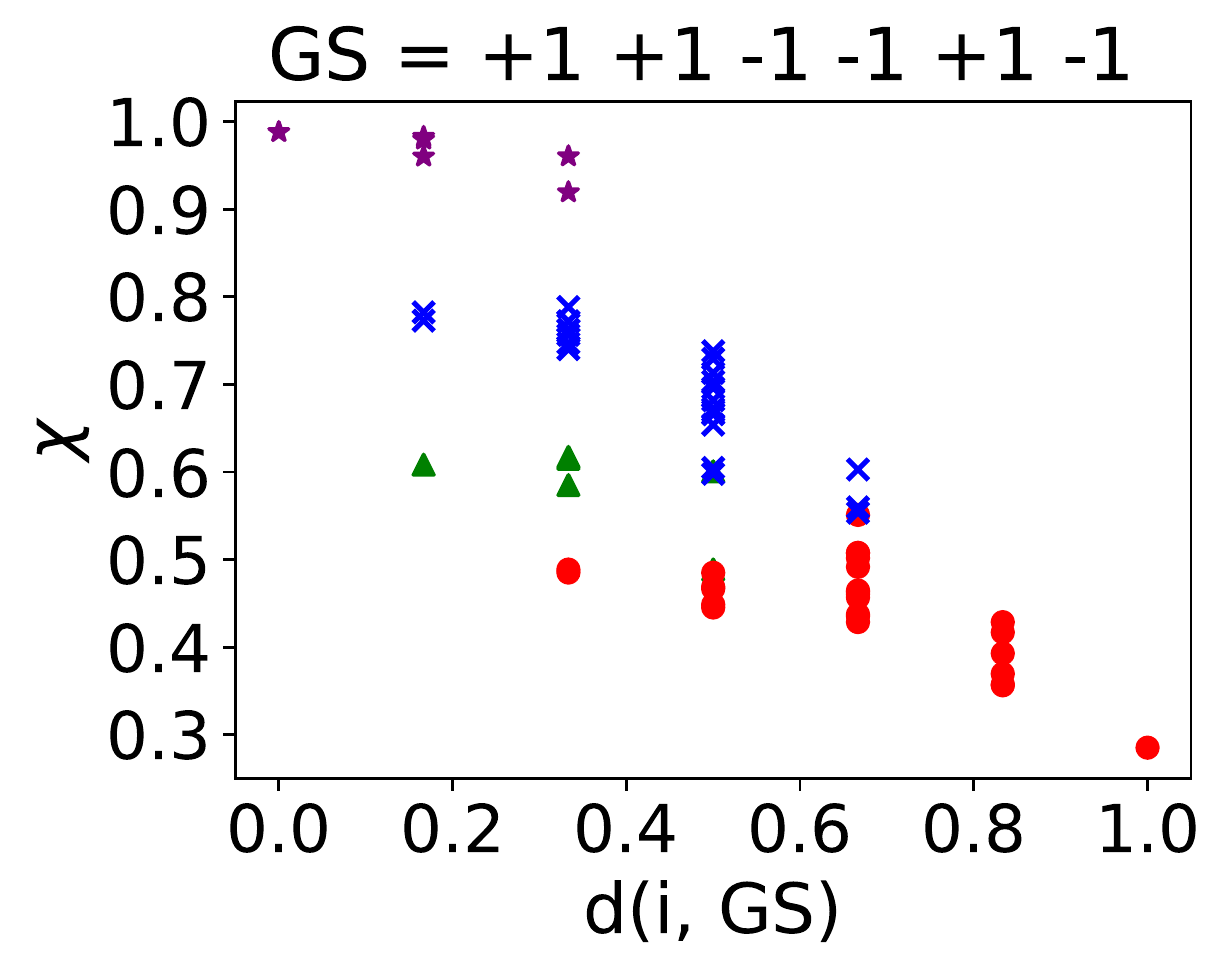}
    \includegraphics[width=0.24\textwidth]{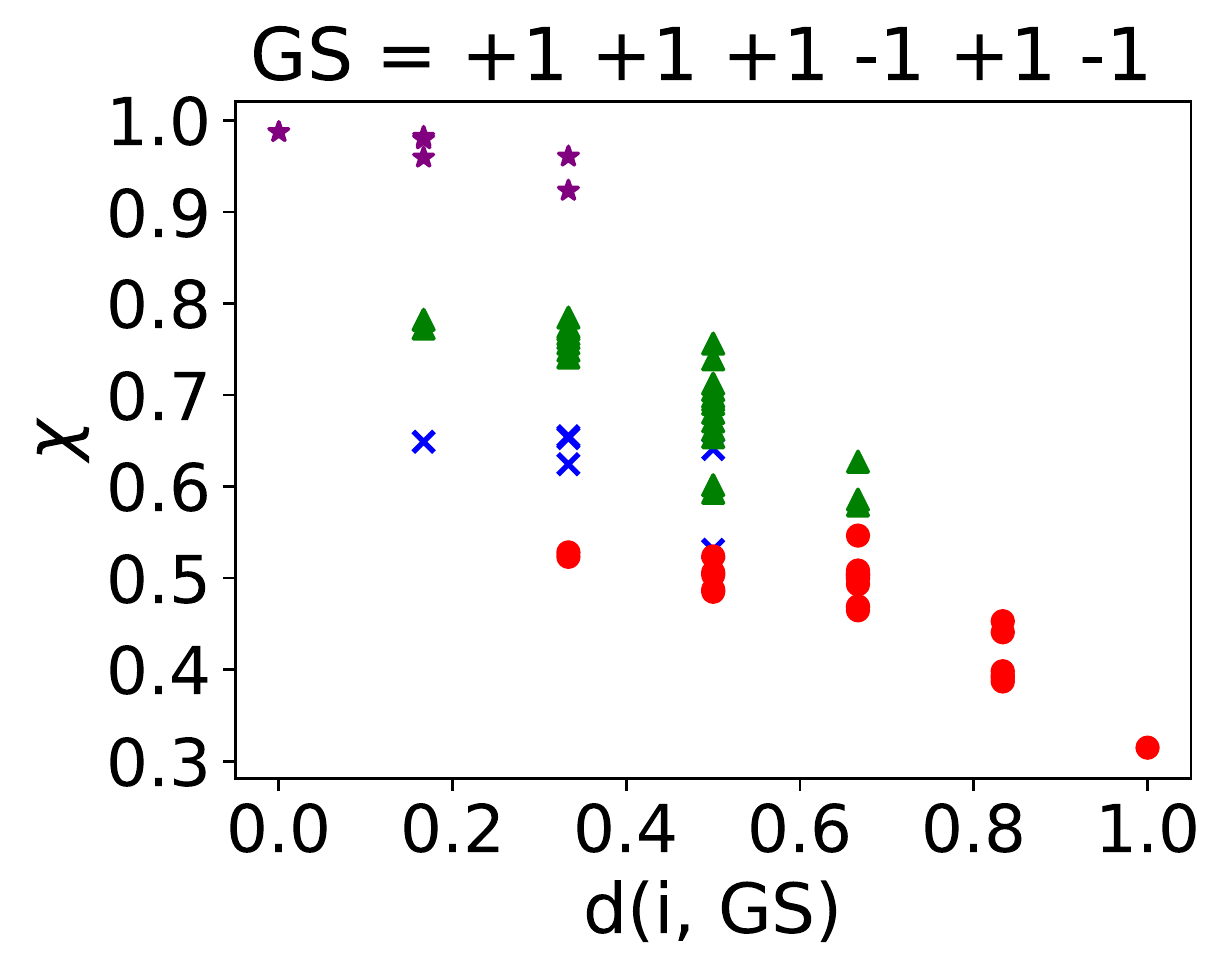}\\
    \includegraphics[width=0.24\textwidth]{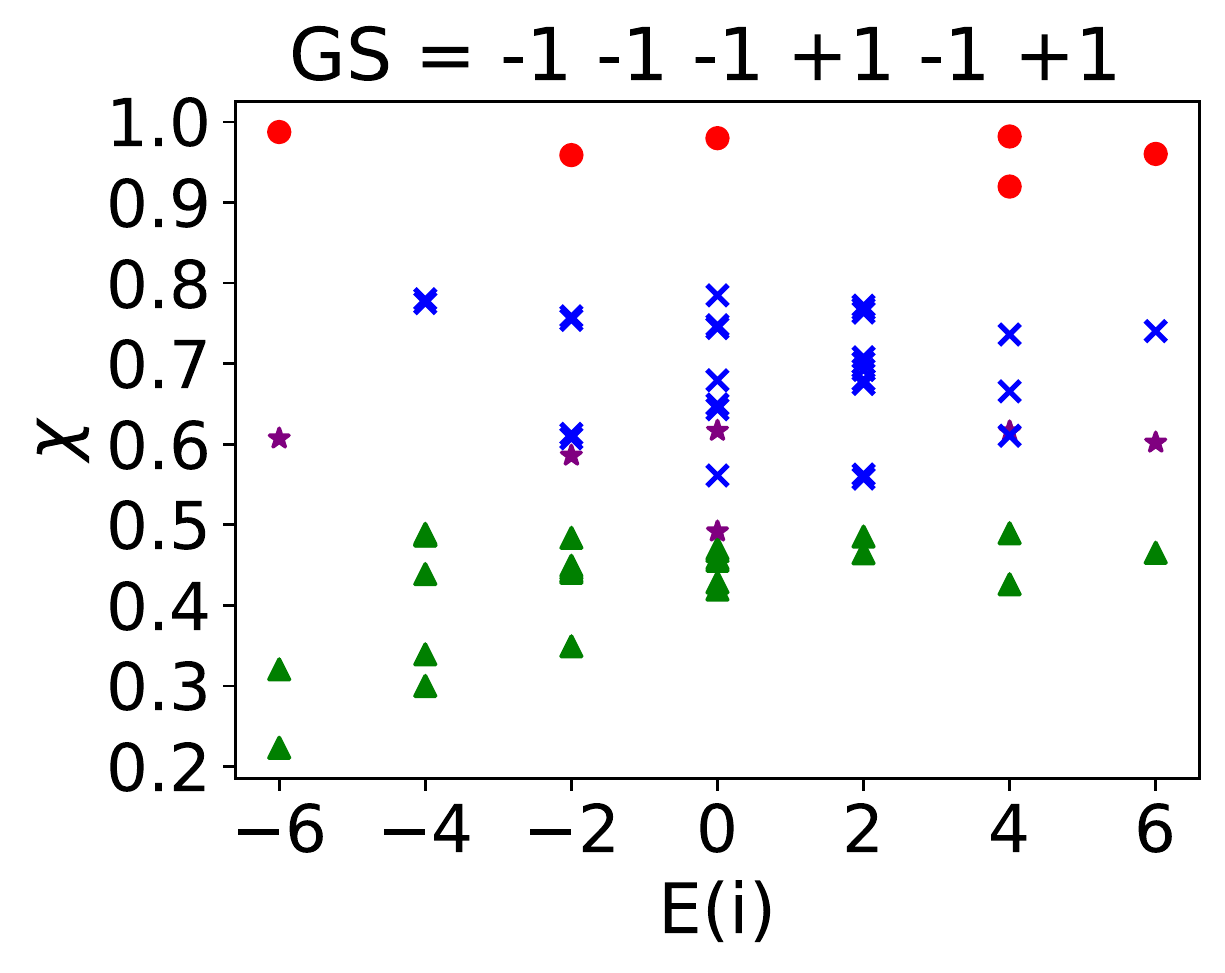}
    \includegraphics[width=0.24\textwidth]{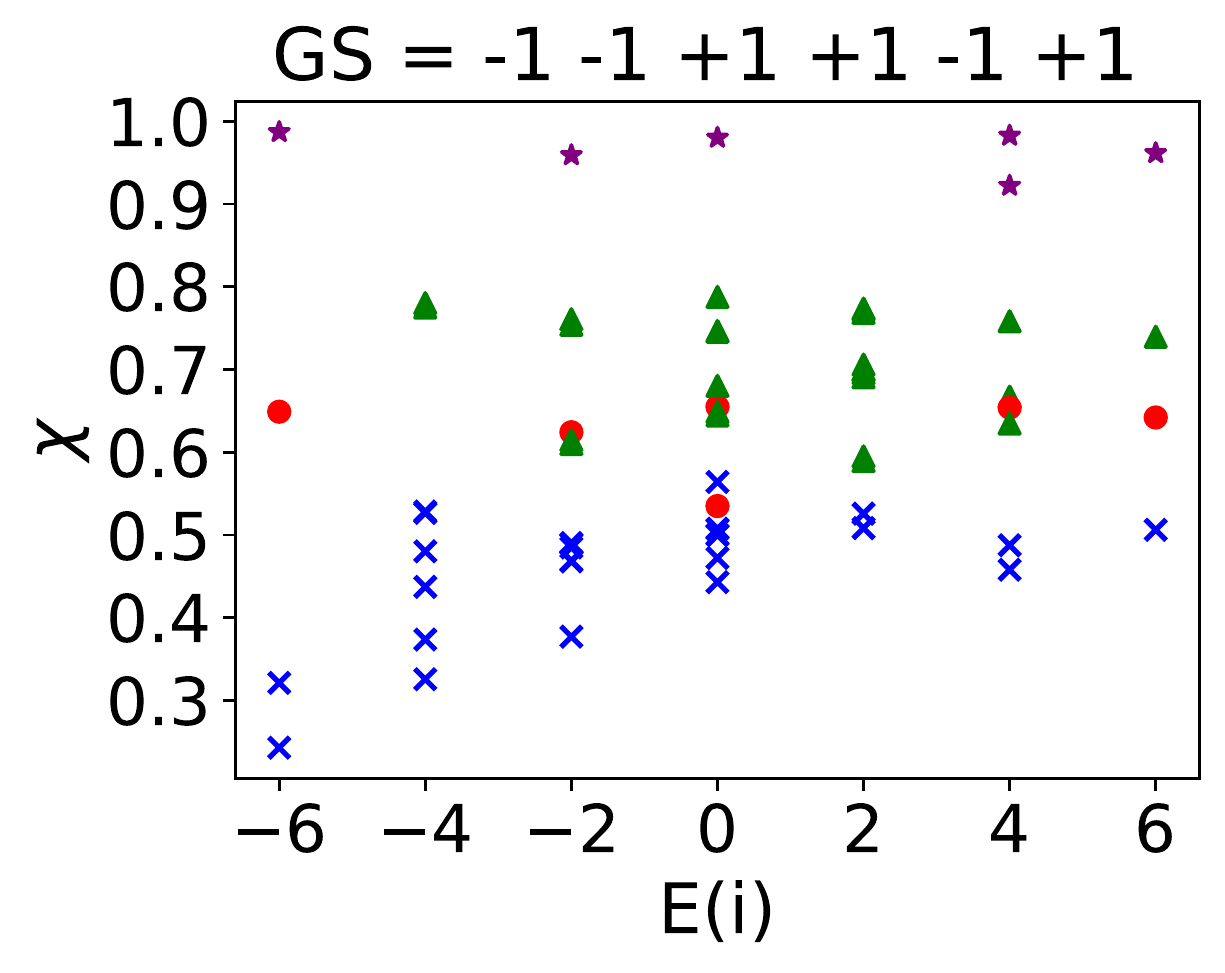}
    \includegraphics[width=0.24\textwidth]{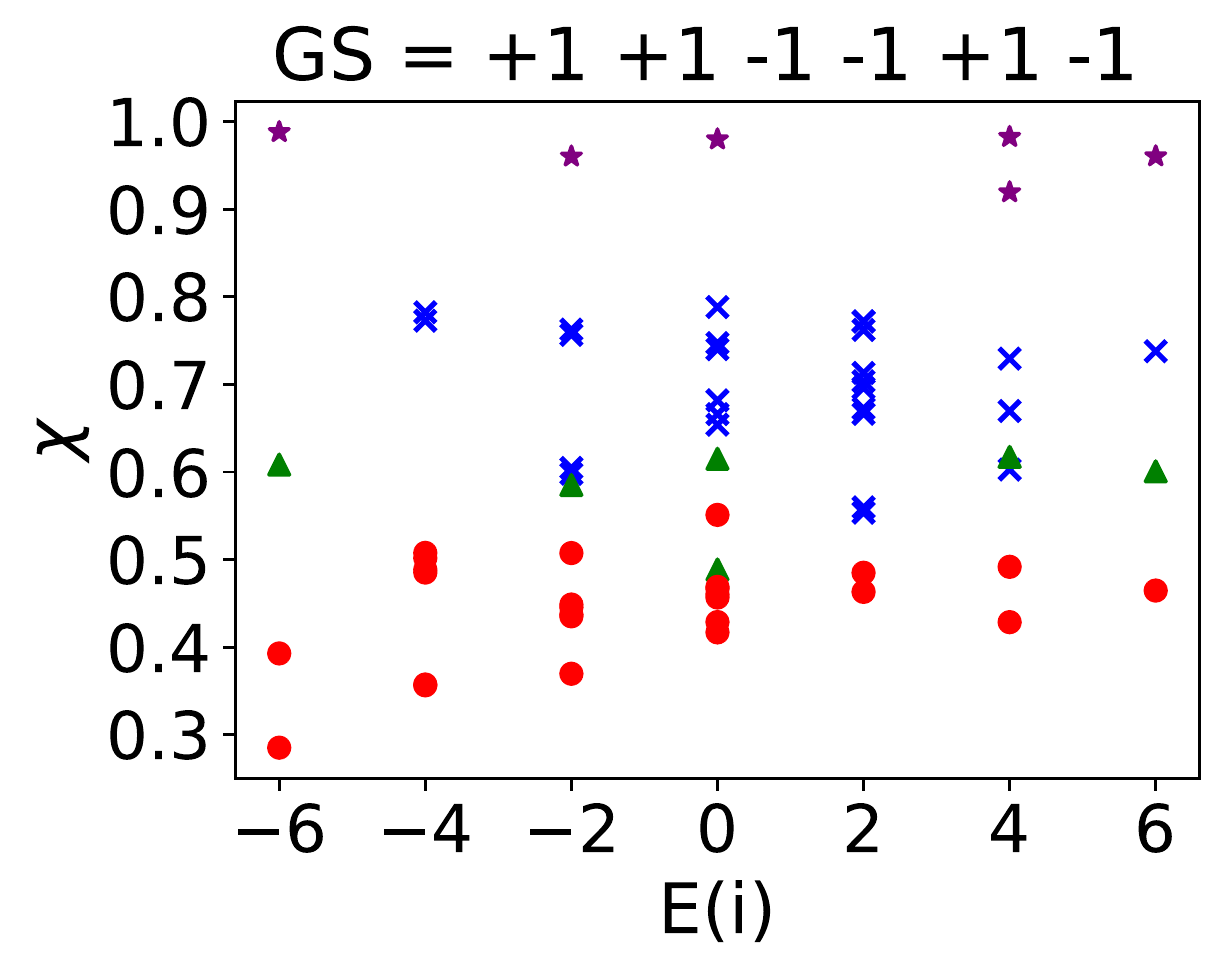}
    \includegraphics[width=0.24\textwidth]{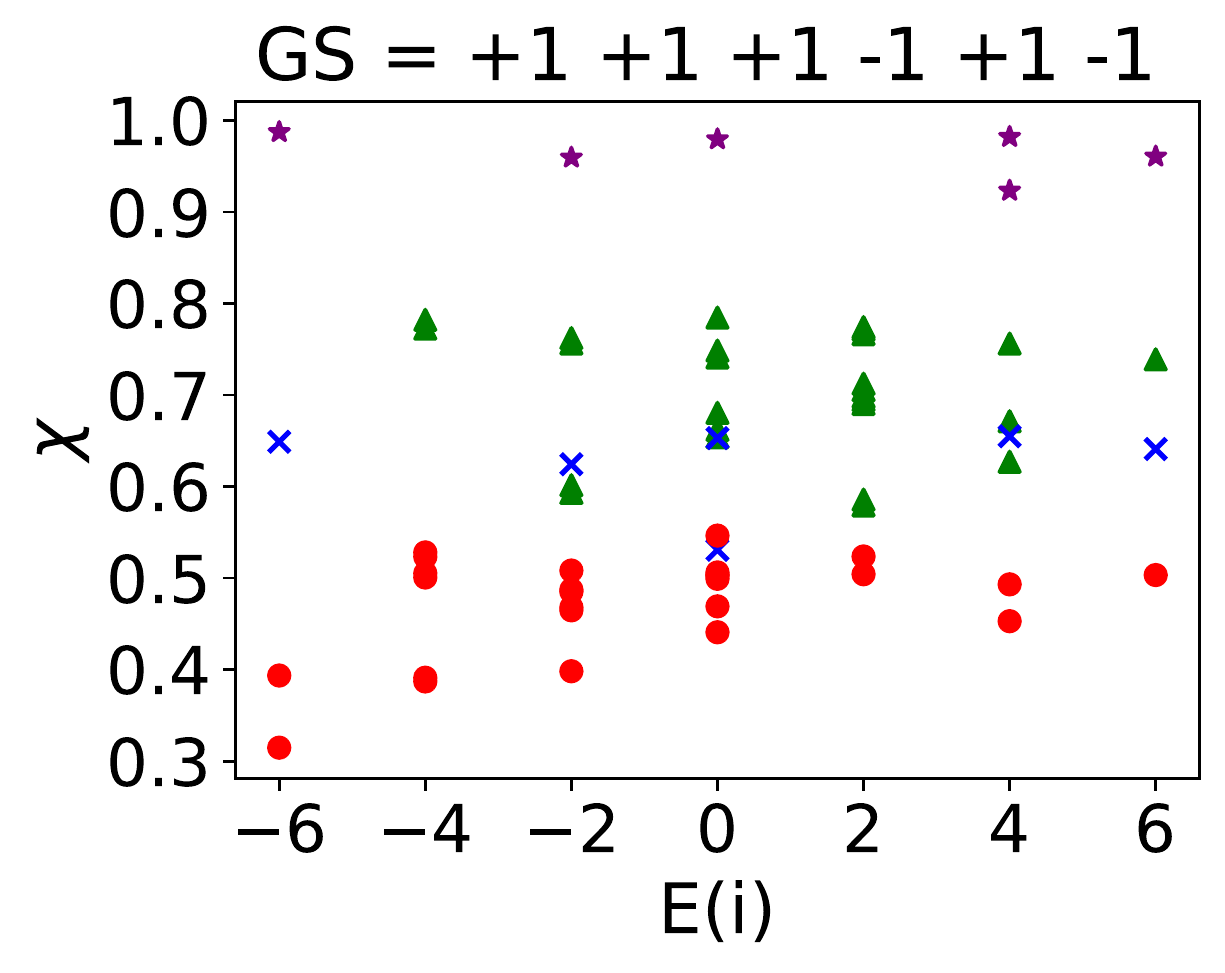}\\
    \includegraphics[width=0.24\textwidth]{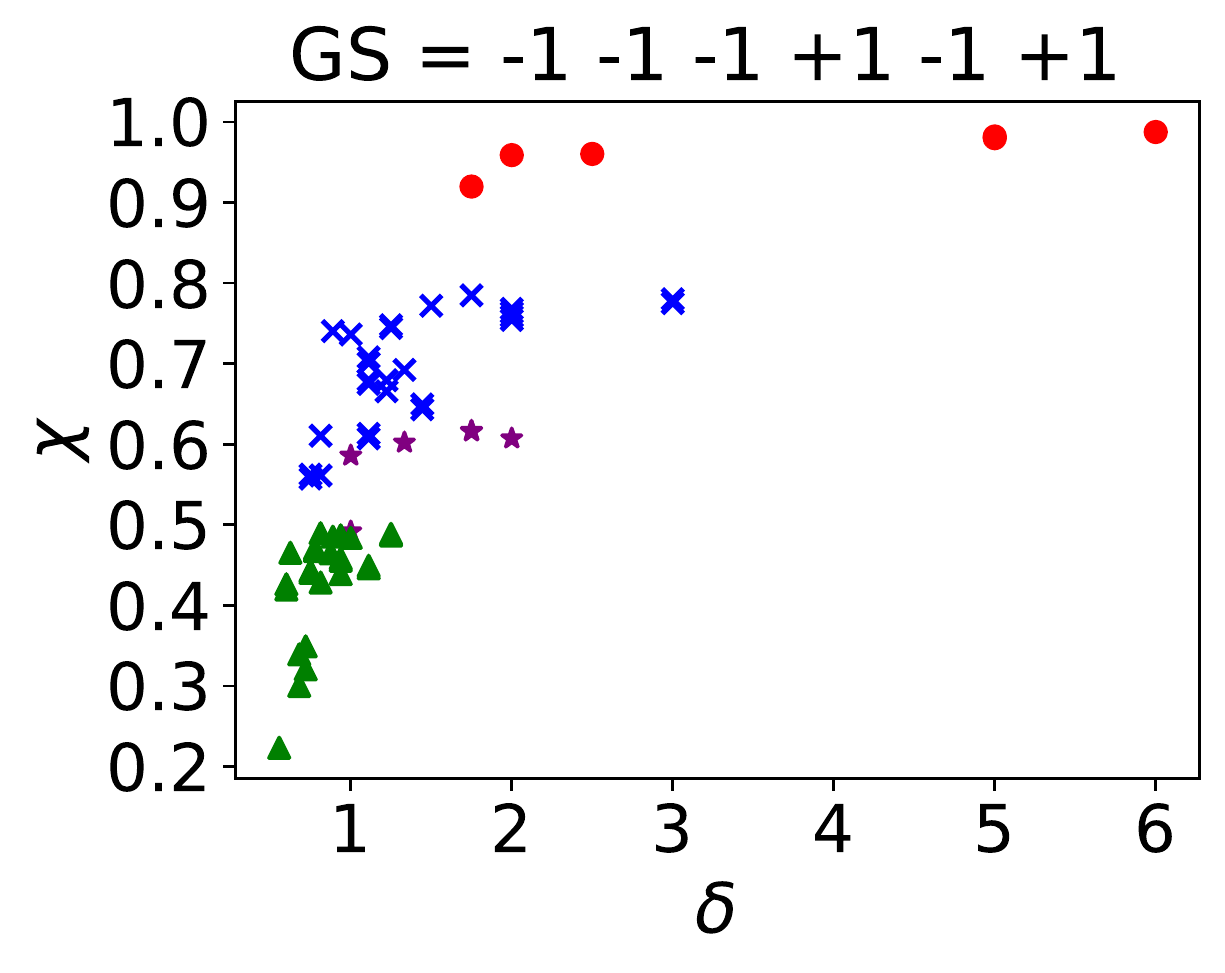}
    \includegraphics[width=0.24\textwidth]{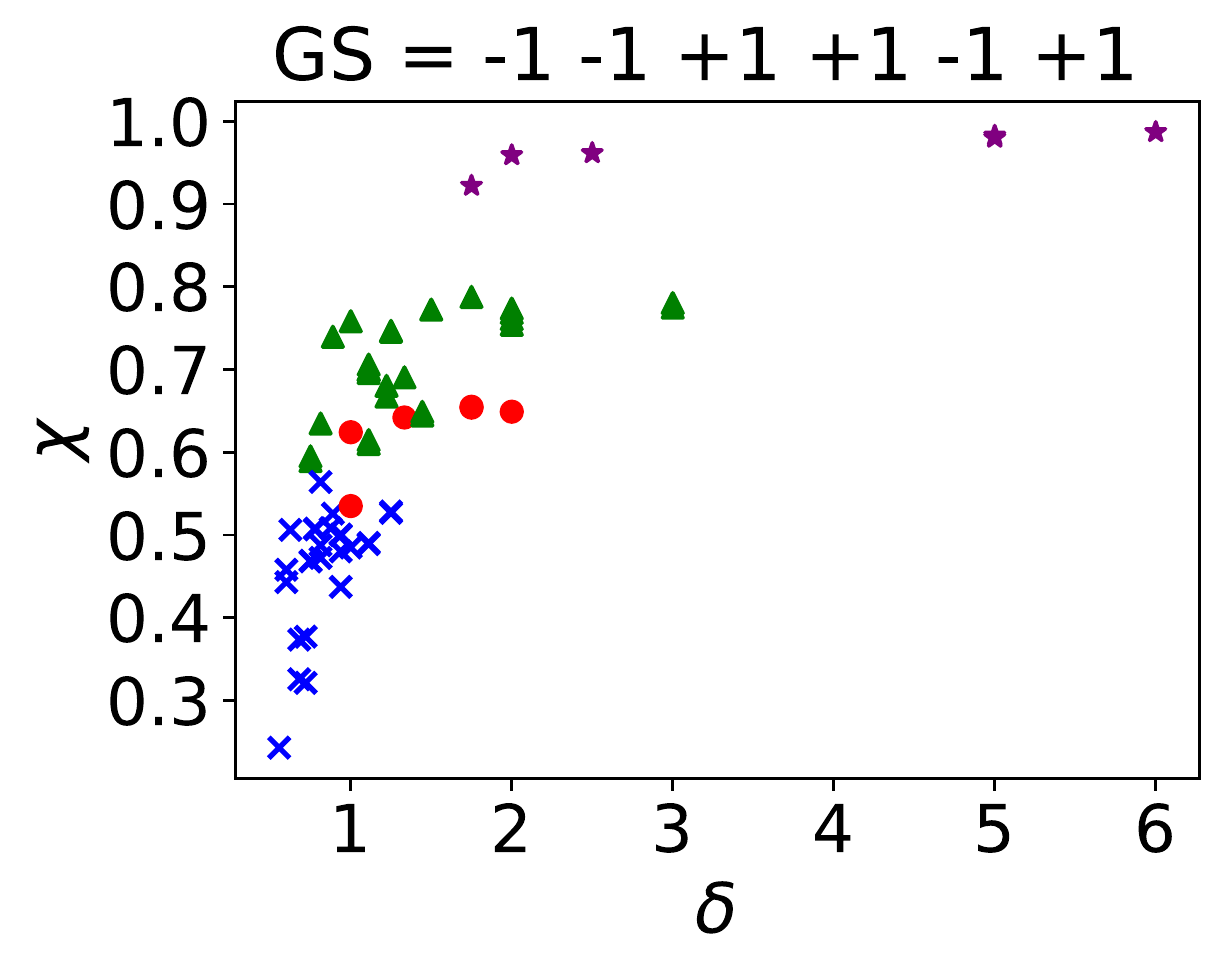}
    \includegraphics[width=0.24\textwidth]{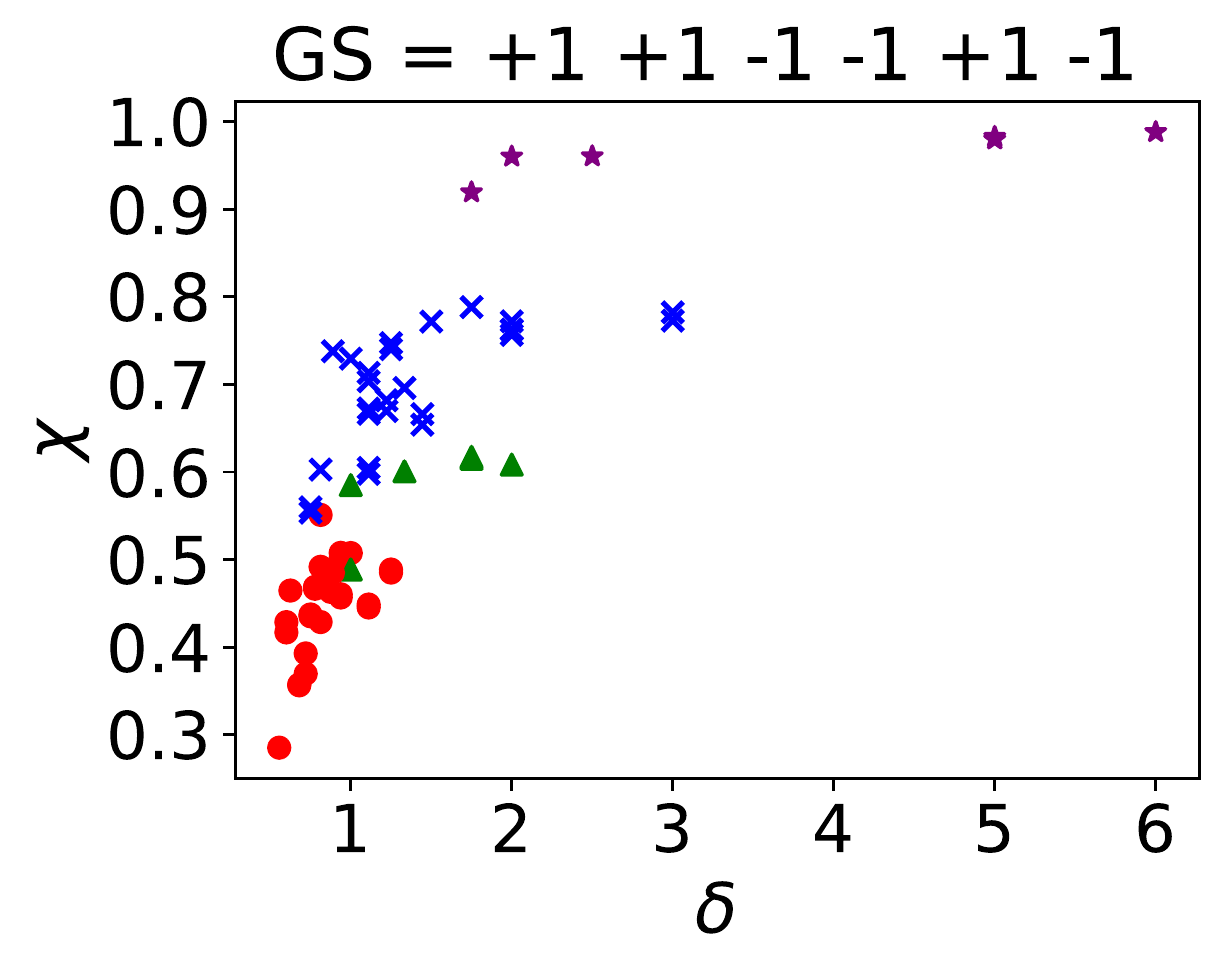}
    \includegraphics[width=0.24\textwidth]{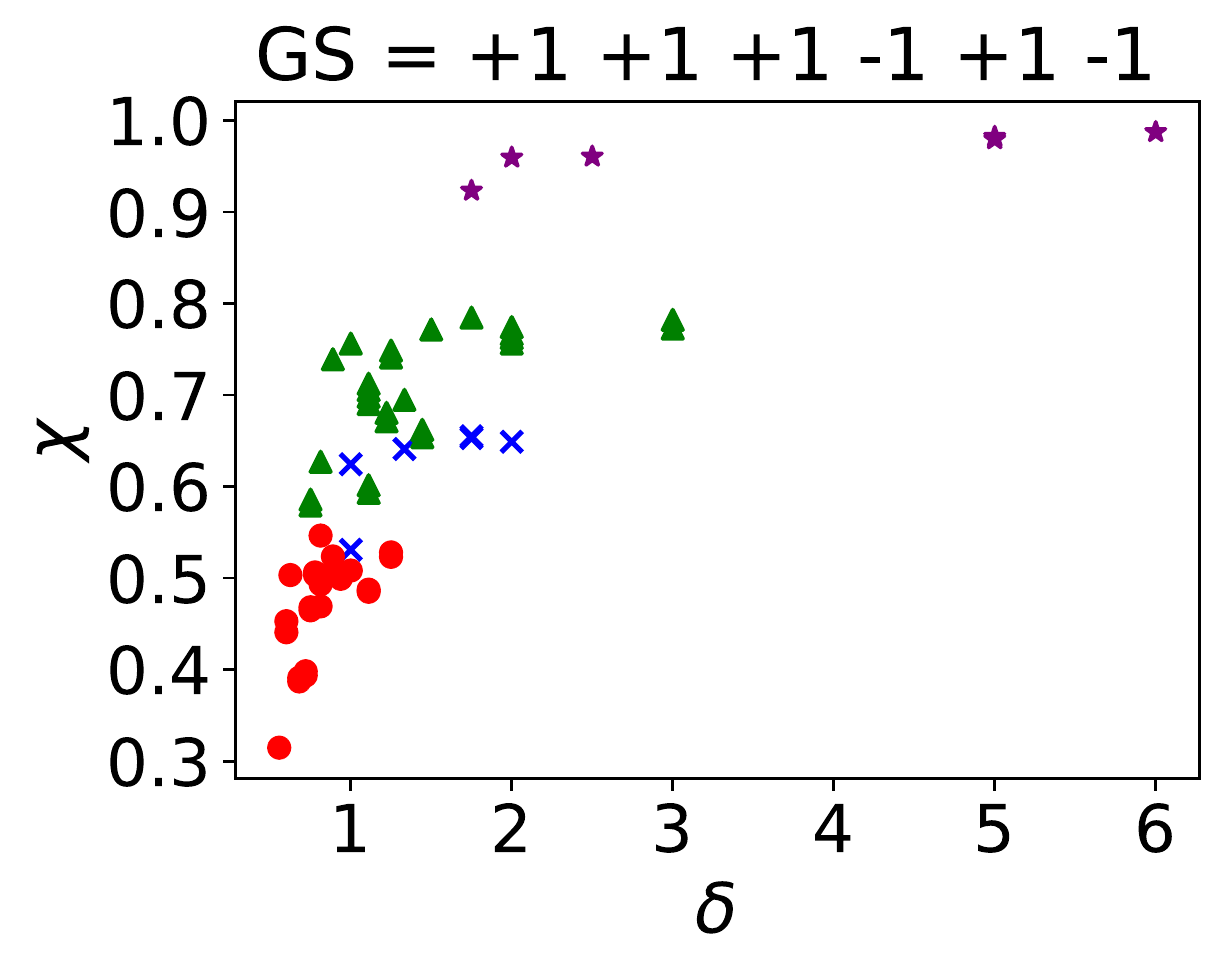}
    \caption{Summary metric plots for the $N_6$ Ising. The $4$ columns correspond to each of the $4$ ground states; the titles of each sub plot are the exact optimal solution vectors. The three rows correspond to three different initial state metrics on the x-axis, the y-axis of each sub-plot is the susceptibility metric $\chi$. The first row has x-axis which are the hamming distance between the ground state and the specific initial state $i$. The second row has x-axis showing the energy of the initial state $i$ evaluated on the $N_6$ Ising. The third row has x-axis showing the $\delta$ metric for each initial state.  }
    \label{fig:summary_metrics_n6}
\end{figure}

In order to help differentiate some of the different observed h-gain response curves, \emph{unsupervised spectral clustering} is performed on the vectors of the h-gain response curves. Specifically we cluster the vectors for each collection of h-gain response curves that are transitioning the annealer to a single ground state; this clustering is then repeated for each of the other ground states in the Ising. We cluster the data into $4$ clusters across all experiments; this number of clusters is somewhat arbitrary and a similar number of clusters could be utilized however $4$ gives a reasonable balance between having reasonably distinct behaviors in the clusters while also being able to visually present all of the clusters in a reasonable amount of space. The spectral clustering implementation used is from the python library \emph{scikit-learn} \cite{scikit-learn, sklearn_api, 10.1093/imaiai/iay008, knyazev2001toward, stella2003multiclass, von2007tutorial, 868688}. 

A natural question that arises in the state transition data is what intermediate states does the anneal pass through at different slices of $h$ (i.e. the strength of the amplification of the ground state). One way to represent this process is to simply construct a graph which consists of the classical states represented as nodes and edges representing when moving from one $h$ value to another changed the dominant classical state in the readout of the samples. For a single mapping procedure from one classical state to a ground state, this graph would simply be a path; the node representing the starting state is at one end and the node representing the largest classical state found among all anneal at the readout when $h=3$ (which we would expect to typically be the intended ground state). Then if there were any intermediate states that the anneal found across the increasing $h$ strengths, we could connect these together in a linear line thus forming a simple linear nearest neighbors (LNN) path. If there are no intermediate states (for example if the samples immediately were pushed into the ground state) then the path would consist entirely of the two nodes and single edge connecting them. Taking the union of each of these paths across all state mappings of initial state to ground state can them form a coherent state transition network for that problem Ising. This representation is a simplification of the data and the state transition process because it only forms edges for the dominant classical state found at each $h$ step. However, this representation does give a notion of \emph{distance} between the starting states and the ground states in the form of how many other states does the anneal transition to before reaching the ground state. For drawing these graphs the layout used is the \textit{spring layout} in the python Networkx \cite{hagberg2008exploring} library, which in part uses the Fruchterman-Reingold force directed algorithm \cite{fruchterman1991graph}.


\section{Results}
\label{section:results}

In this section we present state transition mapping results for each of the three test Isings. Section \ref{section:results_n6} details results for the $N_6$ Ising, Section \ref{section:results_n7} details results for the $N_7$ Ising, and Section \ref{section:results_n8} shows results for the $N_8$ Ising. Section \ref{section:results_state_transition_networks} details the state transition network representation of the data. Section \ref{section:results_groundstate_to_groundstate} further investigates ground state to ground state transitions in each of the three test Isings. Section \ref{section:results_fair_sampling} analyzes what the h-gain susceptibility metric shows in regards to fair sampling - specifically how similarly the h-gain response curves and susceptibility quantities behave across the multiple ground states for each of the test Isings. Lastly Section \ref{section:results_RA_only} examines the ground state proportion sampling rate of the three test Isings when only reverse annealing is applied in order to determine if there are similarities to the h-gain response curves. 


\subsection{\texorpdfstring{$N_6$} \qquad \space Ising h-gain response curves}
\label{section:results_n6}

\begin{figure}[t!]
    \centering
    \includegraphics[width=0.19\textwidth]{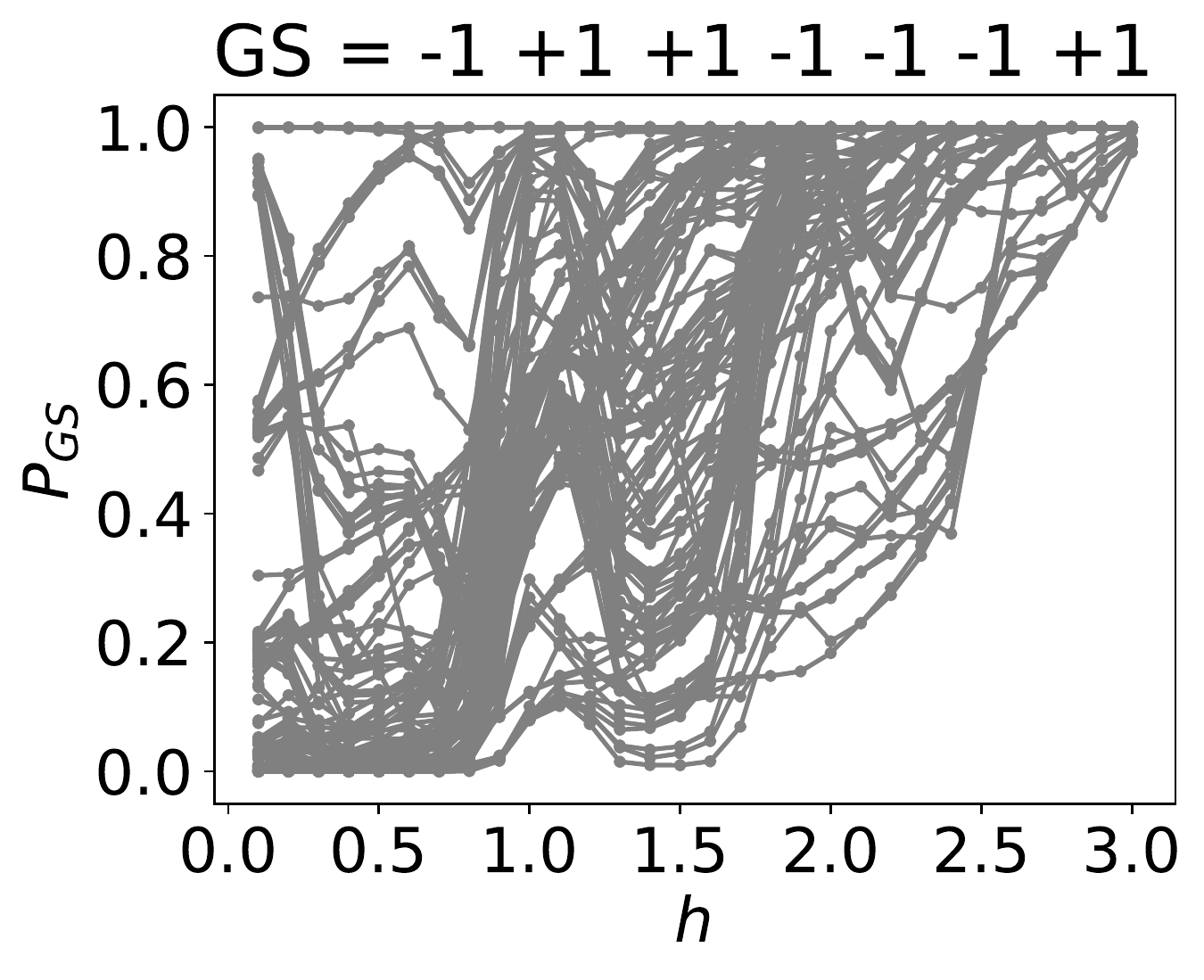}
    \includegraphics[width=0.19\textwidth]{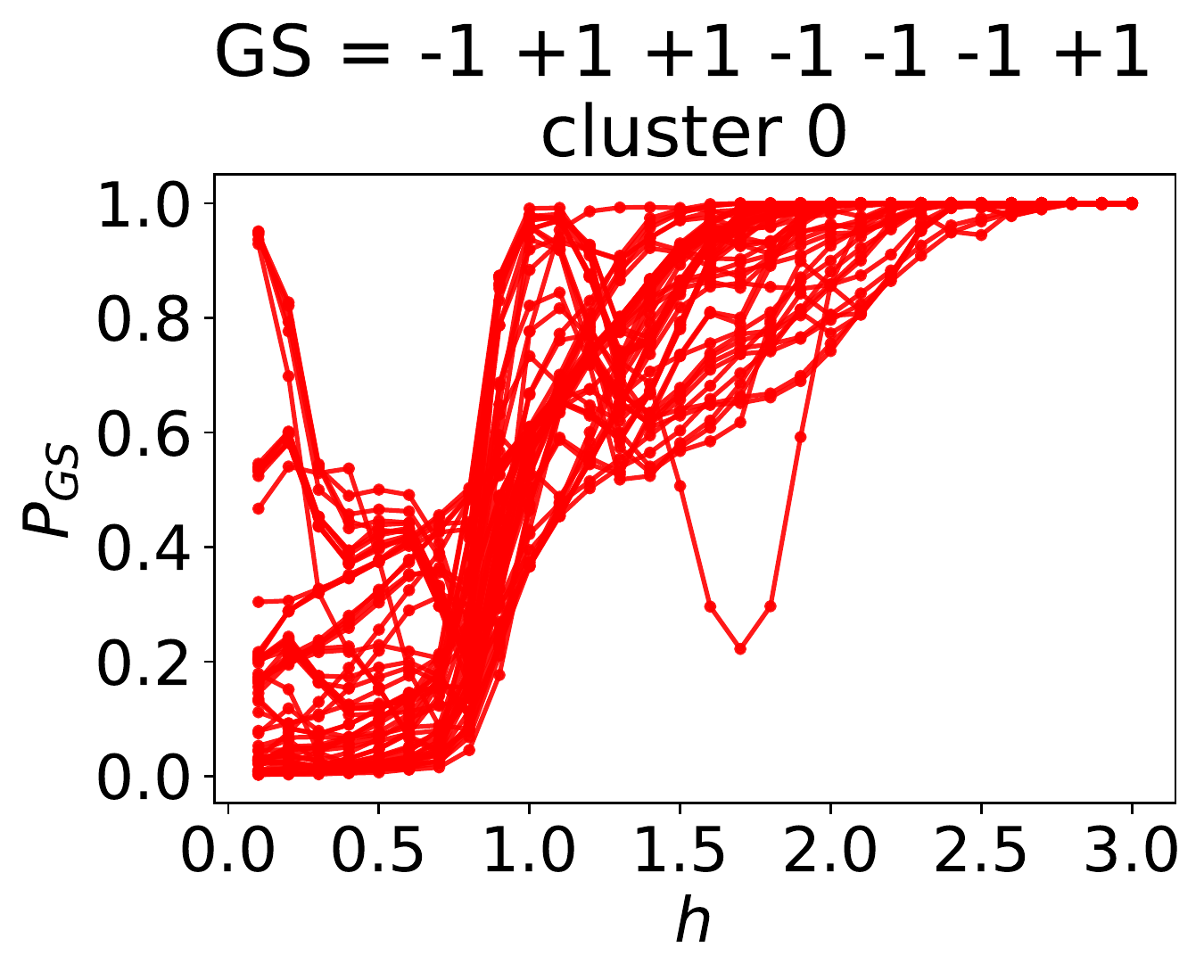}
    \includegraphics[width=0.19\textwidth]{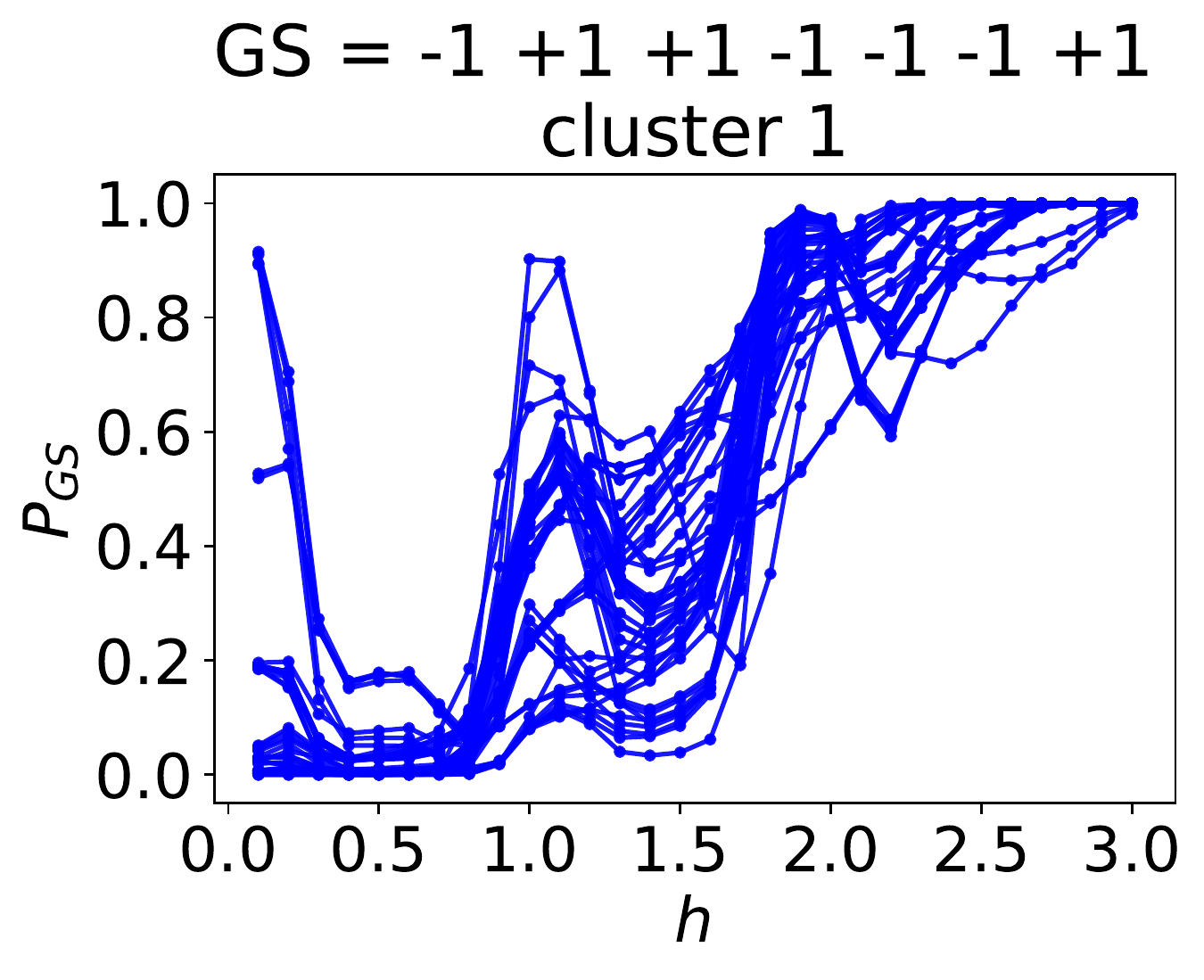}
    \includegraphics[width=0.19\textwidth]{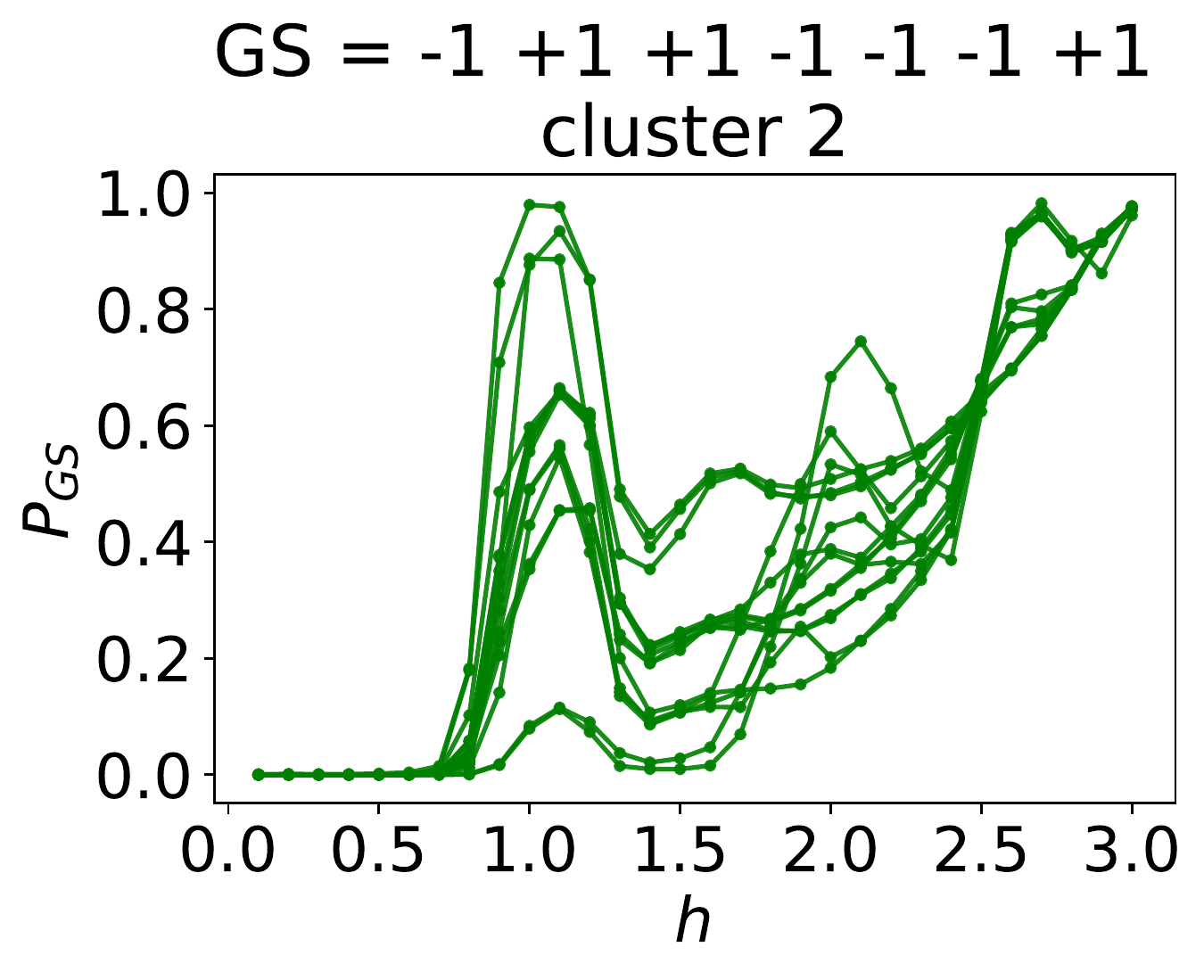}
    \includegraphics[width=0.19\textwidth]{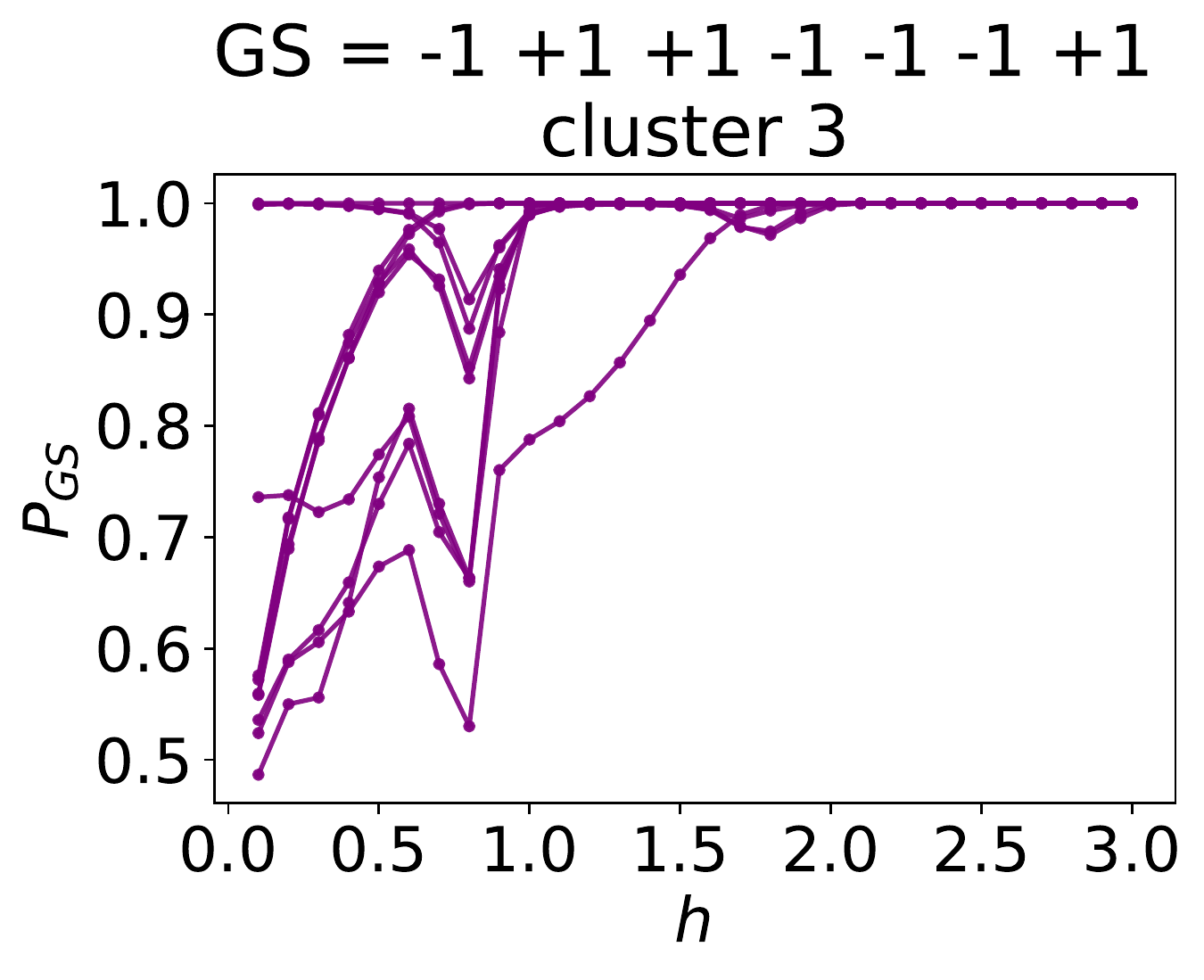}
    \caption{The left hand gray colored plot shows the distribution of $P_{GS}$ as $h$ increases for transitioning all $2^7$ input states into a single ground state for the $N_7$ Ising (the exact ground state is shown in the title of each of the sub-figures). The four right-hand most plots split up this data into $4$ distinct clusters using unsupervised spectral clustering of the vectors of $P_{GS}$ values across the increases $h$ strengths. }
    \label{fig:h_strength_vs_GSP_n7_clustered_GS0}
\end{figure}

Figures \ref{fig:h_strength_vs_GSP_n6_clustered_GS0} and \ref{fig:h_strength_vs_GSP_n6_clustered_GS1} show the clustering results for two of the four ground states of the $N_6$ Ising. An interesting characteristic of the results shown in Figure \ref{fig:h_strength_vs_GSP_n6_clustered_GS0} and \ref{fig:h_strength_vs_GSP_n6_clustered_GS1} is the similarity between them even though they are showing the h-gain response curve for two different ground states and these two ground states are not complements of each other. This suggests that there is quite a bit of symmetry in this problem Ising even. While the $h$ strength parameter consistently is consistently increasing on the x-axis of both Figures \ref{fig:h_strength_vs_GSP_n6_clustered_GS0} and \ref{fig:h_strength_vs_GSP_n6_clustered_GS1}, we see that the h-gain response curves are \emph{not} all monotonically increasing as a function of $h$. This means that this incremental increase of $h$ is transitioning the annealing process into other intermediate states, including in particular other ground states, which would therefore necessarily decrease the success probability. The unsupervised clustering of the h-gain response curves help to differentiate which response curves have a non-monotonic response and which ones do; for example the cluster of red curves in Figure \ref{fig:h_strength_vs_GSP_n6_clustered_GS0} and the cluster of purple curves in Figure \ref{fig:h_strength_vs_GSP_n6_clustered_GS1} have identified a subset of states which required comparably smaller amount of applied $h$ strength in order to transition to the ground state. This state transition process will be investigated in more detail for ground state to ground state transitions in Section \ref{section:results_groundstate_to_groundstate} and in the form of a state transition network in Section \ref{section:results_state_transition_networks}. The h-gain response curves for the other two ground states are not shown here for brevity. 

Figure \ref{fig:HGain_susceptibility_to_groundstate_n6} shows the susceptibility metric across all $2^6$ initial states of the $N_6$ Ising when forcing the anneal into the four distinct ground states. An immediate observation that can be made on this data is that for each ground state; the complement of the ground state is always the minimum susceptibility across all possible states. 

Figure \ref{fig:summary_metrics_n6} shows scatterplots of the three metrics (hamming distance, energy, and $\delta$) outlined in Section \ref{section:methods_metrics} vs $\chi$ for each of the initial states. For each of these three metrics there are a couple of clear observations to be made. First, there is a consistent positive correlation between susceptibility and hamming distance proportion where higher hamming distance proportion leads to lower susceptibility. Second, there is seemingly very little trend between energy and susceptibility; we can see a clear stratification of states with a similar susceptibility being distributed across the entire energy spectrum of the Ising. Third, there does appear to be some correlation between $\delta$ and $\chi$ where a larger $\delta$ corresponds to higher susceptibility. 

Figures \ref{fig:summary_metrics_n6} and \ref{fig:HGain_susceptibility_to_groundstate_n6} use the same coloring schemes from the spectral clustering done on the h-gain response curves, however the actual colors (i.e. index of the cluster) used are arbitrary and randomly assigned by the clustering algorithm. However, all of the figures follow consistent coloring schemes for each of the $4$ different ground states. Note that because two of the h-gain response curves were omitted, the coloring of the other two datasets is not shown in Figures \ref{fig:h_strength_vs_GSP_n6_clustered_GS0} and \ref{fig:h_strength_vs_GSP_n6_clustered_GS1}. For a clear example, the clustering coloring used in Figure \ref{fig:h_strength_vs_GSP_n6_clustered_GS0} is the same coloring used in Figure \ref{fig:HGain_susceptibility_to_groundstate_n6} top left, which is also the same coloring used in the left hand column of Figure \ref{fig:summary_metrics_n6}; all of these plots are in some way representing and analyzing the same dataset where the \emph{h-gain state encoding} was applied to a single one of the four ground states. 

\begin{figure}[t!]
    \centering
    \includegraphics[width=0.49\textwidth]{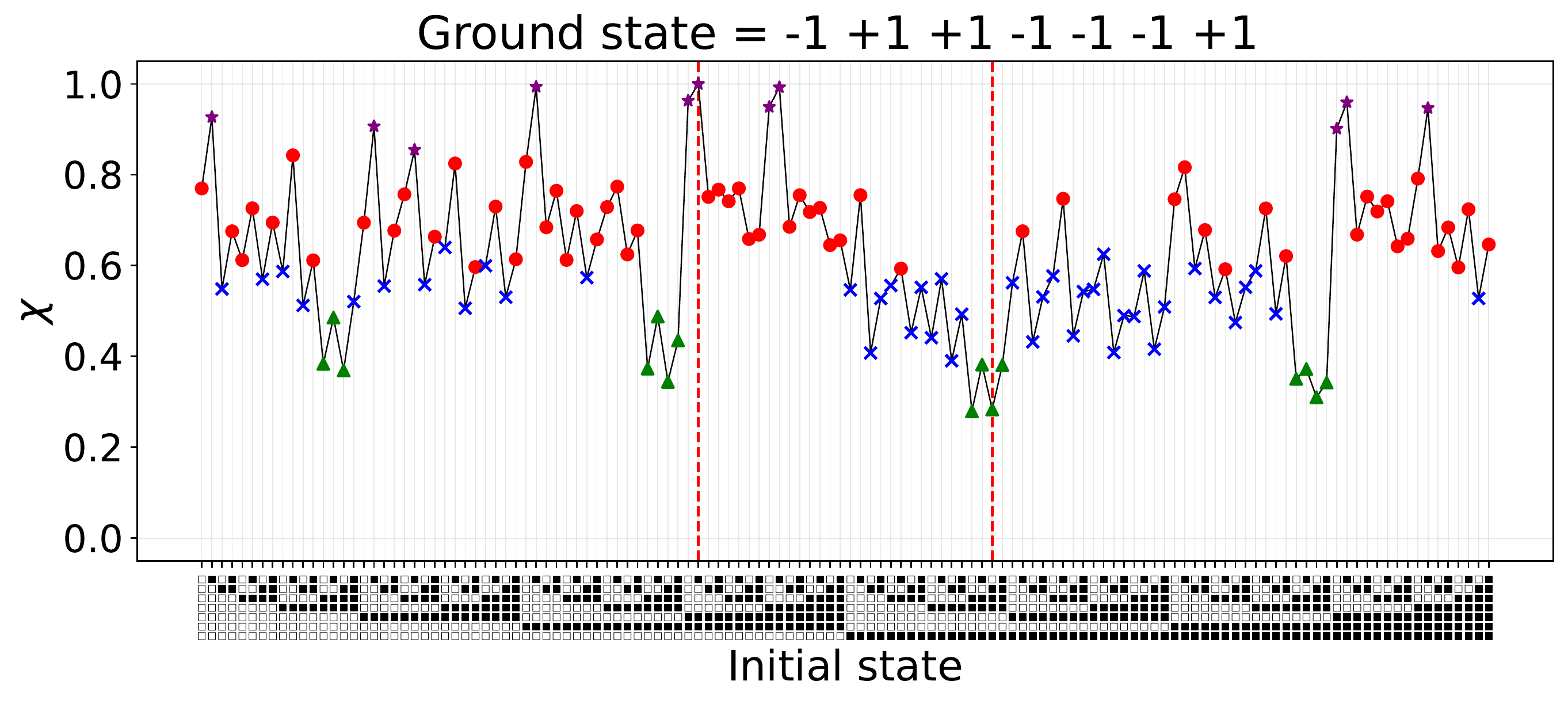}
    \includegraphics[width=0.49\textwidth]{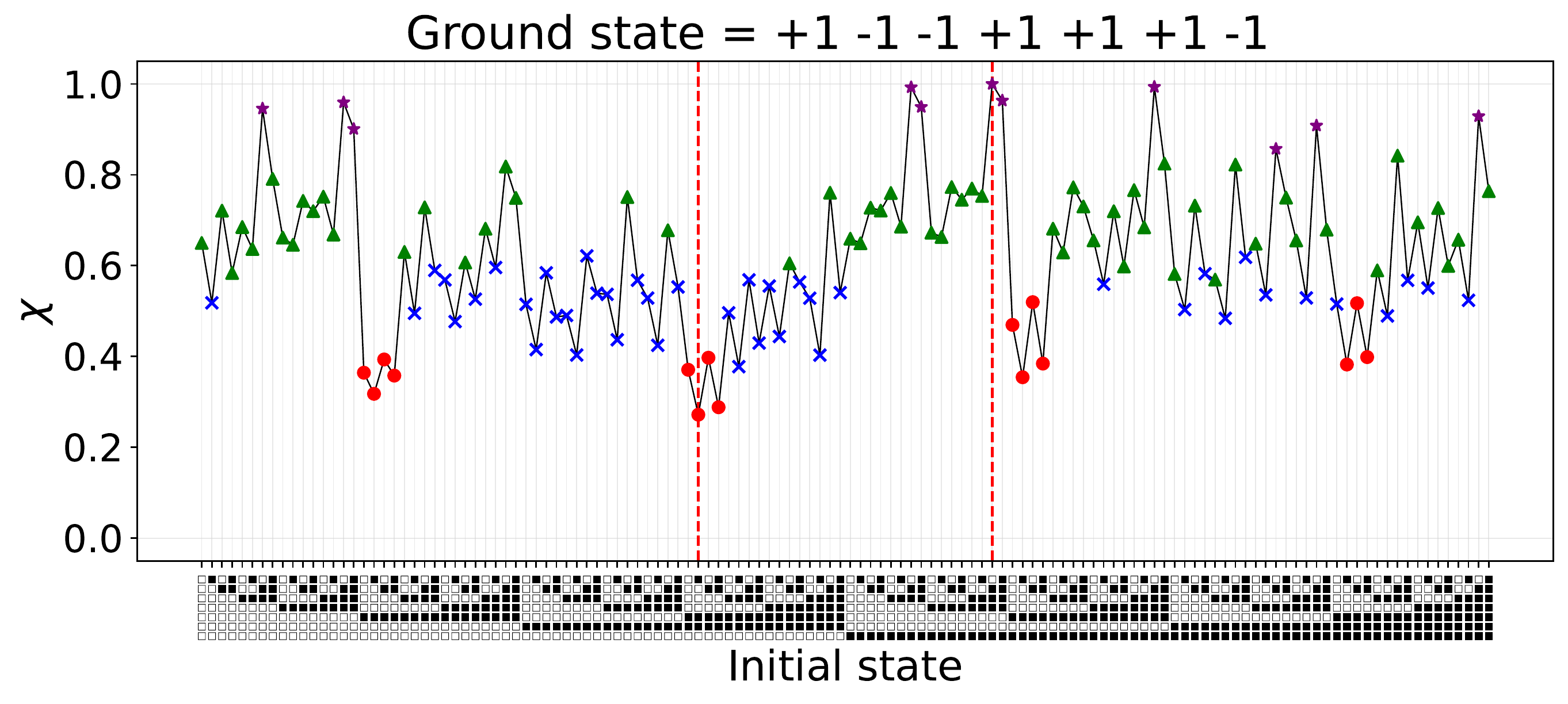}
    \caption{$N_7$ Ising susceptibility across all $2^7$ initial states when applying the h-gain schedule to transition the system into the each of the two ground states. The x-axis encodes these initial states as vectors of vertical blocks where $\blacksquare$ denotes a variable state of $1$ and $\square$ denotes a variable state of $-1$. The initial state vectors are read from bottom to top where the bottom is the first index which corresponds to variable $0$ in the problem Ising. The initial states which are also other ground states are marked with dashed red vertical lines. For each sub-figure, the reflexive ground state mapping (i.e. where the initial state and the intended state are the same ground state) case can be found visually as the state marked with a red dashed vertical line which has the maximum susceptibility measure among all of the initial states. }
    \label{fig:HGain_susceptibility_to_groundstate_n7}
\end{figure}

\begin{figure}[t!]
    \centering
    \includegraphics[width=0.24\textwidth]{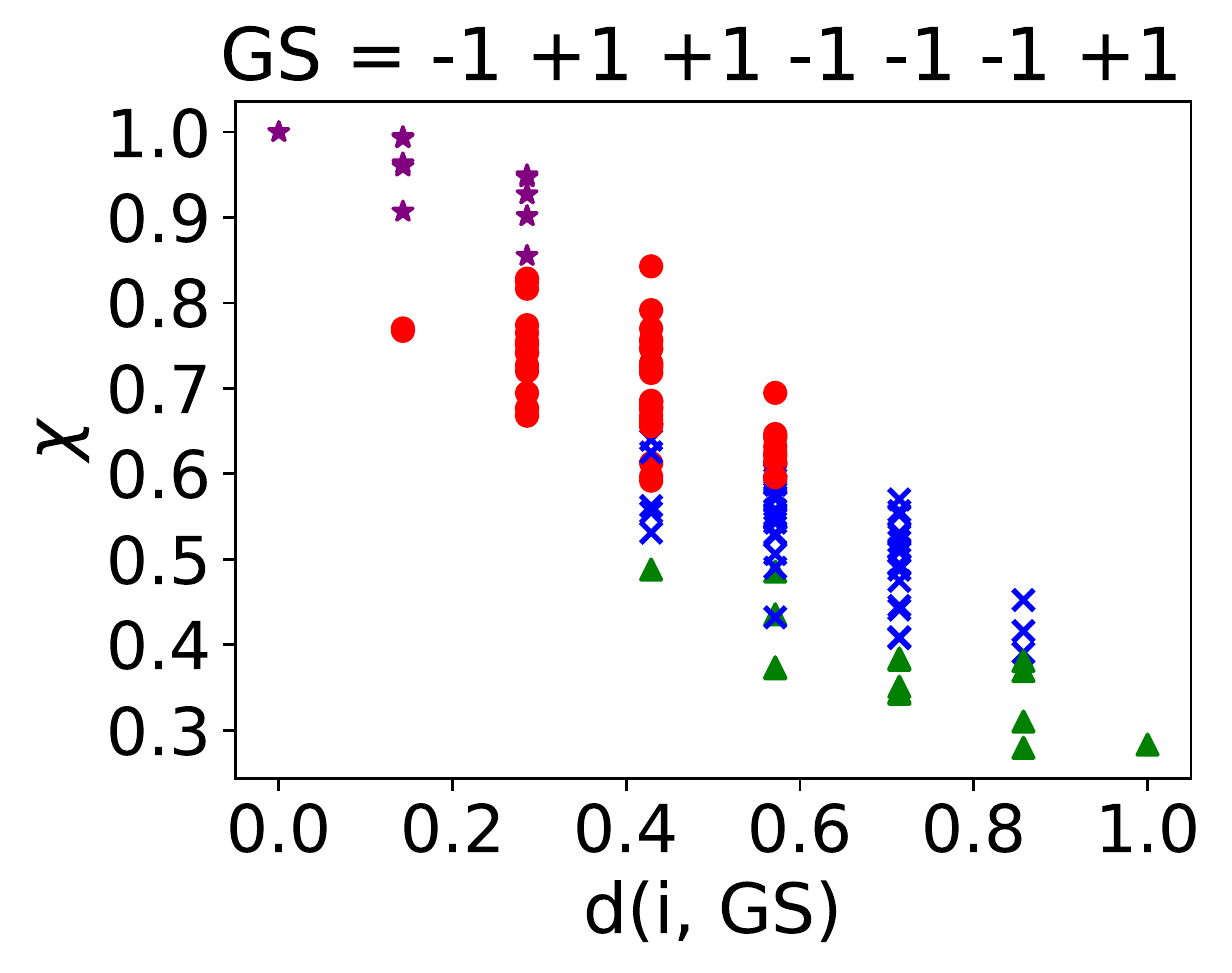}
    \includegraphics[width=0.24\textwidth]{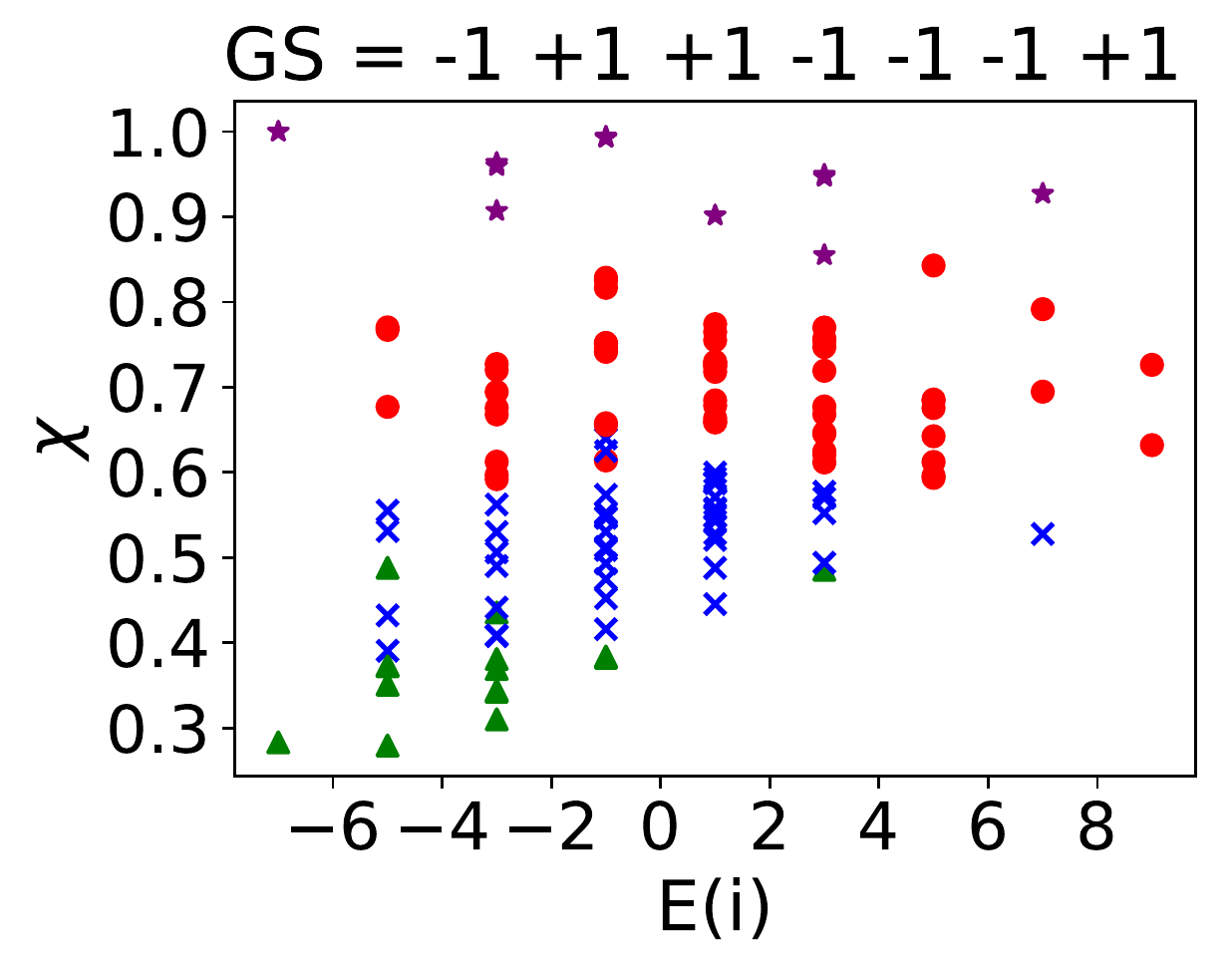}
    \includegraphics[width=0.24\textwidth]{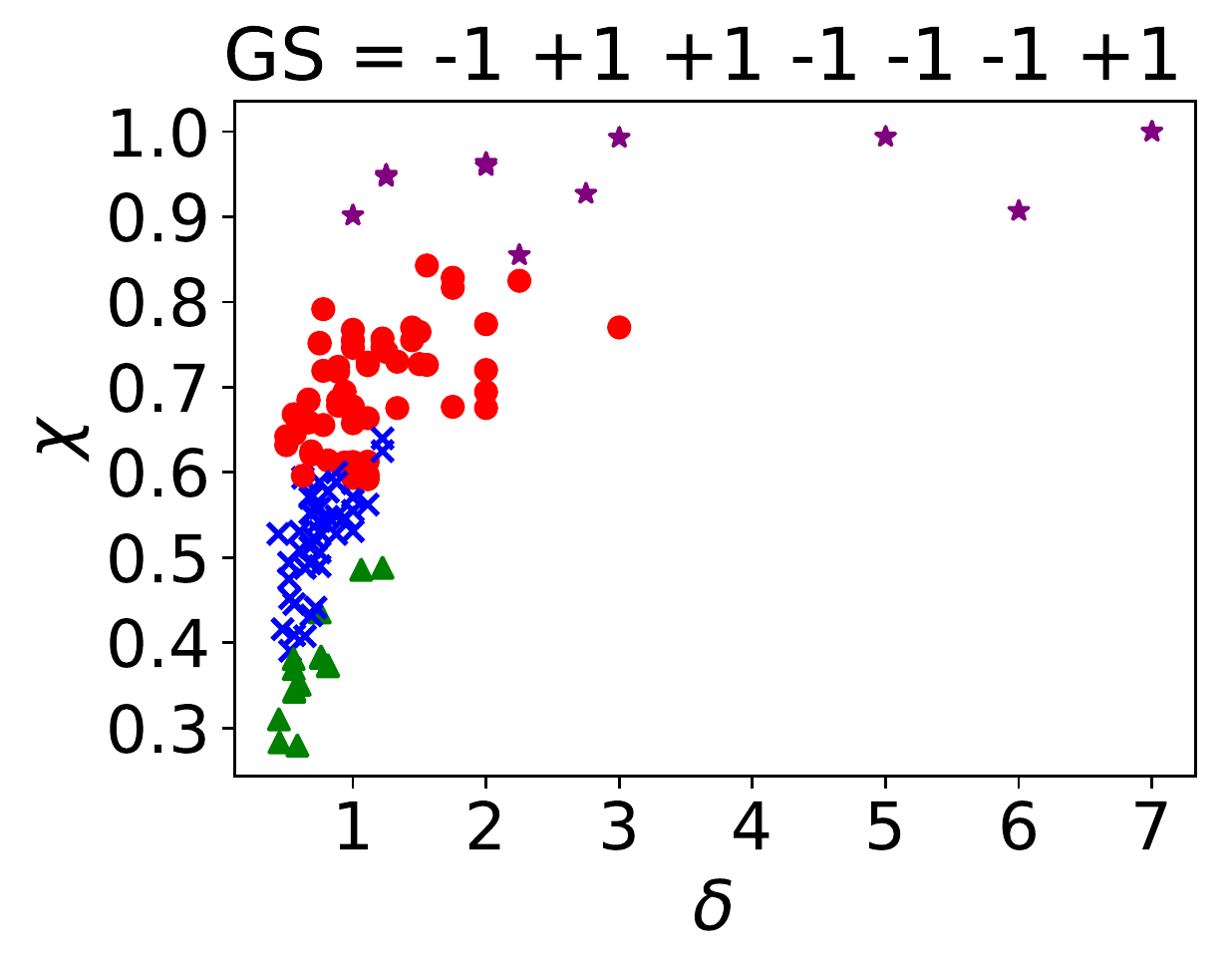}\\
    \includegraphics[width=0.24\textwidth]{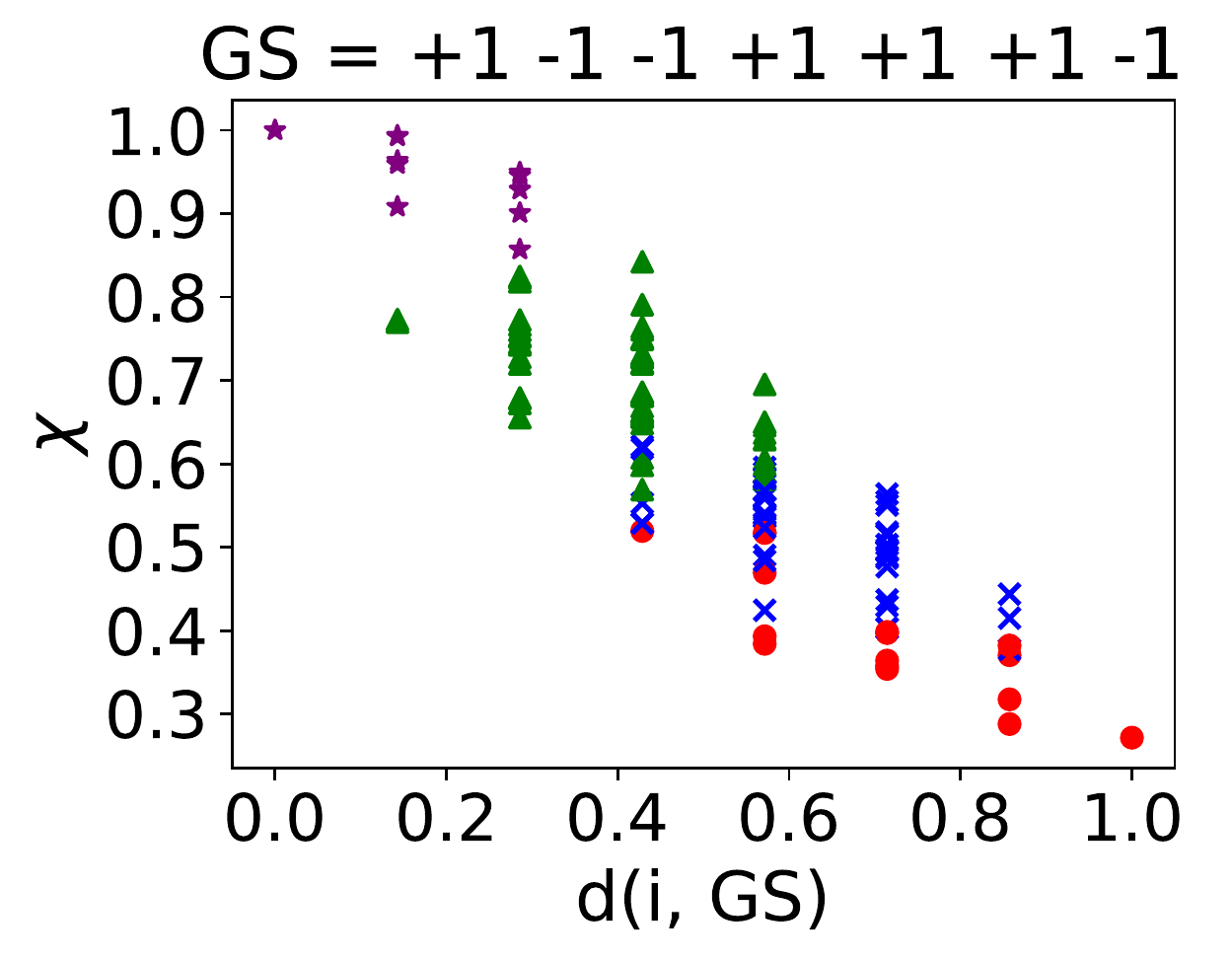}
    \includegraphics[width=0.24\textwidth]{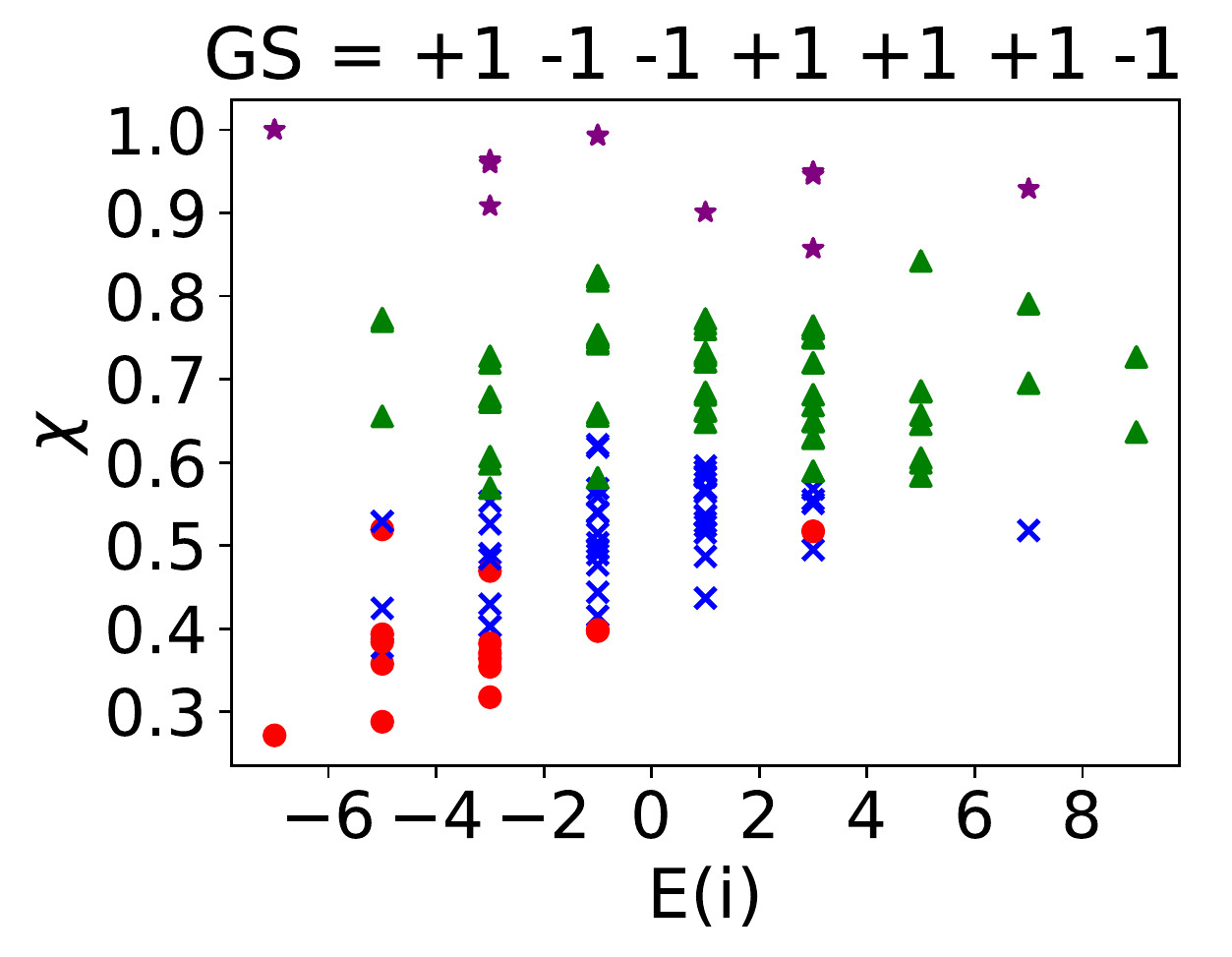}
    \includegraphics[width=0.24\textwidth]{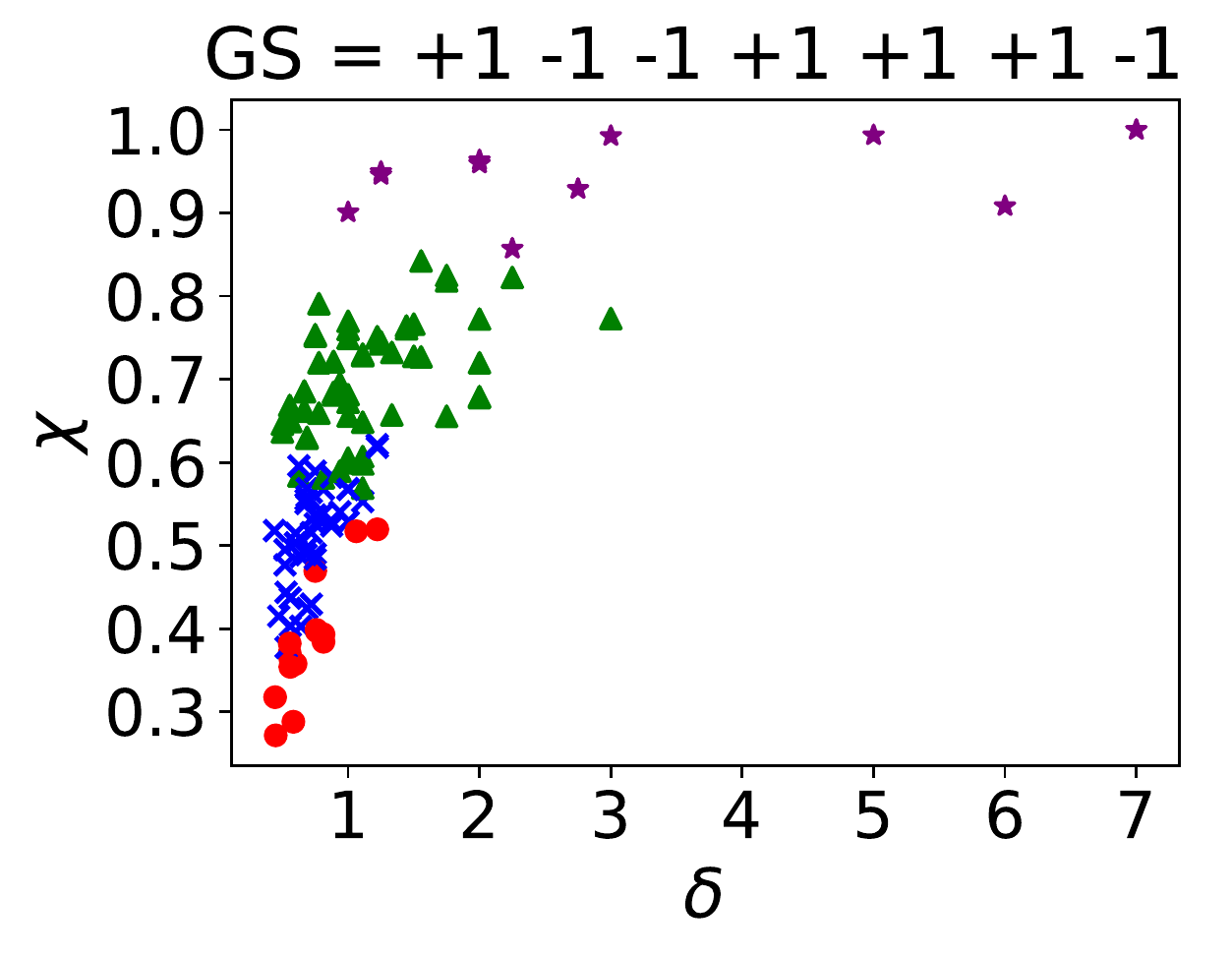}
    \caption{Summary metric plots for the $N_7$ Ising. The $2$ rows correspond to each of the $2$ ground states; the titles of each sub plot are the optimal variable assignment vectors. The three column correspond to three different initial state metrics on the x-axis, the y-axis of each sub-plot is the susceptibility metric $\chi$. The first column has x-axis which are the hamming distance between the ground state and the specific initial state $i$. The second column has x-axis showing the energy of the initial state $i$ evaluated on the $N_7$ Ising. The third column has x-axis showing the $\delta$ metric for each initial state. The node colorings for the top row of sub-figures share the same node coloring as Figure \ref{fig:HGain_susceptibility_to_groundstate_n7} left, and the node colorings for the bottom row of sub-figures share the same node coloring as Figure \ref{fig:HGain_susceptibility_to_groundstate_n7} right. }
    \label{fig:summary_metrics_n7}
\end{figure}

Examining the results from the h-gain response curve clustering (Figures \ref{fig:h_strength_vs_GSP_n6_clustered_GS0} and \ref{fig:h_strength_vs_GSP_n6_clustered_GS1}) in conjunction with Figure \ref{fig:HGain_susceptibility_to_groundstate_n6} and Figure \ref{fig:summary_metrics_n6} we can observe several interesting trends. First, we can clearly see the motivation for developing the $\delta$ metric in Section \ref{section:methods_metrics}. Specifically, in Figure \ref{fig:HGain_susceptibility_to_groundstate_n6} we see that when the initial state we encode using reverse annealing is also the ground state that is specified using \emph{h-gain state encoding}, the susceptibility metric is at a minimum (i.e. is nearly $0$). This is to be expected, however we also see that there are several other states which have a relatively large susceptibility. However, these states are not ground states (ground states are marked in Figure \ref{fig:HGain_susceptibility_to_groundstate_n6} using dashed red vertical lines). If we examine where these states occur in the block of sub plots in Figure \ref{fig:summary_metrics_n6}, which can be done using the fixed clustered coloring scheme and the ground state labels on the titles, we see that there is apparently no correlation with respect to the energy of these states. There is some correlation with respect to the hamming distance proportion, namely that all of the hamming distances are $\leq \frac{2}{3}$. The interesting property of these high susceptibility states is that they are very close to the ground state in terms of hamming distance (i.e. bit flips), but only for variables which are high degree in the graph of the Ising problem; this is the observation that led to the creation of the $\delta$ metric. 

As a specific example, we can examine Figure \ref{fig:HGain_susceptibility_to_groundstate_n6}, upper left hand sub plot. The ground state in this plot has variable assignments $[-1, -1, -1, +1, -1, +1]$ where the index denotes the variable in the original Ising problem. There are a total of $6$ initial states which are colored red in this plot, meaning that their h-gain response curves all behaved similarly and indeed we see these $6$ states have the highest susceptibility metrics among all $2^6$ states. The two states which have the largest susceptibility metric besides the ground state are $[-1, +1, -1, +1, -1, +1]$ and $[+1, -1, -1, +1, -1, +1]$. Now if we examine Figure \ref{fig:logical_isings} for $N_6$ (left hand graph), we see that variables $0$ and $1$ both have a degree of $5$, which is the largest degree in the graph. Examining these two other states which had maximum susceptibility shows that they are both exactly one bit-flip away from the ground state -- specifically if we flip $[-1, +1, -1, +1, -1, +1]$ at index $1$ from a $+1$ to a $-1$, we get the ground state. And if we flip $[+1, -1, -1, +1, -1, +1]$ at index $0$ from a $+1$ to a $-1$ we also arrive at the ground state. If we evaluate these two states using Equation $\ref{eq:delta_metric}$, we get a value of $5$ because these states are exactly one degree $5$ variable away from the ground state. The points corresponding to these two states appear in Figure \ref{fig:summary_metrics_n6} in the bottom left hand corner plot where both $\chi$ and $\delta$ are overlapping causing them to appear as a single point. High degree variables in the problem Ising being easier to induce a bit flip in compared to most other states is consistent with how the physical device implements these problems; namely that these high degree variables have more constraints acting on them compared to the rest of the variables, which then requires smaller h-gain amplification of the linear terms encoding the ground state to change the state of the variable compared to other initial states.


\subsection{\texorpdfstring{$N_7$} \qquad \space Ising h-gain response curves}
\label{section:results_n7}

Figure \ref{fig:h_strength_vs_GSP_n7_clustered_GS0} shows the h-gain response curves for all initial states along with the four groups of clustered response curves for one of the two ground states for the $N_7$ Ising results. As with the $N_6$ h-gain response curves, Figure \ref{fig:h_strength_vs_GSP_n7_clustered_GS0} shows non-monotonic increases of $P_{GS}$ as a function of increasing $h$ for most initial states, again suggesting that the process is getting trapped in other intermediate states besides the intended ground state. For brevity the response curves for the other ground state is omitted. Figure \ref{fig:HGain_susceptibility_to_groundstate_n7} shows the susceptibility metric across all $2^7$ initial states of the $N_7$ Ising when applying the \emph{h-gain state encoding} for each of the two ground states. Figure \ref{fig:summary_metrics_n7} shows scatterplots of the three metrics (hamming distance, energy, and $\delta$) outlined in Section \ref{section:methods_metrics} vs $\chi$ for all of the initial states. Compared to the $N_6$ Ising results in Section \ref{section:results_n6}, these results look similar and have similar patterns overall. For example, the $\delta$ metric still reasonably applies to high susceptibility initial states. The h-gain response curve plot (Figure \ref{fig:h_strength_vs_GSP_n7_clustered_GS0}) is noticeably different than the h-gain response curves of the $N_6$ Ising in Figures \ref{fig:h_strength_vs_GSP_n6_clustered_GS0} and \ref{fig:h_strength_vs_GSP_n6_clustered_GS1}. In Figure \ref{fig:HGain_susceptibility_to_groundstate_n7}, because this Ising has exactly two ground states which are complementary, we can observe a clear symmetry in the data - these two plots are mirrors of each other if reflected about the mid-line of the initial state vectors on the x-axis. This is especially clear because of the two ground states, but a similar symmetry between the pairs of ground states in the $N_6$ Ising can also be seen in Figure \ref{fig:HGain_susceptibility_to_groundstate_n6}. 

Overall, the results for the $N_7$ are consistent with the same trends and observations made on the $N_6$ Ising, despite the additional variable and decreased number of ground states. 

\begin{figure}[t!]
    \centering
    \includegraphics[width=0.19\textwidth]{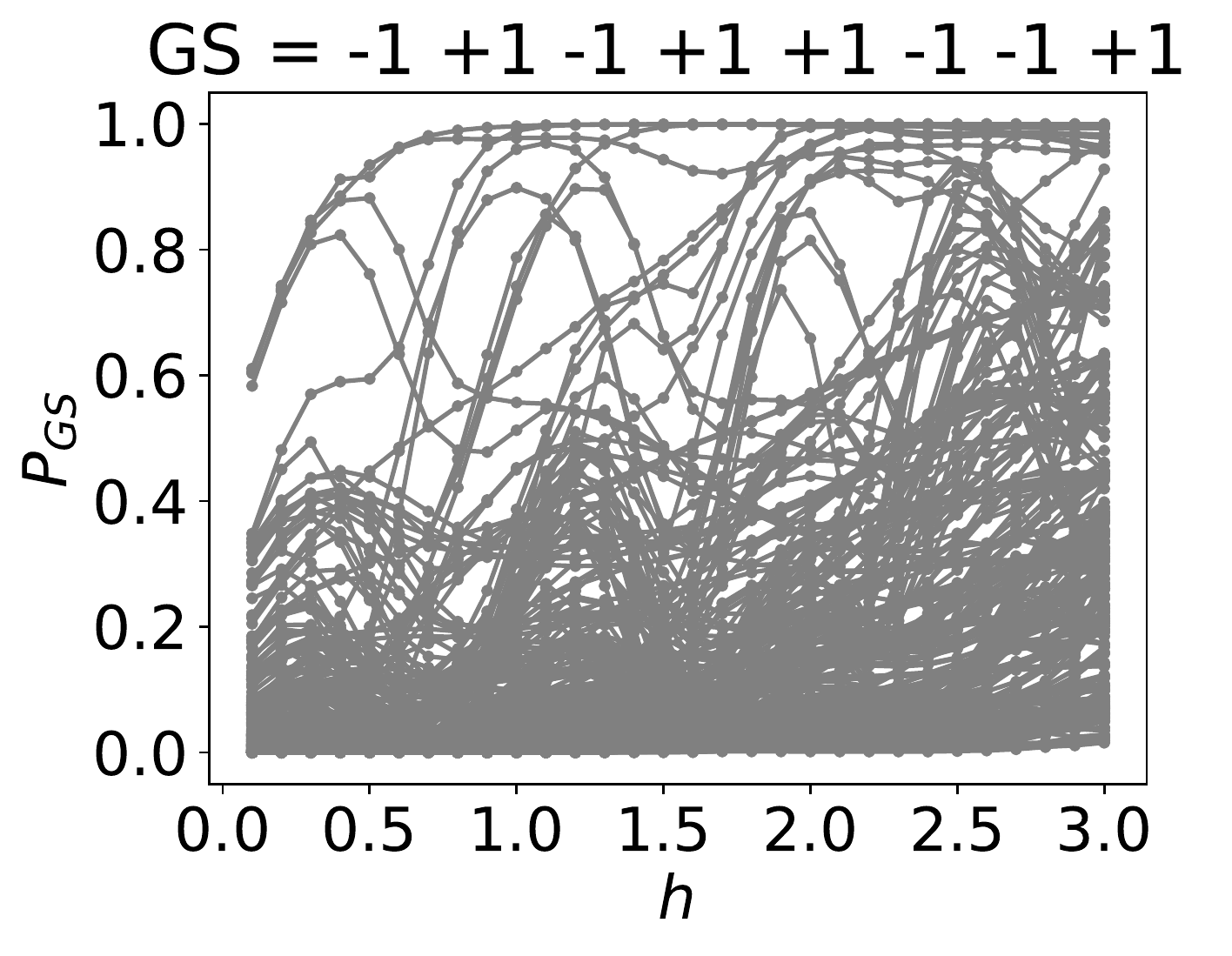}
    \includegraphics[width=0.19\textwidth]{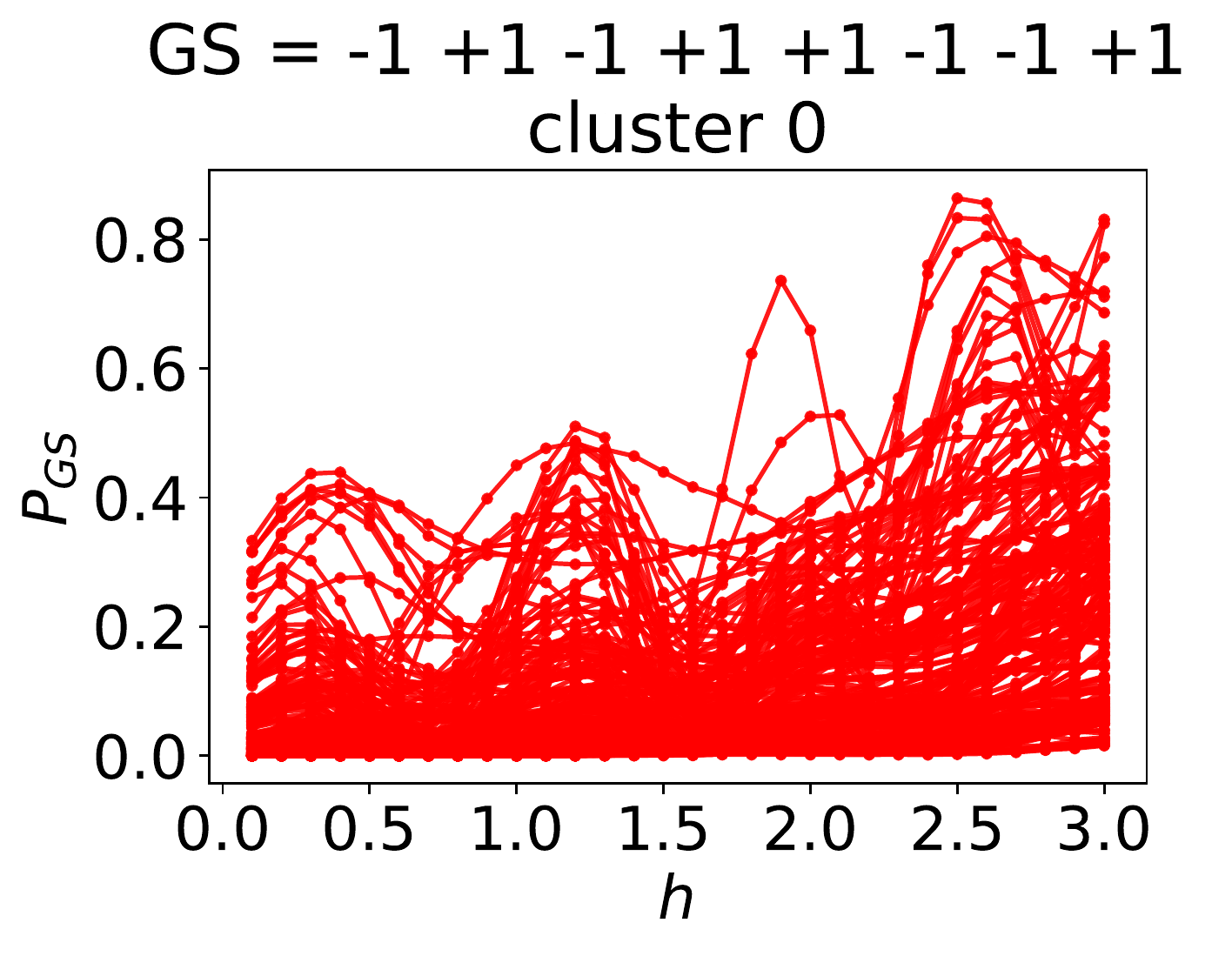}
    \includegraphics[width=0.19\textwidth]{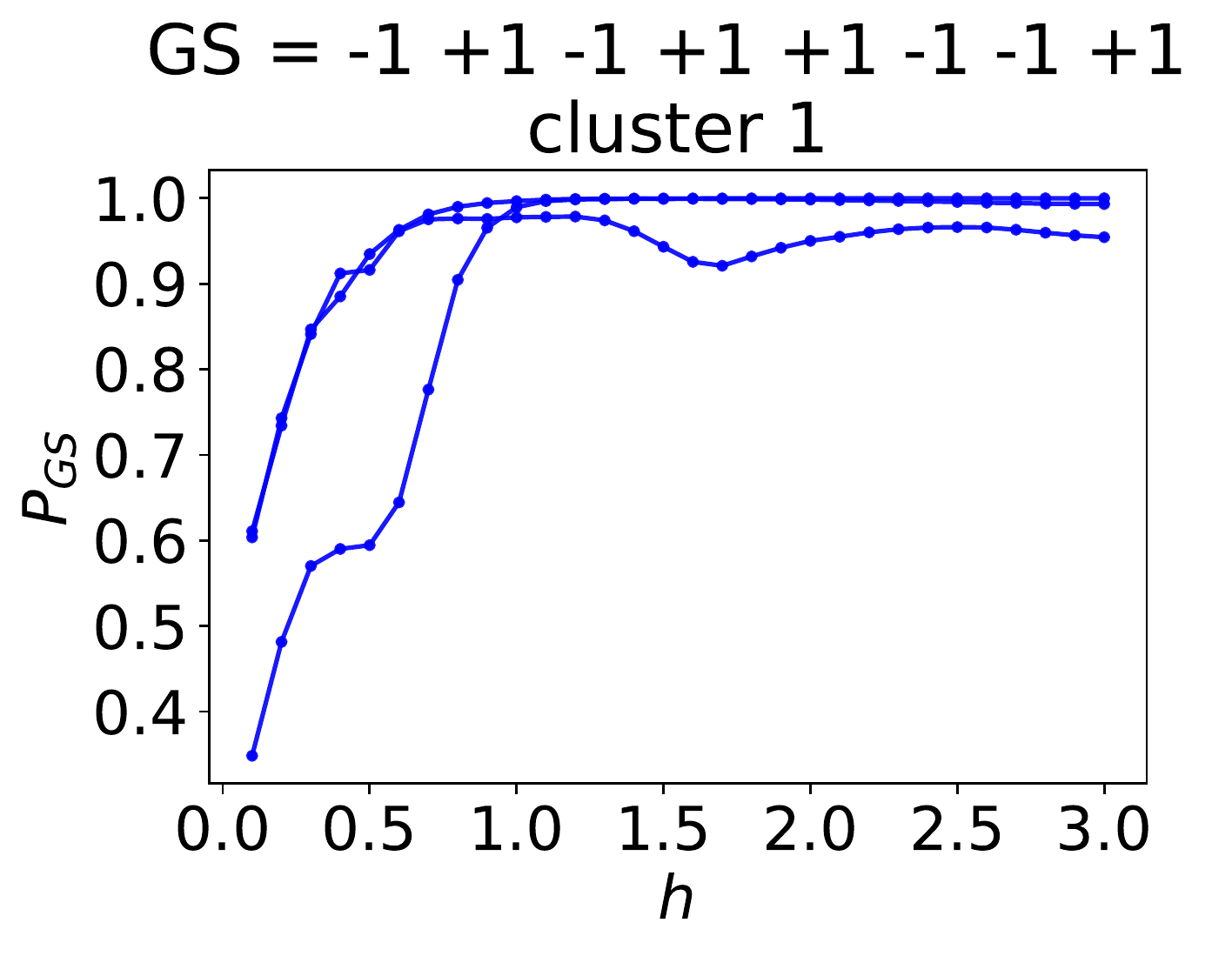}
    \includegraphics[width=0.19\textwidth]{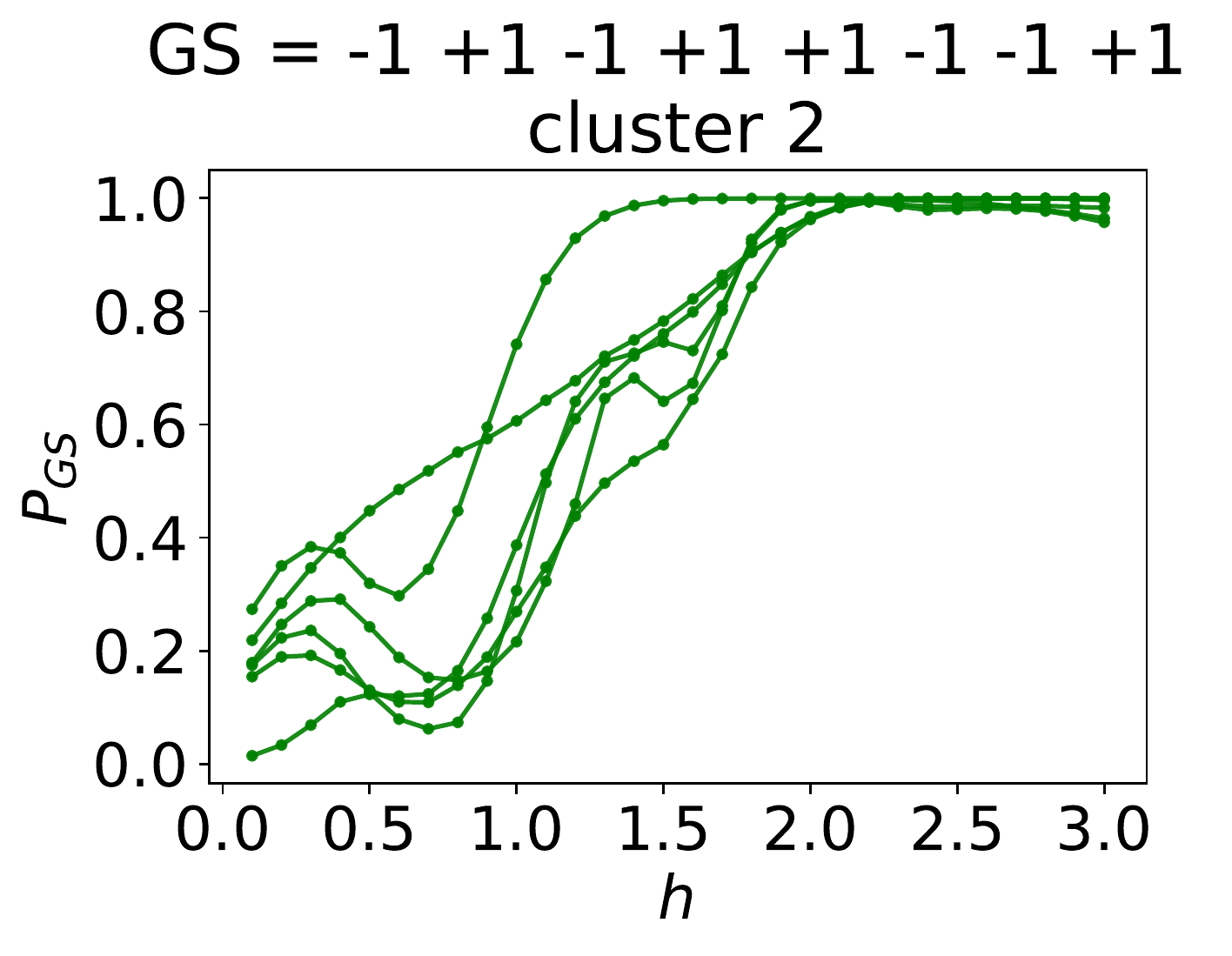}
    \includegraphics[width=0.19\textwidth]{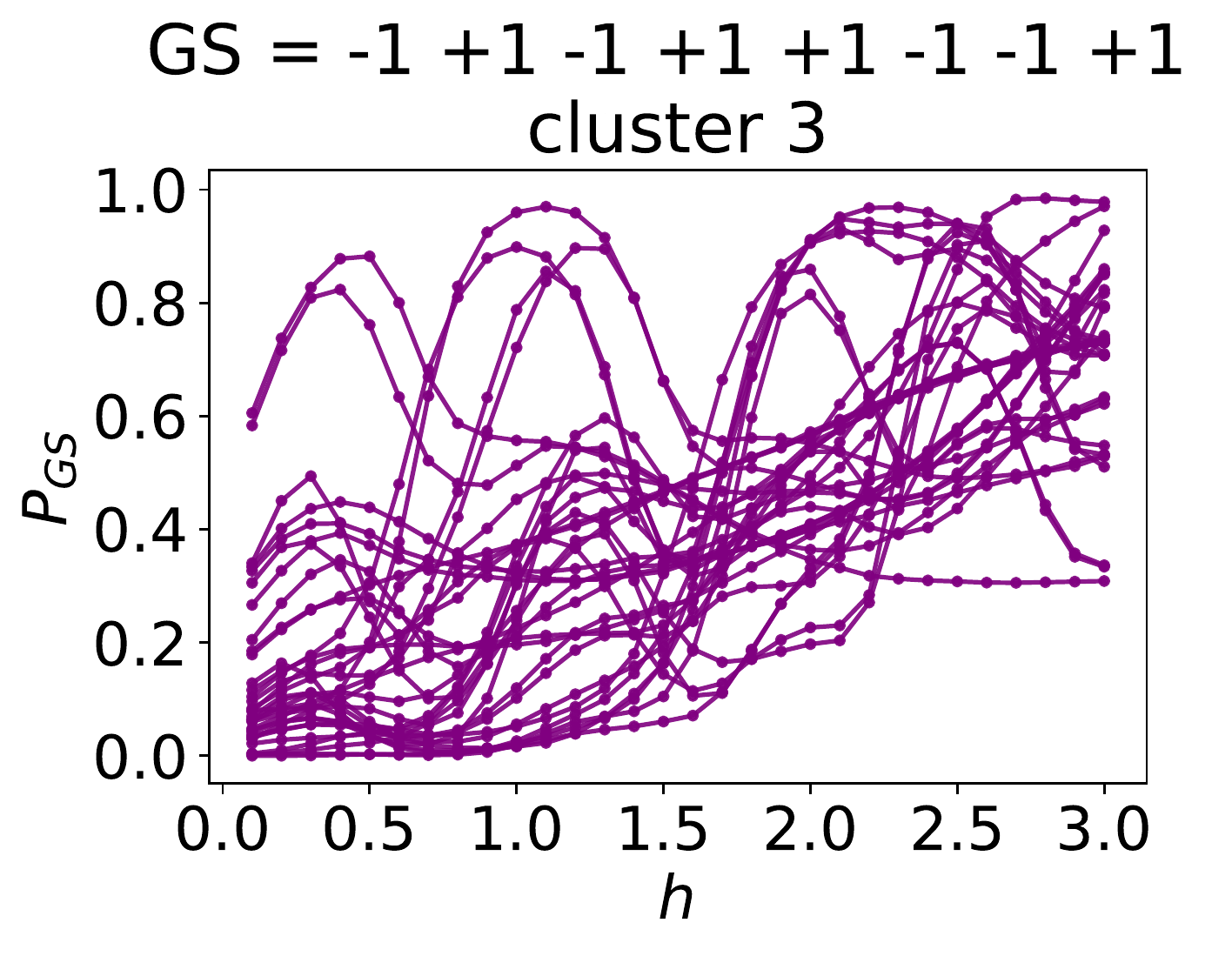}
    \caption{The left hand gray colored plot shows the distribution of $P_{GS}$ as $h$ increases for transitioning all $2^8$ input states into a single ground state for the $N_8$ Ising (the exact ground state is shown in the title of each of the sub-figures). The four right-hand plots split up this data into $4$ distinct clusters using unsupervised spectral clustering of the vectors of $P_{GS}$ values across the increases $h$ strengths. }
    \label{fig:h_strength_vs_GSP_n8_clustered_GS0}
\end{figure}

\begin{figure}[t!]
    \centering
    \includegraphics[width=0.90\textwidth]{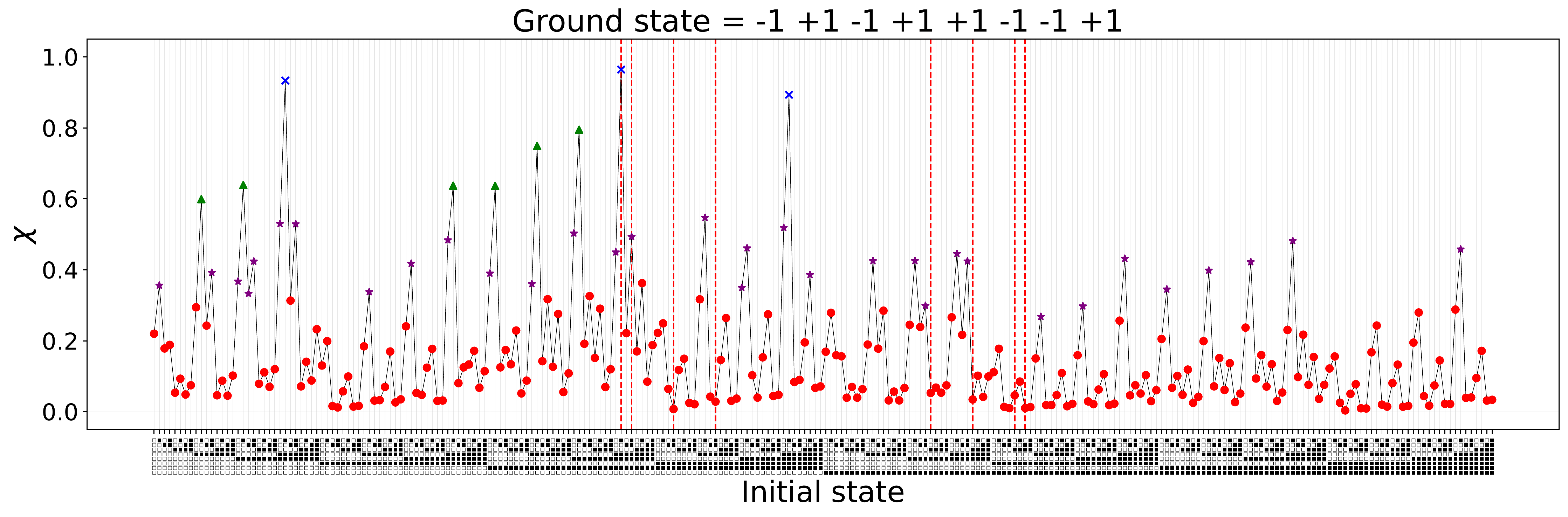}
    \includegraphics[width=0.90\textwidth]{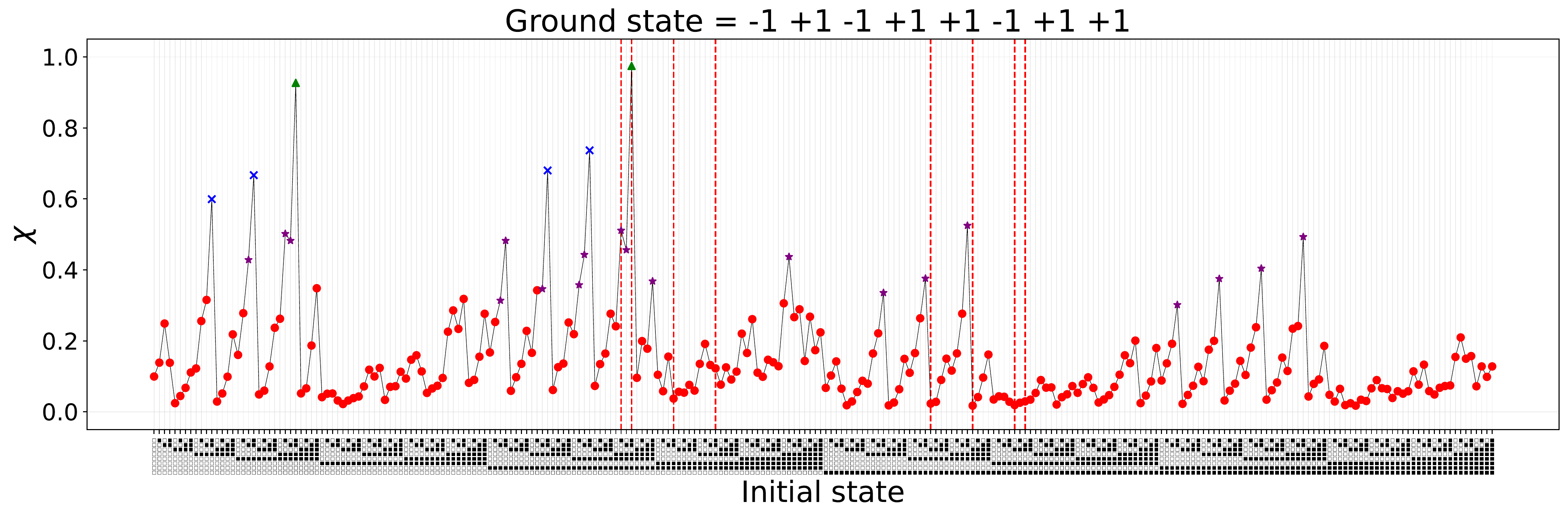}
    \caption{$N_8$ Ising susceptibility across all $2^8$ initial states when applying the h-gain schedule to push the system into two out of the eight ground states where the two sub-figure represents one ground state transition. The other six ground state plots are shown in Figure \ref{fig:appendix_HGain_susceptibility_n8} in Appendix \ref{section:appendix_extra_figures}. The x-axis encodes the initial states as vectors of vertical blocks where $\blacksquare$ denotes a variable state of $+1$ and $\square$ denotes a variable state of $-1$. The initial state vectors are read from bottom to top where the bottom is the first index which corresponds to variable $0$ in the problem Ising. The initial states which are also other ground states are marked with dashed red vertical lines. For each sub-figure, the reflexive ground state mapping (i.e. where the initial state and the intended state are the same ground state) case can be found visually as the state marked with a red vertical line which has the maximum $\chi$ measure among all of the initial states. }
    \label{fig:HGain_susceptibility_to_groundstate_n8}
\end{figure}

\begin{figure}[t!]
    \centering
    \includegraphics[width=0.32\textwidth]{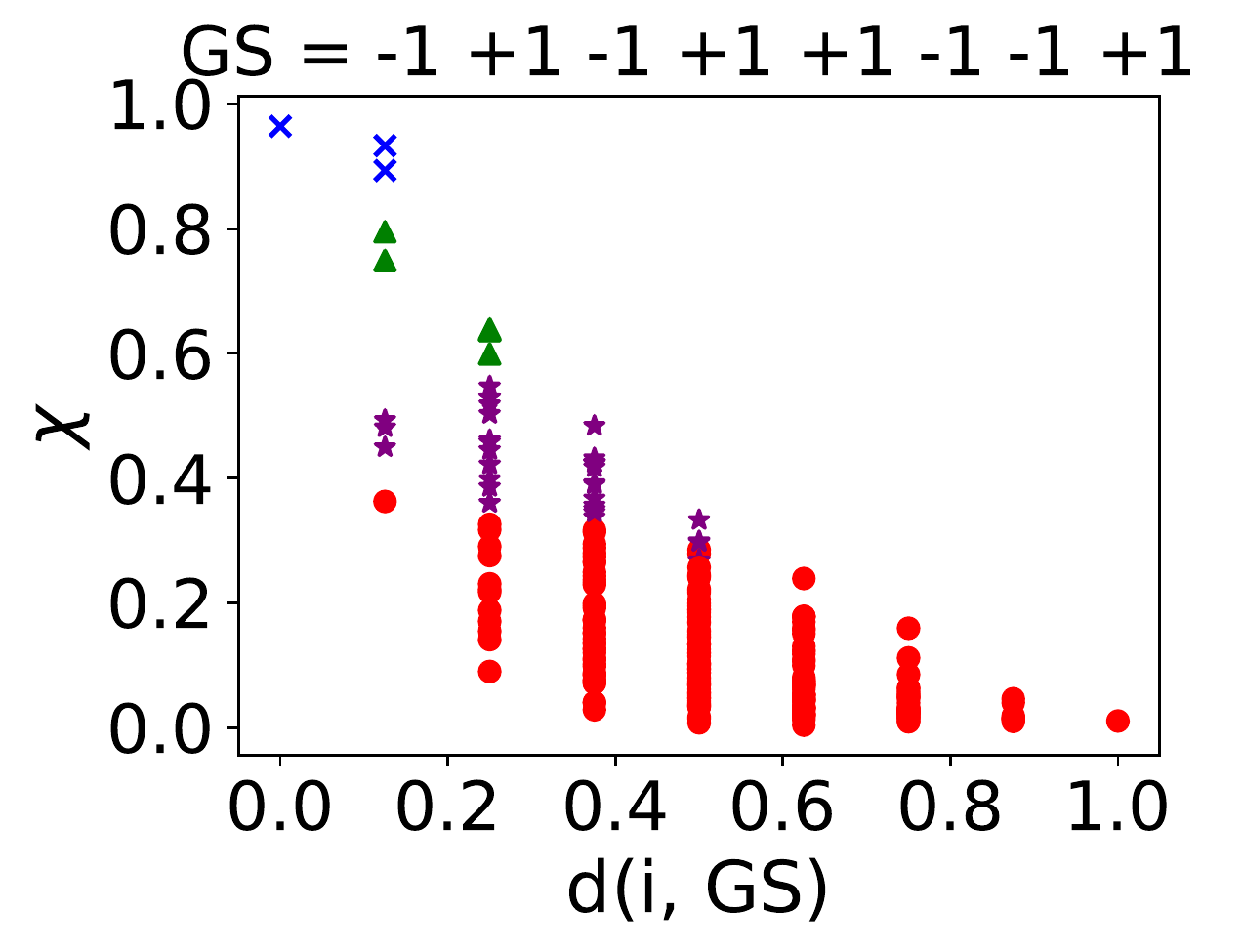}
    \includegraphics[width=0.32\textwidth]{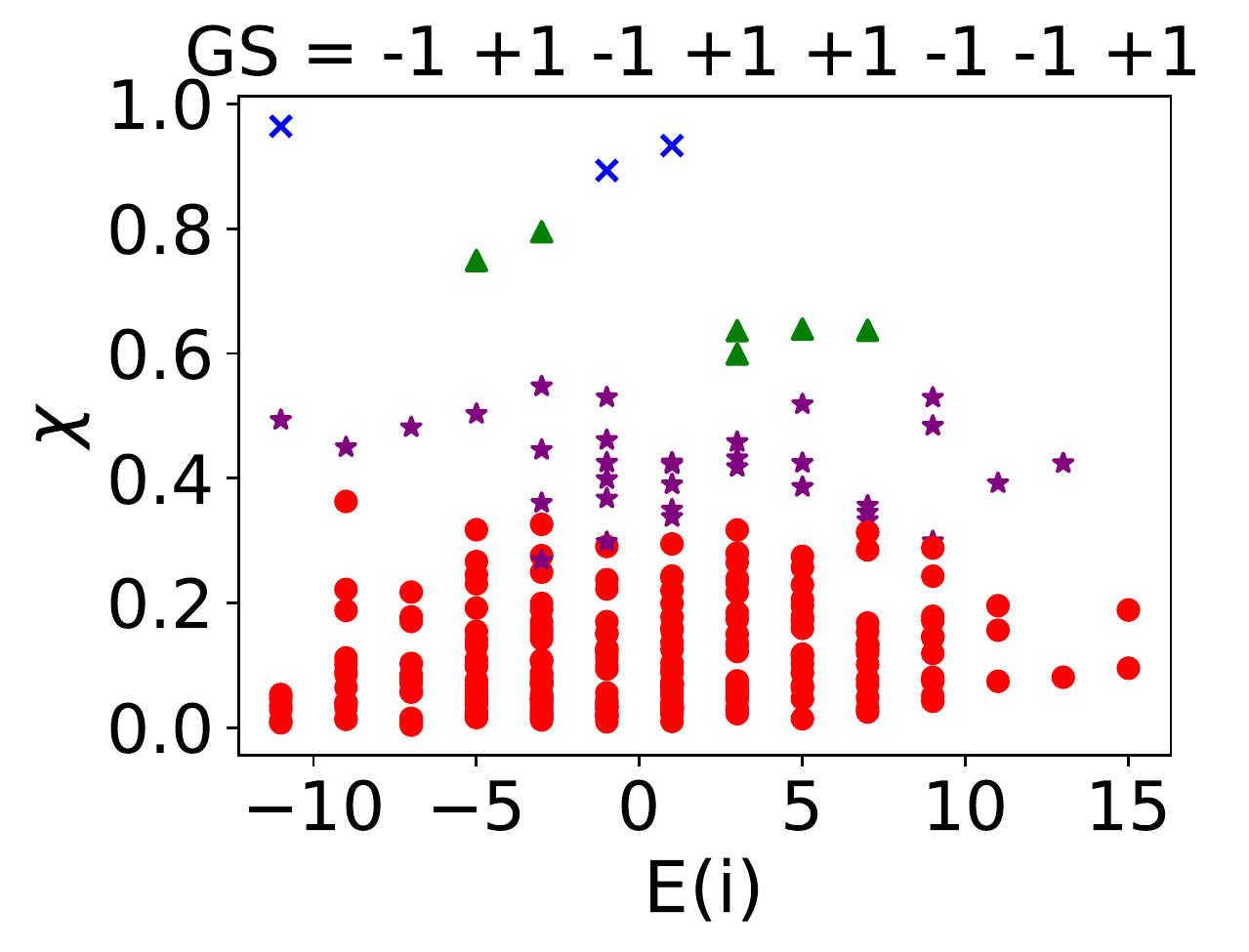}
    \includegraphics[width=0.32\textwidth]{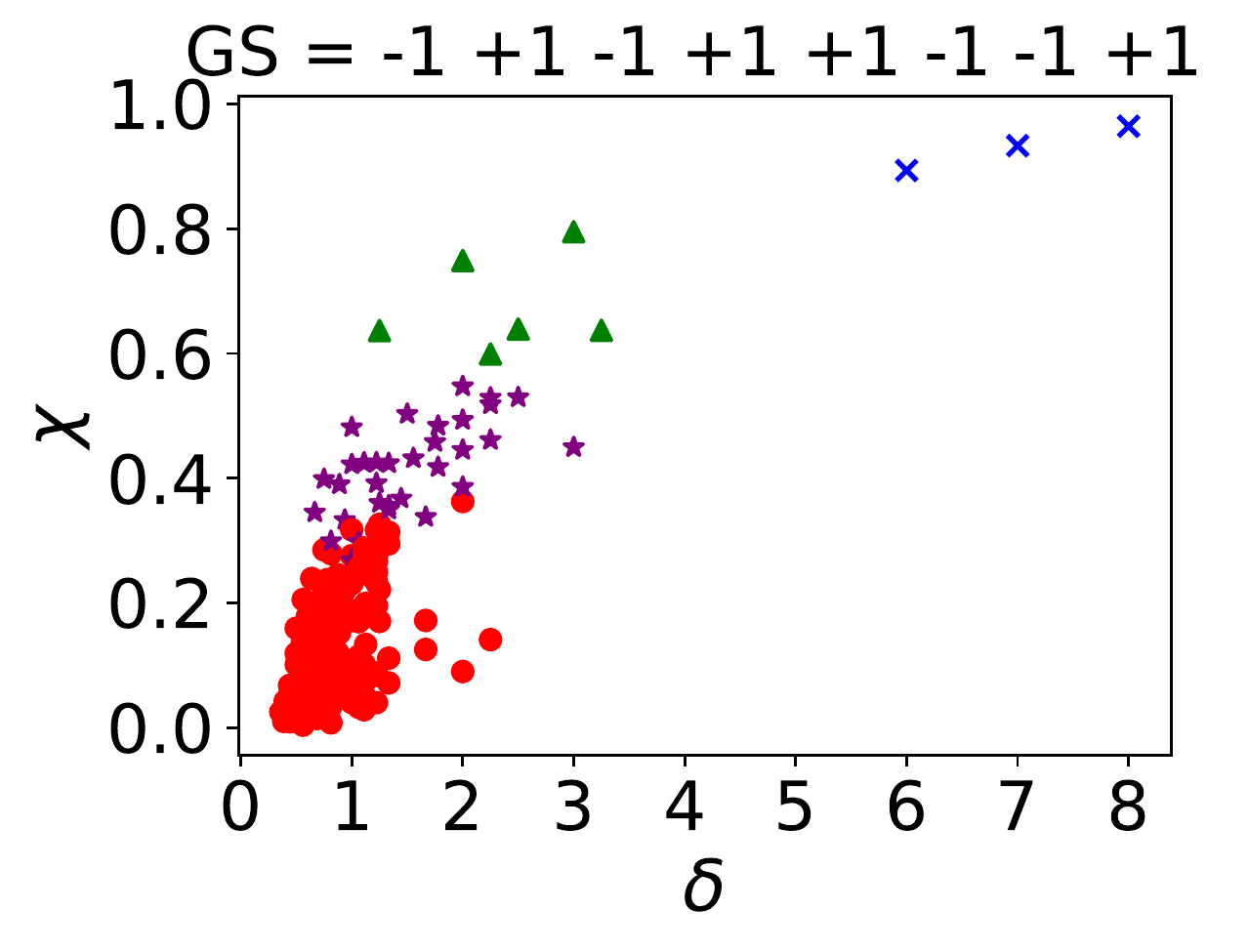}
    \caption{Summary metric plots for the $N_8$ Ising. These three plots show $\chi$ on the y-axis, plotted against hamming distance between the ground state and the specific initial state $i$ (left), the energy of the initial state $i$ evaluated on the $N_8$ Ising (middle), and the $\delta$ metric for each initial state (right). This data is only for the h-gain mapping to the first ground state - the mappings to the other $7$ ground states of the $N_8$ Ising are displayed in Figure \ref{fig:appendix_summary_metrics_N8} in Section \ref{section:appendix_extra_figures}. }
    \label{fig:summary_metrics_n8}
\end{figure}


\begin{figure}[t!]
    \centering
    \includegraphics[width=0.32\textwidth]{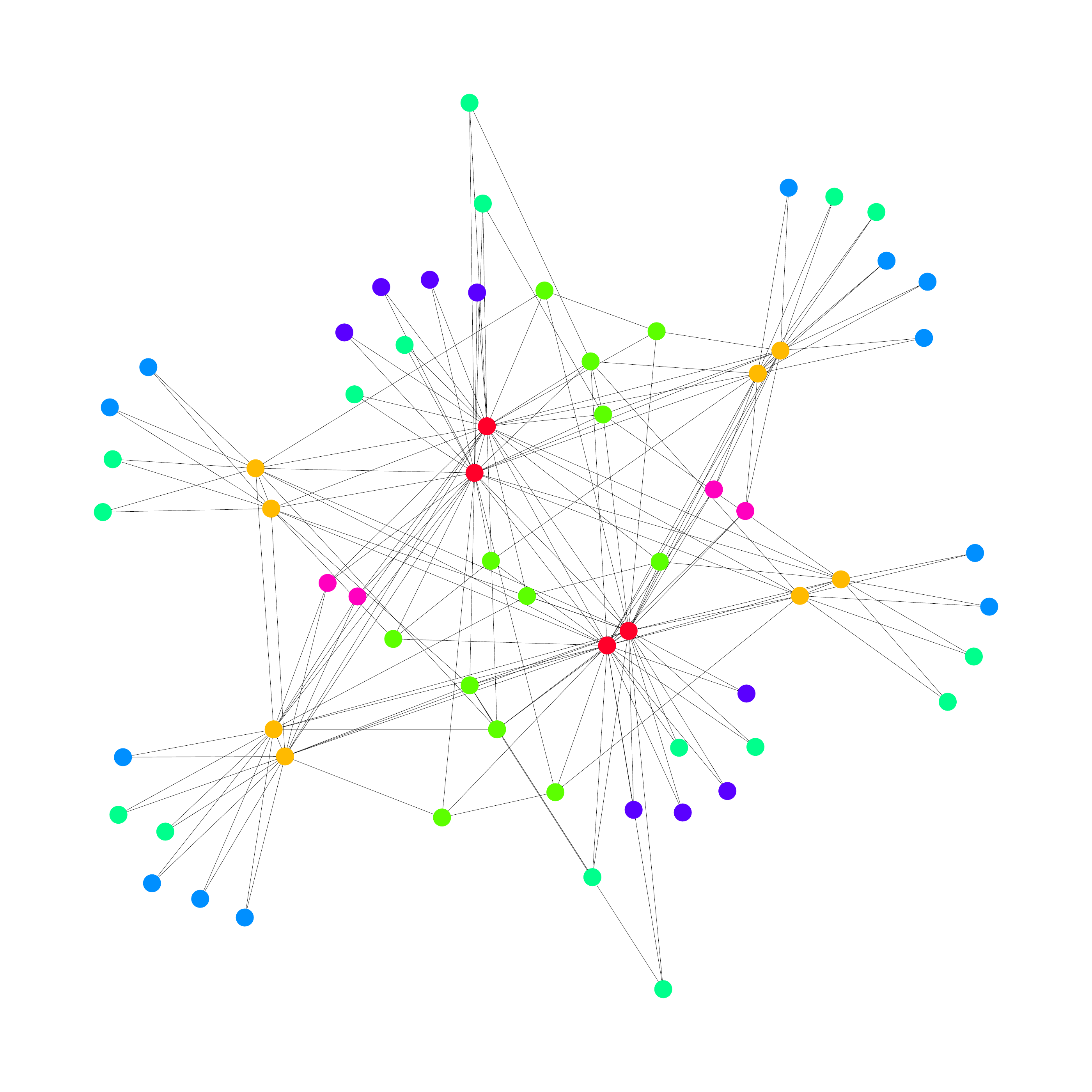}
    \includegraphics[width=0.32\textwidth]{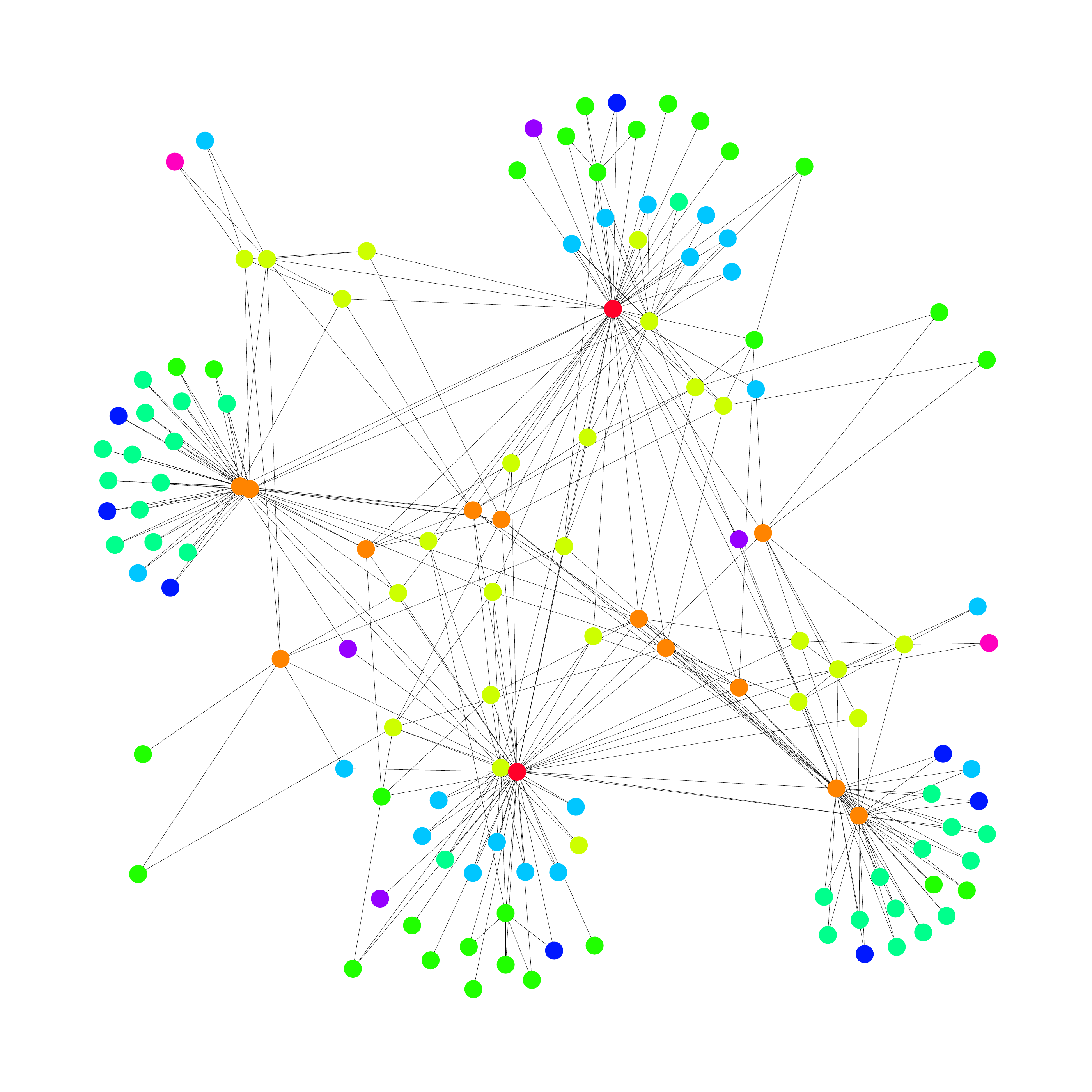}
    \includegraphics[width=0.32\textwidth]{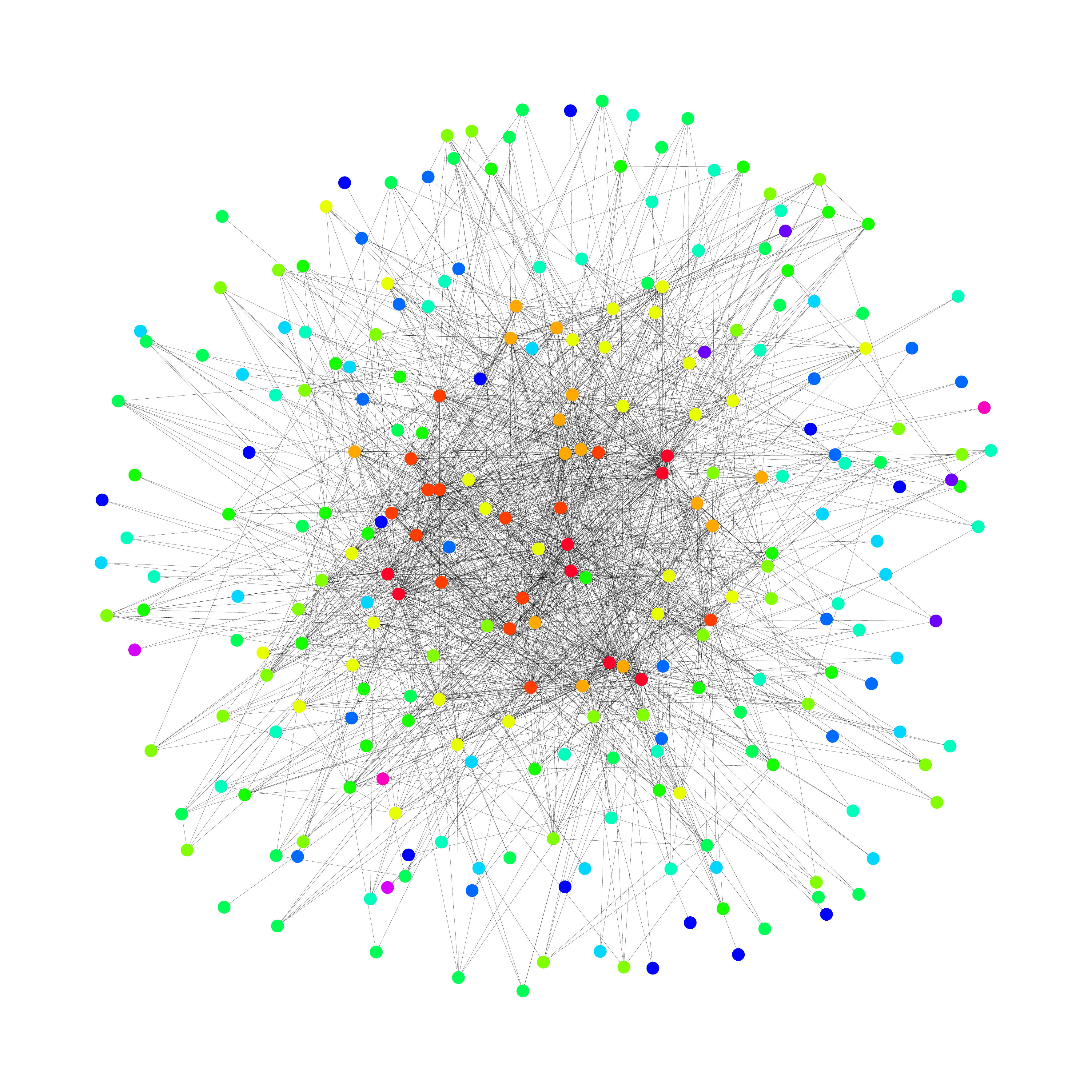}\\
    \includegraphics[width=0.32\textwidth]{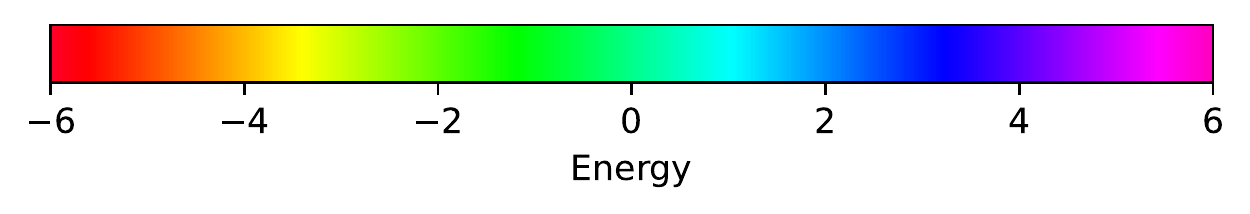}
    \includegraphics[width=0.32\textwidth]{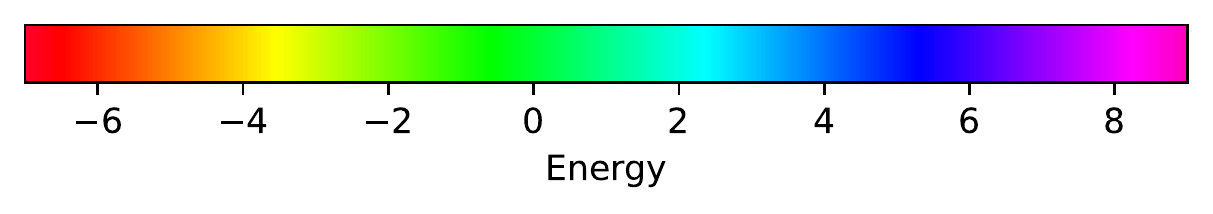}
    \includegraphics[width=0.32\textwidth]{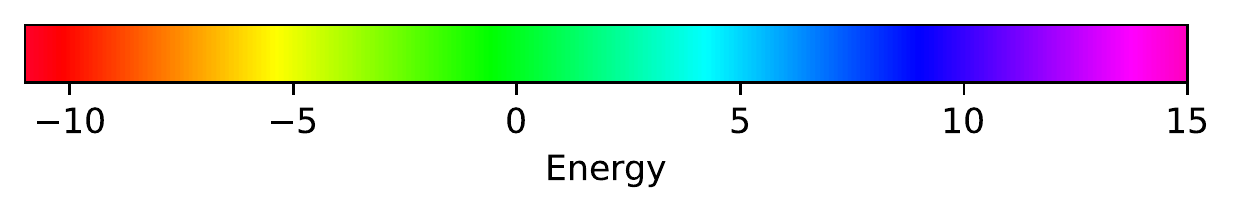}
    \caption{State transition networks for the $N_6$ Ising (left), $N_7$ (middle), and $N_8$ (right) when aggregating the h-gain response curves across all mappings for each ground state. Each node represents one classical state; the $N_6$ graph has $64$ nodes, the $N_7$ graph has $128$ nodes, and the $N_8$ graph has $256$ nodes. The nodes are colored corresponding to the energy (i.e., objective value) of that state evaluated on the Ising, the heatmaps for which are shown below each graph. The red node coloring represents the ground state solutions of the problem Ising and the purple nodes represent the highest energy states of the problem Ising. }
    \label{fig:state_transition_networks}
\end{figure}

\begin{figure}[t!]
    \centering
    \includegraphics[width=0.24\textwidth]{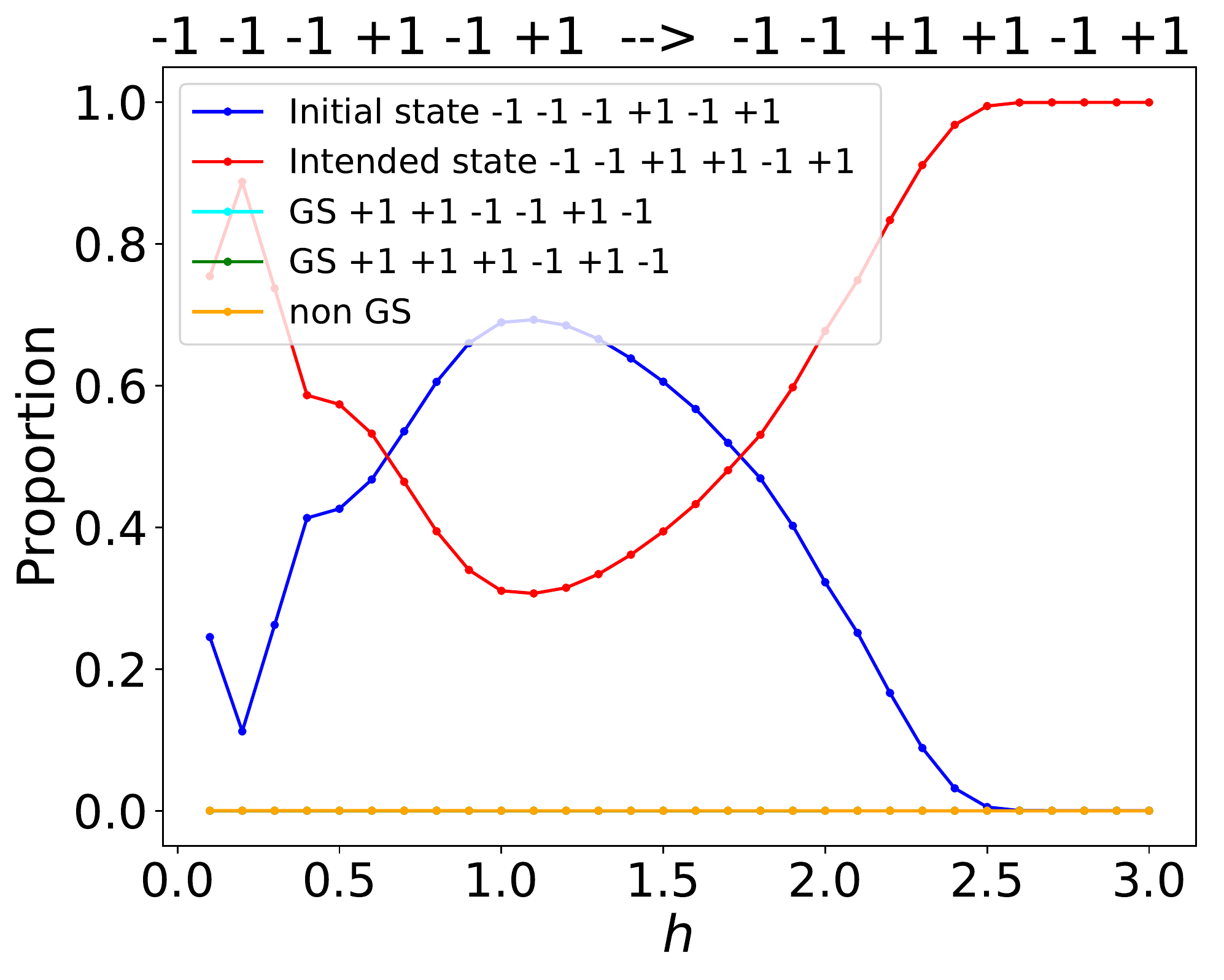}
    \includegraphics[width=0.24\textwidth]{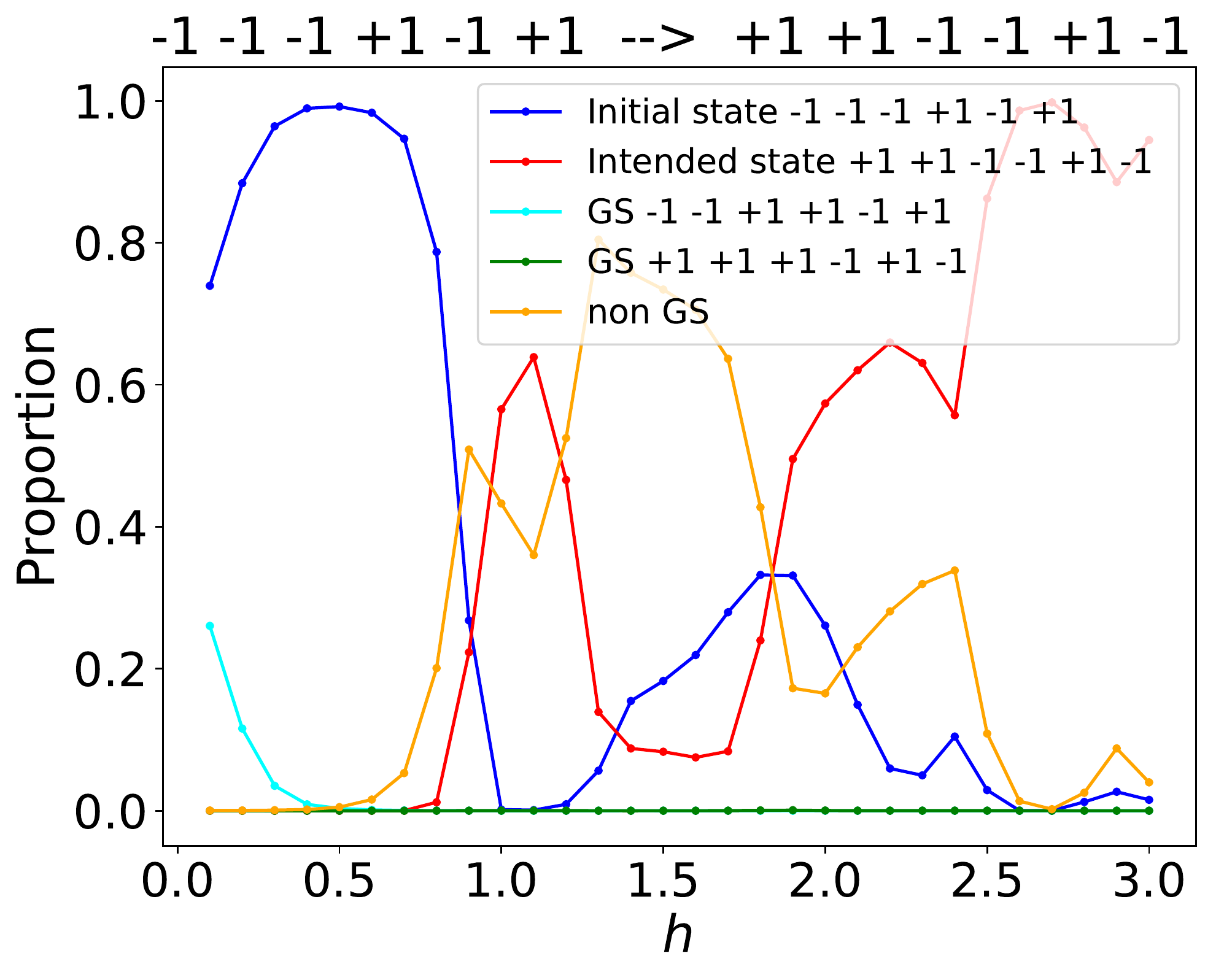}
    \includegraphics[width=0.24\textwidth]{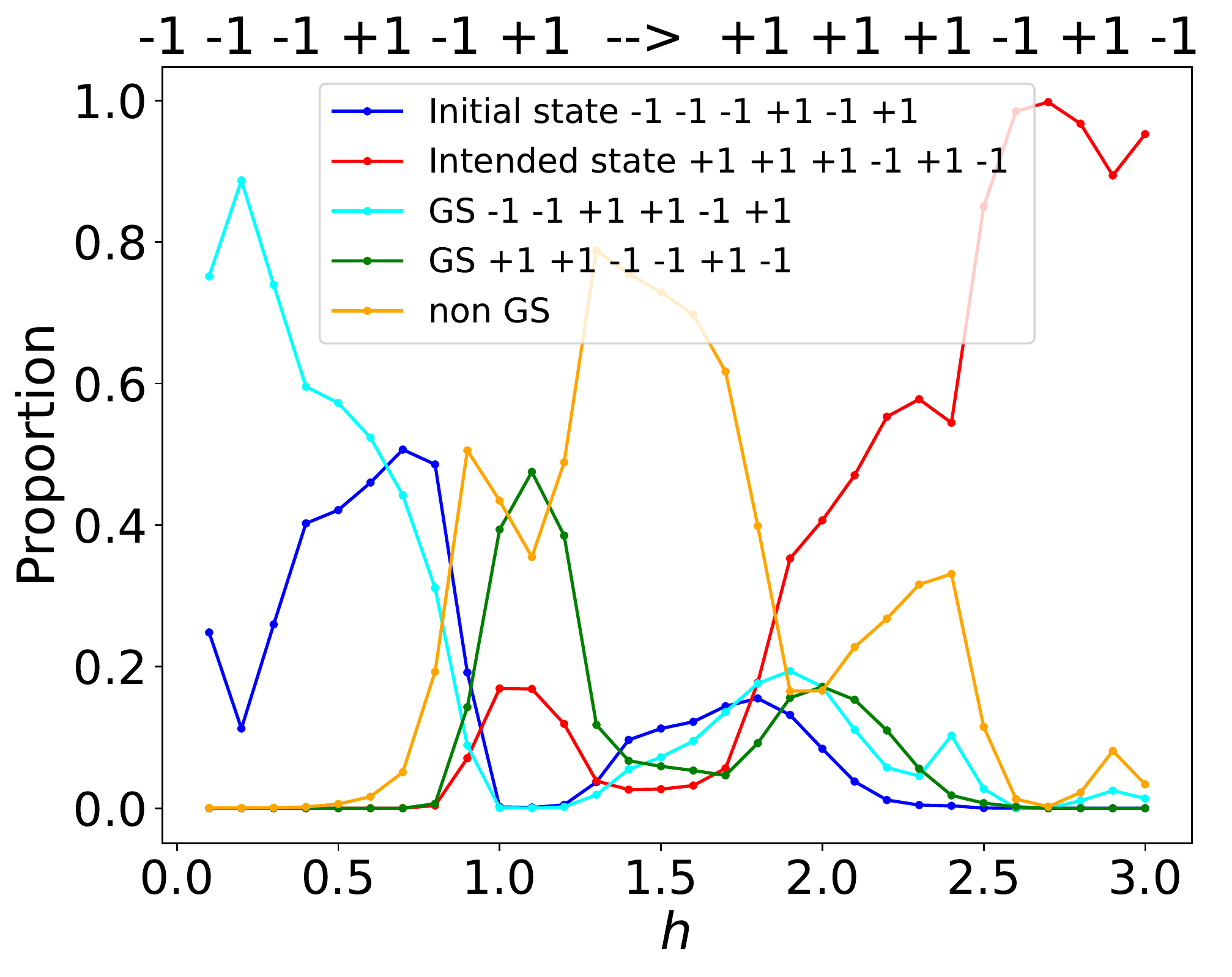}
    \includegraphics[width=0.24\textwidth]{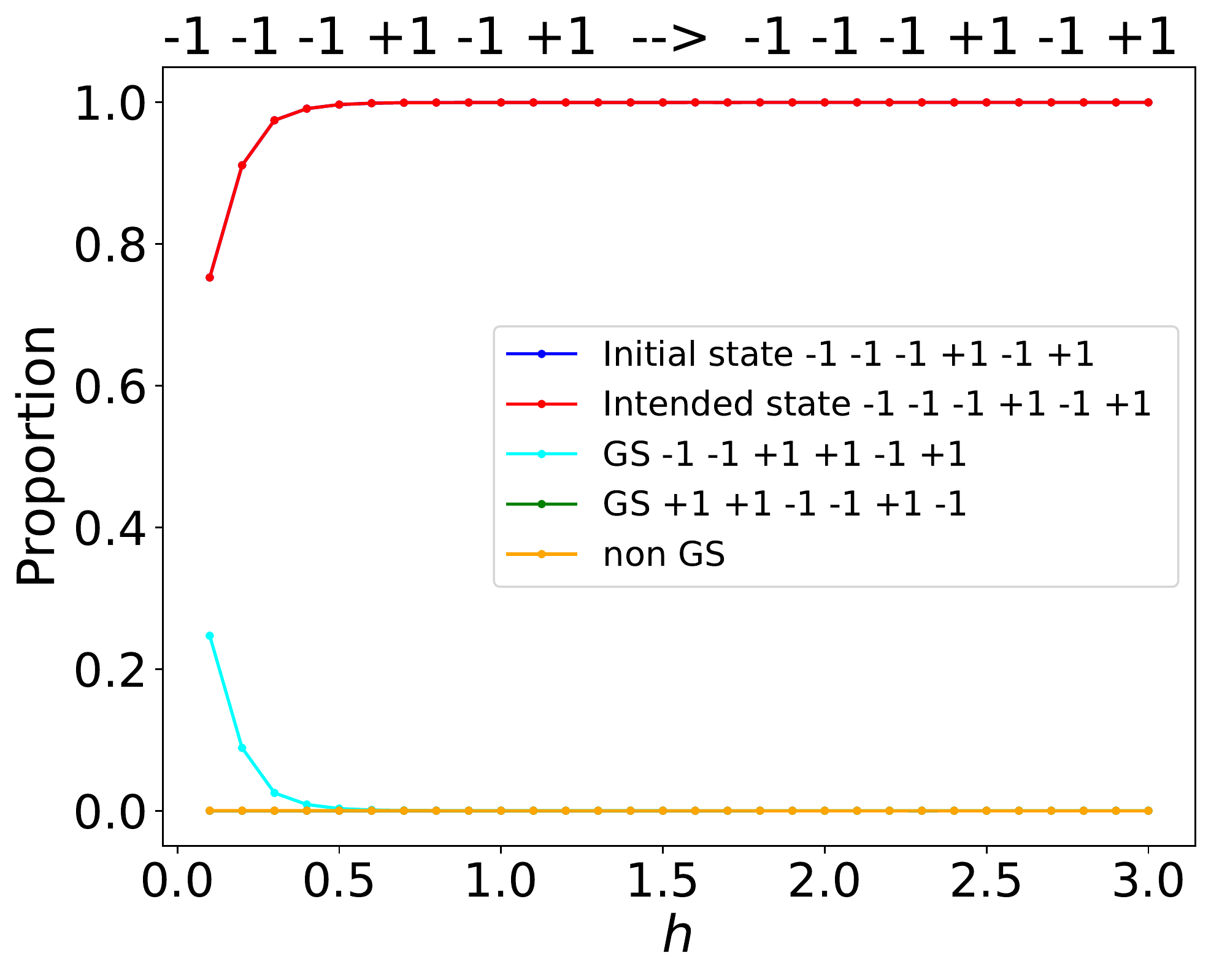}
    \caption{ A more detailed analysis of ground state to ground state h-gain response curves in the $N_6$ Ising. Here the initial state is the same for each of the three plots, and the proportion of the samples at each $h$ increment which were in this initial state is colored blue. The titles of each sub figure show the direction of the state transition. The red lines in the three plots denote the proportion of samples that were the intended ground state. The cyan and green lines denote the proportion of samples which were in the remaining two ground states. The orange colored lines denote all remaining samples which were not in either of the $4$ ground states. Note that, as mentioned in Section \ref{section:methods_problem_isings}, these problems have complementary ground state solutions; the middle-right hand sub-plot shows the state transition from a ground state to its complement. The reflexive self-mapping of ground state to itself is shown in the right hand plot, which shows little state change as a function of $h$.} 
    \label{fig:GS_to_GS}
\end{figure}

\subsection{\texorpdfstring{$N_8$} \qquad \space Ising h-gain response curves}
\label{section:results_n8}

Figure \ref{fig:h_strength_vs_GSP_n8_clustered_GS0} shows the h-gain response curves for all initial states along with the four groups of clustered response curves for one of the two ground states for the $N_8$ Ising results. Figure \ref{fig:HGain_susceptibility_to_groundstate_n8} shows the susceptibility metric across all $2^8$ initial states of the $N_8$ Ising when applying the \emph{h-gain state encoding} for each of the two ground states. Figure \ref{fig:summary_metrics_n8} shows scatterplots of the three metrics (hamming distance, energy, and $\delta$) outlined in Section \ref{section:methods_metrics} vs $\chi$ for all of the initial states. 

The top row of Figure \ref{fig:summary_metrics_n8} has noticeably different characteristics compared to the other hamming distance proportion plots from the $N_6$ and $N_7$ Ising (Figure \ref{fig:summary_metrics_n6} top row and Figure \ref{fig:summary_metrics_n7} top row). The plots for the $N_6$ and $N_7$ have roughy linear relationships between the hamming distance proportion and susceptibility. However, in the corresponding plot for the $N_8$ Ising (Figure \ref{fig:summary_metrics_n8} top row) we see a slightly different trend occurring; the shape of points appears to be have a concave shape facing towards the bottom left hand corner of the plot (and a convex shape facing towards the bottom right hand corner of the plot). This appears to be a result of most of the h-gain response curves having a lower success rate of managing to get the anneal into the intended ground state compared to the smaller $N_6$ and $N_7$ Isings. This could also be a result of the increased number of ground states for this Ising, which could be causing anneals to be trapped in non-intended ground states (this is investigated further in Section \ref{section:results_groundstate_to_groundstate}).

\begin{figure}[t!]
    \centering
    \includegraphics[width=0.32\textwidth]{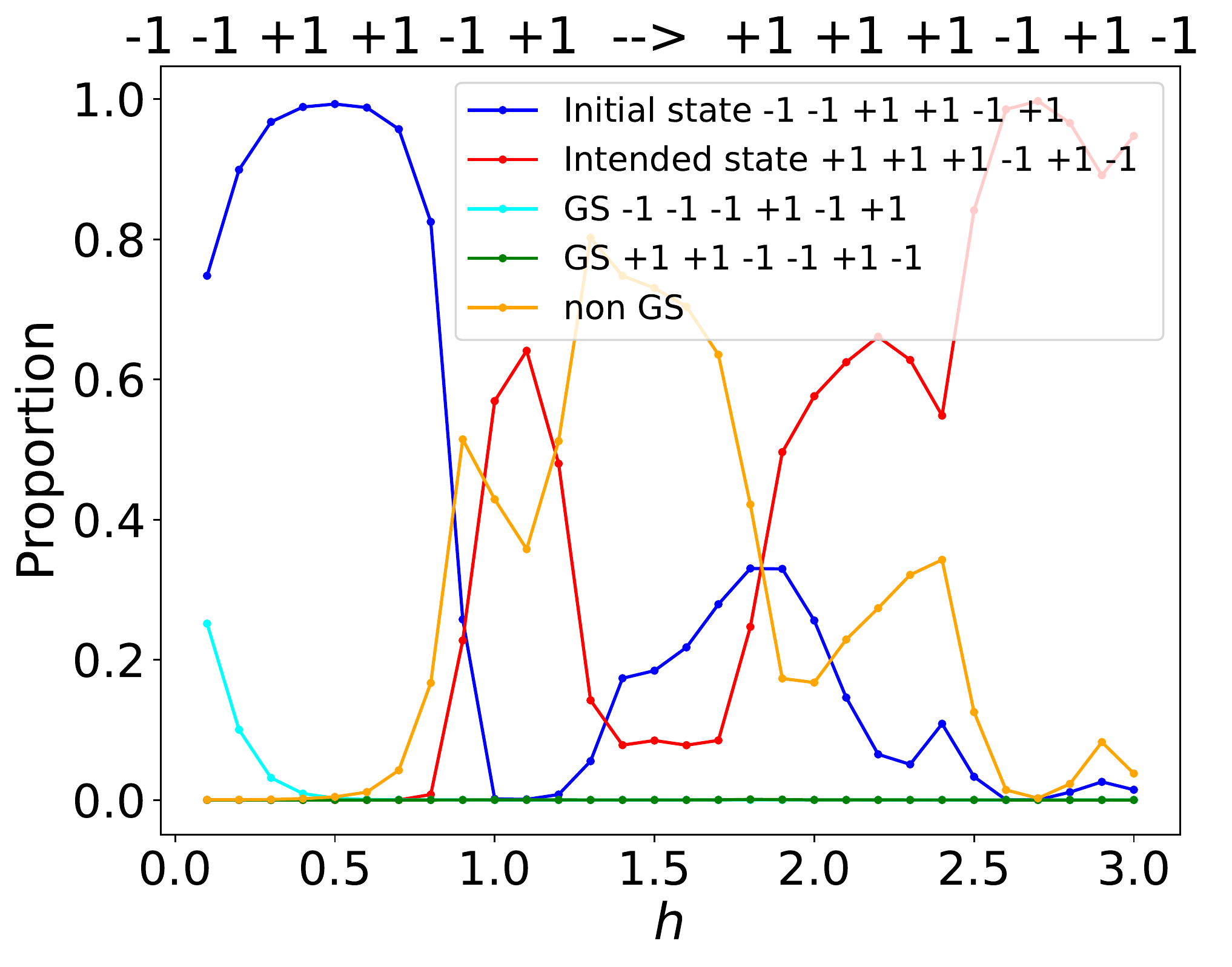}
    \includegraphics[width=0.32\textwidth]{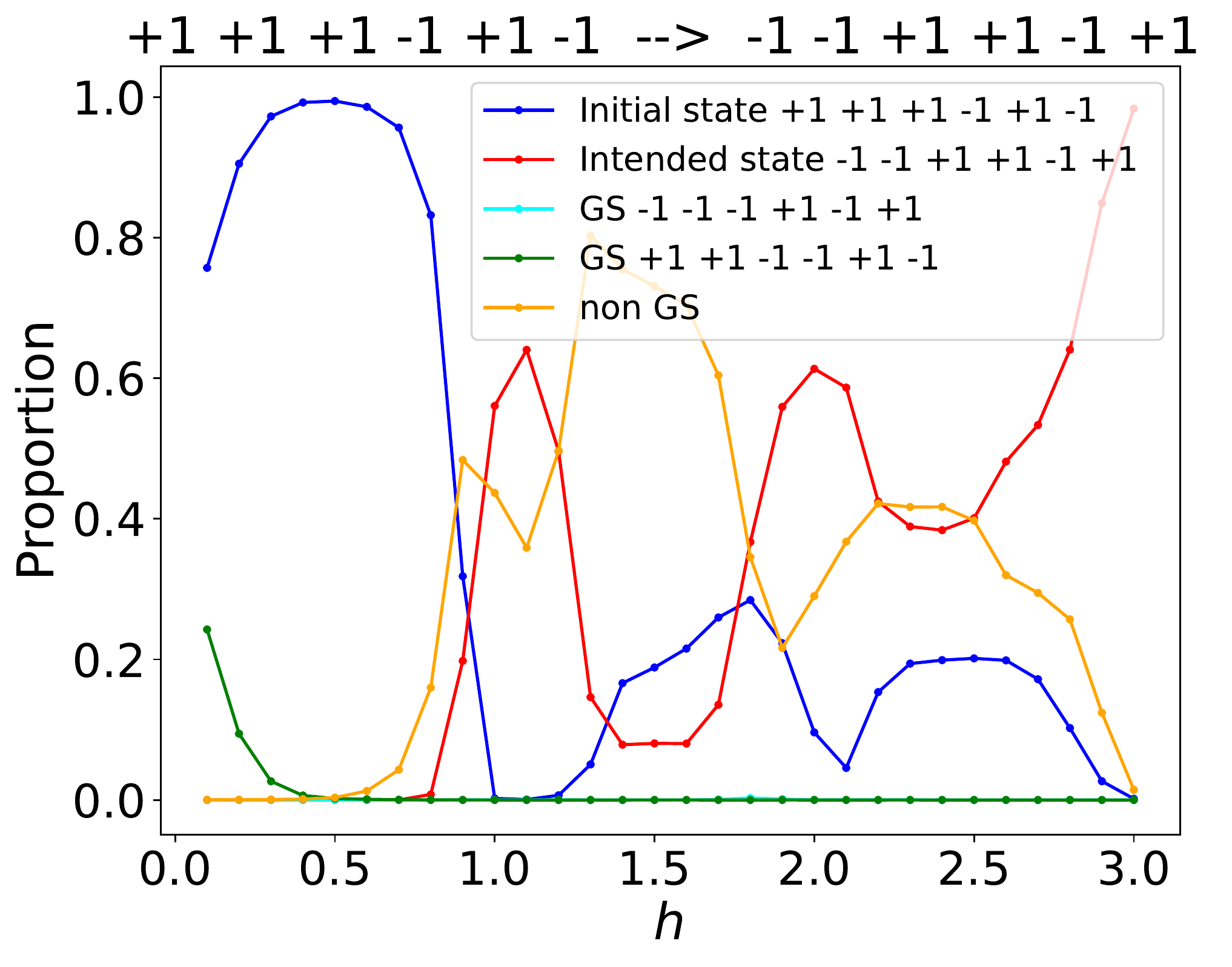}\\
    \includegraphics[width=0.32\textwidth]{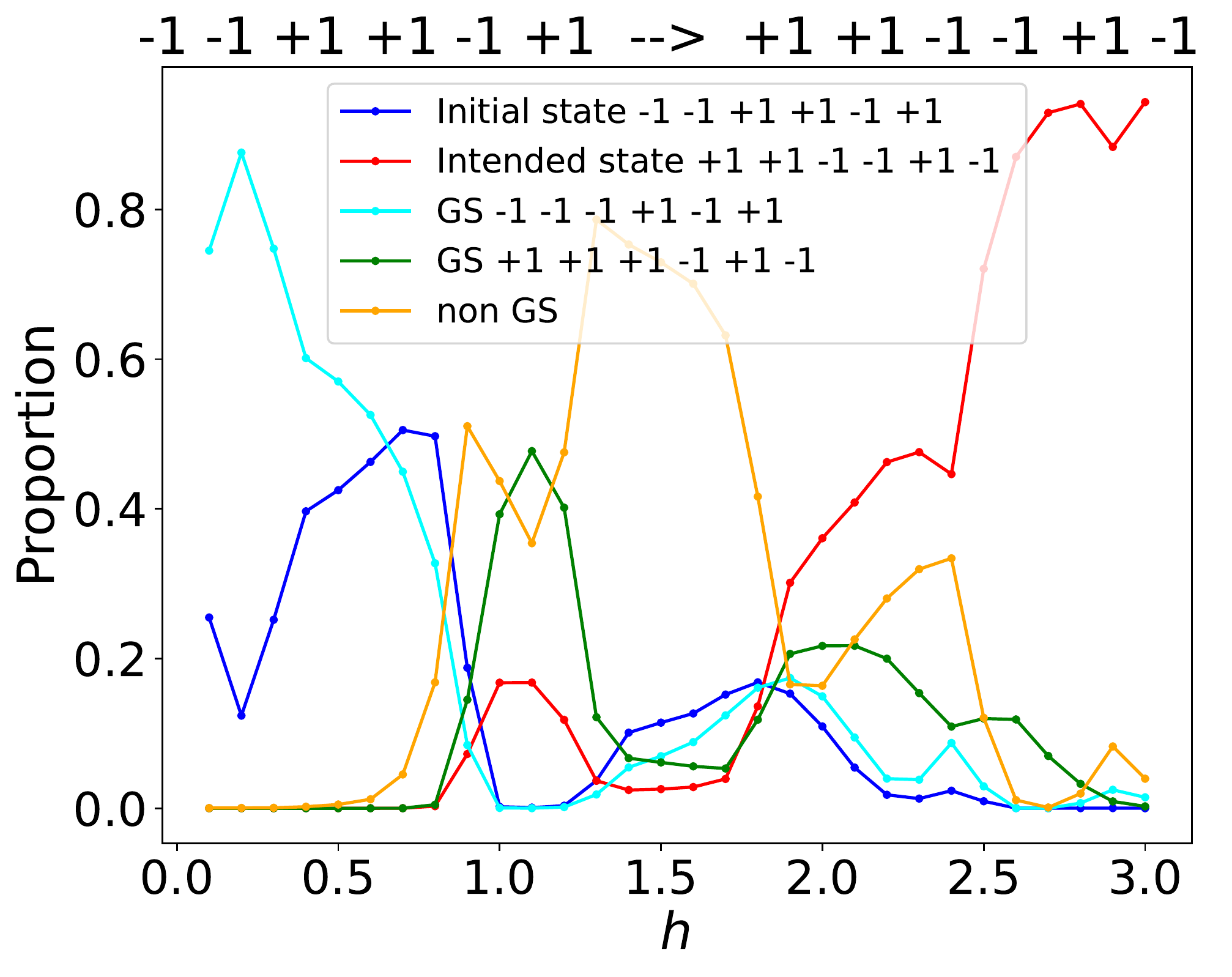}
    \includegraphics[width=0.32\textwidth]{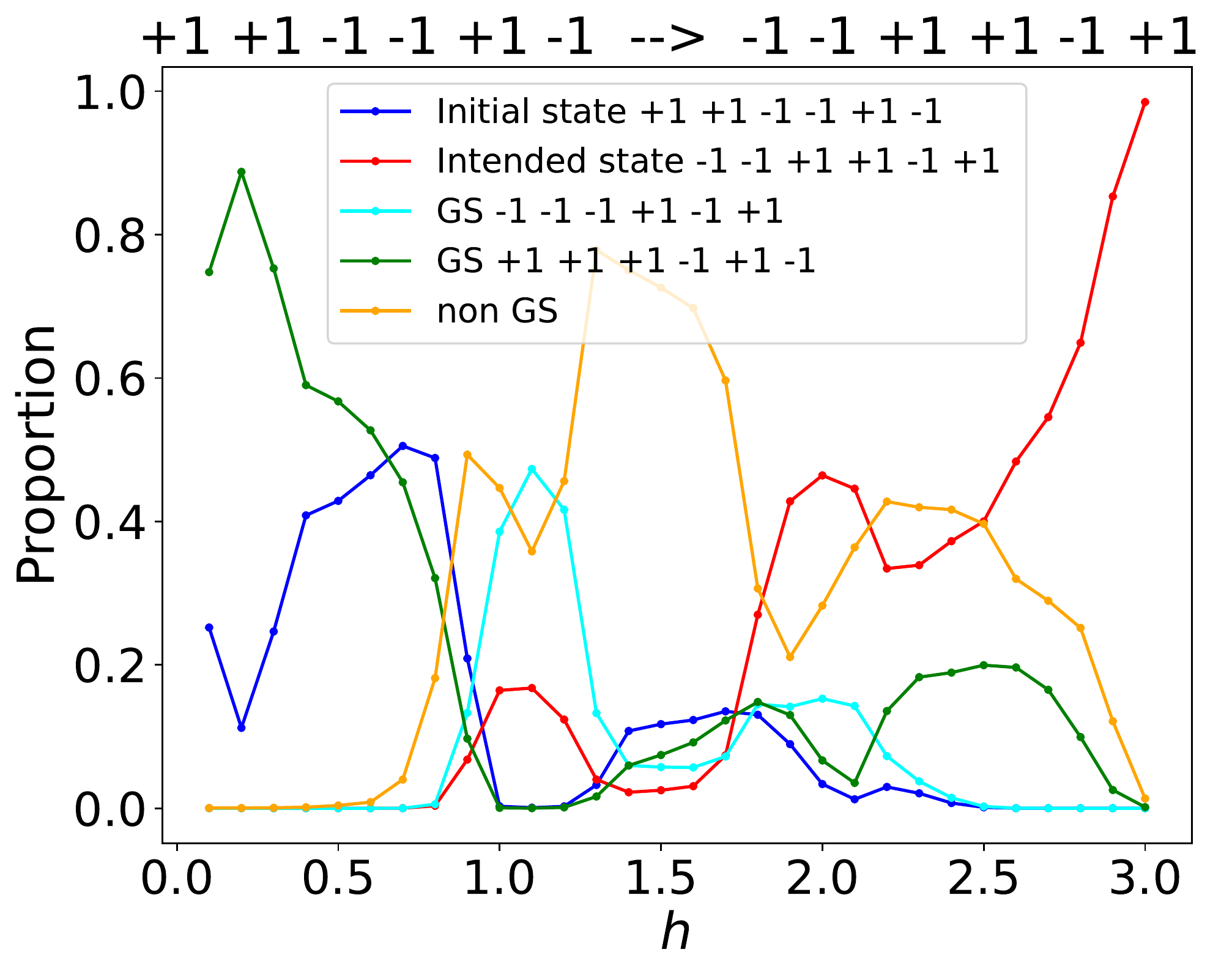}
    \caption{ Two examples of the symmetry of the ground-state to ground-state h-gain response curves in the $N_6$ Ising. The two rows are showing two different symmetric state mappings (e.g., $A \rightarrow B$ and then $B \rightarrow A$). The noticeable property of these plots is that the response curves have a clear symmetry, where when the two ground states are swapped a nearly identical response curve for all $4$ ground states emerges (the difference is simply which place each ground state takes in the plots). Notice as well that the states in the bottom row are complementary ground states, whereas the ground states in the top row are not complementary ground states. }
    \label{fig:GS_to_GS_symmetry}
\end{figure}

As with the other two test Isings, there were not any non-symmetric results across the different ground states - either all of the ground states behaved roughly the same for a given metric, or sets of ground states displayed clear symmetries.

\subsection{State transition networks}
\label{section:results_state_transition_networks}

Figure \ref{fig:state_transition_networks} shows graphical representations of the state transition networks for each of the three test Isings. These state transition networks contain several notable characteristics. First, the nodes which represent the ground states are the highest degree nodes in the graphs, which is expected since they are the intended end point of the state transitions. Second, it is not the case that all non ground state nodes have an edge directly connecting them to the ground state nodes. Such a direct edge connection would mean that the susceptibility of moving the anneal into the intended ground state did not pass through other intermediate states. Instead, we see that for some states the path to the ground states could contain several other nodes. Third, there are some states which are consistent intermediate states for clusters of nodes; in the left hand most graph these nodes are the $8$ orange colored nodes which are persistent intermediate states for small groups of classical states. Fourth, there are clear symmetries in these graphs with respect to the local communities of nodes that are connected similarly, or identically, in other parts of the graph. 

The $N_8$ state transition graph in Figure \ref{fig:state_transition_networks} has fewer discernible trends compared to the state transition graphs for the $N_6$ and $N_7$ Isings. This is largely due to the increase in the number of nodes and edges present in the graph, which is both due to the increase in the number of classical states ($256$ in total) as well as the increase in the number of ground states ($8$). What is clear from the state transition graph in Figure \ref{fig:state_transition_networks} is that there are many intermediate states during the transition from one state to a ground state. 

\begin{figure}[t!]
    \centering
    \includegraphics[width=0.29\textwidth]{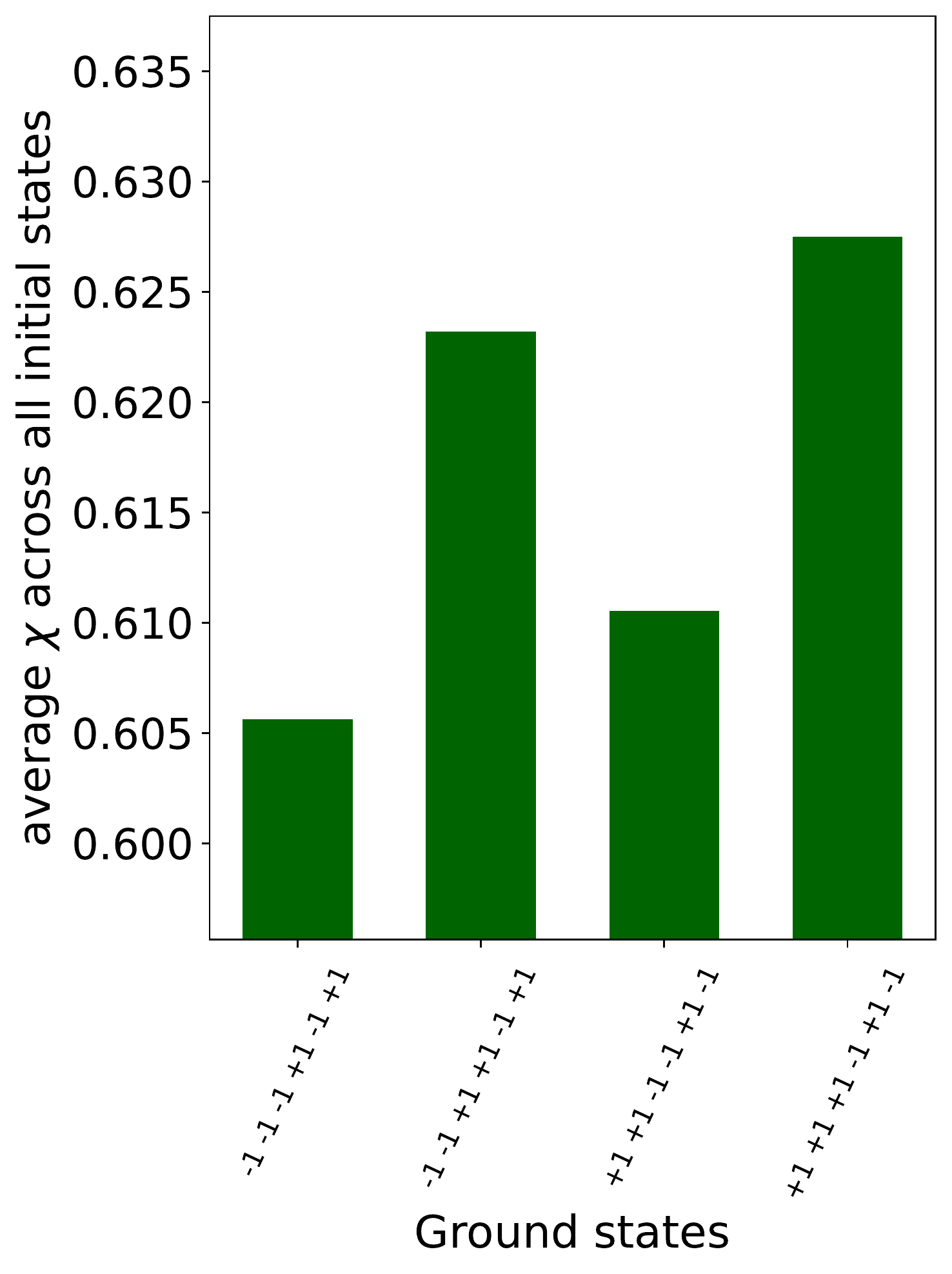}
    \includegraphics[width=0.29\textwidth]{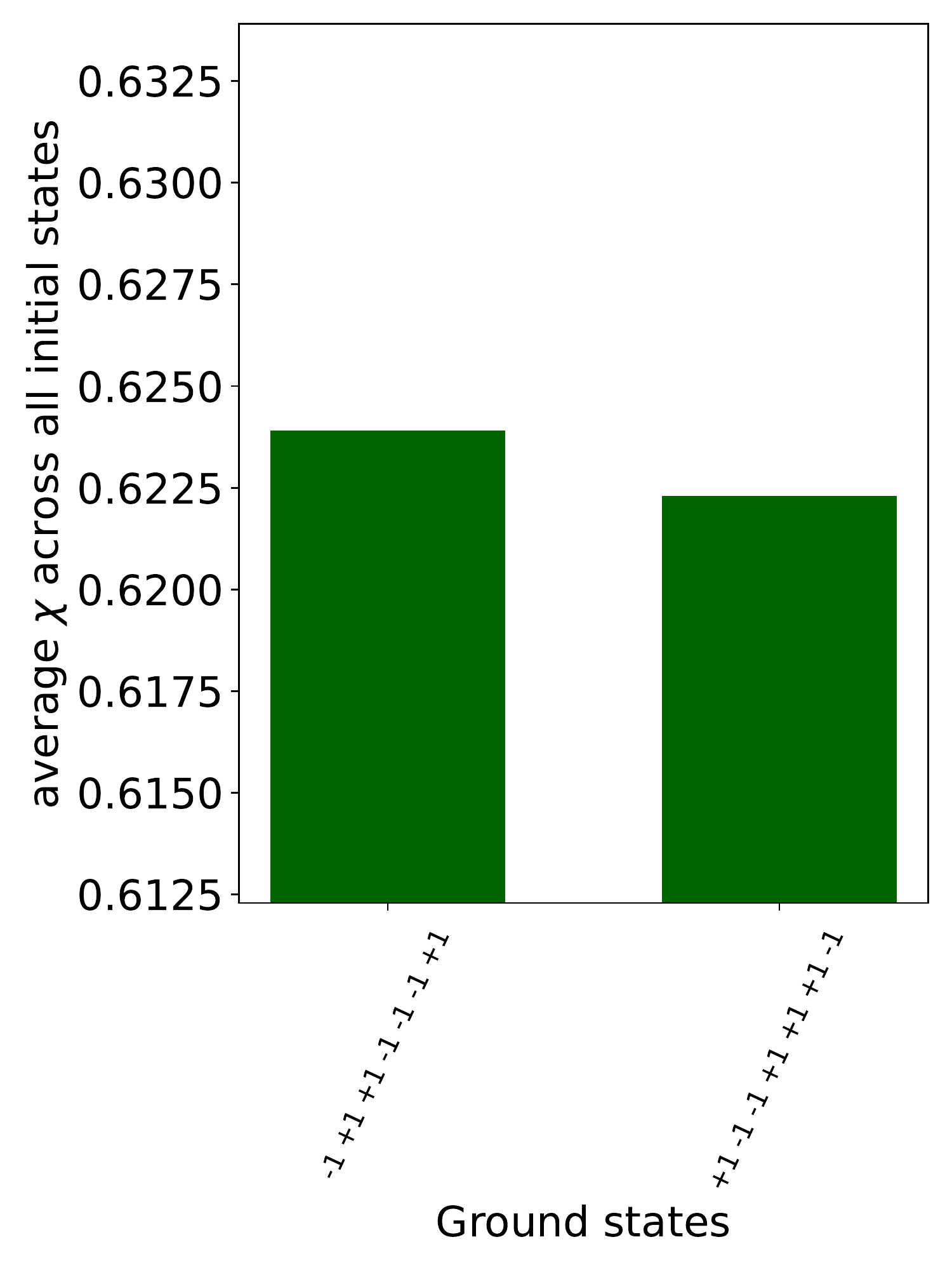}
    \includegraphics[width=0.29\textwidth]{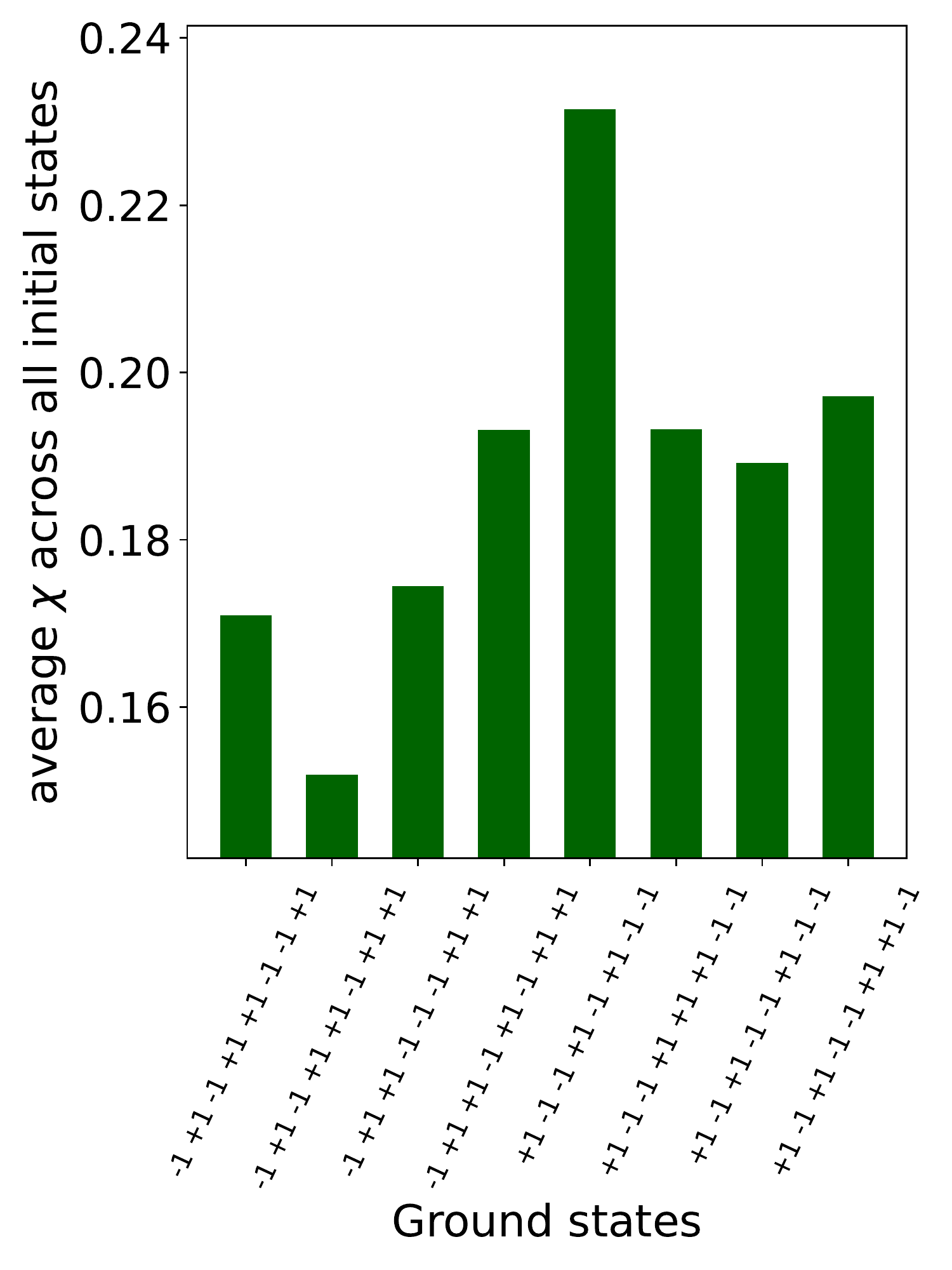}
    \caption{ Distribution of the average susceptibility (y-axis) across all initial states when transitioning the annealer into different ground states (x-axis) for the $N_6$ (left), $N_7$ (middle), $N_8$ (right) Isings. }
    \label{fig:fair_sampling_susceptibility_averages}
\end{figure}

\begin{figure}[t!]
    \centering
    \includegraphics[width=0.29\textwidth]{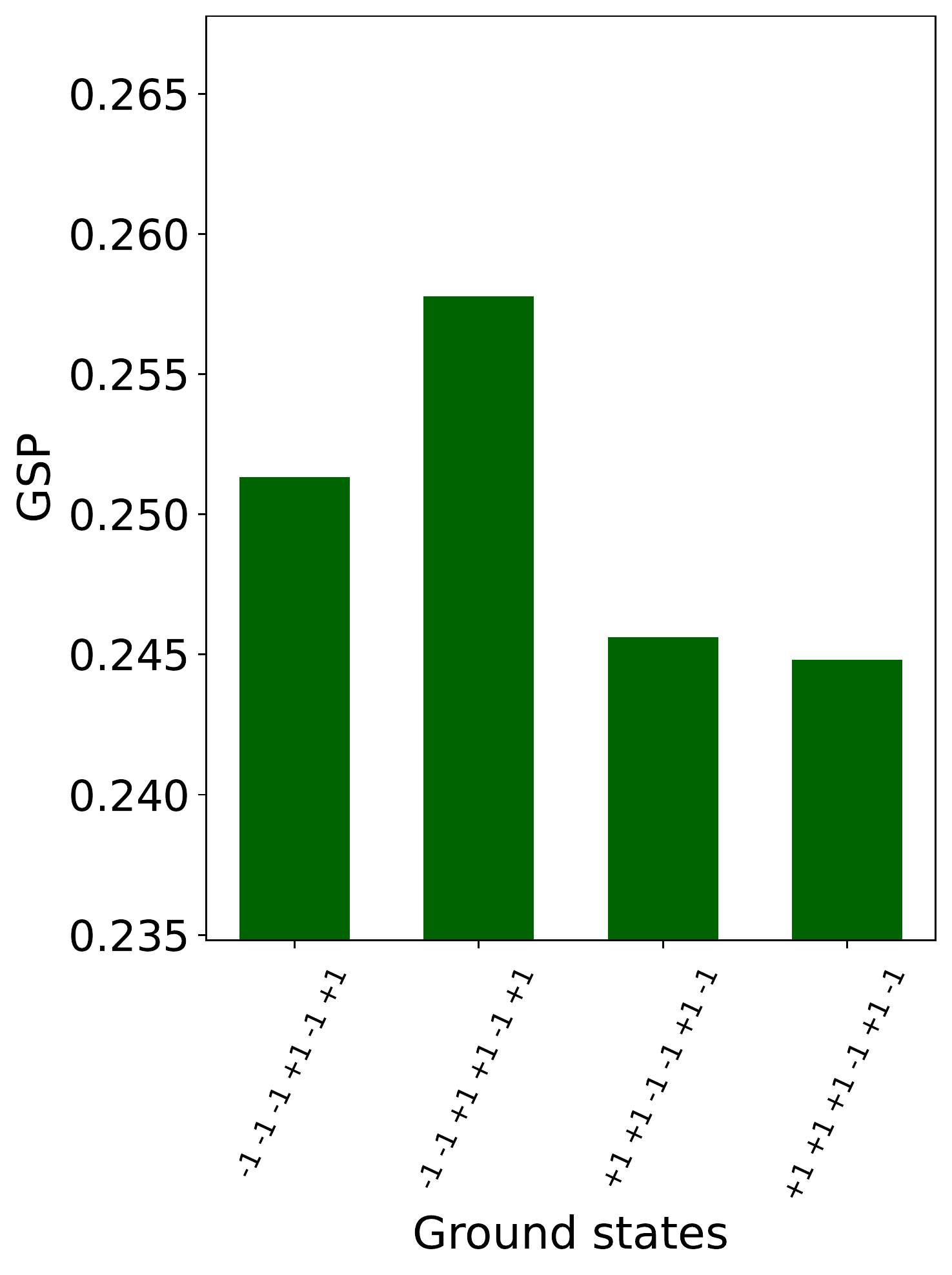}
    \includegraphics[width=0.29\textwidth]{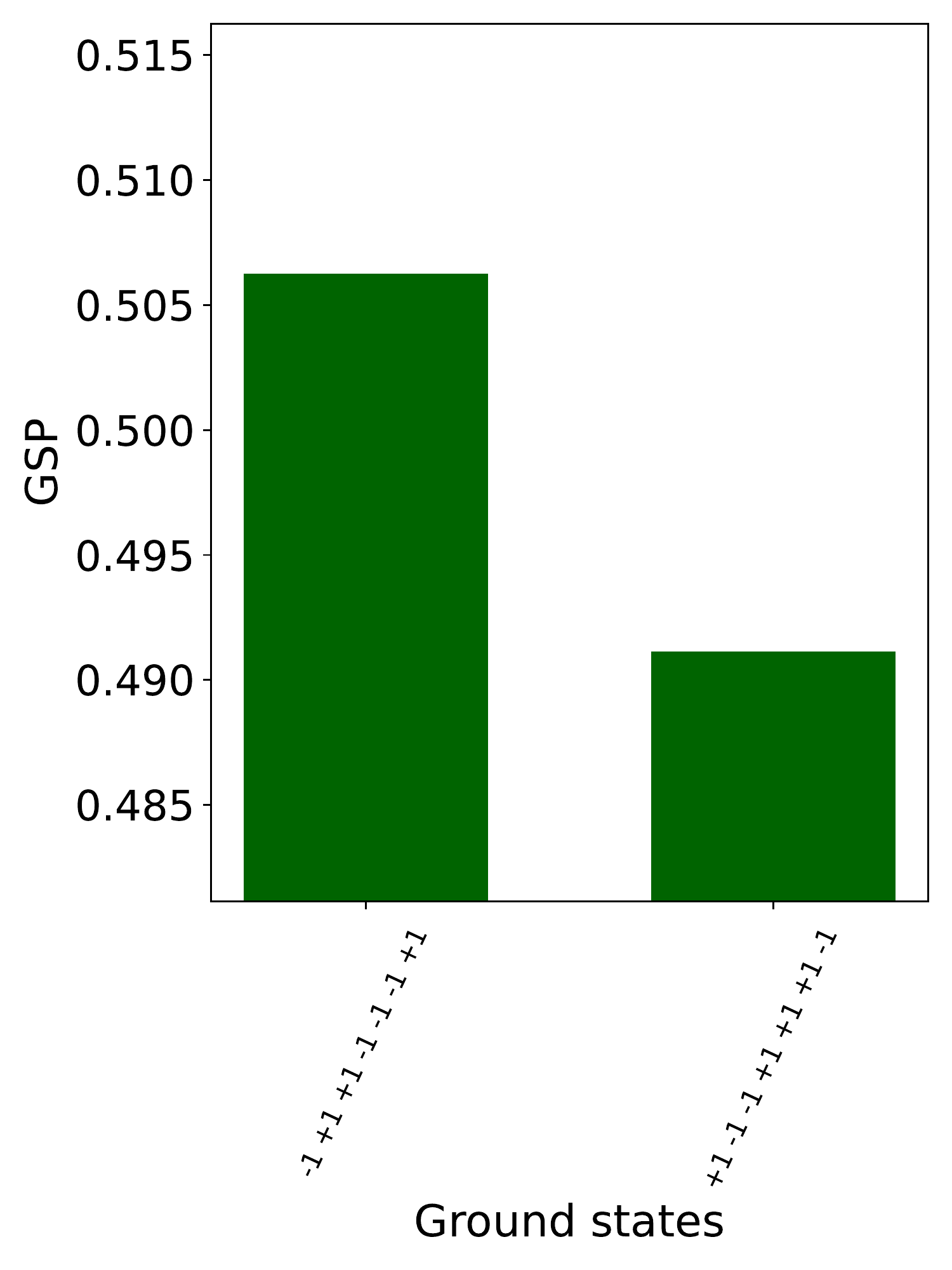}
    \includegraphics[width=0.29\textwidth]{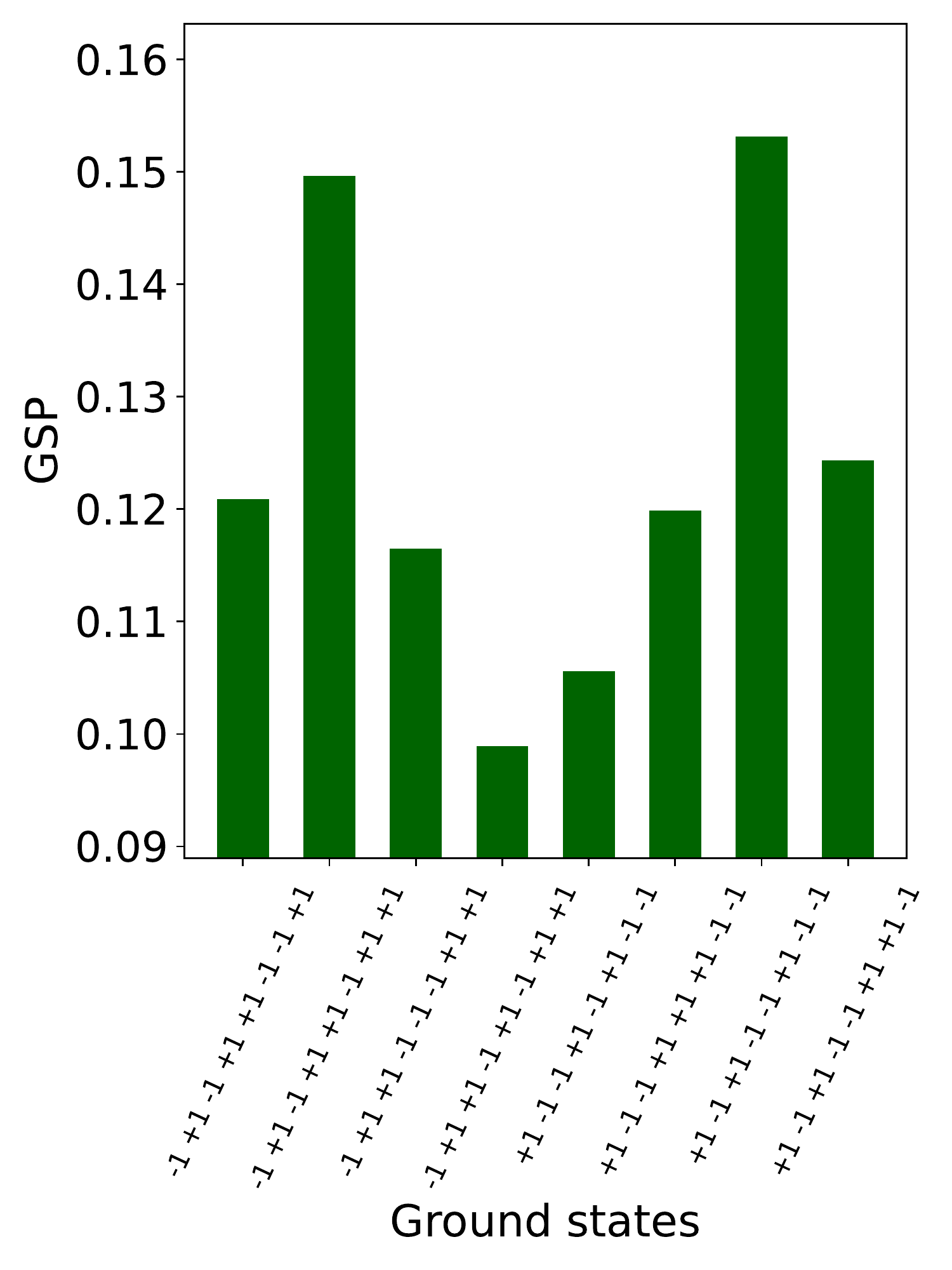}\\
    \caption{ Ground state (x-axis) vs ground state proportion distribution (y-axis) for the $N_6$ (left), $N_7$ (middle), $N_8$ (right) Isings having been executed on using forward annealing with no schedule modification and an annealing time of $100$ microseconds. }
    \label{fig:fair_sampling_forward_QA}
\end{figure}

\subsection{Ground state to ground state transitions}
\label{section:results_groundstate_to_groundstate}
Here we specifically highlight some of the state transitions which are of particular interest - those being ground state to ground state transitions and what intermediate states are found during the transition from one ground state to another. These response curves provide some intuition as to how quantum annealing works, at least in the reverse annealing mode. Specifically, moving from one ground state to another ground state does not require zero amount of h-gain strength. Indeed, as seen in Section \ref{section:results_n6} the complementary ground states in the $N_6$ Ising had \emph{maximal} susceptibility when switching from the initial ground state to its complement. 

Figure \ref{fig:GS_to_GS} shows a detailed analysis of a subset ground state to ground state transition curves for the $N_6$ Ising in terms of the proportion of samples which were each of the $4$ ground states as well as higher energy states. The h-gain response curves (Figures \ref{fig:h_strength_vs_GSP_n6_clustered_GS0}, \ref{fig:h_strength_vs_GSP_n6_clustered_GS1}, \ref{fig:h_strength_vs_GSP_n7_clustered_GS0}, \ref{fig:h_strength_vs_GSP_n8_clustered_GS0}) and the state transition networks in Figure \ref{fig:state_transition_networks} have all shown that during this state transition induced by the increasing $h$ strength, the samples are not immediately flipped from the initial state into the intended ground state. Instead, the variable states are each flipped at different rates in response to the increase in applied $h$; the $\delta$ metric is a clear example where a single variable can be more susceptible to flipping its state compared to other variables due to properties of the problem Ising. Figure \ref{fig:GS_to_GS} shows for each increase of $h$ what other states the samples end up in, which clearly shows that transitioning from one state to another flips the variable states at different rates causing the anneal to end up in other intermediate states. 

Because of the h-gain initial state encoding scheme \cite{pelofske2020advanced}, the optimal solution of the Ising (the problem Ising is a union of the problem Ising quadratic terms and the complement of the intended ground state) is simply the intended ground state. As we increase $h$ the objective function evaluation of that optimal solution decreases. Therefore, it is a notable observation that in Figures \ref{fig:GS_to_GS} and \ref{fig:GS_to_GS_symmetry} we do not see a monotonic increase in the proportion of the intended ground state - indeed it seems to fluctuate and at roughly $h=1.5$ there a local minima across all of the plots (at least for the non reflexive ground state mappings). This shows, as observed in the other data analysis especially in Section \ref{section:results_state_transition_networks}, there are paths being traversed in the search space of this reverse annealing procedure which are not always in an optimal solution.

\subsection{Fair sampling analysis}
\label{section:results_fair_sampling}
Given that quantum annealing in the transverse field Ising model is known to not sample sample ground states fairly, it is natural to ask the question of whether these state mapping enumeration results show a bias for a particular ground state or set of ground states. In particular, for a set of state transition mappings from all initial states to a ground state, the question is if the average susceptibility across all of those initial states is very different when comparing the different ground state mappings. Figure \ref{fig:fair_sampling_susceptibility_averages} shows this comparison across the three test Isings. Interestingly, the average susceptibility distributions for the $N_6$ and $N_8$ Ising are not uniform across the ground states, although in the $N_7$ Ising case the two ground states appear to have very close average susceptibility. 

Another interesting question could be if the non-uniform distribution in Figure \ref{fig:fair_sampling_susceptibility_averages} is also present when applying forward annealing to the test Isings. To this end, a small set of experiments with forward annealing was performed on the three test Isings. The device parameters were $10,000$ anneals, and $100$ microseconds of annealing time for consistency with the state mapping experiments, readout\_thermalization of $0$ microseconds, programming\_thermalization of $0$ microseconds. The same device, D-Wave Advantage\_system4.1, was used for these experiments. The same parallel disjoint embeddings outlined in Section \ref{section:methods_problem_isings} were utilized. Figure \ref{fig:fair_sampling_forward_QA} shows the ground state proportion (GSP) from these forward annealing experiments across the three test Isings'. The forward annealing GSP results clearly show a consistent bias for some ground states which is in agreement with previous D-Wave quantum annealing fair sampling experiments \cite{pelofske2021sampling, mandra2017exponentially, konz2019uncertain}. In particular for the $N_8$ Ising, although the differences are less significant especially for the $N_7$ Ising. Note that the sum of the ground state proportions in each plot is always equal to $1$, as opposed to Figure \ref{fig:fair_sampling_susceptibility_averages} where the y-axis is simply average $\chi$. Nevertheless, these plots can be compared to see if there are consistencies; for example if the states which are broadly more difficult to transition into during the state mapping procedure are also sampled with a lower success proportion during forward annealing. Such a trend does not appear in the data - even though both the distributions of forward annealing ground state proportions and average susceptibility are non uniform, there does not appear to be consistency on the distributions. Susceptibility measurements using \emph{h-gain state encoding} for forward annealing might provide more insight into the behavior of unfair sampling in quantum annealers. 

\subsection{Reverse annealing only}
\label{section:results_RA_only}

\begin{figure}[h!]
    \centering
    \includegraphics[width=0.49\textwidth]{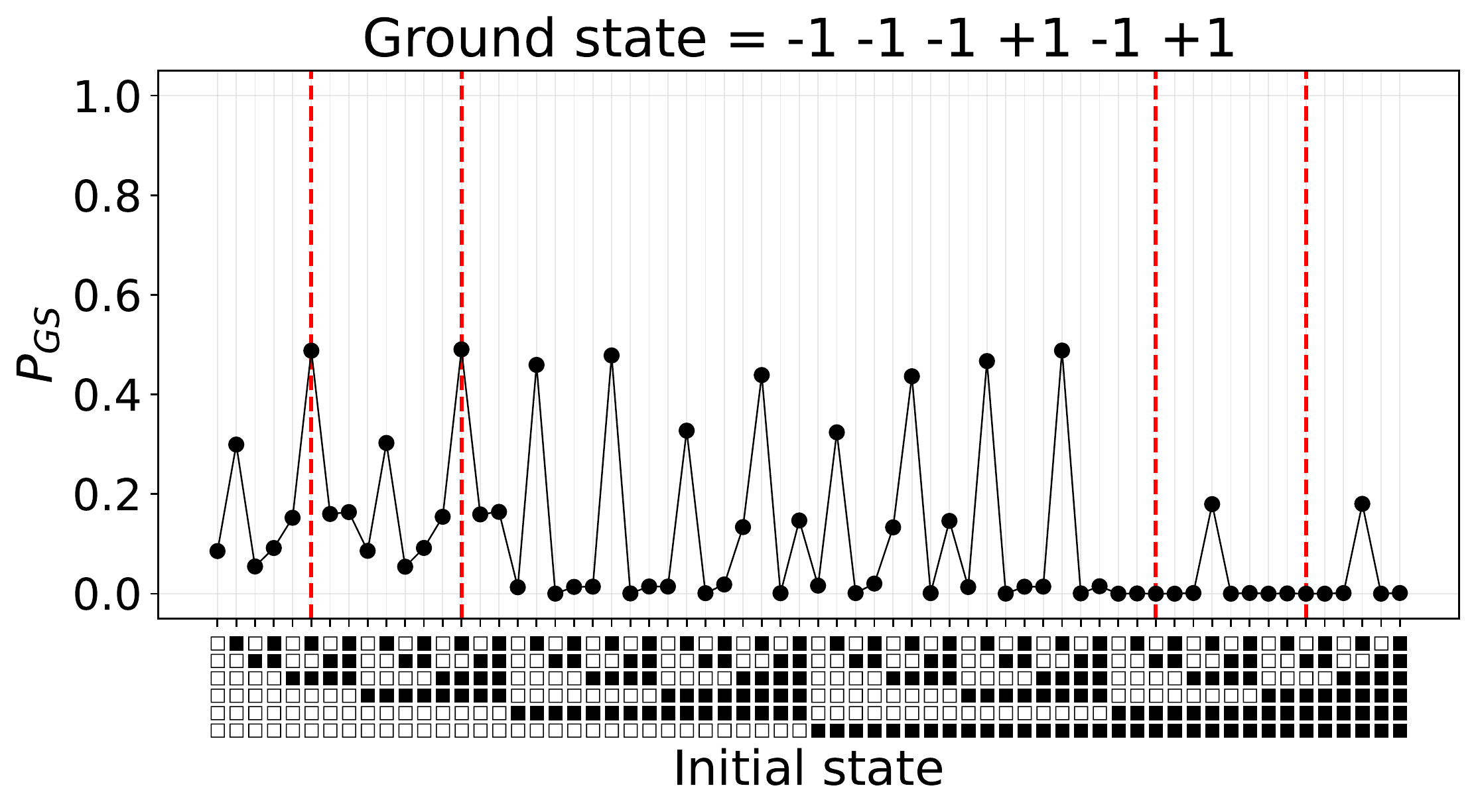}
    \includegraphics[width=0.49\textwidth]{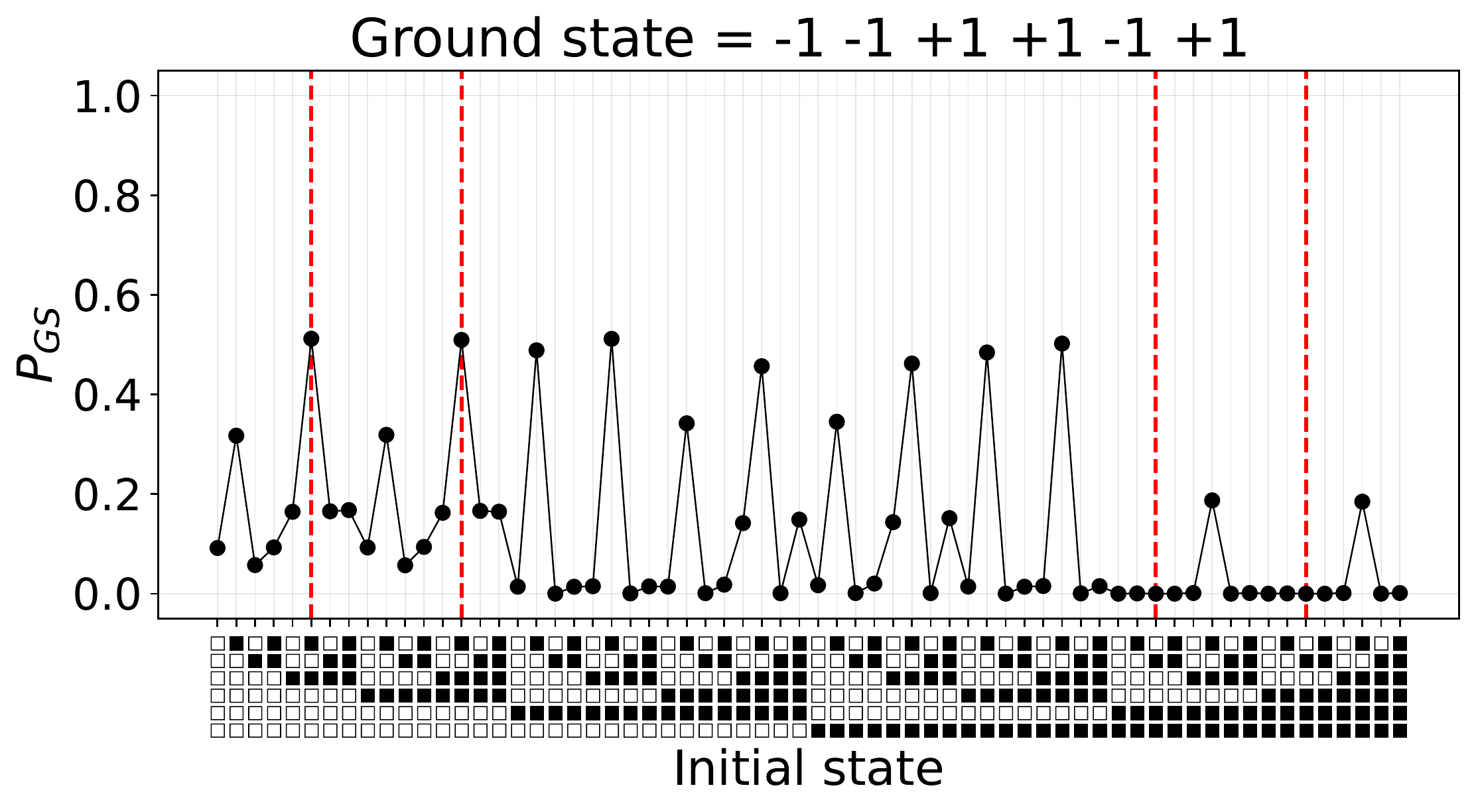}\\
    \includegraphics[width=0.49\textwidth]{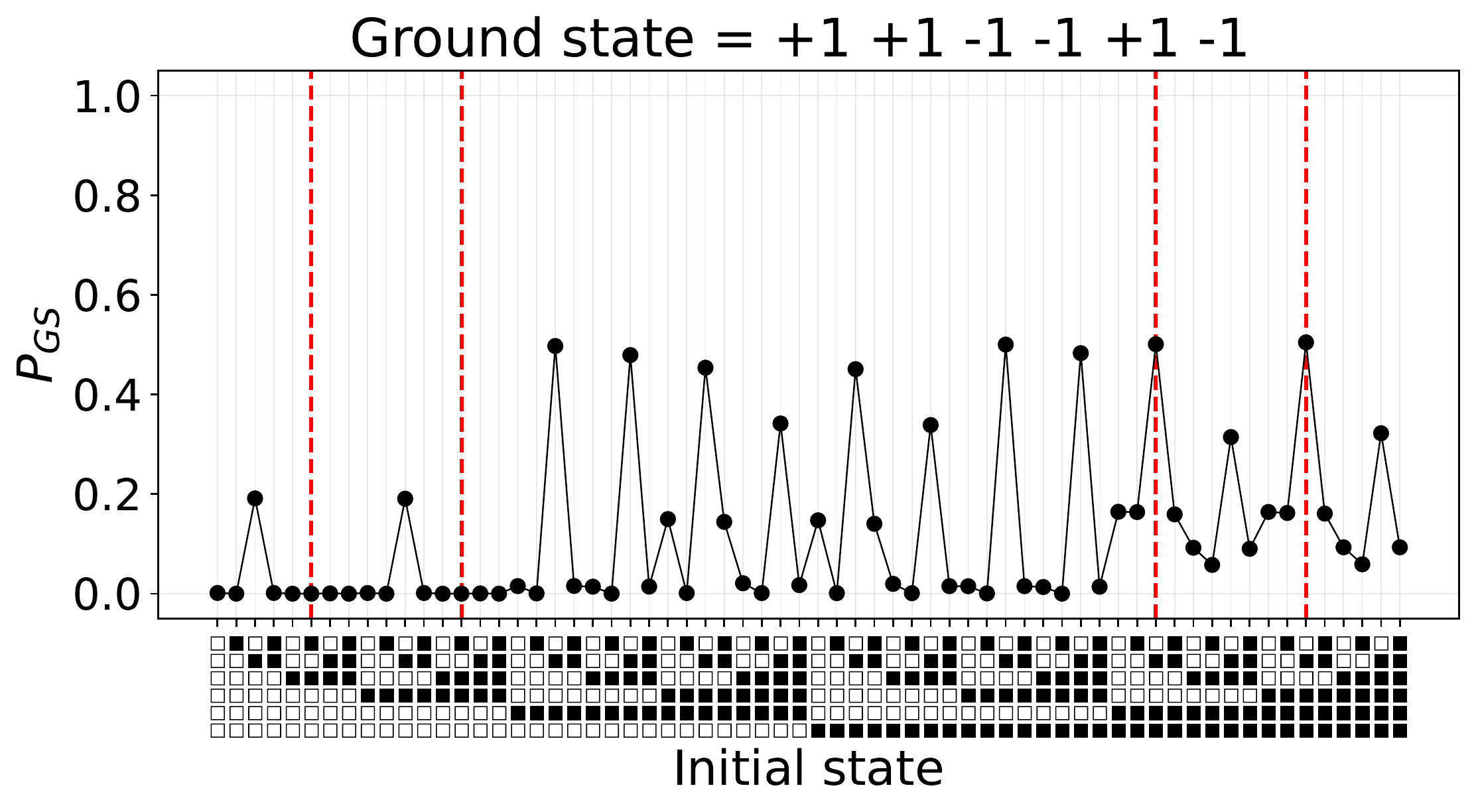}
    \includegraphics[width=0.49\textwidth]{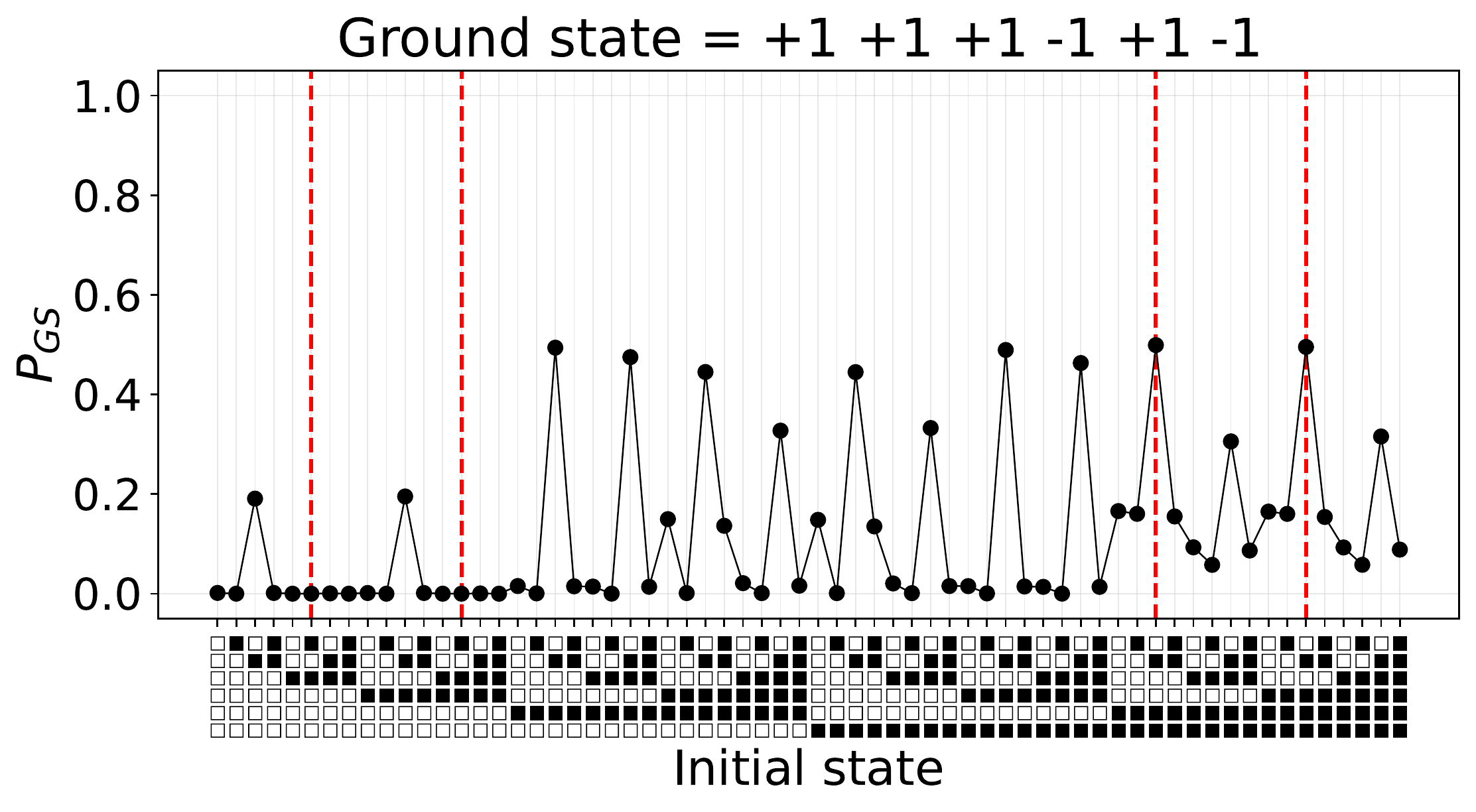}
    \caption{Ground state success probability for reverse annealing only, applied on the $N_6$ Ising. The x-axis encodes the RA initial states as vectors of vertical blocks where $\blacksquare$ denotes a variable state of $+1$ and $\square$ denotes a variable state of $-1$. The initial state vectors are read from bottom to top where the bottom is the first index which corresponds to variable $0$ in the problem Ising. The initial states which are also other ground states are marked with dashed red vertical lines. There are clearly some initial states which cause the reverse annealing $P_{GS}$ to be much higher compared to other initial states. }
    \label{fig:RA_only_N6}
\end{figure}

\begin{figure}[h!]
    \centering
    \includegraphics[width=0.49\textwidth]{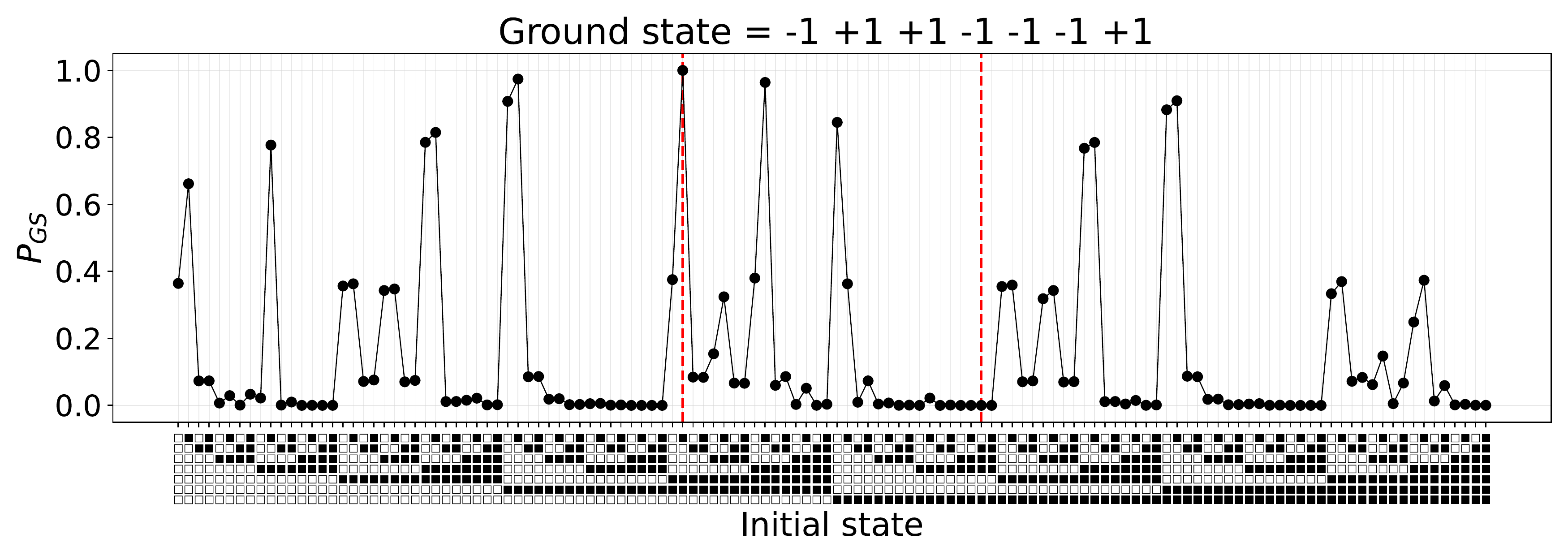}
    \includegraphics[width=0.49\textwidth]{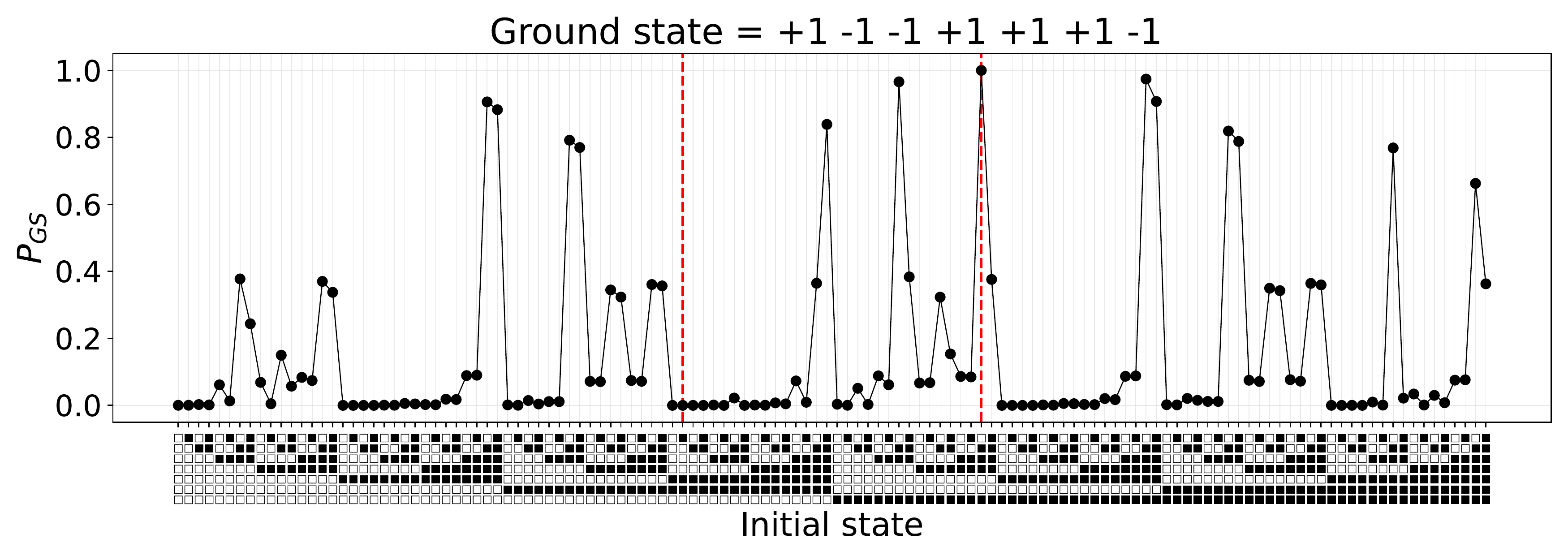}
    \caption{Ground state success probability for reverse annealing only, applied on the $N_7$ Ising. The x-axis encodes the RA initial states as vectors of vertical blocks where $\blacksquare$ denotes a variable state of $+1$ and $\square$ denotes a variable state of $-1$. The initial state vectors are read from bottom to top where the bottom is the first index which corresponds to variable $0$ in the problem Ising. The initial states which are also other ground states are marked with dashed red vertical lines. Because there are exactly two ground states, we can visually see a clear reflected symmetry between these two sub-figures. }
    \label{fig:RA_only_N7}
\end{figure}

\begin{figure}[h!]
    \centering
    \includegraphics[width=0.90\textwidth]{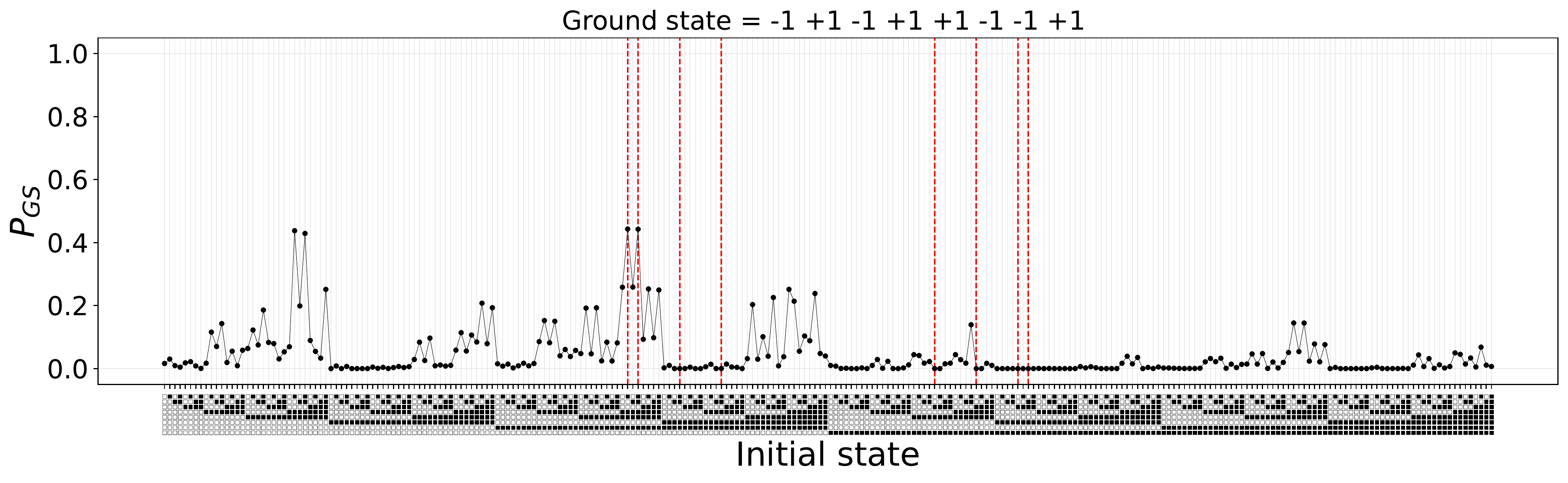}\\
    \includegraphics[width=0.90\textwidth]{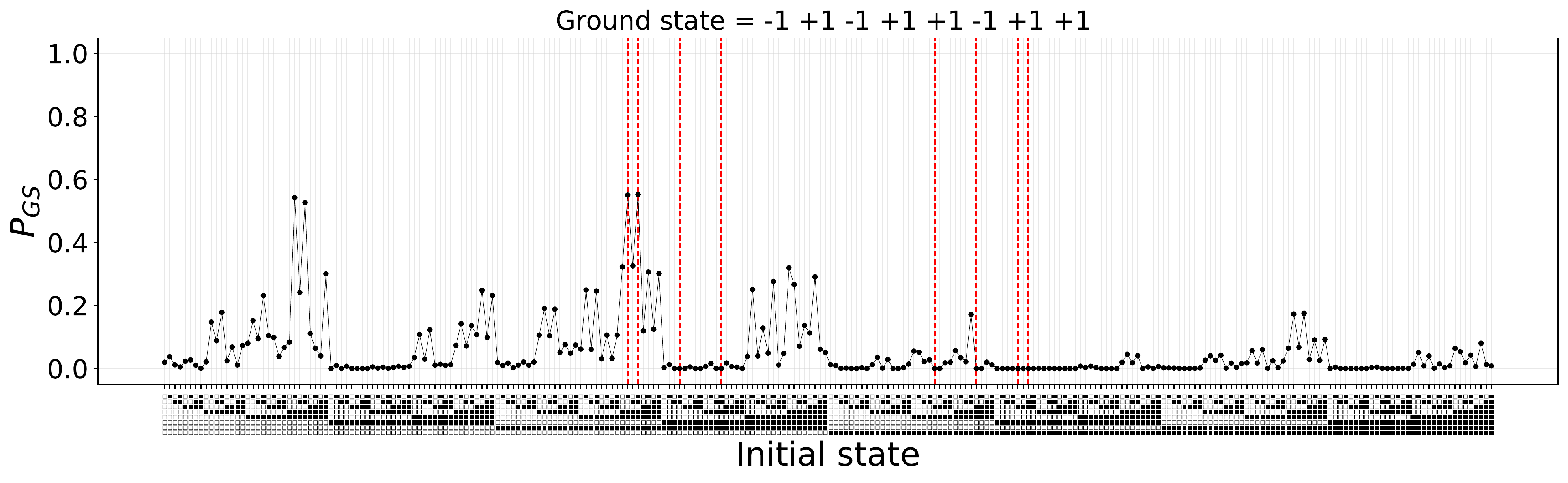}
    \caption{Ground state success probability for reverse annealing only, applied on the $N_8$ Ising. These two plots correspond to two out of the eight ground states of the $N_8$ Ising - the plots for the other six ground states are given in Figure \ref{fig:appendix_RA_only_N8} in Appendix \ref{section:appendix_extra_figures}. The x-axis encodes the RA initial states as vectors of vertical blocks where $\blacksquare$ denotes a variable state of $+1$ and $\square$ denotes a variable state of $-1$. The initial state vectors are read from bottom to top where the bottom is the first index which corresponds to variable $0$ in the problem Ising. The initial states which are also other ground states are marked with dashed red vertical lines.  }
    \label{fig:RA_only_N8}
\end{figure}

\begin{figure}[h!]
    \centering
    \includegraphics[width=0.32\textwidth]{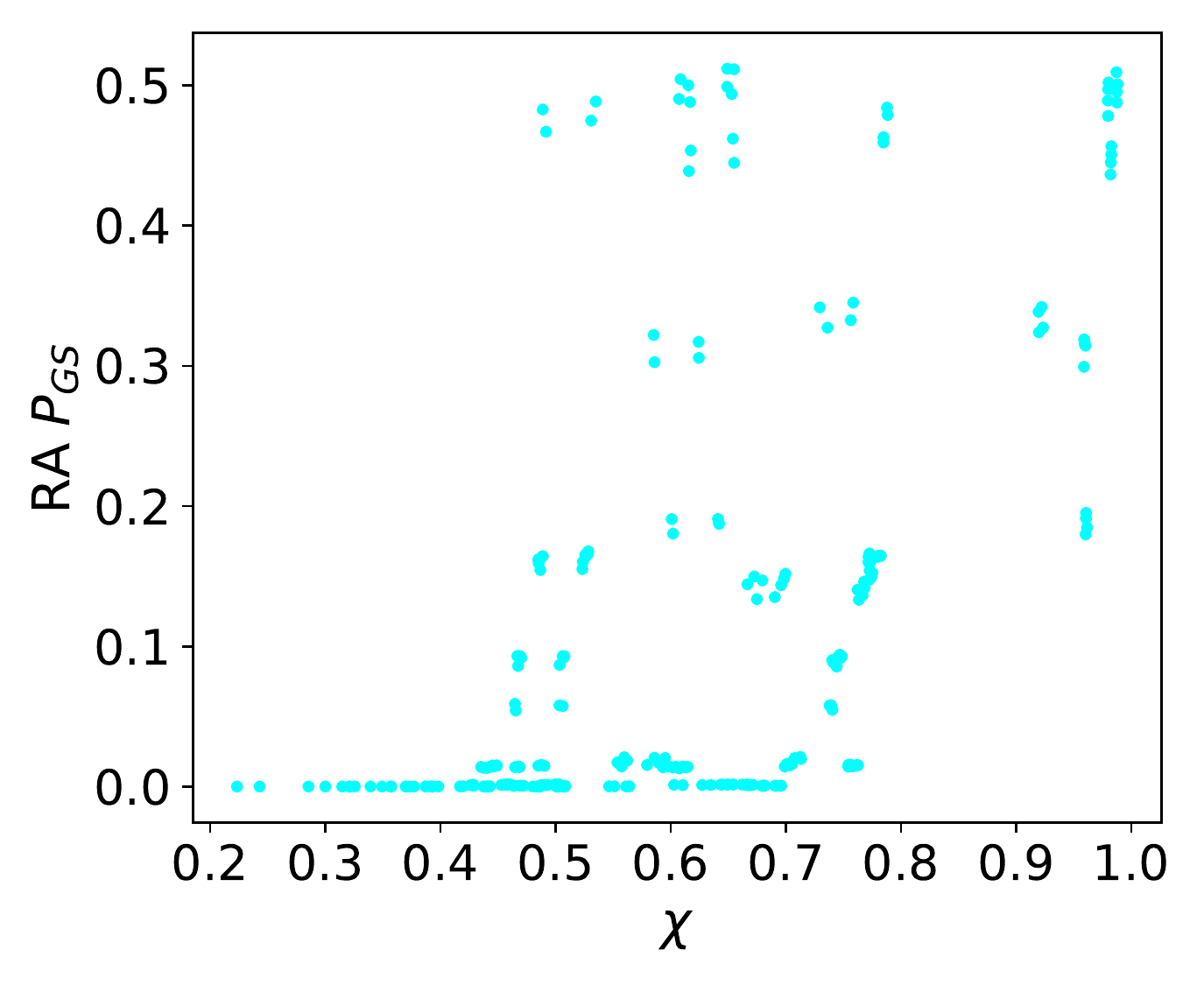}
    \includegraphics[width=0.32\textwidth]{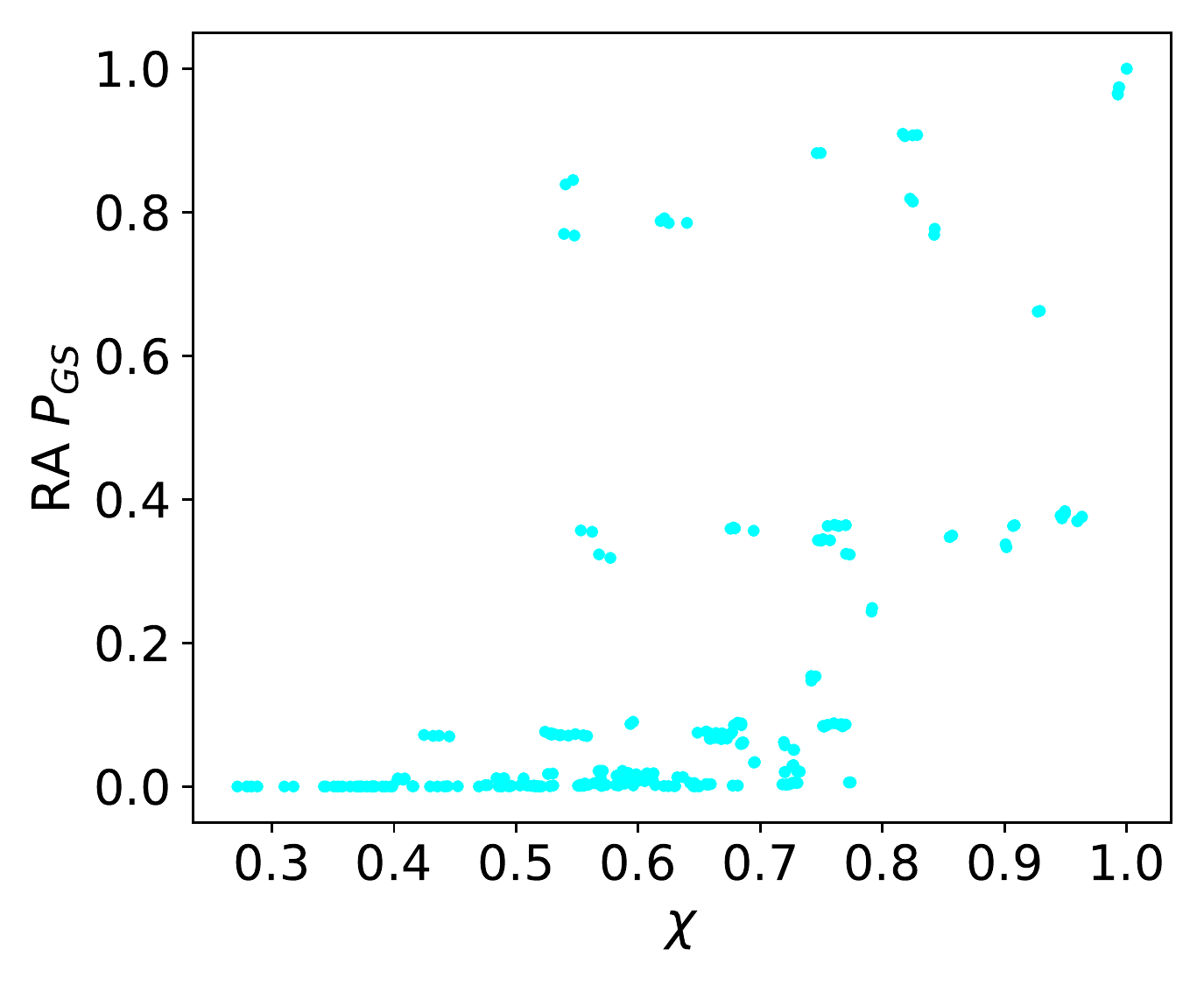}
    \includegraphics[width=0.32\textwidth]{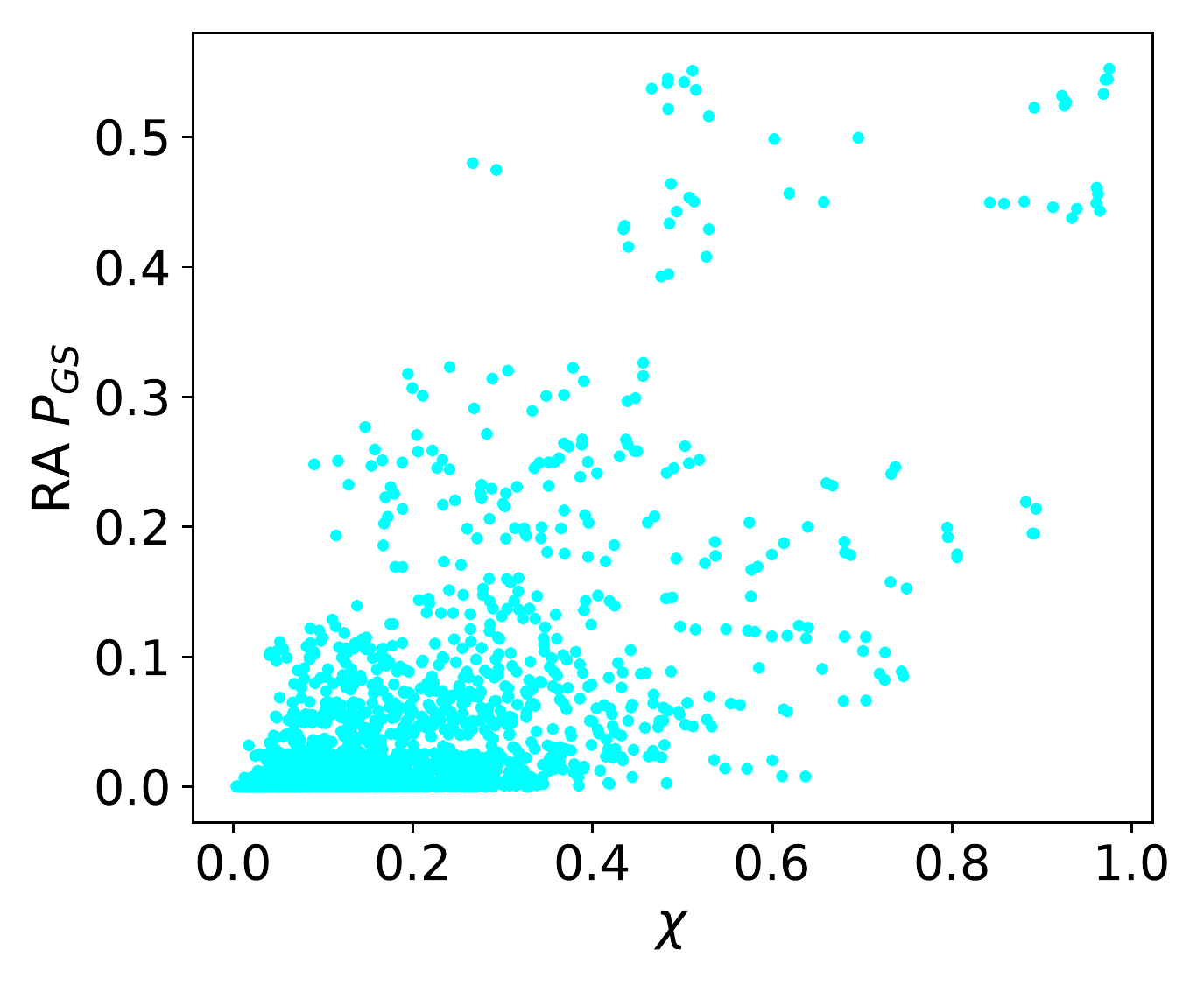}
    \caption{Scatterplots of $\chi$ (x-axis) vs reverse annealing $P_{GS}$ (y-axis) for each of the three Isings. $N_6$ Ising (left) with Pearson correlation coefficient of 0.553, $N_7$ Ising (middle) with a Pearson correlation coefficient of $0.575$, and $N_8$ Ising (right) with a Pearson correlation coefficient of $0.648$. }
    \label{fig:RA_chi_scatter}
\end{figure}

A natural question here is how reverse annealing by itself samples these optimization problems. And along with how reverse annealing (RA) samples these optimization problems for each of the initial states, another relevant question is how the success probability of sampling the optimal solution in reverse annealing correlates with $\chi$. It would make sense for these two metrics to correlated such that higher success probability for an initial state reaching some groundstate in reverse annealing corresponds to higher susceptibility of mapping that initial state to that ground state with RA and h-gain. 

Because of the dynamics observed in the previous section (in particular Section \ref{section:results_state_transition_networks}), we know that in this RA setup states will transition into intermediate states as $h$ increases while the anneal is minimizing the state towards the programmed optimal solution. Therefore these intermediate states may cause the $\chi$ metric to not directly correspond to the plain RA success probability. On the other hand it would be expected that there is correlation between the RA ground state probability and $\chi$, which would confirm that in RA (with some initial state) some states are much easier to transition into than others. 

Figures \ref{fig:RA_only_N6}, \ref{fig:RA_only_N7}, and \ref{fig:RA_only_N8} show the RA ground state proportions across all initial classical states. Notice that for this data no unsupervised clustering of the data was applied, and therefore all of the nodes are colored the same. The dynamics of these figures are different from the RA and h-gain state transition mapping procedure plots, namely Figures \ref{fig:HGain_susceptibility_to_groundstate_n6}, \ref{fig:HGain_susceptibility_to_groundstate_n7}, and \ref{fig:HGain_susceptibility_to_groundstate_n8}. However, there are some similarities. First, the different ground state plots are still symmetric for complementary groundstates. Second, the reflexive ground state mappings have the highest success rate, whereas the complementary ground states have the lowest ground state success probability. 

Figure \ref{fig:RA_chi_scatter} shows the correlation between $\chi$ and RA $P_{GS}$ for the three test Isings. Here the data is not being separated based on the ground state the mapping was applied to, and instead all of the results are aggregated together into a single dataset. The Pearson correlation coefficient, computed with scipy \cite{2020SciPy-NMeth} shows a linear relationship between RA $P_{GS}$ and $\chi$, as expected.

\section{Discussion and Conclusion}
\label{section:discussion}
In this article a method for quantifying the hardness, analogous to the susceptibility, for transitioning a quantum annealing system from one classical start at the beginning of the anneal to a classical ground state of a problem Ising at the end of the anneal. A full enumeration across all initial states was executed on three small test Isings embedded in parallel on the D-Wave \texttt{Advantage\_system4.1} quantum annealing processor. Importantly, this state transition susceptibility does not necessarily generally apply to forward annealing because the methodology specifically uses reverse annealing to encode an classical initial state. However, this state susceptibility enumeration does allow for an examination of interesting dynamics in the quantum annealer, in particular it allows us to develop a notion of \emph{distance}, or hardness, for moving between two classical states in quantum annealer. This proposed state mapping procedure may be able to help in examining quantum annealing dynamics, when sampling ground states, by viewing the internal (not directly measurable) QA dynamics as a linear combination of the full $2^n$ initial state mapping. Section \ref{section:results_fair_sampling} showed that the $2^n$ state mappings do not provide a mechanism to explain fair sampling QA dynamics assuming a uniform linear combination of the state mappings, but perhaps more advanced models would be able to better analyze QA behavoir using these state mappings.  

The set of data across the $N_6$, $N_7$, and $N_8$ Isings are notable because there is a clear finding that, at least for reverse annealing with an anneal fraction of $s=0.65$, low energy states are not necessarily easily accessible from all input states. For the purposes of improving reverse annealing, selecting states which are near to the ground state with respect to certain metrics, such as hamming distance or the $\delta$ measure, could be beneficial, as opposed to selecting states which simply low in energy, which could actually be a large distance away from the ground state in terms of the quantum annealing process. The observation that the success probability of reverse annealing is largely determined by the hamming distance of the initial state compared to the ground state has been quantified since the inception of reverse annealing \cite{perdomo2011study}. This study is therefore consistent with previous reverse annealing findings, but also presents a more detailed mapping of the state transition process. 

Utilizing this mapping technique across input states and having some states substantially more accessible as defined by the $\delta$ metric could allow for improved design of test instances with planted solutions for reverse annealing. In particular because one could construct problems with initial states that are far away from the optimal solution by objective function evaluation but are easy for the annealer to transition to the optimal solution. Typically problems with known planted solutions \cite{kowalsky20223, hen2015probing, perera2020computational, raymond2022hybrid, perera2020chook, pang2020structure, albash2018advantage, denchev2016computational, king2019quantum, king2015performance, king2015benchmarking, zdeborova2016statistical} are intended to serve as a benchmark for how accurate quantum annealing and other heuristic computation tools are for problem sizes for which the optimal solution is difficult to compute exactly in reasonable time. However, specifically constructing planted solutions and specific initial states to begin the optimization at is not a tool that has been developed for reverse annealing. Applying the full classical state enumeration that is utilized in this article is costly because it scales exponentially as $2^n$ states for $n$ variables. Therefore for practical reasons this type of state mapping will need to be restricted to small problem instances. 

Examining how initial state choices impact the optimization process is not only relevant for quantum annealing, but is also a relevant question in warm-start QAOA \cite{https://doi.org/10.48550/arxiv.2207.05089, Egger_2021}. Because the Quantum Alternating Operator Ansatz (QAOA) algorithm \cite{Hadfield_2019} is effectively a Trotterization of quantum annealing, understanding the dynamics of variable state changes in quantum annealing could inform QAOA algorithm development. For example, it has been observed that in specific cases warm start QAOA for non-optimal low energy initial solutions gets stuck in local minima \cite{https://doi.org/10.48550/arxiv.2207.05089}. 

There are several open research questions that can be investigated:

\begin{itemize}
    \item The observation that it requires less h-gain strength to flip high degree variables into a neighboring ground state, which lead to the creation of the $\delta$ metric, could be further evaluated for Isings of comparable number of variables but which have uniform variable degrees in their graph. In particular, following the observations made in this article it would be expected that Isings with uniform degrees would have more uniform $\chi$ across all input states. 
    \item What are the quantum annealing dynamics are when forcing the system into a ground state during normal forward annealing. This can be accomplished by specifying the ground state using linear terms and an h-gain schedule which gradually increases from $0$ up to some specified strength. This could allow a for determination as to whether some ground states have a higher susceptibility than others under the D-Wave implementation of the transverse field Ising model of quantum annealing, and if this correlates with the unfair sampling of ground states. 
    \item To what extent can the reverse annealing and h-gain state-to-state mapping presented in this article be applied for mapping to states which are not ground states. In principle if the h-gain field is applied continuously over the duration of the anneal and the state freezes out before the effect of the h-gain field subsides or is intentionally reduced then the anneal could remain in a non ground state at readout. If this could be implemented, a full Markov transition matrix for every classical state to every other classical state, having been passed through a quantum annealer, could be developed (including which intermediate states are found during the transition from one state to another). 
    \item Expanding the results presented in this article to include other anneal fractions to pause the anneal at, more complex reverse annealing schedules, or more complex h-gain schedules. 
    \item D-Wave Advantage\_system4.1 has a maximum h-gain schedule strength of $3$, but other D-Wave quantum annealers have other allowed maximum h-gain schedule values (for example Advantage\_system6.1 allows up to an h-gain schedule strength of $4$). Utilizing stronger h-gain schedule amplifications could facilitate finding higher success proportions for getting the anneal into an intended ground state, in particular for the $N_8$ Ising. 
    \item The h-gain field can also be programmed to have a \emph{negative} bias, as opposed to a positive bias. Applying both the negative and positive h-gain fields to state mapping or phase transition experiments could be useful for the purpose of determining if there is an asymmetry in the responses to the applied h-gain field. 
    \item The h-gain state encoding method, unlike reverse annealing which requires all active qubits to have an initial state specified, allows selective encoding of an initial state for some subset of the qubits being used for a problem Ising. Therefore, in a slightly different context than what is described in this paper, the h-gain state encoding method could be used for methods where selective variable encoding is important. For example, this could allow one to encode the states of weak or strong variable persistencies of an optimization problem \cite{hammer1984roof, billionnet1992persistency, windheuser2012generalized}, which can be efficiently computed classically and describe the state of variables which always take this state in optimum solutions (strong persistencies) or variables which take a specific state in at least one optimum solution if there are multiple optimal solutions (weak persistencies). 
    \item These systems are small enough that quantum annealing simulators could be applied to compare against the experimental results. The limitation is that such a simulator would need to include both a reverse annealing schedule and an h-gain schedule capability. 
\end{itemize}


\section{Acknowledgments}
\label{section:acknowledgments}
The author thanks Carleton Coffrin and Stephan Eidenbenz for providing feedback on the experimental results and versions of the manuscript at various stages of the research. The author thanks Denny Dahl for discussions on the h-gain feature which eventually led to this idea. This work was supported by the U.S. Department of Energy through the Los Alamos National Laboratory. Los Alamos National Laboratory is operated by Triad National Security, LLC, for the National Nuclear Security Administration of U.S. Department of Energy (Contract No. 89233218CNA000001). The research presented in this article was supported by the Laboratory Directed Research and Development program of Los Alamos National Laboratory under project number 20220656ER. This research used resources provided by the Los Alamos National Laboratory Institutional Computing Program. 

\noindent
LA-UR-22-31301

\setlength\bibitemsep{0pt}
\printbibliography

@article{Kadowaki_1998,
	doi = {10.1103/physreve.58.5355},
  
	url = {https://doi.org/10.1103%2Fphysreve.58.5355},
  
	year = 1998,
	month = {nov},
  
	publisher = {American Physical Society ({APS})},
  
	volume = {58},
  
	number = {5},
  
	pages = {5355--5363},
  
	author = {Tadashi Kadowaki and Hidetoshi Nishimori},
  
	title = {Quantum annealing in the transverse Ising model},
	journal = {Physical Review E}
}

@article{PhysRevApplied.17.044005,
  title = {Customized Quantum Annealing Schedules},
  author = {Khezri, Mostafa and Dai, Xi and Yang, Rui and Albash, Tameem and Lupascu, Adrian and Lidar, Daniel A.},
  journal = {Phys. Rev. Applied},
  volume = {17},
  issue = {4},
  pages = {044005},
  numpages = {19},
  year = {2022},
  month = {Apr},
  publisher = {American Physical Society},
  doi = {10.1103/PhysRevApplied.17.044005},
  url = {https://link.aps.org/doi/10.1103/PhysRevApplied.17.044005}
}

@article{pelofske2022quantum,
  title={Quantum annealing algorithms for Boolean tensor networks},
  author={Pelofske, Elijah and Hahn, Georg and O’Malley, Daniel and Djidjev, Hristo N and Alexandrov, Boian S},
  journal={Scientific Reports},
  volume={12},
  number={1},
  pages={1--19},
  year={2022},
  publisher={Nature Publishing Group},
  doi = {10.1038/s41598-022-12611-9},
}

@article{pelofske2022parallel,
  title={Parallel quantum annealing},
  author={Pelofske, Elijah and Hahn, Georg and Djidjev, Hristo N},
  journal={Scientific Reports},
  volume={12},
  number={1},
  pages={1--11},
  year={2022},
  publisher={Nature Publishing Group},
  doi={10.1038/s41598-022-08394-8}
}

@inproceedings{pelofske2020advanced,  
author={Pelofske, Elijah and Hahn, Georg and Djidjev, Hristo N.},  booktitle={2020 IEEE International Conference on Quantum Computing and Engineering (QCE)},   title={Advanced anneal paths for improved quantum annealing},   year={2020},  volume={},  number={},  pages={256-266},  doi={10.1109/QCE49297.2020.00040}}

@article{PhysRevA.105.022410,
  title = {Perturbed ferromagnetic chain: Tunable test of hardness in the transverse-field Ising model},
  author = {O'Connor, D. T. and Fry-Bouriaux, L. and Warburton, P. A.},
  journal = {Phys. Rev. A},
  volume = {105},
  issue = {2},
  pages = {022410},
  numpages = {12},
  year = {2022},
  month = {Feb},
  publisher = {American Physical Society},
  doi = {10.1103/PhysRevA.105.022410},
  url = {https://link.aps.org/doi/10.1103/PhysRevA.105.022410}
}

@article{vert2021benchmarking,
  title={Benchmarking Quantum Annealing Against “Hard” Instances of the Bipartite Matching Problem},
  author={Vert, Daniel and Sirdey, Renaud and Louise, St{\'e}phane},
  journal={SN Computer Science},
  volume={2},
  number={2},
  pages={1--12},
  year={2021},
  publisher={Springer}
}

@article{das2008colloquium,
  title={Colloquium: Quantum annealing and analog quantum computation},
  author={Das, Arnab and Chakrabarti, Bikas K},
  journal={Reviews of Modern Physics},
  volume={80},
  number={3},
  pages={1061},
  year={2008},
  publisher={APS}
}

@article{johnson2011quantum,
  title={Quantum annealing with manufactured spins},
  author={Johnson, Mark W and Amin, Mohammad HS and Gildert, Suzanne and Lanting, Trevor and Hamze, Firas and Dickson, Neil and Harris, Richard and Berkley, Andrew J and Johansson, Jan and Bunyk, Paul and others},
  journal={Nature},
  volume={473},
  number={7346},
  pages={194--198},
  year={2011},
  publisher={Nature Publishing Group}
}

@article{hauke2020perspectives,
  title={Perspectives of quantum annealing: Methods and implementations},
  author={Hauke, Philipp and Katzgraber, Helmut G and Lechner, Wolfgang and Nishimori, Hidetoshi and Oliver, William D},
  journal={Reports on Progress in Physics},
  volume={83},
  number={5},
  pages={054401},
  year={2020},
  publisher={IOP Publishing}
}

@article{morita2008mathematical,
  title={Mathematical foundation of quantum annealing},
  author={Morita, Satoshi and Nishimori, Hidetoshi},
  journal={Journal of Mathematical Physics},
  volume={49},
  number={12},
  pages={125210},
  year={2008},
  publisher={American Institute of Physics}
}

@article{boixo2014evidence,
  title={Evidence for quantum annealing with more than one hundred qubits},
  author={Boixo, Sergio and R{\o}nnow, Troels F and Isakov, Sergei V and Wang, Zhihui and Wecker, David and Lidar, Daniel A and Martinis, John M and Troyer, Matthias},
  journal={Nature physics},
  volume={10},
  number={3},
  pages={218--224},
  year={2014},
  publisher={Nature Publishing Group}
}

@article{boixo2013experimental,
  title={Experimental signature of programmable quantum annealing},
  author={Boixo, Sergio and Albash, Tameem and Spedalieri, Federico M and Chancellor, Nicholas and Lidar, Daniel A},
  journal={Nature communications},
  volume={4},
  number={1},
  pages={1--8},
  year={2013},
  publisher={Nature Publishing Group}
}

@inproceedings{pelofske2021sampling,
  title={Sampling on NISQ Devices:" Who’s the Fairest One of All?"},
  author={Pelofske, Elijah and Golden, John and B{\"a}rtschi, Andreas and O’Malley, Daniel and Eidenbenz, Stephan},
  booktitle={2021 IEEE International Conference on Quantum Computing and Engineering (QCE)},
  pages={207--217},
  year={2021},
  organization={IEEE}
}

@article{konz2019uncertain,
  title={Uncertain fate of fair sampling in quantum annealing},
  author={K{\"o}nz, Mario S and Mazzola, Guglielmo and Ochoa, Andrew J and Katzgraber, Helmut G and Troyer, Matthias},
  journal={Physical Review A},
  volume={100},
  number={3},
  pages={030303},
  year={2019},
  publisher={APS}
}

@article{kumar2020achieving,
  title={Achieving fair sampling in quantum annealing},
  author={Kumar, Vaibhaw and Tomlin, Casey and Nehrkorn, Curt and O'Malley, Daniel and Dulny III, Joseph},
  journal={arXiv preprint arXiv:2007.08487},
  year={2020}
}

@article{mandra2017exponentially,
  title = {Exponentially Biased Ground-State Sampling of Quantum Annealing Machines with Transverse-Field Driving Hamiltonians},
  author = {Mandr\`a, Salvatore and Zhu, Zheng and Katzgraber, Helmut G.},
  journal = {Phys. Rev. Lett.},
  volume = {118},
  issue = {7},
  pages = {070502},
  numpages = {6},
  year = {2017},
  month = {Feb},
  publisher = {American Physical Society},
  doi = {10.1103/PhysRevLett.118.070502},
  url = {https://link.aps.org/doi/10.1103/PhysRevLett.118.070502}
}

@article{yamamoto2020fair,
  title={Fair sampling by simulated annealing on quantum annealer},
  author={Yamamoto, Masayuki and Ohzeki, Masayuki and Tanaka, Kazuyuki},
  journal={Journal of the Physical Society of Japan},
  volume={89},
  number={2},
  pages={025002},
  year={2020},
  publisher={The Physical Society of Japan},
  doi={10.7566/JPSJ.89.025002}
}

@article{zhou2021experimental,
  title = {Experimental realization of classical ${\mathbb{Z}}_{2}$ spin liquids in a programmable quantum device},
  author = {Zhou, Shiyu and Green, Dmitry and Dahl, Edward D. and Chamon, Claudio},
  journal = {Phys. Rev. B},
  volume = {104},
  issue = {8},
  pages = {L081107},
  numpages = {6},
  year = {2021},
  month = {Aug},
  publisher = {American Physical Society},
  doi = {10.1103/PhysRevB.104.L081107},
  url = {https://link.aps.org/doi/10.1103/PhysRevB.104.L081107}
}

@article{king2021qubit,
  title={Qubit spin ice},
  author={King, Andrew D and Nisoli, Cristiano and Dahl, Edward D and Poulin-Lamarre, Gabriel and Lopez-Bezanilla, Alejandro},
  journal={Science},
  volume={373},
  number={6554},
  pages={576--580},
  year={2021},
  publisher={American Association for the Advancement of Science},
  doi={10.1126/science.abe2824}
}

@inproceedings{10.1145/2482767.2482797,
author = {McGeoch, Catherine C. and Wang, Cong},
title = {Experimental Evaluation of an Adiabiatic Quantum System for Combinatorial Optimization},
year = {2013},
isbn = {9781450320535},
publisher = {Association for Computing Machinery},
address = {New York, NY, USA},
url = {https://doi.org/10.1145/2482767.2482797},
doi = {10.1145/2482767.2482797},
booktitle = {Proceedings of the ACM International Conference on Computing Frontiers},
articleno = {23},
numpages = {11},
location = {Ischia, Italy},
series = {CF '13}
}

@article{finnila1994quantum,
  title={Quantum annealing: A new method for minimizing multidimensional functions},
  author={Finnila, Aleta Berk and Gomez, MA and Sebenik, C and Stenson, Catherine and Doll, Jimmie D},
  journal={Chemical physics letters},
  volume={219},
  number={5-6},
  pages={343--348},
  year={1994},
  publisher={Elsevier}
}

@article{king2021scaling,
  title={Scaling advantage over path-integral Monte Carlo in quantum simulation of geometrically frustrated magnets},
  author={King, Andrew D and Raymond, Jack and Lanting, Trevor and Isakov, Sergei V and Mohseni, Masoud and Poulin-Lamarre, Gabriel and Ejtemaee, Sara and Bernoudy, William and Ozfidan, Isil and Smirnov, Anatoly Yu and others},
  journal={Nature communications},
  volume={12},
  number={1},
  pages={1--6},
  year={2021},
  publisher={Nature Publishing Group}
}

@article{titiloye2011quantum,
  title={Quantum annealing of the graph coloring problem},
  author={Titiloye, Olawale and Crispin, Alan},
  journal={Discrete Optimization},
  volume={8},
  number={2},
  pages={376--384},
  year={2011},
  publisher={Elsevier}
}

@article{perdomo2015quantum,
  title={A quantum annealing approach for fault detection and diagnosis of graph-based systems},
  author={Perdomo-Ortiz, Alejandro and Fluegemann, Joseph and Narasimhan, Sriram and Biswas, Rupak and Smelyanskiy, Vadim N},
  journal={The European Physical Journal Special Topics},
  volume={224},
  number={1},
  pages={131--148},
  year={2015},
  publisher={Springer}
}

@article{stollenwerk2019quantum,
  title={Quantum annealing applied to de-conflicting optimal trajectories for air traffic management},
  author={Stollenwerk, Tobias and O’Gorman, Bryan and Venturelli, Davide and Mandra, Salvatore and Rodionova, Olga and Ng, Hokkwan and Sridhar, Banavar and Rieffel, Eleanor Gilbert and Biswas, Rupak},
  journal={IEEE transactions on intelligent transportation systems},
  volume={21},
  number={1},
  pages={285--297},
  year={2019},
  publisher={IEEE}
}

@article{silva2021mapping,
  title={Mapping a logical representation of TSP to quantum annealing},
  author={Silva, Carla and Aguiar, Ana and Lima, Priscila and Dutra, In{\^e}s},
  journal={Quantum Information Processing},
  volume={20},
  number={12},
  pages={1--21},
  year={2021},
  publisher={Springer}
}

@article{yarkoni2021quantum,
  title={Quantum Annealing for Industry Applications: Introduction and Review},
  author={Yarkoni, Sheir and Raponi, Elena and Schmitt, Sebastian and B{\"a}ck, Thomas},
  journal={arXiv preprint arXiv:2112.07491},
  year={2021}
}

@article{chapuis2019finding,
  title={Finding maximum cliques on the d-wave quantum annealer},
  author={Chapuis, Guillaume and Djidjev, Hristo and Hahn, Georg and Rizk, Guillaume},
  journal={Journal of Signal Processing Systems},
  volume={91},
  number={3},
  pages={363--377},
  year={2019},
  publisher={Springer}
}

@article{negre2020detecting,
  title={Detecting multiple communities using quantum annealing on the D-Wave system},
  author={Negre, Christian FA and Ushijima-Mwesigwa, Hayato and Mniszewski, Susan M},
  journal={Plos one},
  volume={15},
  number={2},
  pages={e0227538},
  year={2020},
  publisher={Public Library of Science San Francisco, CA USA}
}

@inproceedings{pelofske2019solving,
  title={Solving large minimum vertex cover problems on a quantum annealer},
  author={Pelofske, Elijah and Hahn, Georg and Djidjev, Hristo},
  booktitle={Proceedings of the 16th ACM International Conference on Computing Frontiers},
  pages={76--84},
  year={2019}
}

@article{Marshall_2022,
	doi = {10.1103/physreva.105.022615},
  
	url = {https://doi.org/10.1103%2Fphysreva.105.022615},
  
	year = 2022,
	month = {feb},
  
	publisher = {American Physical Society ({APS})},
  
	volume = {105},
  
	number = {2},
  
	author = {Jeffrey Marshall and Gianni Mossi and Eleanor G. Rieffel},
  
	title = {Perils of embedding for quantum sampling},
  
	journal = {Physical Review A}
}

@inproceedings{ushijima2017graph,
  title={Graph partitioning using quantum annealing on the d-wave system},
  author={Ushijima-Mwesigwa, Hayato and Negre, Christian FA and Mniszewski, Susan M},
  booktitle={Proceedings of the Second International Workshop on Post Moores Era Supercomputing},
  pages={22--29},
  year={2017}
}

@inproceedings{novotny2016spanning,
  title={Spanning tree calculations on D-Wave 2 machines},
  author={Novotny, MA and Hobl, Q L and Hall, JS and Michielsen, Kristel},
  booktitle={Journal of Physics: Conference Series},
  volume={681},
  number={1},
  pages={012005},
  year={2016},
  organization={IOP Publishing}
}

@article{dattani2019pegasus,
  title={Pegasus: The second connectivity graph for large-scale quantum annealing hardware},
  author={Dattani, Nike and Szalay, Szilard and Chancellor, Nick},
  journal={arXiv preprint arXiv:1901.07636},
  year={2019}
}

@inproceedings{zbinden2020embedding,
  title={Embedding algorithms for quantum annealers with chimera and pegasus connection topologies},
  author={Zbinden, Stefanie and B{\"a}rtschi, Andreas and Djidjev, Hristo and Eidenbenz, Stephan},
  booktitle={International Conference on High Performance Computing},
  pages={187--206},
  year={2020},
  organization={Springer}
}

@article{boothby2020next,
  title={Next-generation topology of d-wave quantum processors},
  author={Boothby, Kelly and Bunyk, Paul and Raymond, Jack and Roy, Aidan},
  journal={arXiv preprint arXiv:2003.00133},
  year={2020}
}

@misc{https://doi.org/10.48550/arxiv.1406.2741,
  doi = {10.48550/ARXIV.1406.2741},
  
  url = {https://arxiv.org/abs/1406.2741},
  
  author = {Cai, Jun and Macready, William G. and Roy, Aidan},
  
  title = {A practical heuristic for finding graph minors},
  
  publisher = {arXiv},
  
  year = {2014},
  
  copyright = {arXiv.org perpetual, non-exclusive license}
}

@inproceedings{pelofske2021boolean,
  title={Boolean hierarchical tucker networks on quantum annealers},
  author={Pelofske, Elijah and Hahn, Georg and O’Malley, Daniel and Djidjev, Hristo N and Alexandrov, Boian S},
  booktitle={International Conference on Large-Scale Scientific Computing},
  pages={351--358},
  year={2021},
  organization={Springer}
}

@inproceedings{o2020tucker,
  title={Tucker-1 Boolean Tensor Factorization with Quantum Annealers},
  author={O’Malley, Daniel and Djidjev, Hristo N and Alexandrov, Boian S},
  booktitle={2020 International Conference on Rebooting Computing (ICRC)},
  pages={58--65},
  year={2020},
  organization={IEEE}
}

@inproceedings{barbosa2020optimizing,
  title={Optimizing embedding-related quantum annealing parameters for reducing hardware bias},
  author={Barbosa, Aaron and Pelofske, Elijah and Hahn, Georg and Djidjev, Hristo N},
  booktitle={International Symposium on Parallel Architectures, Algorithms and Programming},
  pages={162--173},
  year={2020},
  organization={Springer}
}

@article{pelofske2020inferring,
  title={Inferring the Dynamics of the State Evolution During Quantum Annealing},
  author={Pelofske, Elijah and Hahn, Georg and Djidjev, Hristo},
  journal={IEEE Transactions on Parallel and Distributed Systems},
  volume={33},
  number={2},
  pages={310--321},
  year={2020},
  publisher={IEEE}
}

@ARTICLE{868688,  author={Jianbo Shi and Malik, J.},  journal={IEEE Transactions on Pattern Analysis and Machine Intelligence},   title={Normalized cuts and image segmentation},   year={2000},  volume={22},  number={8},  pages={888-905},  doi={10.1109/34.868688}}

@article{von2007tutorial,
  title={A tutorial on spectral clustering},
  author={Von Luxburg, Ulrike},
  journal={Statistics and computing},
  volume={17},
  number={4},
  pages={395--416},
  year={2007},
  publisher={Springer}
}

@inproceedings{stella2003multiclass,  author={Stella, X Yu and Shi, Jianbo},  booktitle={Proceedings Ninth IEEE International Conference on Computer Vision},   title={Multiclass spectral clustering},   year={2003},  volume={},  number={},  pages={313-319 vol.1},  doi={10.1109/ICCV.2003.1238361}}

@article{knyazev2001toward,
  title={Toward the optimal preconditioned eigensolver: Locally optimal block preconditioned conjugate gradient method},
  author={Knyazev, Andrew V},
  journal={SIAM journal on scientific computing},
  volume={23},
  number={2},
  pages={517--541},
  year={2001},
  publisher={SIAM}
}

@article{10.1093/imaiai/iay008,
    author = {Damle, Anil and Minden, Victor and Ying, Lexing},
    title = "{Simple, direct and efficient multi-way spectral clustering}",
    journal = {Information and Inference: A Journal of the IMA},
    volume = {8},
    number = {1},
    pages = {181-203},
    year = {2018},
    month = {06},
    issn = {2049-8772},
    doi = {10.1093/imaiai/iay008},
    url = {https://doi.org/10.1093/imaiai/iay008},
    eprint = {https://academic.oup.com/imaiai/article-pdf/8/1/181/28053156/iay008.pdf},
}

@article{scikit-learn,
 title={Scikit-learn: Machine Learning in {P}ython},
 author={Pedregosa, F. and Varoquaux, G. and Gramfort, A. and Michel, V.
         and Thirion, B. and Grisel, O. and Blondel, M. and Prettenhofer, P.
         and Weiss, R. and Dubourg, V. and Vanderplas, J. and Passos, A. and
         Cournapeau, D. and Brucher, M. and Perrot, M. and Duchesnay, E.},
 journal={Journal of Machine Learning Research},
 volume={12},
 pages={2825--2830},
 year={2011}
}

@inproceedings{sklearn_api,
  author    = {Lars Buitinck and Gilles Louppe and Mathieu Blondel and
               Fabian Pedregosa and Andreas Mueller and Olivier Grisel and
               Vlad Niculae and Peter Prettenhofer and Alexandre Gramfort
               and Jaques Grobler and Robert Layton and Jake VanderPlas and
               Arnaud Joly and Brian Holt and Ga{\"{e}}l Varoquaux},
  title     = {{API} design for machine learning software: experiences from the scikit-learn
               project},
  booktitle = {ECML PKDD Workshop: Languages for Data Mining and Machine Learning},
  year      = {2013},
  pages = {108--122},
}

@article{PhysRevE.99.063314,
  title = {Fair sampling of ground-state configurations of binary optimization problems},
  author = {Zhu, Zheng and Ochoa, Andrew J. and Katzgraber, Helmut G.},
  journal = {Phys. Rev. E},
  volume = {99},
  issue = {6},
  pages = {063314},
  numpages = {6},
  year = {2019},
  month = {Jun},
  publisher = {American Physical Society},
  doi = {10.1103/PhysRevE.99.063314},
  url = {https://link.aps.org/doi/10.1103/PhysRevE.99.063314}
}

@article{katzgraber2014glassy,
  title={Glassy chimeras could be blind to quantum speedup: Designing better benchmarks for quantum annealing machines},
  author={Katzgraber, Helmut G and Hamze, Firas and Andrist, Ruben S},
  journal={Physical Review X},
  volume={4},
  number={2},
  pages={021008},
  year={2014},
  publisher={APS},
  doi = {10.1103/physrevx.4.021008}
}

@article{PRXQuantum.2.030317,
  title = {Quantum Annealing Simulation of Out-of-Equilibrium Magnetization in a Spin-Chain Compound},
  author = {King, Andrew D. and Batista, Cristian D. and Raymond, Jack and Lanting, Trevor and Ozfidan, Isil and Poulin-Lamarre, Gabriel and Zhang, Hao and Amin, Mohammad H.},
  journal = {PRX Quantum},
  volume = {2},
  issue = {3},
  pages = {030317},
  numpages = {11},
  year = {2021},
  month = {Jul},
  publisher = {American Physical Society},
  doi = {10.1103/PRXQuantum.2.030317},
  url = {https://link.aps.org/doi/10.1103/PRXQuantum.2.030317}
}

@article{abel2021quantum,
  title={Quantum computing for quantum tunneling},
  author={Abel, Steven and Chancellor, Nicholas and Spannowsky, Michael},
  journal={Physical Review D},
  volume={103},
  number={1},
  pages={016008},
  year={2021},
  publisher={APS}
}

@ARTICLE{9485068,  author={Chen, Jie and Stollenwerk, Tobias and Chancellor, Nicholas},  journal={IEEE Transactions on Quantum Engineering},   title={Performance of Domain-Wall Encoding for Quantum Annealing},   year={2021},  volume={2},  number={},  pages={1-14},  doi={10.1109/TQE.2021.3094280}}

@misc{berwald2021understanding,
  doi = {10.48550/ARXIV.2108.12004},
  
  url = {https://arxiv.org/abs/2108.12004},
  
  author = {Berwald, Jesse and Chancellor, Nicholas and Dridi, Raouf},
  
  title = {Understanding domain-wall encoding theoretically and experimentally},
  
  publisher = {arXiv},
  
  year = {2021},
  
  copyright = {arXiv.org perpetual, non-exclusive license}
}

@article{chancellor2019domain,
  title={Domain wall encoding of discrete variables for quantum annealing and QAOA},
  author={Chancellor, Nicholas},
  journal={Quantum Science and Technology},
  volume={4},
  number={4},
  pages={045004},
  year={2019},
  publisher={IOP Publishing}
}

@article{Abel_2021,
	doi = {10.1103/prxquantum.2.010349},
  
	url = {https://doi.org/10.1103%2Fprxquantum.2.010349},
  
	year = 2021,
	month = {mar},
  
	publisher = {American Physical Society ({APS})},
  
	volume = {2},
  
	number = {1},
  
	author = {Steven Abel and Michael Spannowsky},
  
	title = {Quantum-Field-Theoretic Simulation Platform for Observing the Fate of the False Vacuum},
  
	journal = {{PRX} Quantum}
}

@article{hammer1984roof,
  title={Roof duality, complementation and persistency in quadratic 0--1 optimization},
  author={Hammer, Peter L and Hansen, Pierre and Simeone, Bruno},
  journal={Mathematical programming},
  volume={28},
  number={2},
  pages={121--155},
  year={1984},
  publisher={Springer}
}

@article{billionnet1992persistency,
  title={Persistency in quadratic 0--1 optimization},
  author={Billionnet, Alain and Sutter, Alain},
  journal={Mathematical programming},
  volume={54},
  number={1},
  pages={115--119},
  year={1992},
  publisher={Springer}
}

@inproceedings{windheuser2012generalized,
  title={Generalized roof duality for multi-label optimization: Optimal lower bounds and persistency},
  author={Windheuser, Thomas and Ishikawa, Hiroshi and Cremers, Daniel},
  booktitle={European Conference on Computer Vision},
  pages={400--413},
  year={2012},
  organization={Springer}
}

@inproceedings{pelofske2021reducing,
author = {Pelofske, Elijah and Hahn, Georg and Djidjev, Hristo N.},
title = {Reducing Quantum Annealing Biases for Solving the Graph Partitioning Problem},
year = {2021},
isbn = {9781450384049},
publisher = {Association for Computing Machinery},
address = {New York, NY, USA},
url = {https://doi.org/10.1145/3457388.3458672},
doi = {10.1145/3457388.3458672},
booktitle = {Proceedings of the 18th ACM International Conference on Computing Frontiers},
pages = {133–139},
numpages = {7},
location = {Virtual Event, Italy},
series = {CF '21}
}

@article{kowalsky20223,
  title={3-regular three-XORSAT planted solutions benchmark of classical and quantum heuristic optimizers},
  author={Kowalsky, Matthew and Albash, Tameem and Hen, Itay and Lidar, Daniel A},
  journal={Quantum Science and Technology},
  volume={7},
  number={2},
  pages={025008},
  year={2022},
  publisher={IOP Publishing}
}

@article{hen2015probing,
  title={Probing for quantum speedup in spin-glass problems with planted solutions},
  author={Hen, Itay and Job, Joshua and Albash, Tameem and R{\o}nnow, Troels F and Troyer, Matthias and Lidar, Daniel A},
  journal={Physical Review A},
  volume={92},
  number={4},
  pages={042325},
  year={2015},
  publisher={APS}
}

@article{raymond2022hybrid,
  title={Hybrid quantum annealing for larger-than-QPU lattice-structured problems},
  author={Raymond, Jack and Stevanovic, Radomir and Bernoudy, William and Boothby, Kelly and McGeoch, Catherine and Berkley, Andrew J and Farr{\'e}, Pau and King, Andrew D},
  journal={arXiv preprint arXiv:2202.03044},
  year={2022}
}

@article{perera2020computational,
  title={Computational hardness of spin-glass problems with tile-planted solutions},
  author={Perera, Dilina and Hamze, Firas and Raymond, Jack and Weigel, Martin and Katzgraber, Helmut G},
  journal={Physical Review E},
  volume={101},
  number={2},
  pages={023316},
  year={2020},
  publisher={APS}
}

@article{perera2020chook,
  title={Chook--A comprehensive suite for generating binary optimization problems with planted solutions},
  author={Perera, Dilina and Akpabio, Inimfon and Hamze, Firas and Mandra, Salvatore and Rose, Nathan and Aramon, Maliheh and Katzgraber, Helmut G},
  journal={arXiv preprint arXiv:2005.14344},
  year={2020}
}

@article{pang2020structure,
  author="Pang, Yuchen and Coffrin, Carleton and Lokhov, Andrey Y. and Vuffray, Marc",
  title="The Potential of Quantum Annealing for Rapid Solution Structure Identification",
  booktitle="Integration of Constraint Programming, Artificial Intelligence, and Operations Research",
  editor="Hebrard, Emmanuel and Musliu, Nysret",
  year="2020",
  publisher="Springer International Publishing"
}

@article{albash2018advantage,
  title = {Demonstration of a Scaling Advantage for a Quantum Annealer over Simulated Annealing},
  author = {Albash, Tameem and Lidar, Daniel A.},
  journal = {Phys. Rev. X},
  volume = {8},
  issue = {3},
  pages = {031016},
  numpages = {26},
  year = {2018},
  month = {Jul},
  publisher = {American Physical Society},
  doi = {10.1103/PhysRevX.8.031016},
  url = {https://link.aps.org/doi/10.1103/PhysRevX.8.031016}
}

@article{denchev2016computational,
  title={What is the Computational Value of Finite-Range Tunneling?},
  author={Denchev, Vasil S and Boixo, Sergio and Isakov, Sergei V and Ding, Nan and Babbush, Ryan and Smelyanskiy, Vadim and Martinis, John and Neven, Hartmut},
  journal={Physical Review X},
  volume={6},
  number={3},
  pages={031015},
  year={2016},
  publisher={APS}
}

@article{king2019quantum,
  title={Quantum annealing amid local ruggedness and global frustration},
  author={King, James and Yarkoni, Sheir and Raymond, Jack and Ozfidan, Isil and King, Andrew D and Nevisi, Mayssam Mohammadi and Hilton, Jeremy P and McGeoch, Catherine C},
  journal={Journal of the Physical Society of Japan},
  volume={88},
  number={6},
  pages={061007},
  year={2019},
  publisher={The Physical Society of Japan}
}

@article{king2015performance,
  title={Performance of a quantum annealer on range-limited constraint satisfaction problems},
  author={King, Andrew D and Lanting, Trevor and Harris, Richard},
  journal={arXiv preprint arXiv:1502.02098},
  year={2015}
}

@article{king2015benchmarking,
  title={Benchmarking a quantum annealing processor with the time-to-target metric},
  author={King, James and Yarkoni, Sheir and Nevisi, Mayssam M and Hilton, Jeremy P and McGeoch, Catherine C},
  journal={arXiv preprint arXiv:1508.05087},
  year={2015}
}

@article{zdeborova2016statistical,
  author = {Zdeborova, Lenka and Krzakala, Florent},
  title = {Statistical physics of inference: thresholds and algorithms},
  journal = {Advances in Physics},
  volume = {65},
  number = {5},
  pages = {453-552},
  year = {2016},
  doi = {10.1080/00018732.2016.1211393},
  URL = {http://dx.doi.org/10.1080/00018732.2016.1211393}
}

@article{matplotlib,
  Author    = {Hunter, J. D.},
  Title     = {Matplotlib: A 2D graphics environment},
  Journal   = {Computing in Science \& Engineering},
  Volume    = {9},
  Number    = {3},
  Pages     = {90--95},
  publisher = {IEEE COMPUTER SOC},
  doi       = {10.1109/MCSE.2007.55},
  year      = 2007
}

@software{thomas_a_caswell_2022_6513224,
  author       = {Thomas A Caswell and
                  Michael Droettboom and
                  Antony Lee and
                  Elliott Sales de Andrade and
                  Tim Hoffmann and
                  Jody Klymak and
                  John Hunter and
                  Eric Firing and
                  David Stansby and
                  Nelle Varoquaux and
                  Jens Hedegaard Nielsen and
                  Benjamin Root and
                  Ryan May and
                  Phil Elson and
                  Jouni K. Seppänen and
                  Darren Dale and
                  Jae-Joon Lee and
                  Damon McDougall and
                  Andrew Straw and
                  Paul Hobson and
                  hannah and
                  Christoph Gohlke and
                  Adrien F. Vincent and
                  Tony S Yu and
                  Eric Ma and
                  Steven Silvester and
                  Charlie Moad and
                  Nikita Kniazev and
                  Elan Ernest and
                  Paul Ivanov},
  title        = {matplotlib/matplotlib: REL: v3.5.2},
  month        = may,
  year         = 2022,
  publisher    = {Zenodo},
  version      = {v3.5.2},
  doi          = {10.5281/zenodo.6513224},
  url          = {https://doi.org/10.5281/zenodo.6513224}
}

@techreport{hagberg2008exploring,
  title={Exploring network structure, dynamics, and function using NetworkX},
  author={Hagberg, Aric and Swart, Pieter and S Chult, Daniel},
  year={2008},
  institution={Los Alamos National Lab.(LANL), Los Alamos, NM (United States)}
}

@article{fruchterman1991graph,
  title={Graph drawing by force-directed placement},
  author={Fruchterman, Thomas MJ and Reingold, Edward M},
  journal={Software: Practice and experience},
  volume={21},
  number={11},
  pages={1129--1164},
  year={1991},
  publisher={Wiley Online Library}
}

@INPROCEEDINGS{pelofske2019peering,  
author={Pelofske, Elijah and Hahn, Georg and Djidjev, Hristo},  booktitle={2019 20th International Conference on Parallel and Distributed Computing, Applications and Technologies (PDCAT)},   title={Peering Into the Anneal Process of a Quantum Annealer},   year={2019},  volume={},  number={},  pages={184-189},  doi={10.1109/PDCAT46702.2019.00043}}

@article{preskill2018quantum,
  title={Quantum computing in the NISQ era and beyond},
  author={Preskill, John},
  journal={Quantum},
  volume={2},
  pages={79},
  year={2018},
  publisher={Verein zur F{\"o}rderung des Open Access Publizierens in den Quantenwissenschaften}
}

@article{jiang2018quantum,
  title={Quantum annealing for prime factorization},
  author={Jiang, Shuxian and Britt, Keith A and McCaskey, Alexander J and Humble, Travis S and Kais, Sabre},
  journal={Scientific reports},
  volume={8},
  number={1},
  pages={1--9},
  year={2018},
  publisher={Nature Publishing Group}
}

@article{PRXQuantum.2.040322,
  title = {Embedding Overhead Scaling of Optimization Problems in Quantum Annealing},
  author = {K\"onz, Mario S. and Lechner, Wolfgang and Katzgraber, Helmut G. and Troyer, Matthias},
  journal = {PRX Quantum},
  volume = {2},
  issue = {4},
  pages = {040322},
  numpages = {11},
  year = {2021},
  month = {Nov},
  publisher = {American Physical Society},
  doi = {10.1103/PRXQuantum.2.040322},
  url = {https://link.aps.org/doi/10.1103/PRXQuantum.2.040322}
}

@article{boyda2017deploying,
  title={Deploying a quantum annealing processor to detect tree cover in aerial imagery of California},
  author={Boyda, Edward and Basu, Saikat and Ganguly, Sangram and Michaelis, Andrew and Mukhopadhyay, Supratik and Nemani, Ramakrishna R},
  journal={PloS one},
  volume={12},
  number={2},
  pages={e0172505},
  year={2017},
  publisher={Public Library of Science San Francisco, CA USA}
}

@article{boothby2016fast,
  title={Fast clique minor generation in Chimera qubit connectivity graphs},
  author={Boothby, Tomas and King, Andrew D and Roy, Aidan},
  journal={Quantum Information Processing},
  volume={15},
  number={1},
  pages={495--508},
  year={2016},
  publisher={Springer}
}

@article{dridi2017prime,
  title={Prime factorization using quantum annealing and computational algebraic geometry},
  author={Dridi, Raouf and Alghassi, Hedayat},
  journal={Scientific reports},
  volume={7},
  number={1},
  pages={1--10},
  year={2017},
  publisher={Nature Publishing Group}
}

@article{peng2019factoring,
  title={Factoring larger integers with fewer qubits via quantum annealing with optimized parameters},
  author={Peng, WangChun and Wang, BaoNan and Hu, Feng and Wang, YunJiang and Fang, XianJin and Chen, XingYuan and Wang, Chao},
  journal={SCIENCE CHINA Physics, Mechanics \& Astronomy},
  volume={62},
  number={6},
  pages={1--8},
  year={2019},
  publisher={Springer}
}

@article{warren2019factoring,
  title={Factoring on a quantum annealing computer},
  author={Warren, Richard H},
  journal={Quantum Information \& Computation},
  volume={19},
  number={3-4},
  pages={252--261},
  year={2019},
  publisher={Rinton Press, Incorporated Paramus, NJ}
}

@incollection{grant2020adiabatic,
  title={Adiabatic quantum computing and quantum annealing},
  author={Grant, Erica K and Humble, Travis S},
  booktitle={Oxford Research Encyclopedia of Physics},
  year={2020}
}

@misc{https://doi.org/10.48550/arxiv.2207.05089,
  doi = {10.48550/ARXIV.2207.05089},
  
  url = {https://arxiv.org/abs/2207.05089},
  
  author = {Cain, Madelyn and Farhi, Edward and Gutmann, Sam and Ranard, Daniel and Tang, Eugene},
  
  title = {The QAOA gets stuck starting from a good classical string},
  
  publisher = {arXiv},
  
  year = {2022},
  
  copyright = {Creative Commons Attribution 4.0 International}
}

@article{Egger_2021,
	doi = {10.22331/q-2021-06-17-479},
  
	url = {https://doi.org/10.22331%2Fq-2021-06-17-479},
  
	year = 2021,
	month = {jun},
  
	publisher = {Verein zur Forderung des Open Access Publizierens in den Quantenwissenschaften},
  
	volume = {5},
  
	pages = {479},
  
	author = {Daniel J. Egger and Jakub Mare{\v{c}
}ek and Stefan Woerner},
  
	title = {Warm-starting quantum optimization},
  
	journal = {Quantum}
}

@article{Hadfield_2019,
	doi = {10.3390/a12020034},
  
	url = {https://doi.org/10.3390%2Fa12020034},
  
	year = 2019,
	month = {feb},
  
	publisher = {{MDPI} {AG}
},
  
	volume = {12},
  
	number = {2},
  
	pages = {34},
  
	author = {Stuart Hadfield and Zhihui Wang and Bryan O{\textquotesingle}Gorman and Eleanor Rieffel and Davide Venturelli and Rupak Biswas},
  
	title = {From the Quantum Approximate Optimization Algorithm to a Quantum Alternating Operator Ansatz},
  
	journal = {Algorithms}
}

@article{grant2022benchmarking,
  title={Benchmarking embedded chain breaking in quantum annealing},
  author={Grant, Erica and Humble, Travis S},
  journal={Quantum Science and Technology},
  volume={7},
  number={2},
  pages={025029},
  year={2022},
  publisher={IOP Publishing}
}

@article{lucas2014ising,
  title={Ising formulations of many NP problems},
  author={Lucas, Andrew},
  journal={Frontiers in physics},
  pages={5},
  year={2014},
  publisher={Frontiers}
}

@article{papalitsas2019qubo,
  title={A QUBO model for the traveling salesman problem with time windows},
  author={Papalitsas, Christos and Andronikos, Theodore and Giannakis, Konstantinos and Theocharopoulou, Georgia and Fanarioti, Sofia},
  journal={Algorithms},
  volume={12},
  number={11},
  pages={224},
  year={2019},
  publisher={MDPI}
}

@article{zaman2021pyqubo,
  title={PyQUBO: Python library for mapping combinatorial optimization problems to QUBO form},
  author={Zaman, Mashiyat and Tanahashi, Kotaro and Tanaka, Shu},
  journal={IEEE Transactions on Computers},
  volume={71},
  number={4},
  pages={838--850},
  year={2021},
  publisher={IEEE}
}

@misc{https://doi.org/10.48550/arxiv.1811.11538,
  doi = {10.48550/ARXIV.1811.11538},
  
  url = {https://arxiv.org/abs/1811.11538},
  
  author = {Glover, Fred and Kochenberger, Gary and Du, Yu},
  
  title = {A Tutorial on Formulating and Using QUBO Models},
  
  publisher = {arXiv},
  
  year = {2018},
  
  copyright = {arXiv.org perpetual, non-exclusive license}
}

@article{PhysRevX.5.031040,
  title = {Quantum Optimization of Fully Connected Spin Glasses},
  author = {Venturelli, Davide and Mandr\`a, Salvatore and Knysh, Sergey and O'Gorman, Bryan and Biswas, Rupak and Smelyanskiy, Vadim},
  journal = {Phys. Rev. X},
  volume = {5},
  issue = {3},
  pages = {031040},
  numpages = {8},
  year = {2015},
  month = {Sep},
  publisher = {American Physical Society},
  doi = {10.1103/PhysRevX.5.031040},
  url = {https://link.aps.org/doi/10.1103/PhysRevX.5.031040}
}

@article{santoro2006optimization,
  title={Optimization using quantum mechanics: quantum annealing through adiabatic evolution},
  author={Santoro, Giuseppe E and Tosatti, Erio},
  journal={Journal of Physics A: Mathematical and General},
  volume={39},
  number={36},
  pages={R393},
  year={2006},
  publisher={IOP Publishing},
  doi={}
}

@article{barahona1982computational,
  title={On the computational complexity of Ising spin glass models},
  author={Barahona, Francisco},
  journal={Journal of Physics A: Mathematical and General},
  volume={15},
  number={10},
  pages={3241},
  year={1982},
  publisher={IOP Publishing}
}

@misc{https://doi.org/10.48550/arxiv.2202.03044,
  doi = {10.48550/ARXIV.2202.03044},
  
  url = {https://arxiv.org/abs/2202.03044},
  
  author = {Raymond, Jack and Stevanovic, Radomir and Bernoudy, William and Boothby, Kelly and McGeoch, Catherine and Berkley, Andrew J. and Farré, Pau and King, Andrew D.},
  
  title = {Hybrid quantum annealing for larger-than-QPU lattice-structured problems},
  
  publisher = {arXiv},
  
  year = {2022},
  
  copyright = {arXiv.org perpetual, non-exclusive license}
}

@article{PhysRevB.95.184416,
  title = {Nonstoquastic Hamiltonians and quantum annealing of an Ising spin glass},
  author = {Hormozi, Layla and Brown, Ethan W. and Carleo, Giuseppe and Troyer, Matthias},
  journal = {Phys. Rev. B},
  volume = {95},
  issue = {18},
  pages = {184416},
  numpages = {9},
  year = {2017},
  month = {May},
  publisher = {American Physical Society},
  doi = {10.1103/PhysRevB.95.184416},
  url = {https://link.aps.org/doi/10.1103/PhysRevB.95.184416}
}

@article{mishra2016performance,
  title={Performance of two different quantum annealing correction codes},
  author={Mishra, Anurag and Albash, Tameem and Lidar, Daniel A},
  journal={Quantum Information Processing},
  volume={15},
  number={2},
  pages={609--636},
  year={2016},
  publisher={Springer}
}

@article{pudenz2015quantum,
  title={Quantum annealing correction for random Ising problems},
  author={Pudenz, Kristen L and Albash, Tameem and Lidar, Daniel A},
  journal={Physical Review A},
  volume={91},
  number={4},
  pages={042302},
  year={2015},
  publisher={APS}
}

@article{vinci2015quantum,
  title={Quantum annealing correction with minor embedding},
  author={Vinci, Walter and Albash, Tameem and Paz-Silva, Gerardo and Hen, Itay and Lidar, Daniel A},
  journal={Physical Review A},
  volume={92},
  number={4},
  pages={042310},
  year={2015},
  publisher={APS}
}

@misc{https://doi.org/10.48550/arxiv.2104.01941,
  doi = {10.48550/ARXIV.2104.01941},
  
  url = {https://arxiv.org/abs/2104.01941},
  
  author = {Mizuno, Yuta and Komatsuzaki, Tamiki},
  
  title = {A Note on Enumeration by Fair Sampling},
  
  publisher = {arXiv},
  
  year = {2021},
  
  copyright = {arXiv.org perpetual, non-exclusive license}
}

@article{weaver2012satisfiability,
  title={Satisfiability-based set membership filters},
  author={Weaver, Sean A and Ray, Katrina J and Marek, Victor W and Mayer, Andrew J and Walker, Alden K},
  journal={Journal on Satisfiability, Boolean Modeling and Computation},
  volume={8},
  number={3-4},
  pages={129--148},
  year={2012},
  publisher={IOS Press}
}

@inproceedings{schaefer1978complexity,
  title={The complexity of satisfiability problems},
  author={Schaefer, Thomas J},
  booktitle={Proceedings of the tenth annual ACM symposium on Theory of computing},
  pages={216--226},
  year={1978}
}

@inproceedings{douglass2015constructing,
  title={Constructing SAT filters with a quantum annealer},
  author={Douglass, Adam and King, Andrew D and Raymond, Jack},
  booktitle={International Conference on Theory and Applications of Satisfiability Testing},
  pages={104--120},
  year={2015},
  organization={Springer}
}

@article{azinovic2017assessment,
  title={Assessment of quantum annealing for the construction of satisfiability filters},
  author={Azinovi{\'c}, Marlon and Herr, Daniel and Heim, Bettina and Brown, Ethan and Troyer, Matthias},
  journal={SciPost Physics},
  volume={2},
  number={2},
  pages={013},
  year={2017}
}

@article{jerrum1986random,
  title={Random generation of combinatorial structures from a uniform distribution},
  author={Jerrum, Mark R and Valiant, Leslie G and Vazirani, Vijay V},
  journal={Theoretical computer science},
  volume={43},
  pages={169--188},
  year={1986},
  publisher={Elsevier}
}

@incollection{gomes2021model,
  title={Model counting},
  author={Gomes, Carla P and Sabharwal, Ashish and Selman, Bart},
  booktitle={Handbook of satisfiability},
  pages={993--1014},
  year={2021},
  publisher={IOS press}
}

@article{PhysRevE.99.043306,
  title = {Feeding the multitude: A polynomial-time algorithm to improve sampling},
  author = {Ochoa, Andrew J. and Jacob, Darryl C. and Mandr\`a, Salvatore and Katzgraber, Helmut G.},
  journal = {Phys. Rev. E},
  volume = {99},
  issue = {4},
  pages = {043306},
  numpages = {15},
  year = {2019},
  month = {Apr},
  publisher = {American Physical Society},
  doi = {10.1103/PhysRevE.99.043306},
  url = {https://link.aps.org/doi/10.1103/PhysRevE.99.043306}
}

@article{hinton2002training,
  title={Training products of experts by minimizing contrastive divergence},
  author={Hinton, Geoffrey E},
  journal={Neural computation},
  volume={14},
  number={8},
  pages={1771--1800},
  year={2002},
  publisher={MIT Press}
}

@article{eslami2014shape,
  title={The shape boltzmann machine: a strong model of object shape},
  author={Eslami, SM and Heess, Nicolas and Williams, Christopher KI and Winn, John},
  journal={International Journal of Computer Vision},
  volume={107},
  number={2},
  pages={155--176},
  year={2014},
  publisher={Springer}
}

@article{matsuda2009ground,
  title={Ground-state statistics from annealing algorithms: quantum versus classical approaches},
  author={Matsuda, Yoshiki and Nishimori, Hidetoshi and Katzgraber, Helmut G},
  journal={New Journal of Physics},
  volume={11},
  number={7},
  pages={073021},
  year={2009},
  publisher={IOP Publishing}
}

@article{ohkuwa2018reverse,
  title={Reverse annealing for the fully connected p-spin model},
  author={Ohkuwa, Masaki and Nishimori, Hidetoshi and Lidar, Daniel A},
  journal={Physical Review A},
  volume={98},
  number={2},
  pages={022314},
  year={2018},
  publisher={APS}
}

@article{golden2021reverse,
  title={Reverse annealing for nonnegative/binary matrix factorization},
  author={Golden, John and O’Malley, Daniel},
  journal={Plos one},
  volume={16},
  number={1},
  pages={e0244026},
  year={2021},
  publisher={Public Library of Science San Francisco, CA USA}
}

@article{venturelli2019reverse,
  title={Reverse quantum annealing approach to portfolio optimization problems},
  author={Venturelli, Davide and Kondratyev, Alexei},
  journal={Quantum Machine Intelligence},
  volume={1},
  number={1},
  pages={17--30},
  year={2019},
  publisher={Springer}
}

@article{harris2018phase,
  title={Phase transitions in a programmable quantum spin glass simulator},
  author={Harris, R and Sato, Y and Berkley, AJ and Reis, M and Altomare, F and Amin, MH and Boothby, K and Bunyk, P and Deng, C and Enderud, C and others},
  journal={Science},
  volume={361},
  number={6398},
  pages={162--165},
  year={2018},
  publisher={American Association for the Advancement of Science}
}

@article{wronski2021practical,
  title={Practical solving of discrete logarithm problem over prime fields using quantum annealing},
  author={Wro{\'n}ski, Micha{\l}},
  journal={Cryptology ePrint Archive},
  year={2021}
}

@article{perdomo2011study,
  title={A study of heuristic guesses for adiabatic quantum computation},
  author={Perdomo-Ortiz, Alejandro and Venegas-Andraca, Salvador E and Aspuru-Guzik, Al{\'a}n},
  journal={Quantum Information Processing},
  volume={10},
  number={1},
  pages={33--52},
  year={2011},
  publisher={Springer},
  doi={10.1007/s11128-010-0168-z}
}

@misc{https://doi.org/10.48550/arxiv.quant-ph/0001106,
  doi = {10.48550/ARXIV.QUANT-PH/0001106},
  
  url = {https://arxiv.org/abs/quant-ph/0001106},
  
  author = {Farhi, Edward and Goldstone, Jeffrey and Gutmann, Sam and Sipser, Michael},
  
  title = {Quantum Computation by Adiabatic Evolution},
  
  publisher = {arXiv},
  
  year = {2000},
  
  copyright = {Assumed arXiv.org perpetual, non-exclusive license to distribute this article for submissions made before January 2004}
}

@article{chancellor2017modernizing,
  title={Modernizing quantum annealing using local searches},
  author={Chancellor, Nicholas},
  journal={New Journal of Physics},
  volume={19},
  number={2},
  pages={023024},
  year={2017},
  publisher={IOP Publishing}
}

@article{marshall2019power,
  title={Power of pausing: Advancing understanding of thermalization in experimental quantum annealers},
  author={Marshall, Jeffrey and Venturelli, Davide and Hen, Itay and Rieffel, Eleanor G},
  journal={Physical Review Applied},
  volume={11},
  number={4},
  pages={044083},
  year={2019},
  publisher={APS}
}

@article{PhysRevA.105.032431,
  title = {Standard quantum annealing outperforms adiabatic reverse annealing with decoherence},
  author = {Passarelli, Gianluca and Yip, Ka-Wa and Lidar, Daniel A. and Lucignano, Procolo},
  journal = {Phys. Rev. A},
  volume = {105},
  issue = {3},
  pages = {032431},
  numpages = {12},
  year = {2022},
  month = {Mar},
  publisher = {American Physical Society},
  doi = {10.1103/PhysRevA.105.032431},
  url = {https://link.aps.org/doi/10.1103/PhysRevA.105.032431}
}

@article{PhysRevA.100.052321,
  title = {Dynamics of reverse annealing for the fully connected $p$-spin model},
  author = {Yamashiro, Yu and Ohkuwa, Masaki and Nishimori, Hidetoshi and Lidar, Daniel A.},
  journal = {Phys. Rev. A},
  volume = {100},
  issue = {5},
  pages = {052321},
  numpages = {11},
  year = {2019},
  month = {Nov},
  publisher = {American Physical Society},
  doi = {10.1103/PhysRevA.100.052321},
  url = {https://link.aps.org/doi/10.1103/PhysRevA.100.052321}
}

@misc{https://doi.org/10.48550/arxiv.2012.04470,
  doi = {10.48550/ARXIV.2012.04470},
  
  url = {https://arxiv.org/abs/2012.04470},
  
  author = {Kwok, Julia and Pudenz, Kristen},
  
  title = {Graph Coloring with Quantum Annealing},
  
  publisher = {arXiv},
  
  year = {2020},
  
  copyright = {arXiv.org perpetual, non-exclusive license}
}

@misc{https://doi.org/10.48550/arxiv.2005.02268,
  doi = {10.48550/ARXIV.2005.02268},
  
  url = {https://arxiv.org/abs/2005.02268},
  
  author = {Mengoni, Riccardo and Ottaviani, Daniele and Iorio, Paolino},
  
  title = {Breaking RSA Security With A Low Noise D-Wave 2000Q Quantum Annealer: Computational Times, Limitations And Prospects},
  
  publisher = {arXiv},
  
  year = {2020},
  
  copyright = {arXiv.org perpetual, non-exclusive license}
}

@misc{https://doi.org/10.48550/arxiv.2208.09068,
  doi = {10.48550/ARXIV.2208.09068},
  
  url = {https://arxiv.org/abs/2208.09068},
  
  author = {Morrell, Zachary and Vuffray, Marc and Lokhov, Andrey and Bärtschi, Andreas and Albash, Tameem and Coffrin, Carleton},
  
  title = {Signatures of Open and Noisy Quantum Systems in Single-Qubit Quantum Annealing},
  
  publisher = {arXiv},
  
  year = {2022},
  
  copyright = {arXiv.org perpetual, non-exclusive license}
}

@article{munoz2019double,
  title={A double-slit proposal for quantum annealing},
  author={Munoz-Bauza, Humberto and Chen, Huo and Lidar, Daniel},
  journal={npj Quantum Information},
  volume={5},
  number={1},
  pages={1--11},
  year={2019},
  publisher={Nature Publishing Group}
}

@article{doi:10.1126/science.284.5415.779,
author = {J. Brooke  and D. Bitko  and T. F.  and null Rosenbaum  and G. Aeppli },
title = {Quantum Annealing of a Disordered Magnet},
journal = {Science},
volume = {284},
number = {5415},
pages = {779-781},
year = {1999},
doi = {10.1126/science.284.5415.779},
URL = {https://www.science.org/doi/abs/10.1126/science.284.5415.779},
eprint = {https://www.science.org/doi/pdf/10.1126/science.284.5415.779}}

@ARTICLE{2020SciPy-NMeth,
  author  = {Virtanen, Pauli and Gommers, Ralf and Oliphant, Travis E. and
            Haberland, Matt and Reddy, Tyler and Cournapeau, David and
            Burovski, Evgeni and Peterson, Pearu and Weckesser, Warren and
            Bright, Jonathan and {van der Walt}, St{\'e}fan J. and
            Brett, Matthew and Wilson, Joshua and Millman, K. Jarrod and
            Mayorov, Nikolay and Nelson, Andrew R. J. and Jones, Eric and
            Kern, Robert and Larson, Eric and Carey, C J and
            Polat, {\.I}lhan and Feng, Yu and Moore, Eric W. and
            {VanderPlas}, Jake and Laxalde, Denis and Perktold, Josef and
            Cimrman, Robert and Henriksen, Ian and Quintero, E. A. and
            Harris, Charles R. and Archibald, Anne M. and
            Ribeiro, Ant{\^o}nio H. and Pedregosa, Fabian and
            {van Mulbregt}, Paul and {SciPy 1.0 Contributors}},
  title   = {{{SciPy} 1.0: Fundamental Algorithms for Scientific
            Computing in Python}},
  journal = {Nature Methods},
  year    = {2020},
  volume  = {17},
  pages   = {261--272},
  adsurl  = {https://rdcu.be/b08Wh},
  doi     = {10.1038/s41592-019-0686-2},
}

@article{osti_1498001,
title = {Challenges with Chains: Testing the Limits of a D-Wave Quantum Annealer for Discrete Optimization},
author = {Coffrin, Carleton James},
abstractNote = {},
doi = {10.2172/1498001},
url = {https://www.osti.gov/biblio/1498001}, journal = {},
place = {United States},
year = {2019},
month = {2}
}

@dataset{elijah_pelofske_2023_7676291,
  author       = {Elijah Pelofske},
  title        = {{Dataset for Mapping State Transition 
                   Susceptibility in Quantum Annealing Part 1}},
  month        = feb,
  year         = 2023,
  publisher    = {Zenodo},
  doi          = {10.5281/zenodo.7676291},
  url          = {https://doi.org/10.5281/zenodo.7676291}
}

@dataset{elijah_pelofske_2023_7676350,
  author       = {Elijah Pelofske},
  title        = {{Dataset for Mapping State Transition 
                   Susceptibility in Quantum Annealing Part 2}},
  month        = feb,
  year         = 2023,
  publisher    = {Zenodo},
  doi          = {10.5281/zenodo.7676350},
  url          = {https://doi.org/10.5281/zenodo.7676350}
}

@dataset{elijah_pelofske_2023_7676389,
  author       = {Elijah Pelofske},
  title        = {{Dataset for Mapping State Transition 
                   Susceptibility in Quantum Annealing Part 3}},
  month        = feb,
  year         = 2023,
  publisher    = {Zenodo},
  doi          = {10.5281/zenodo.7676389},
  url          = {https://doi.org/10.5281/zenodo.7676389}
}

@dataset{elijah_pelofske_2023_7677424,
  author       = {Elijah Pelofske},
  title        = {{Dataset for Mapping State Transition 
                   Susceptibility in Quantum Annealing Part 4}},
  month        = feb,
  year         = 2023,
  publisher    = {Zenodo},
  doi          = {10.5281/zenodo.7677424},
  url          = {https://doi.org/10.5281/zenodo.7677424}
}

\appendix
\section{Additional figures}
\label{section:appendix_extra_figures}

\begin{figure}[h!]
    \centering
    \includegraphics[width=0.49\textwidth]{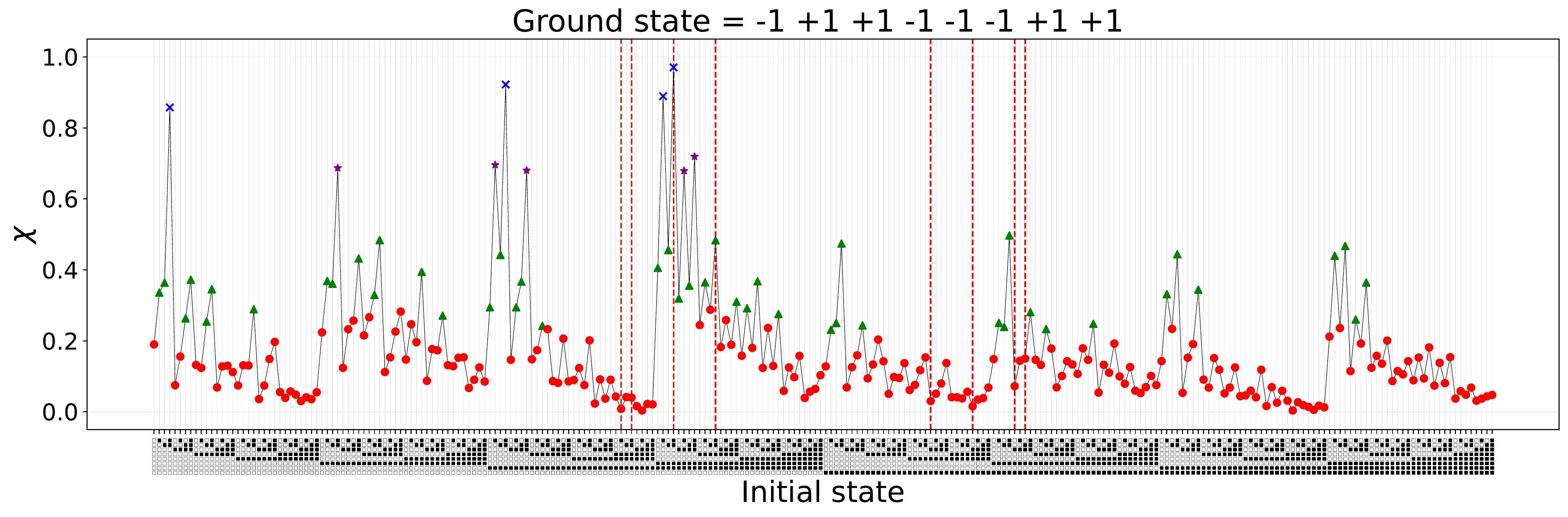}
    \includegraphics[width=0.49\textwidth]{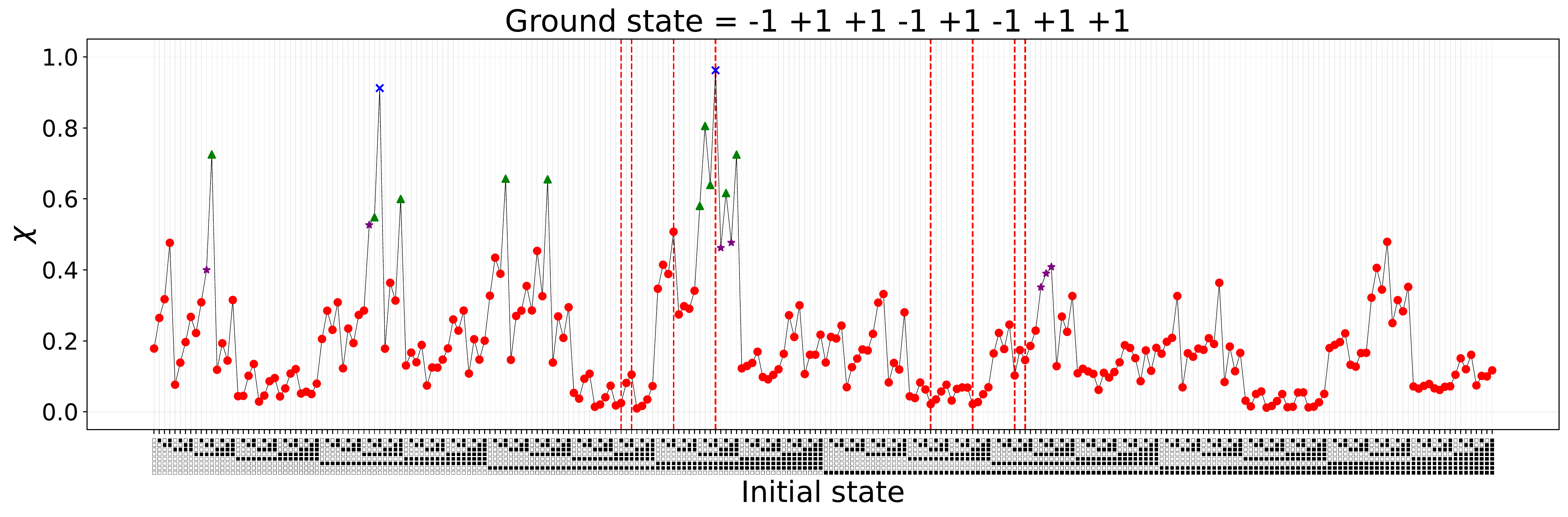}
    \includegraphics[width=0.49\textwidth]{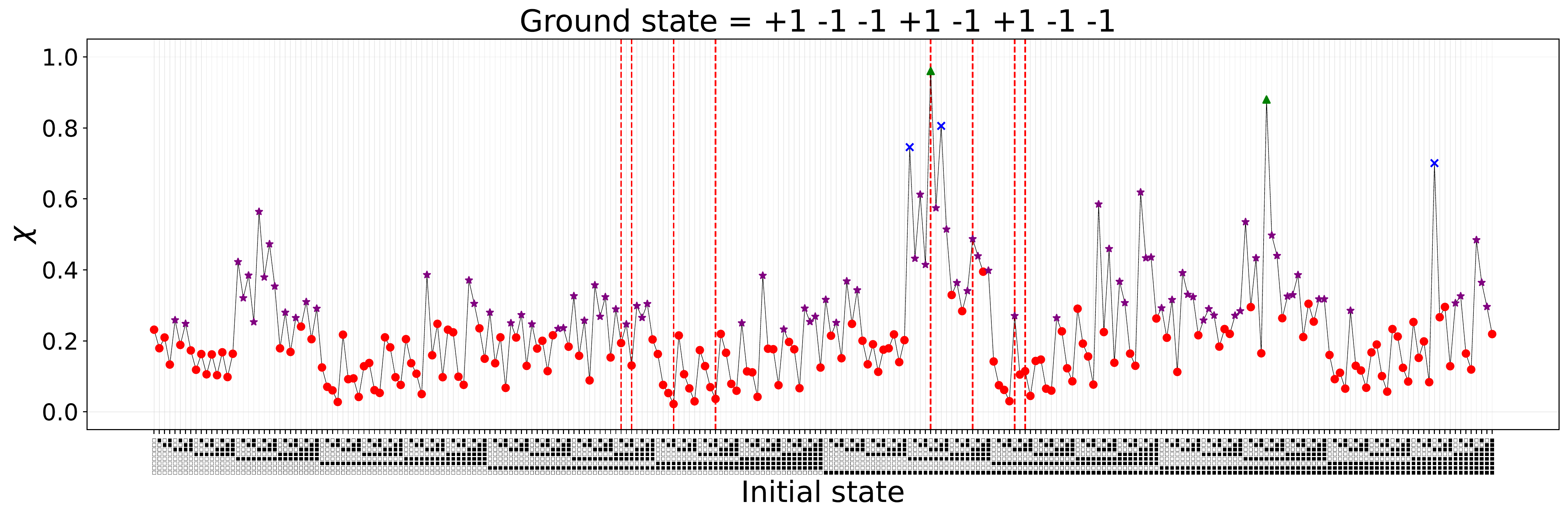}
    \includegraphics[width=0.49\textwidth]{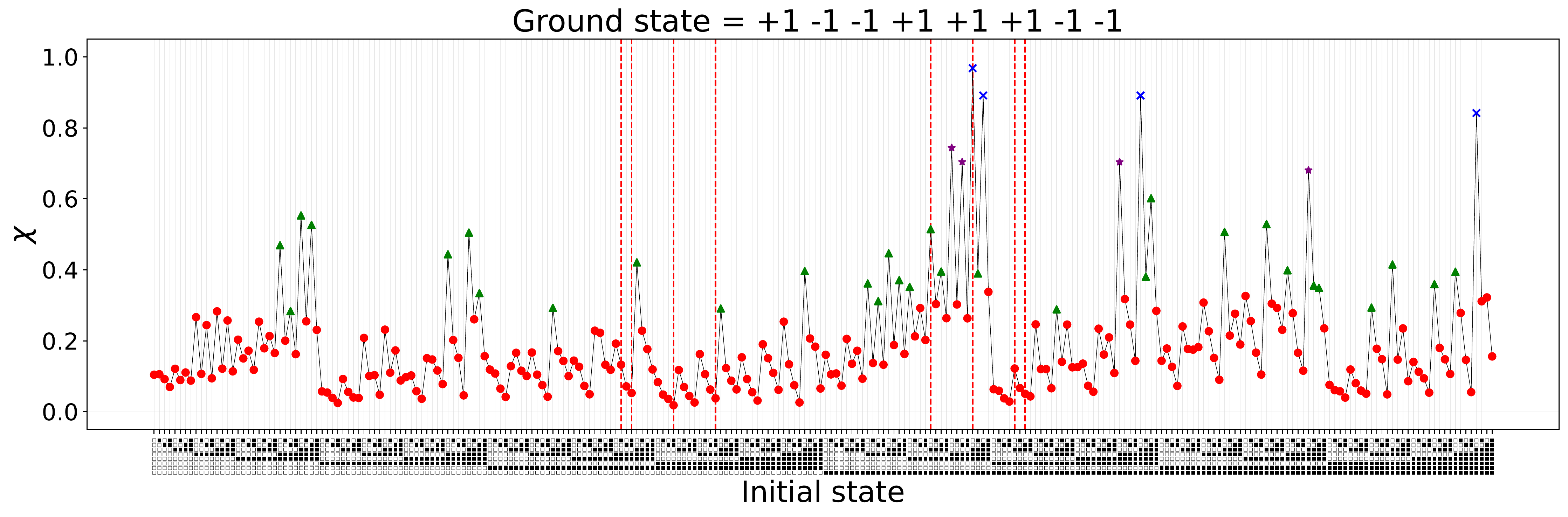}
    \includegraphics[width=0.49\textwidth]{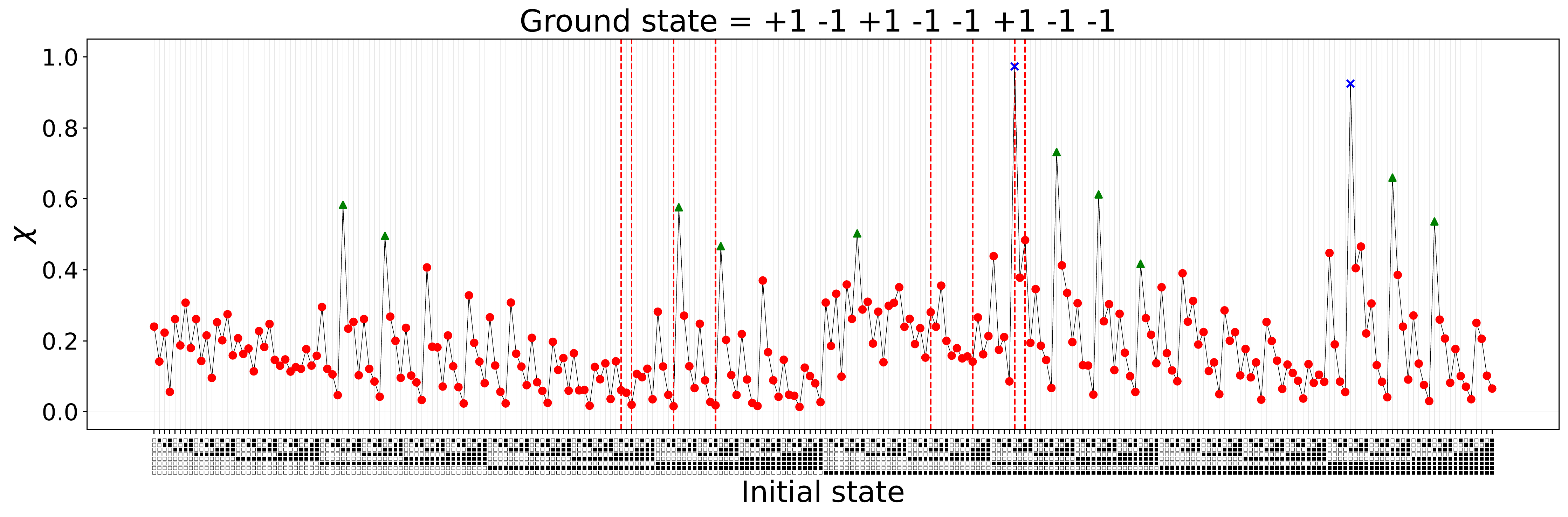}
    \includegraphics[width=0.49\textwidth]{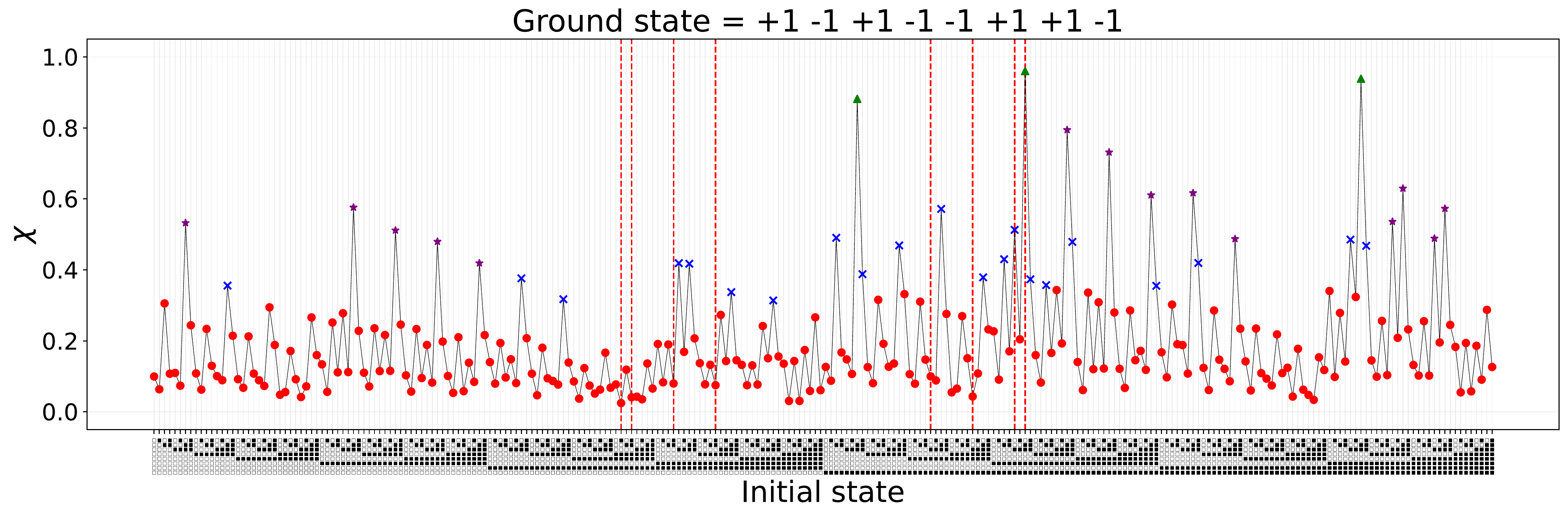}
    \caption{These plots extend Figure \ref{fig:HGain_susceptibility_to_groundstate_n8} with the other six ground states. The x-axis encodes the initial states as vectors of vertical blocks where $\blacksquare$ denotes a variable state of $+1$ and $\square$ denotes a variable state of $-1$. The initial state vectors are read from bottom to top where the bottom is the first index which corresponds to variable $0$ in the problem Ising. The initial states which are also other ground states are marked with dashed red vertical lines. For each sub-figure, the reflexive ground state mapping (i.e. where the initial state and the intended state are the same ground state) case can be found visually as the state marked with a red vertical dashed line which has the maximum susceptibility among all of the initial states. Although this plot is considerably more dense compared to Figures \ref{fig:HGain_susceptibility_to_groundstate_n6} and \ref{fig:HGain_susceptibility_to_groundstate_n7} because of the increase in initial states ($256$) and the increase in ground states ($8$), there are symmetries that can be clearly identified. In particular, complementary ground states show some symmetries which can be identified by the h-gain response curve clustering coloring scheme. For example, the top left hand sub-figure and the middle right hand sub-figure represent complementary ground states and these plots have a symmetry across the vertical midline of each figure; the lowest susceptibility states in these two plots are complements of each other (blue colored nodes in the top left hand sub-figure and green colored nodes in the bottom right hand sub-figure). }
    \label{fig:appendix_HGain_susceptibility_n8}
\end{figure}

\begin{figure}[h!]
    \centering
    \includegraphics[width=0.13\textwidth]{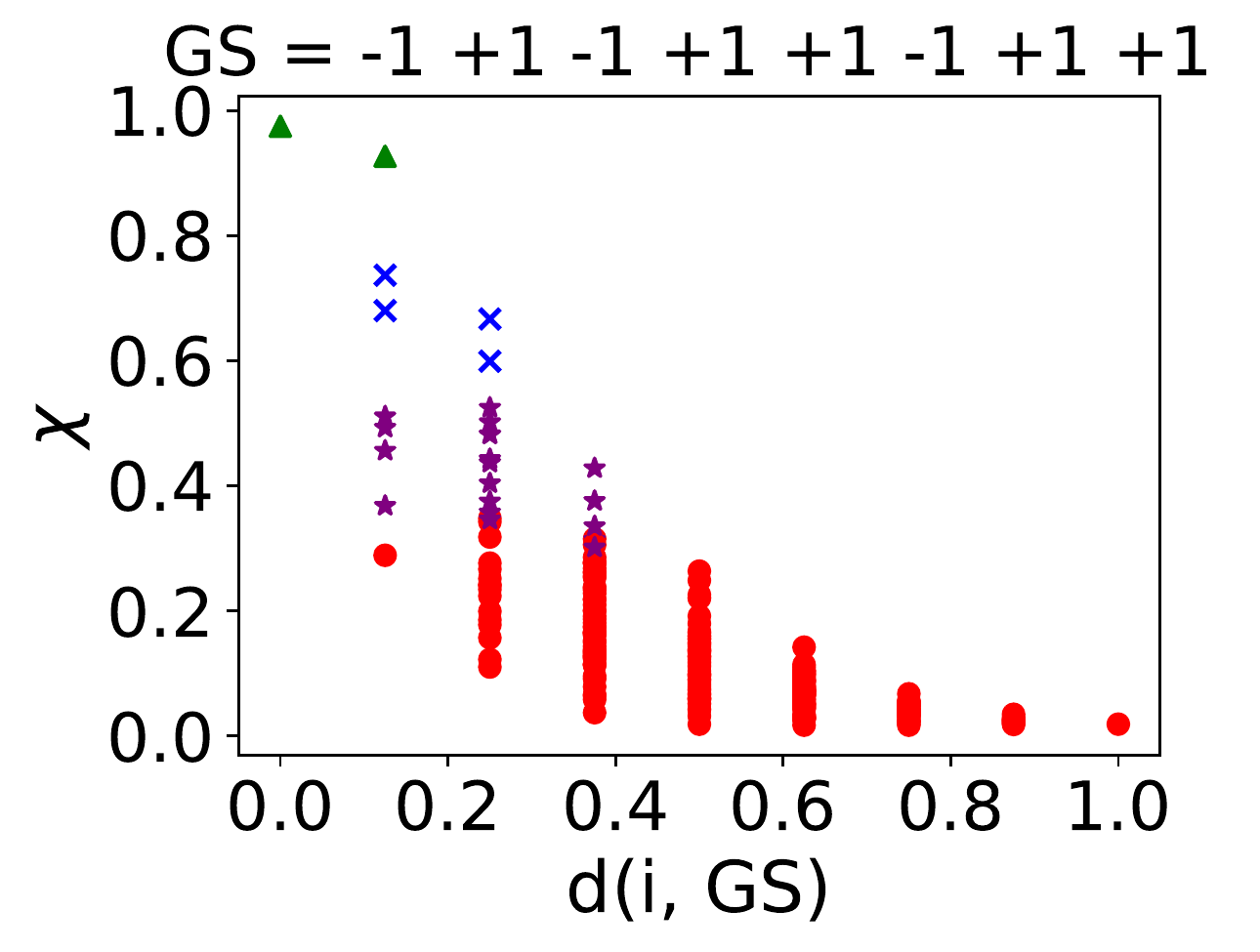}
    \includegraphics[width=0.13\textwidth]{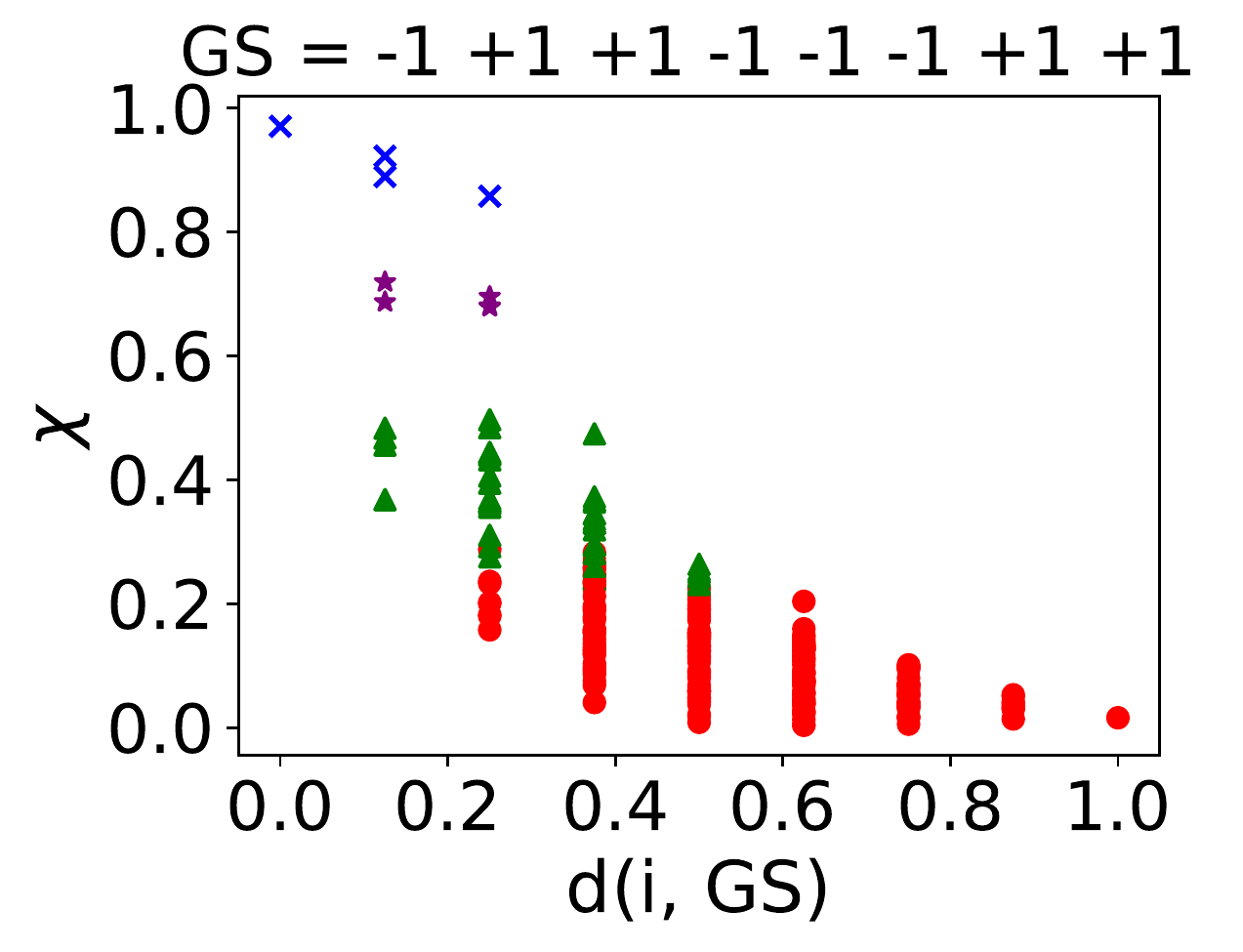}
    \includegraphics[width=0.13\textwidth]{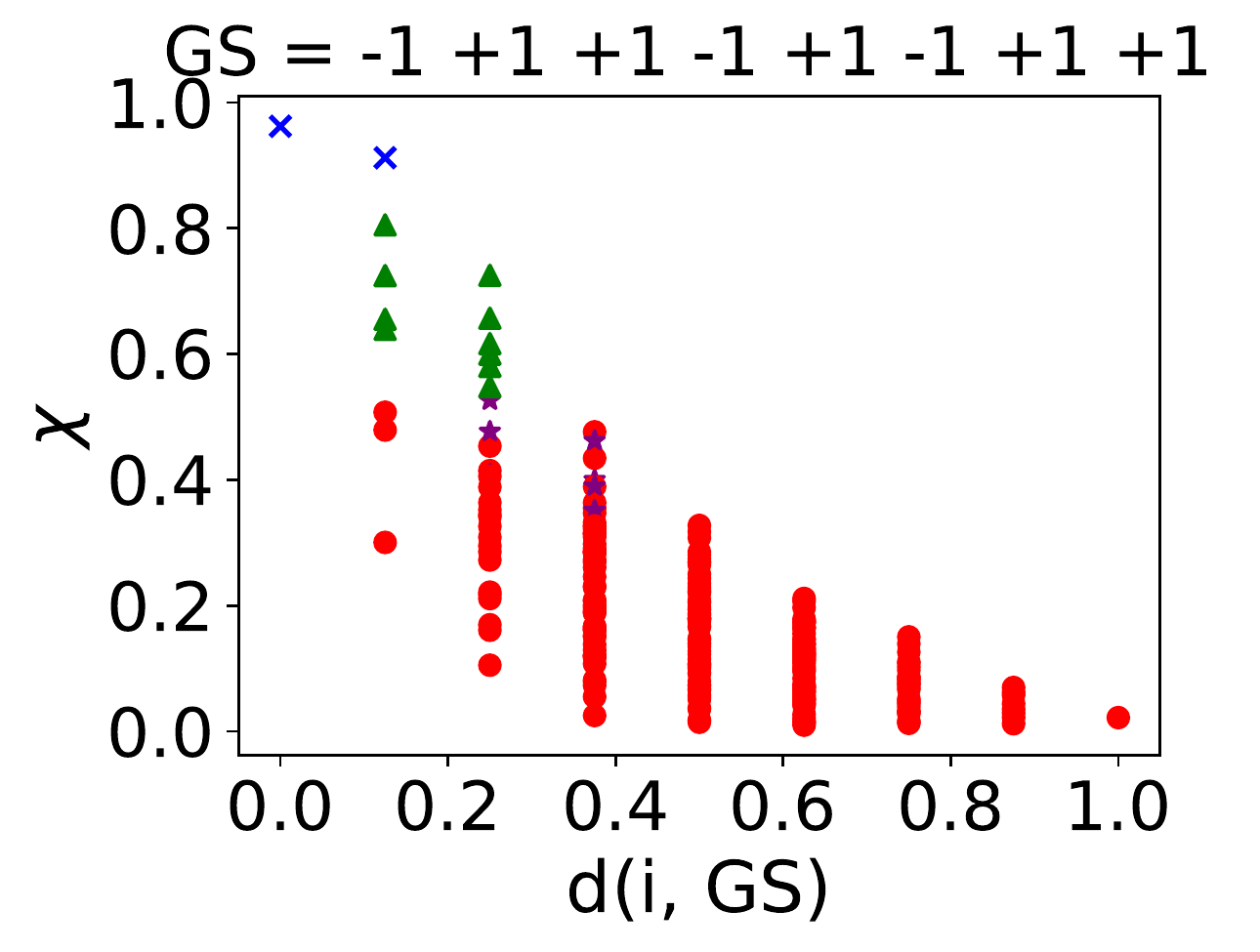}
    \includegraphics[width=0.13\textwidth]{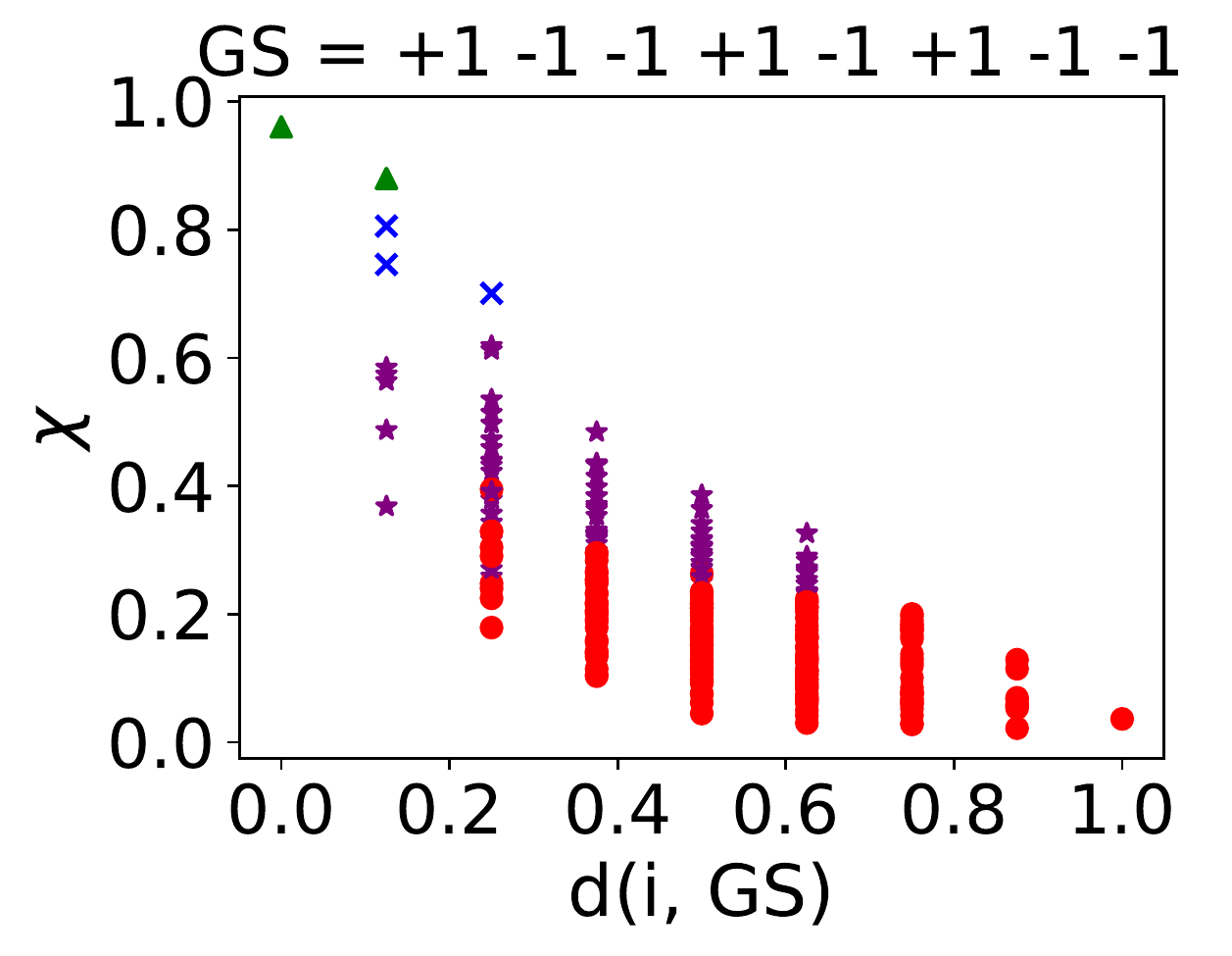}
    \includegraphics[width=0.13\textwidth]{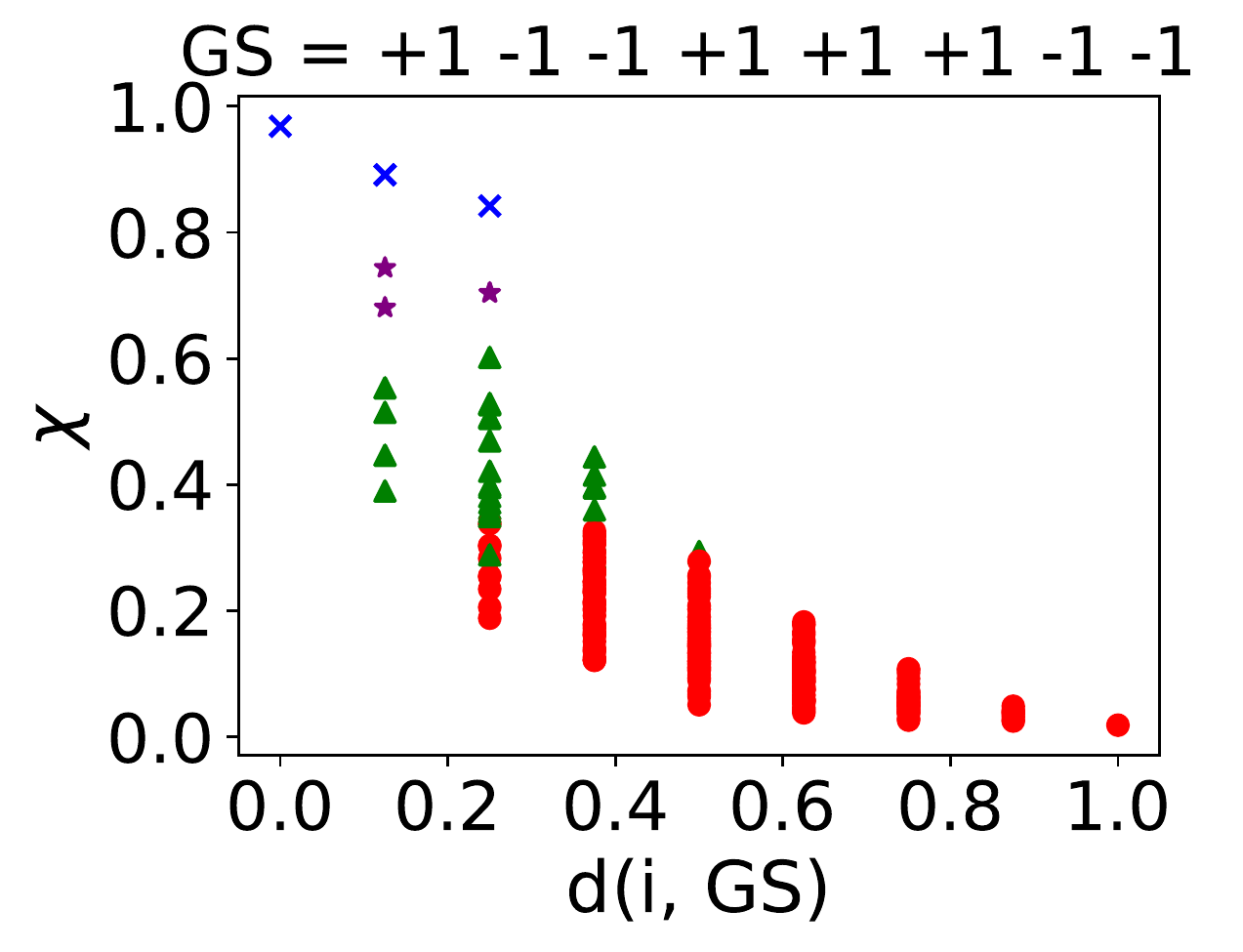}
    \includegraphics[width=0.13\textwidth]{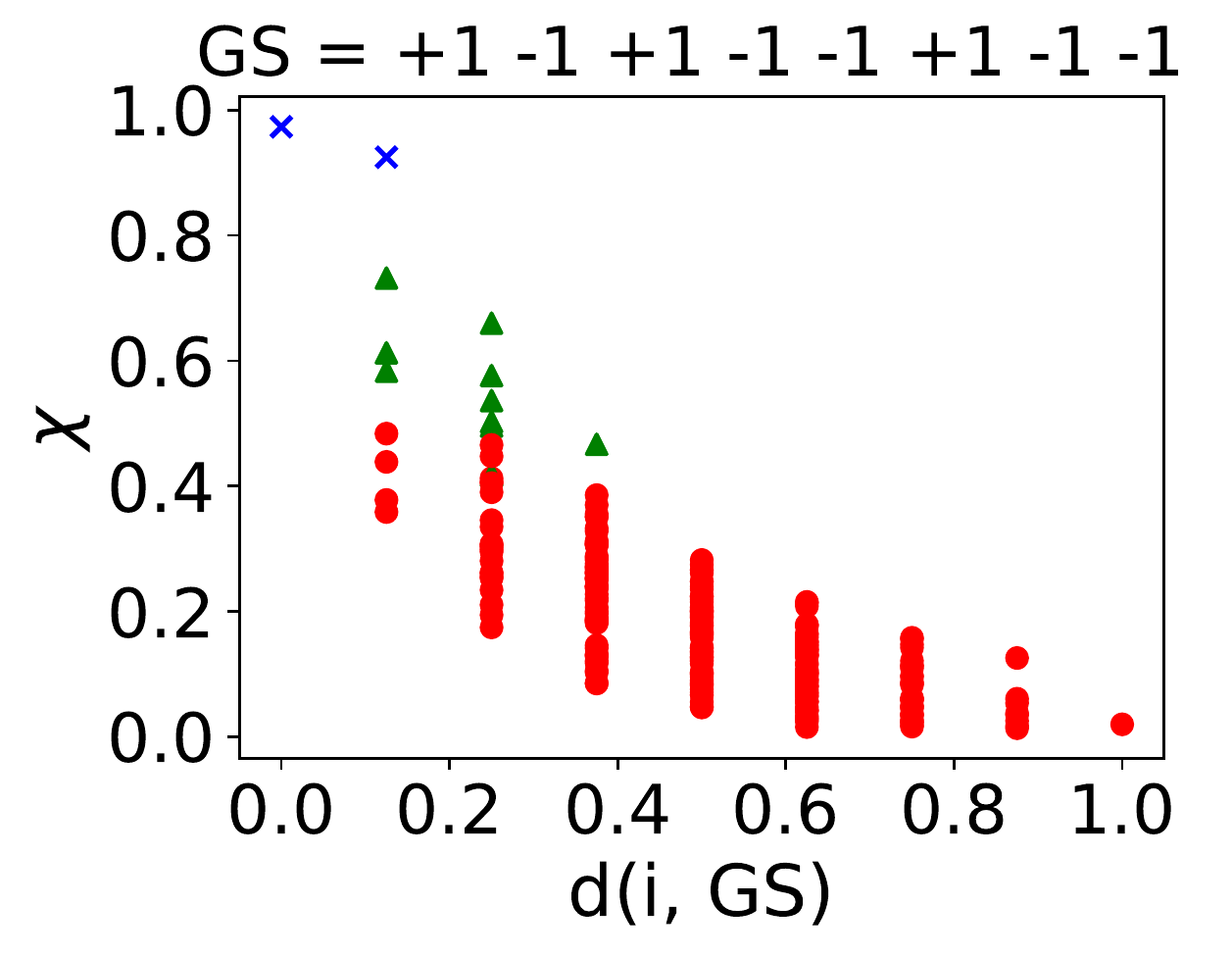}
    \includegraphics[width=0.13\textwidth]{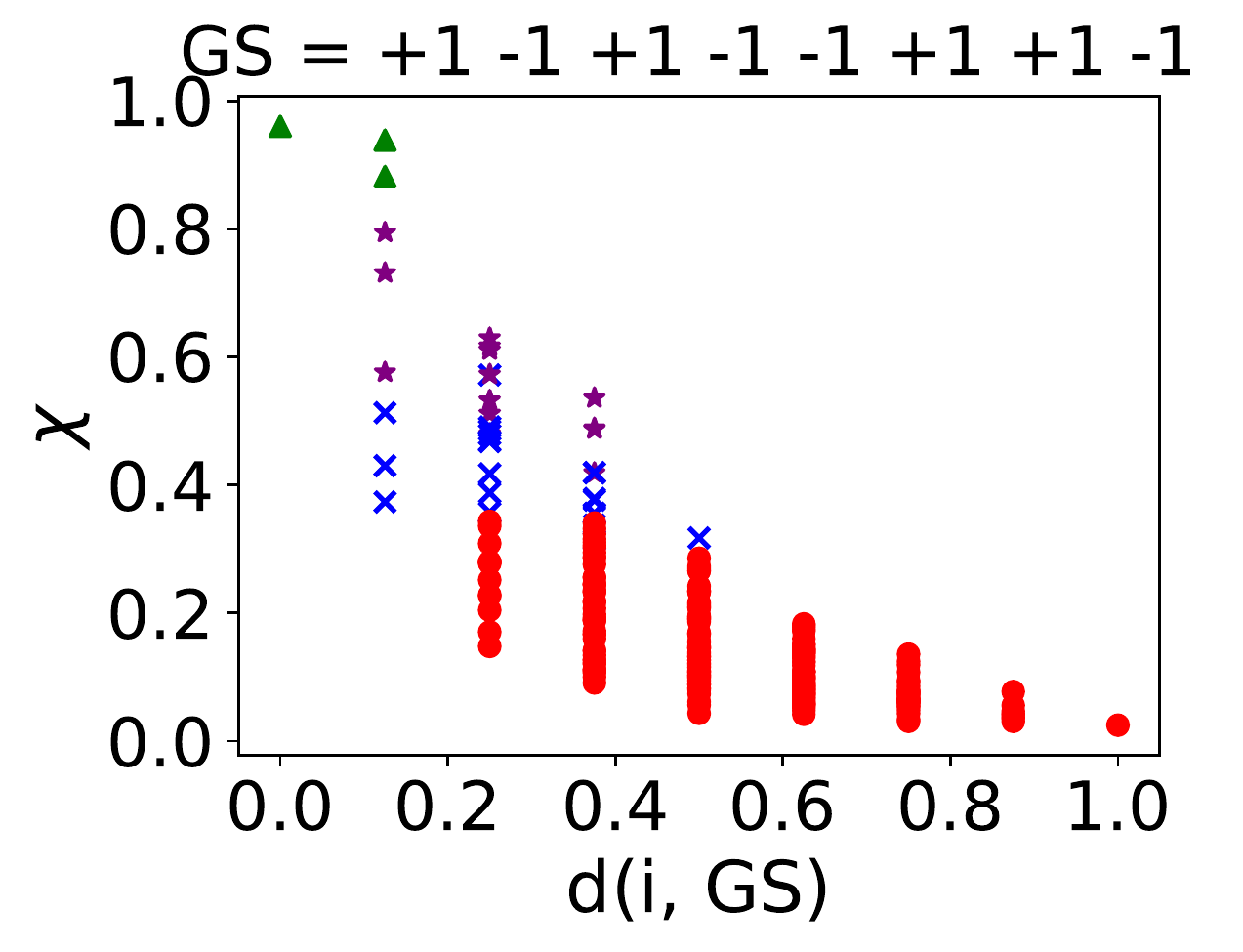}\\
    \includegraphics[width=0.13\textwidth]{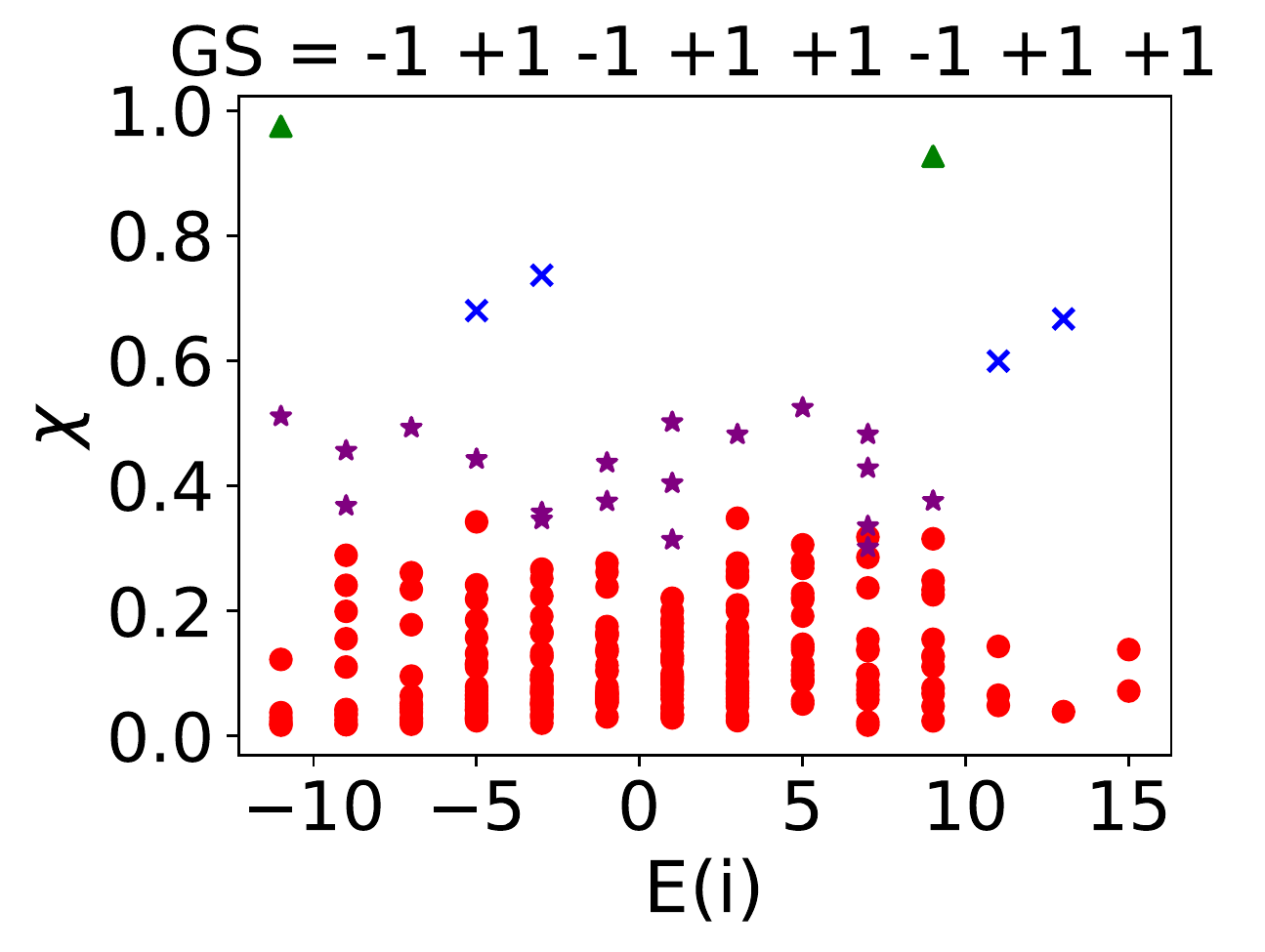}
    \includegraphics[width=0.13\textwidth]{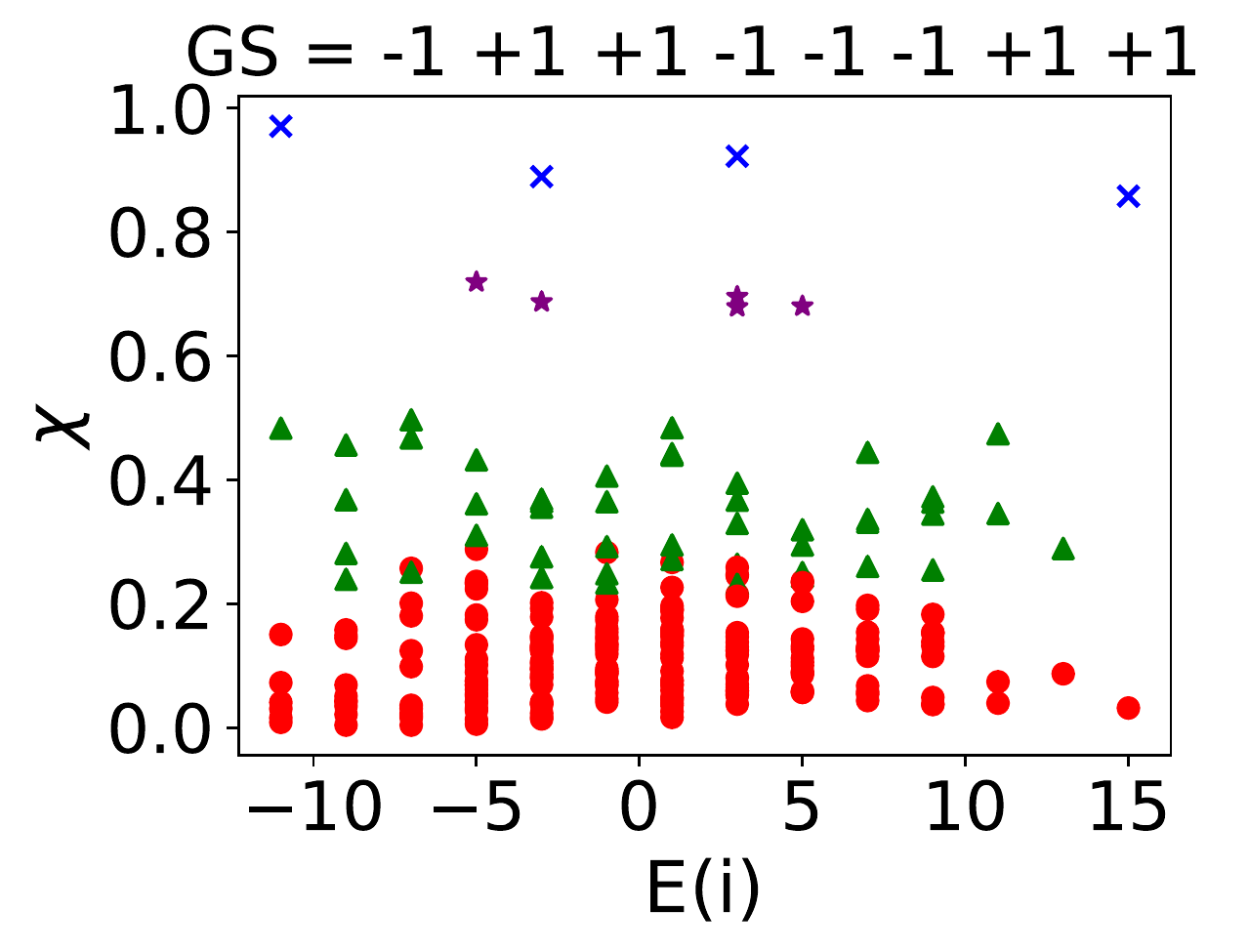}
    \includegraphics[width=0.13\textwidth]{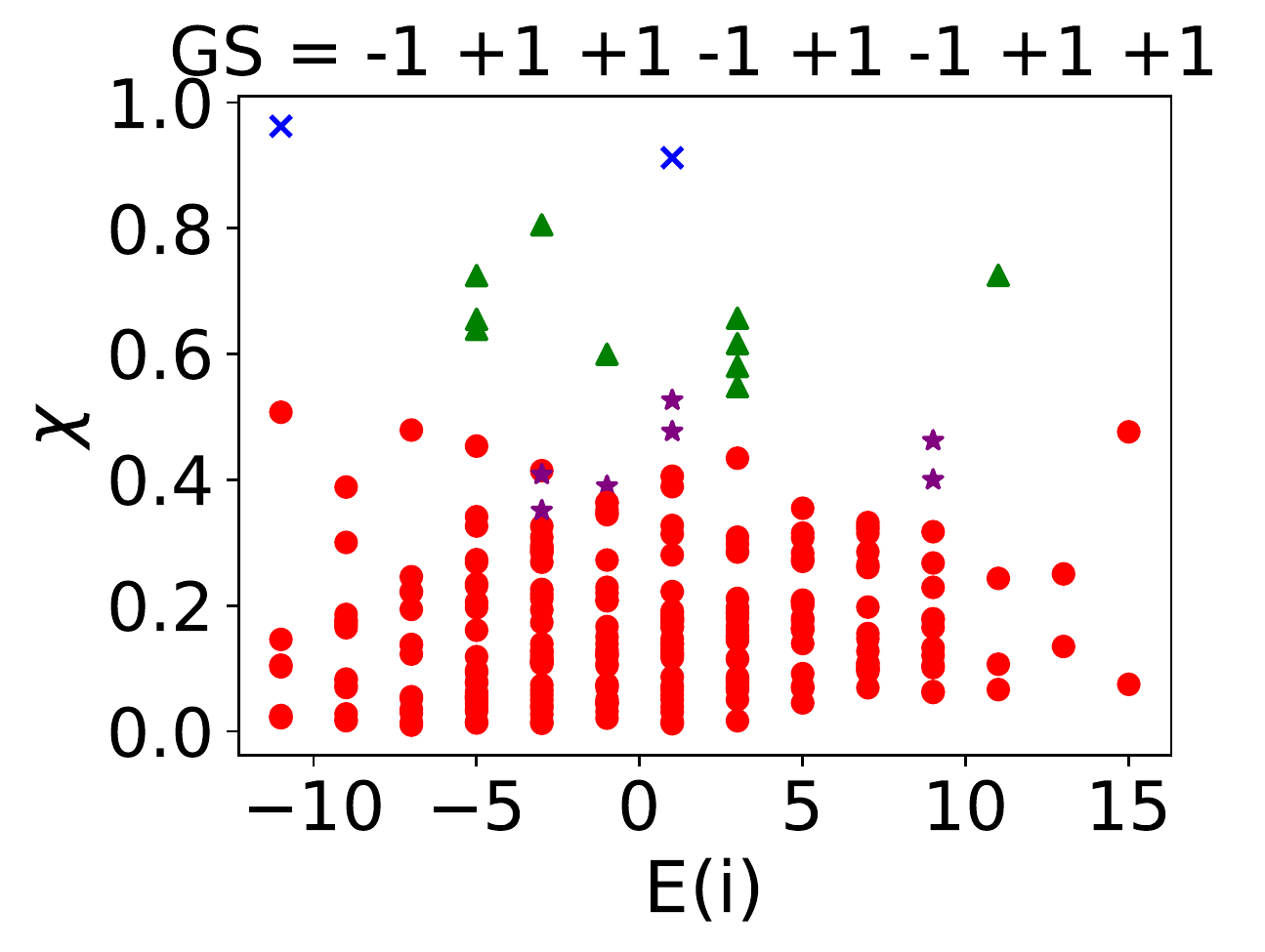}
    \includegraphics[width=0.13\textwidth]{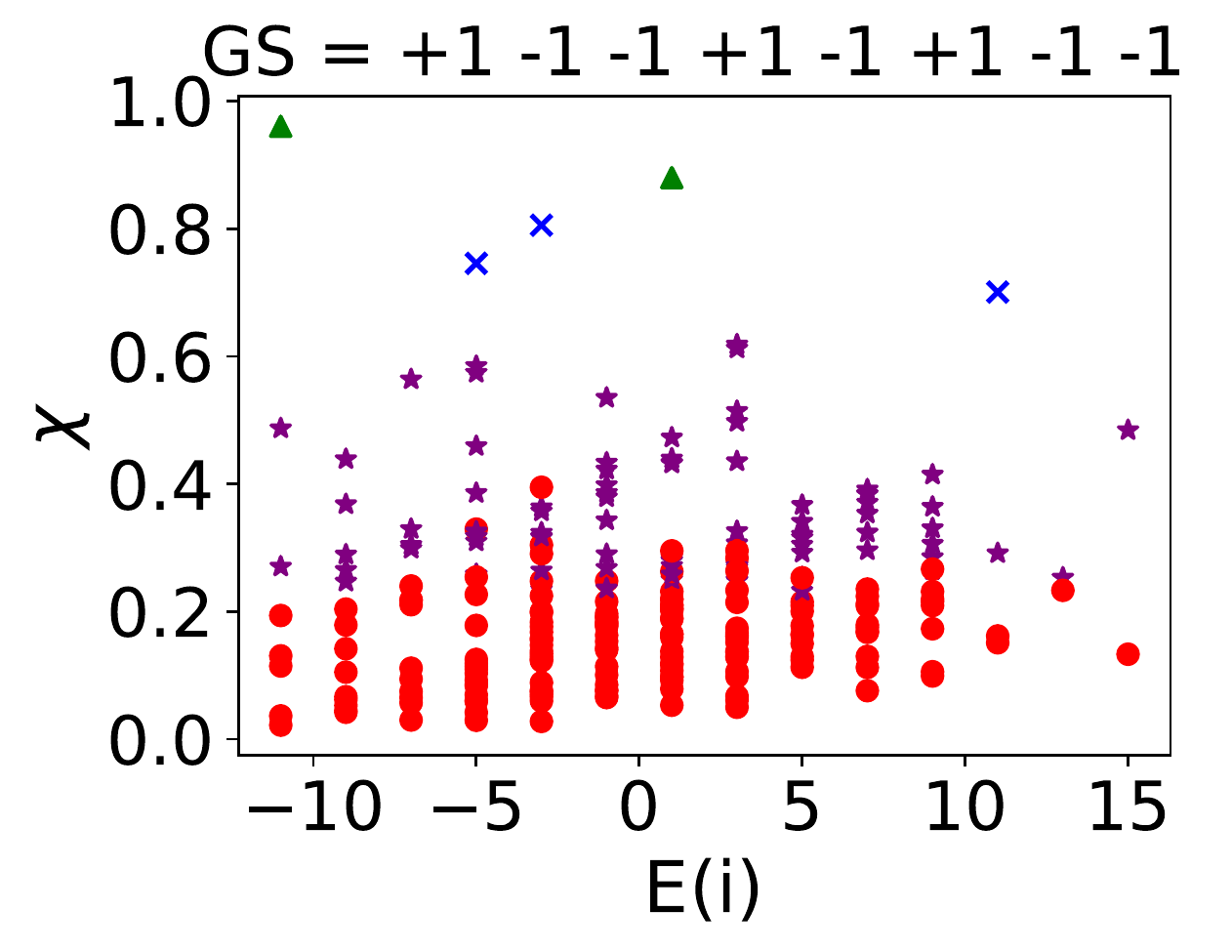}
    \includegraphics[width=0.13\textwidth]{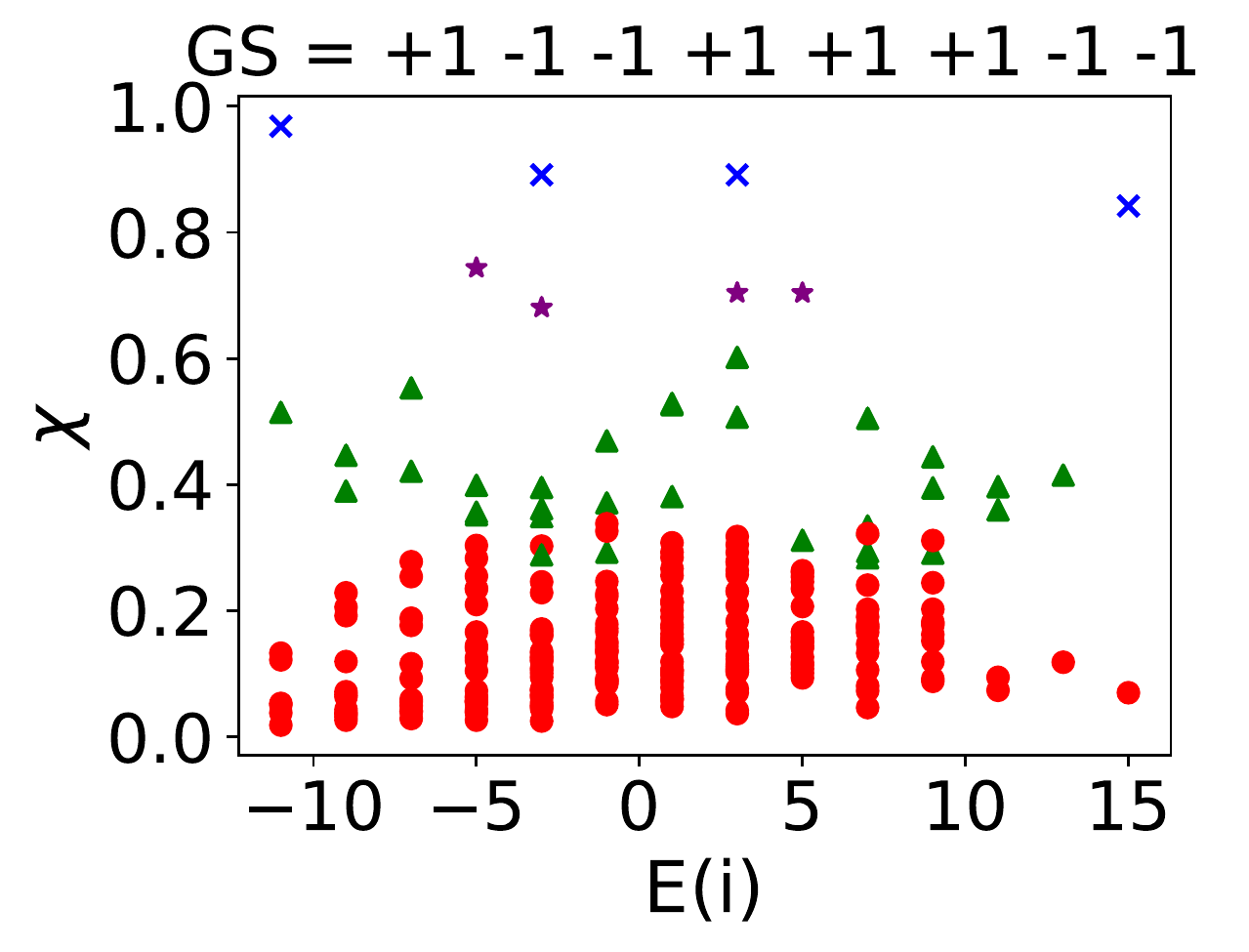}
    \includegraphics[width=0.13\textwidth]{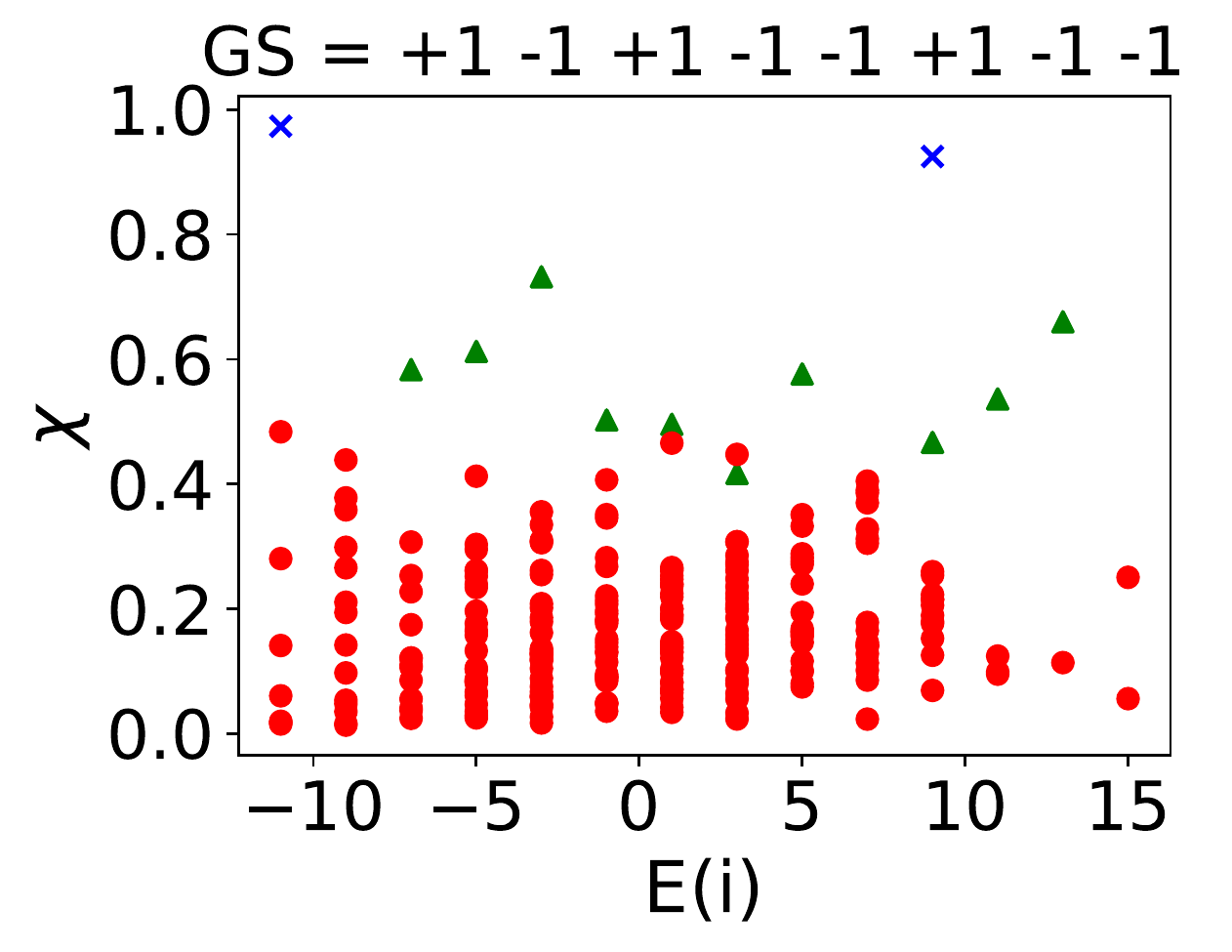}
    \includegraphics[width=0.13\textwidth]{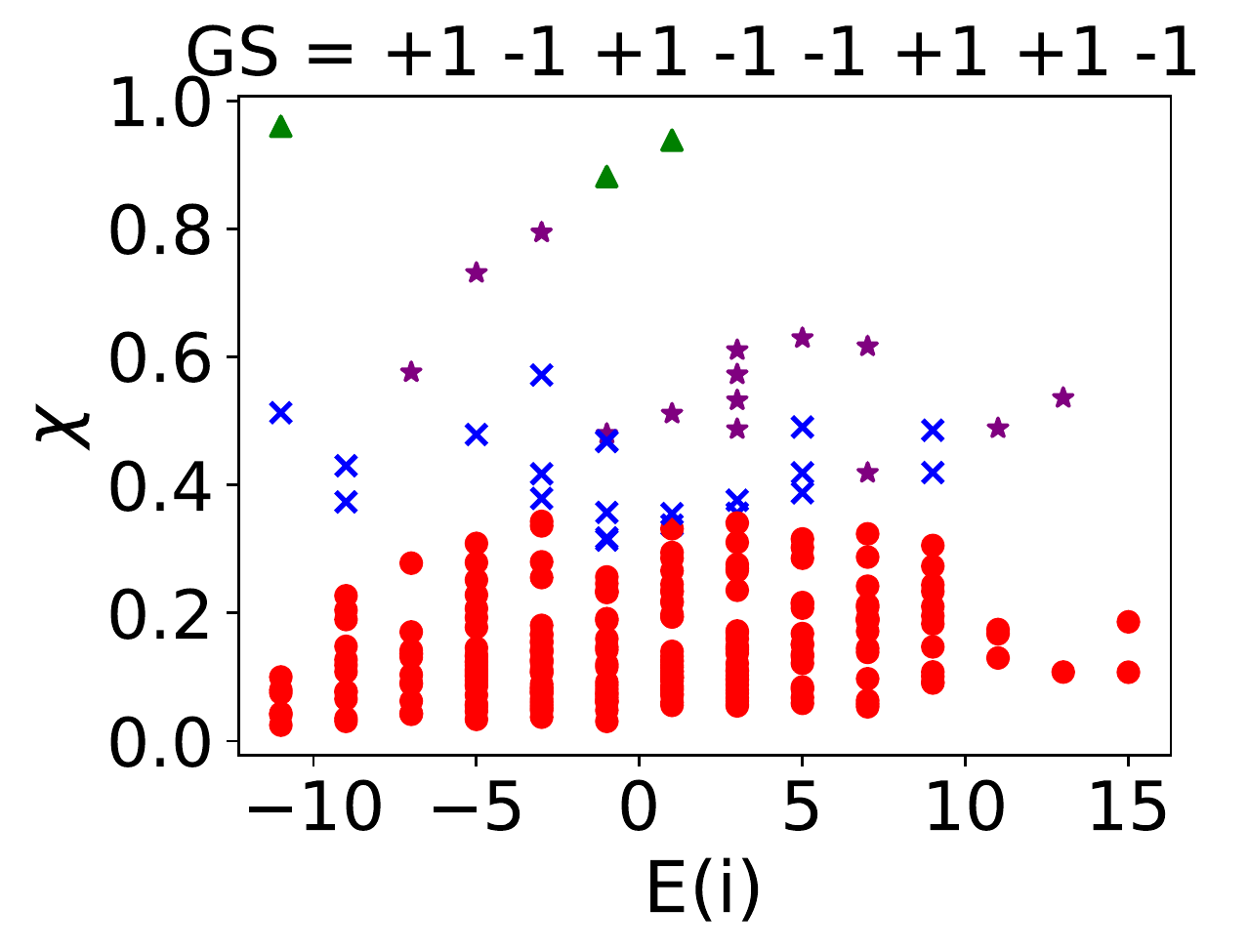}\\
    \includegraphics[width=0.13\textwidth]{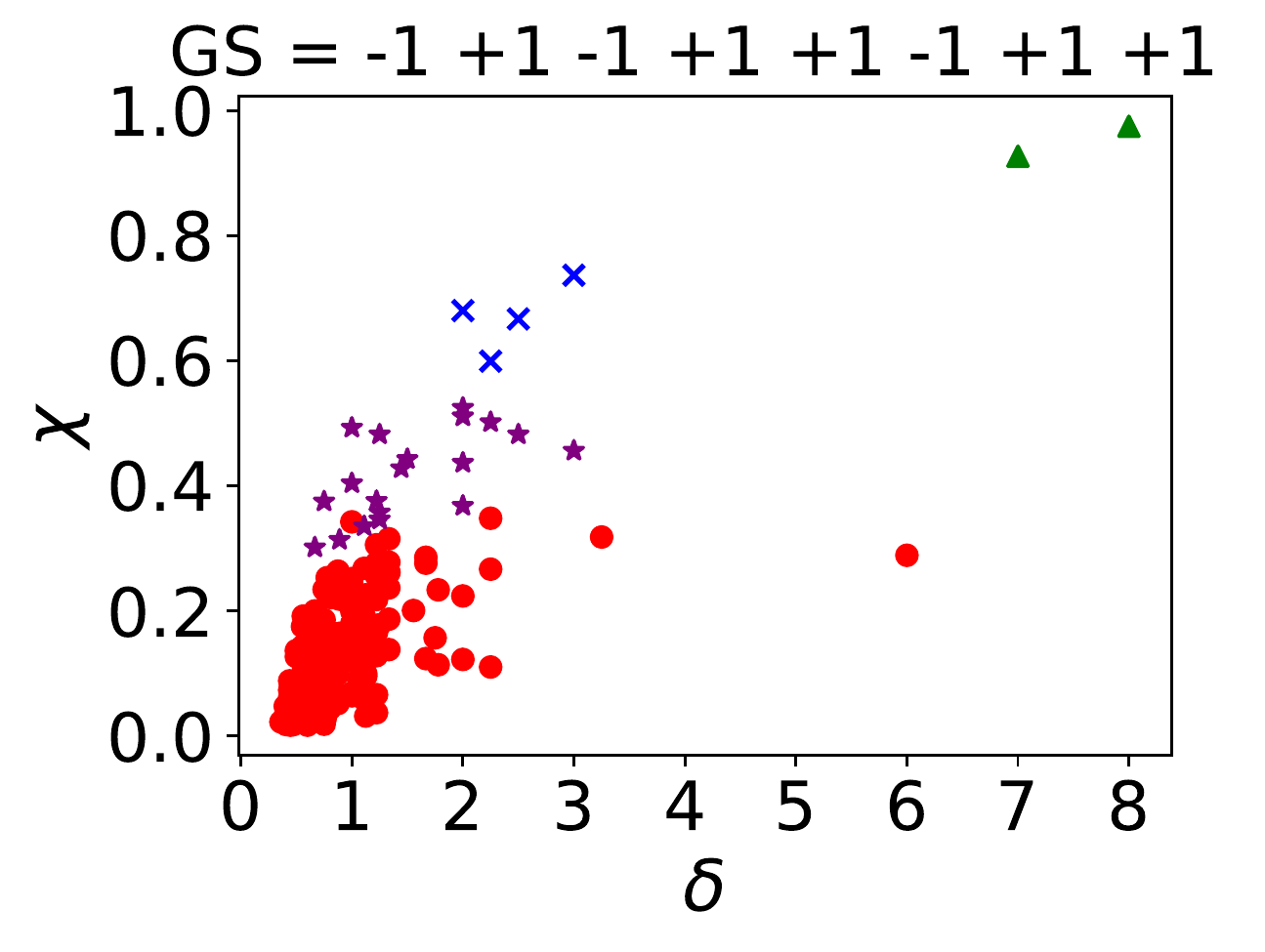}
    \includegraphics[width=0.13\textwidth]{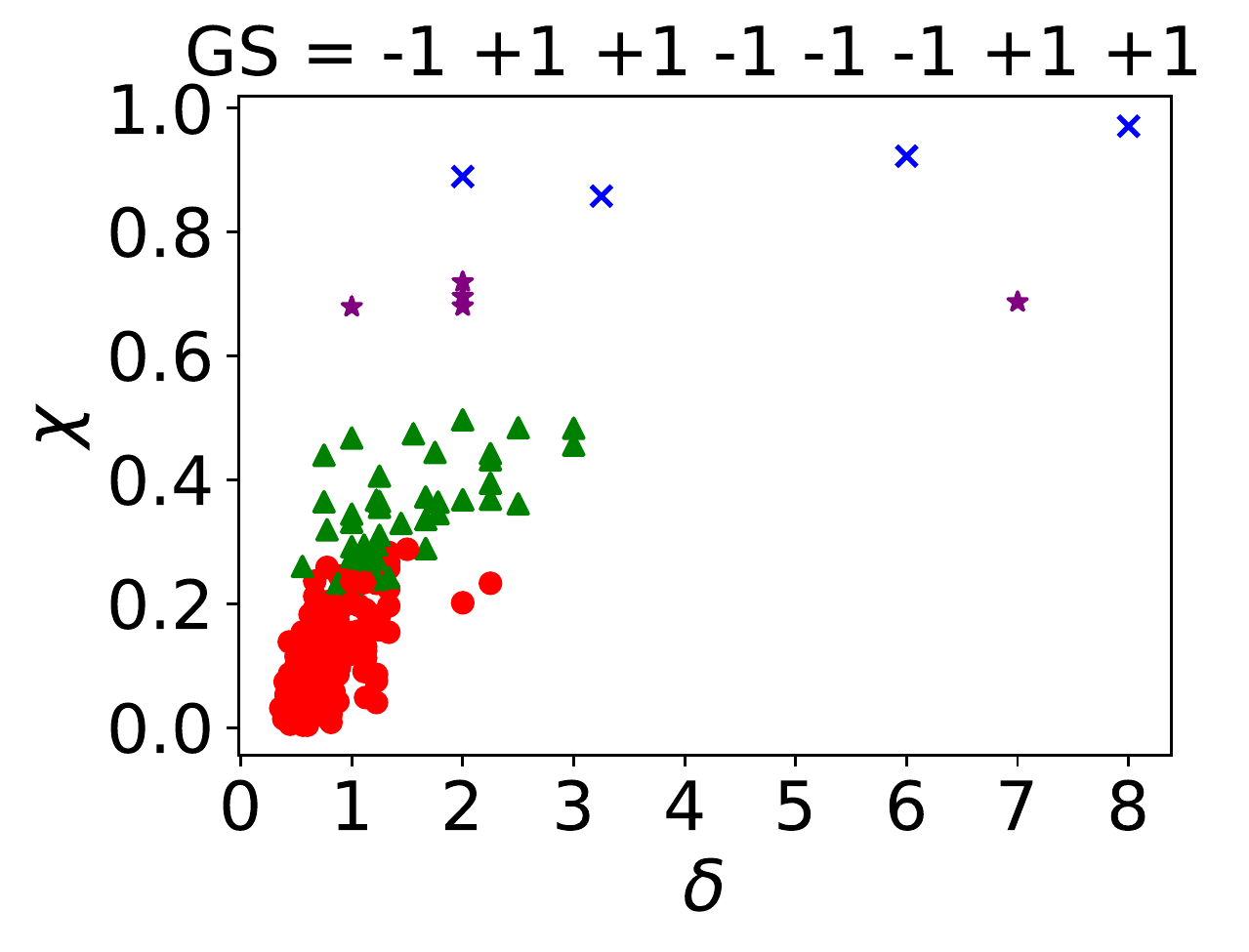}
    \includegraphics[width=0.13\textwidth]{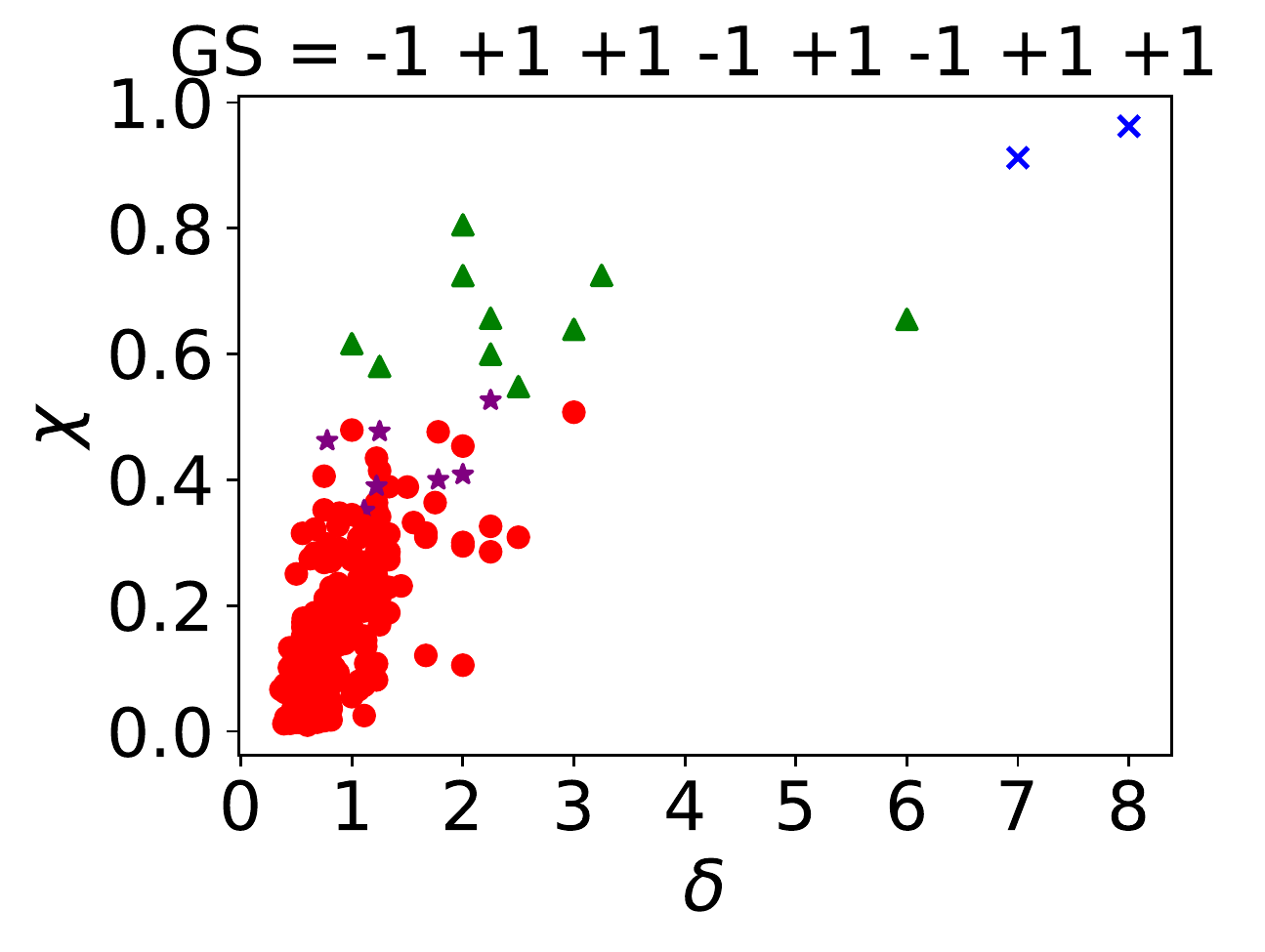}
    \includegraphics[width=0.13\textwidth]{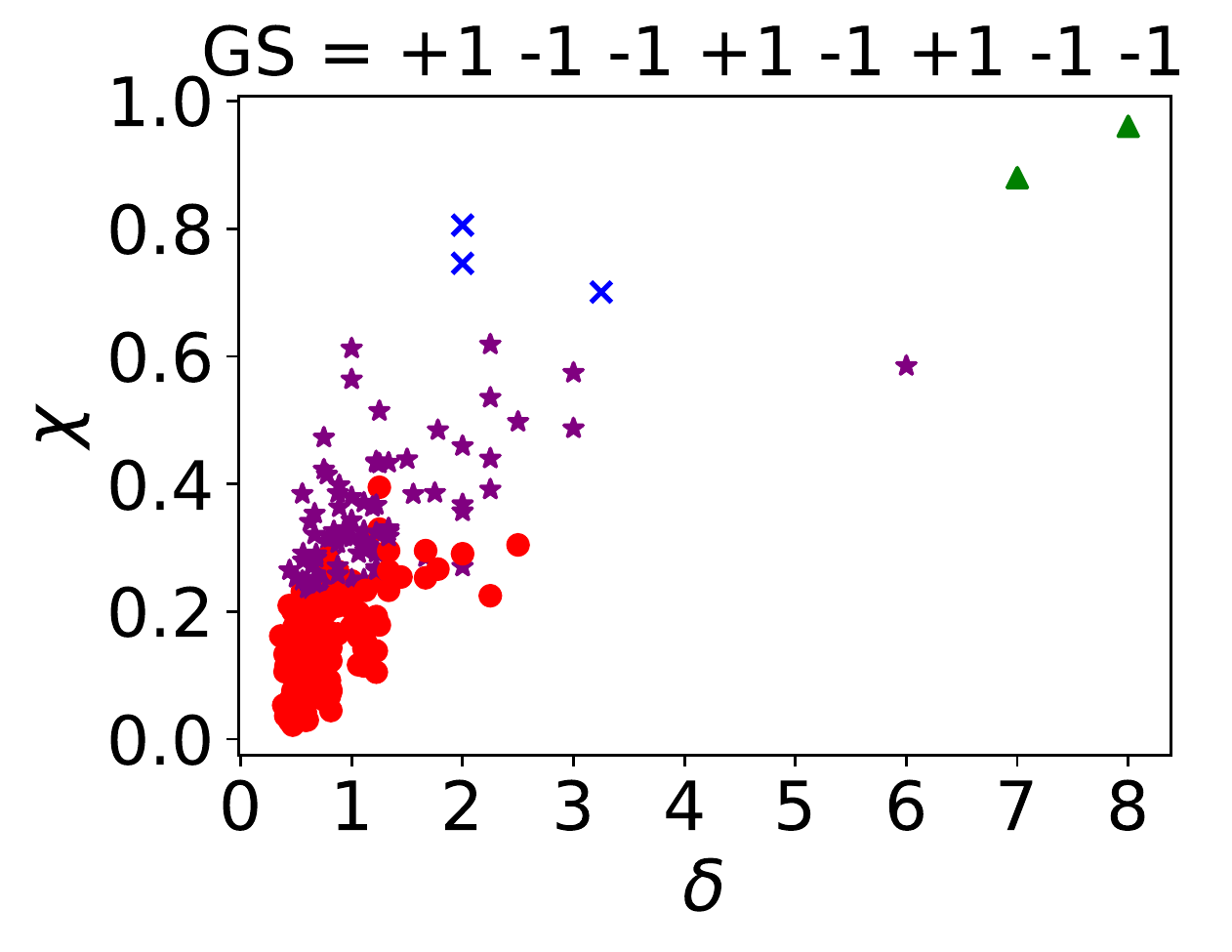}
    \includegraphics[width=0.13\textwidth]{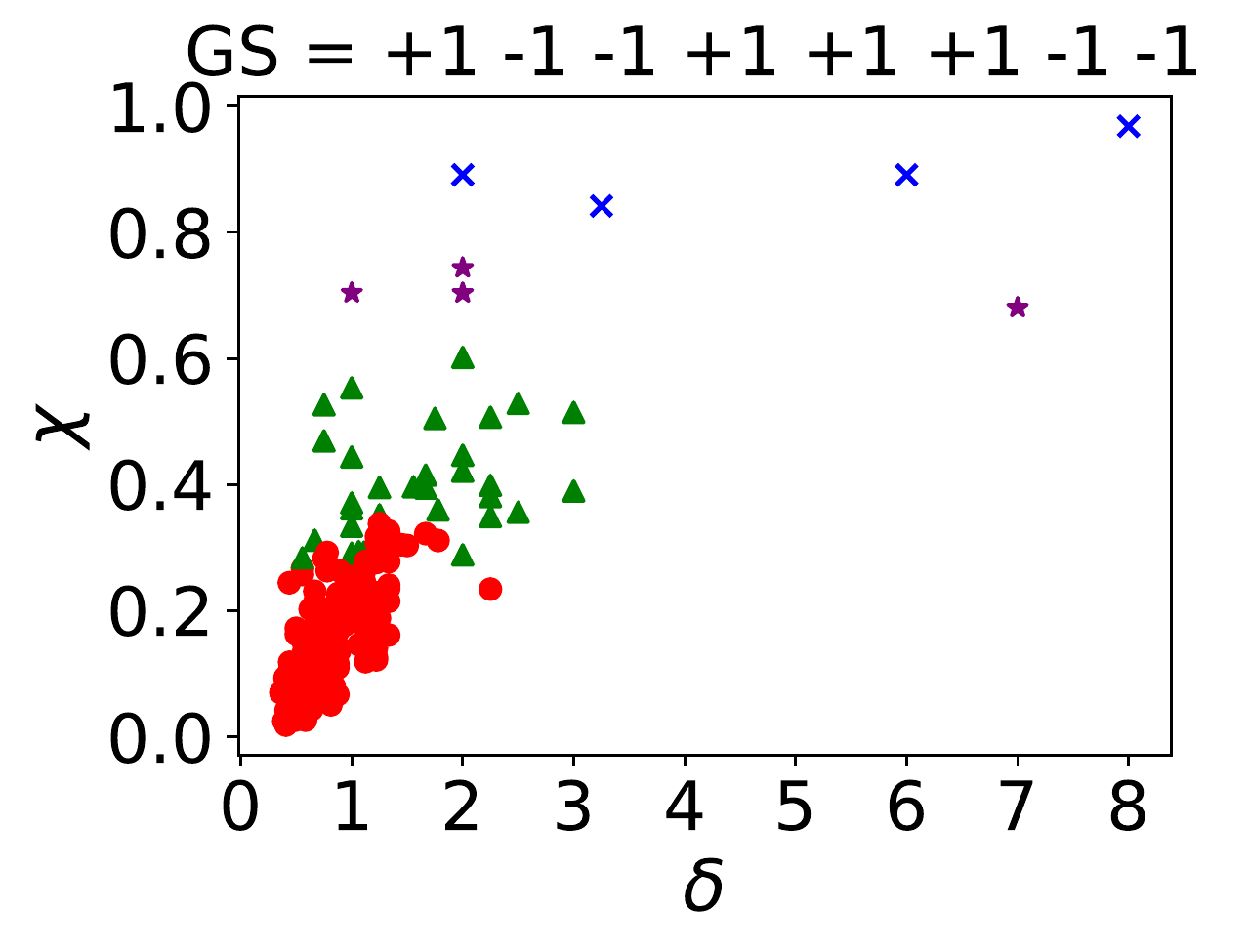}
    \includegraphics[width=0.13\textwidth]{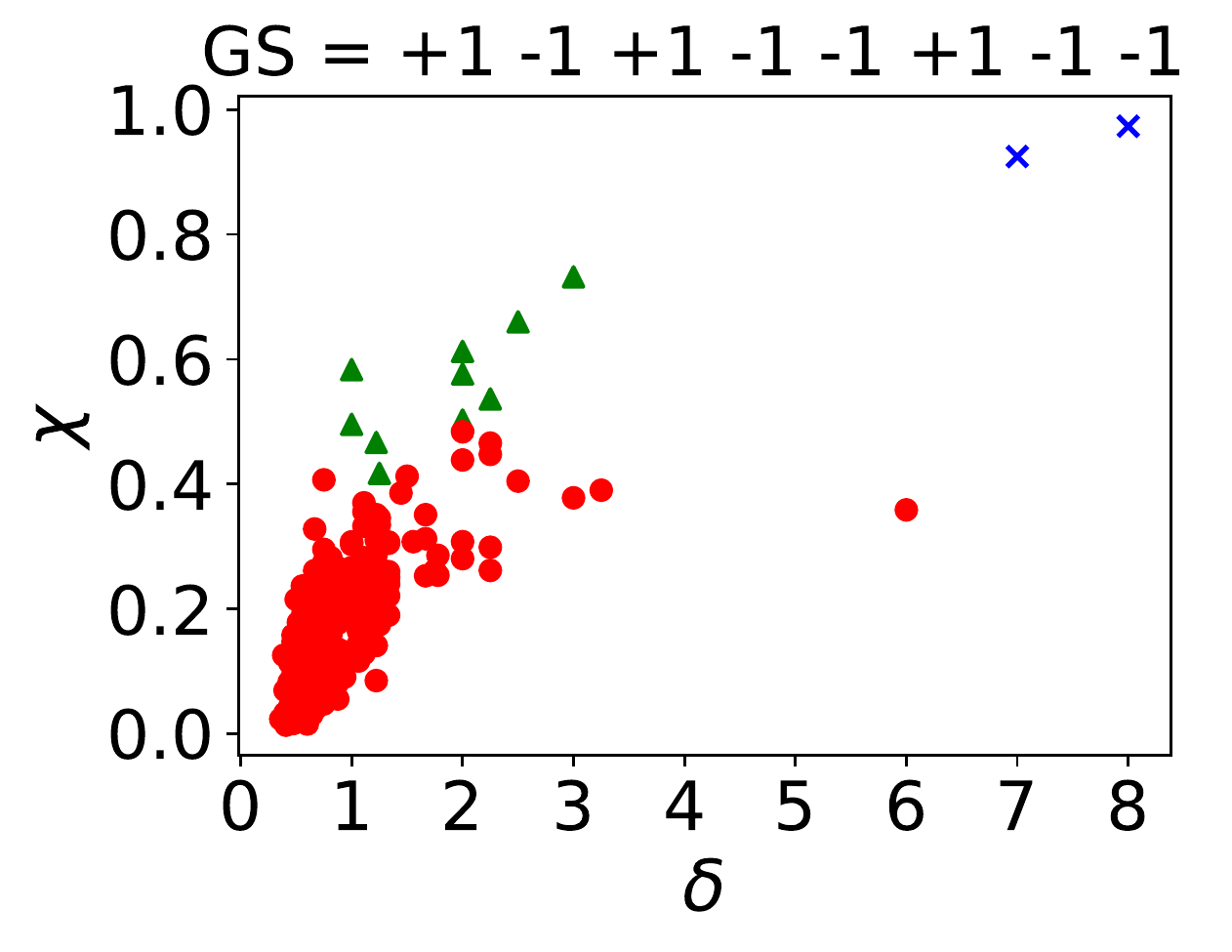}
    \includegraphics[width=0.13\textwidth]{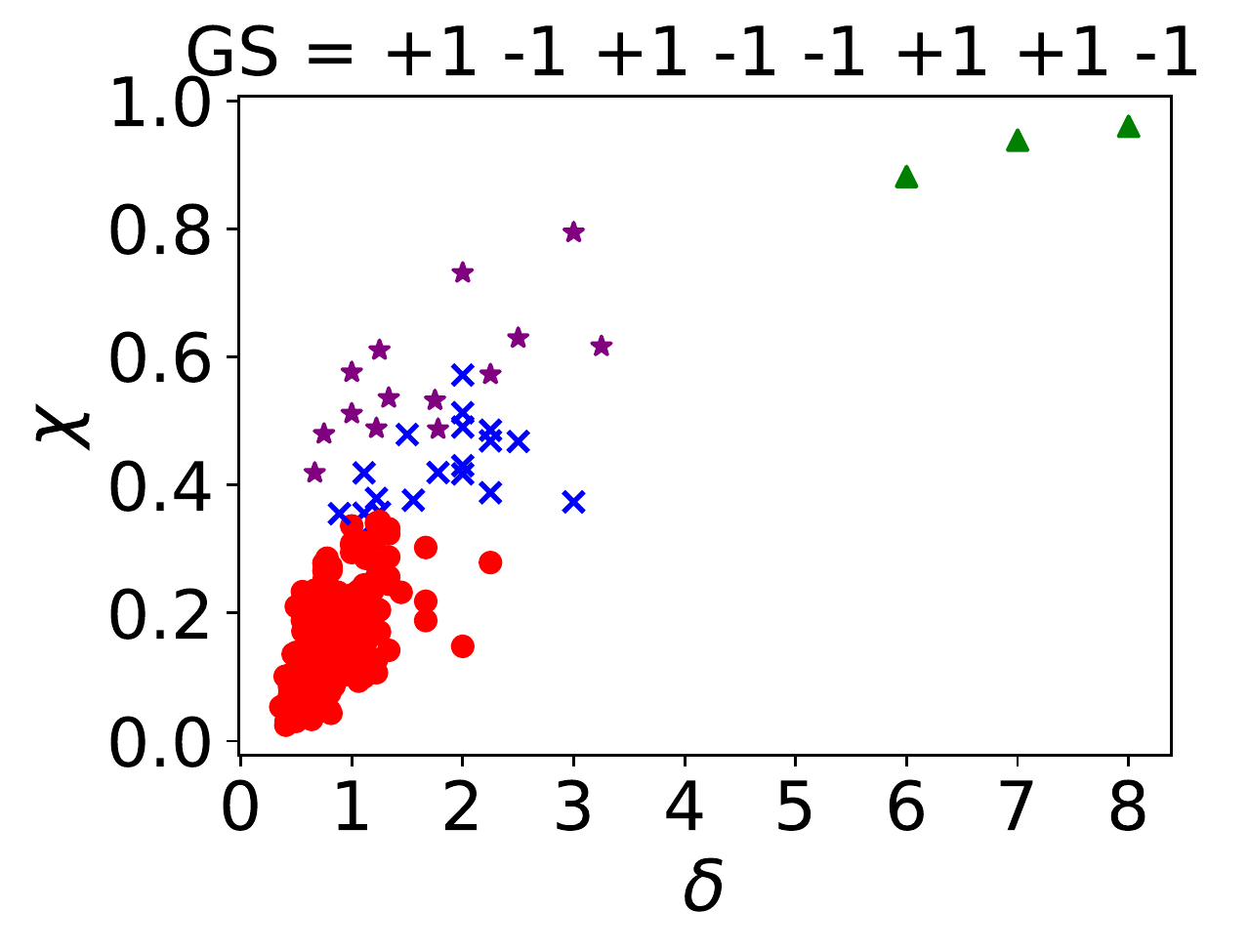}
    \caption{The rest of the data that completes Figure \ref{fig:summary_metrics_n8} with the other $7$ ground state mappings. Summary metric plots for the $N_8$ Ising. The $7$ columns correspond to each of the $8$ ground states; the titles of each sub plot are the exact optimal solution vectors. The three rows correspond to three different initial state metrics on the x-axis, the y-axis of each sub-plot is $\chi$. The first row has x-axis which are the hamming distance between the ground state and the specific initial state $i$. The second row has x-axis showing the energy of the initial state $i$ evaluated on the $N_8$ Ising. The third row has x-axis showing the $\delta$ metric for each initial state. }
    \label{fig:appendix_summary_metrics_N8}
\end{figure}

\begin{figure}[h!]
    \centering
    \includegraphics[width=0.7\textwidth]{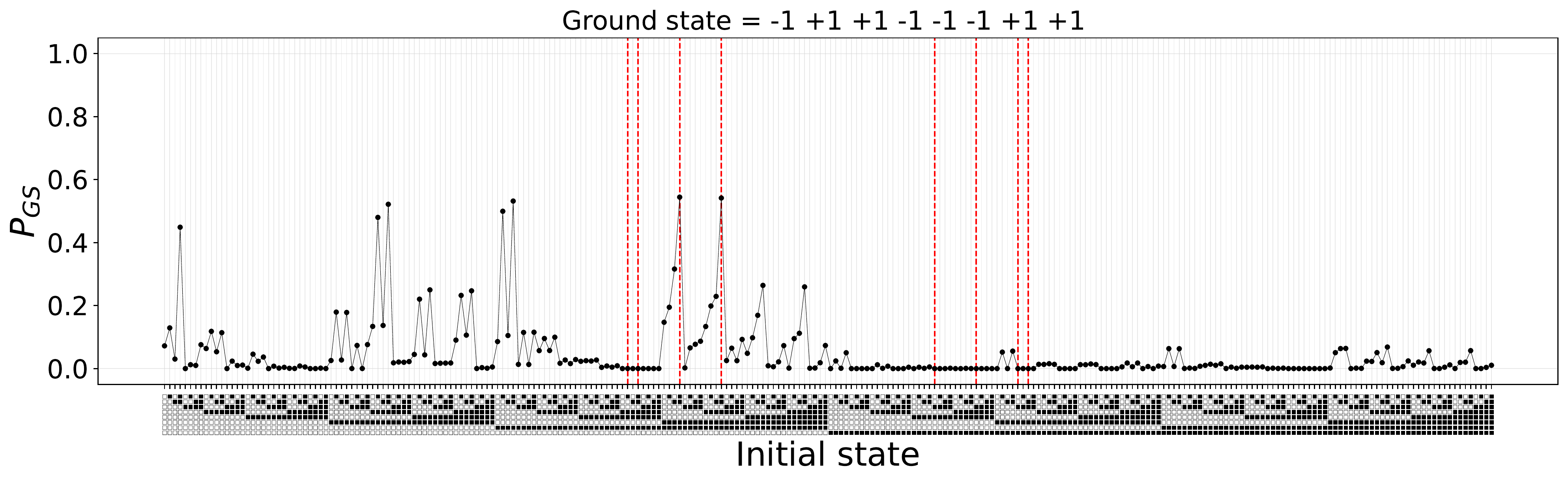}
    \includegraphics[width=0.7\textwidth]{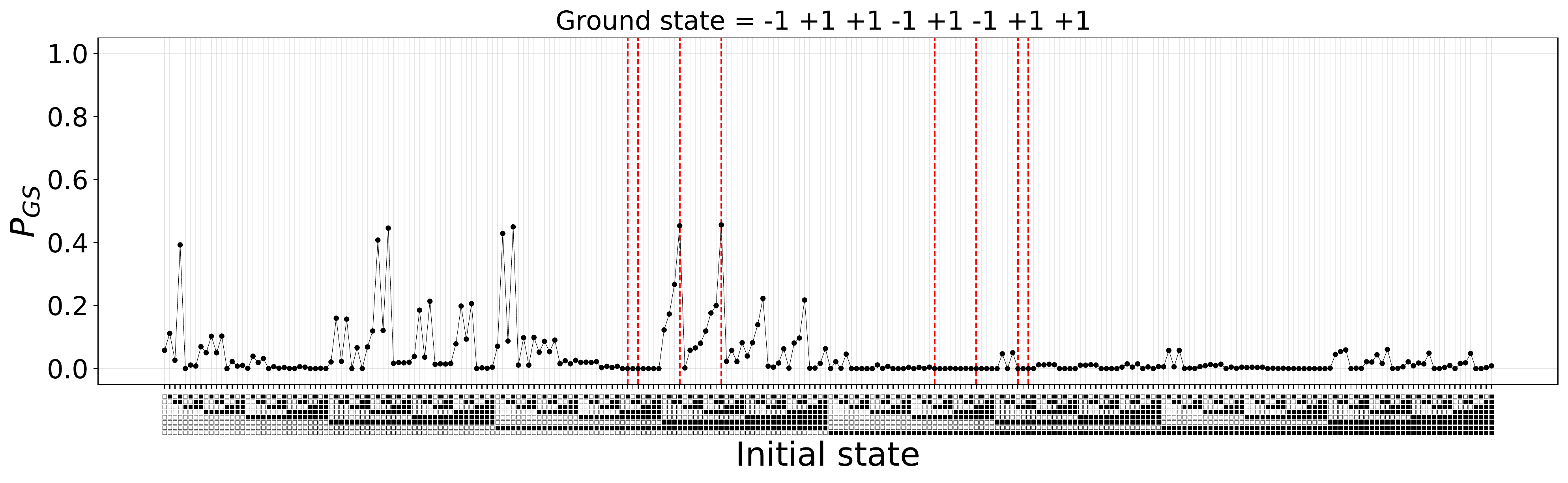}
    \includegraphics[width=0.7\textwidth]{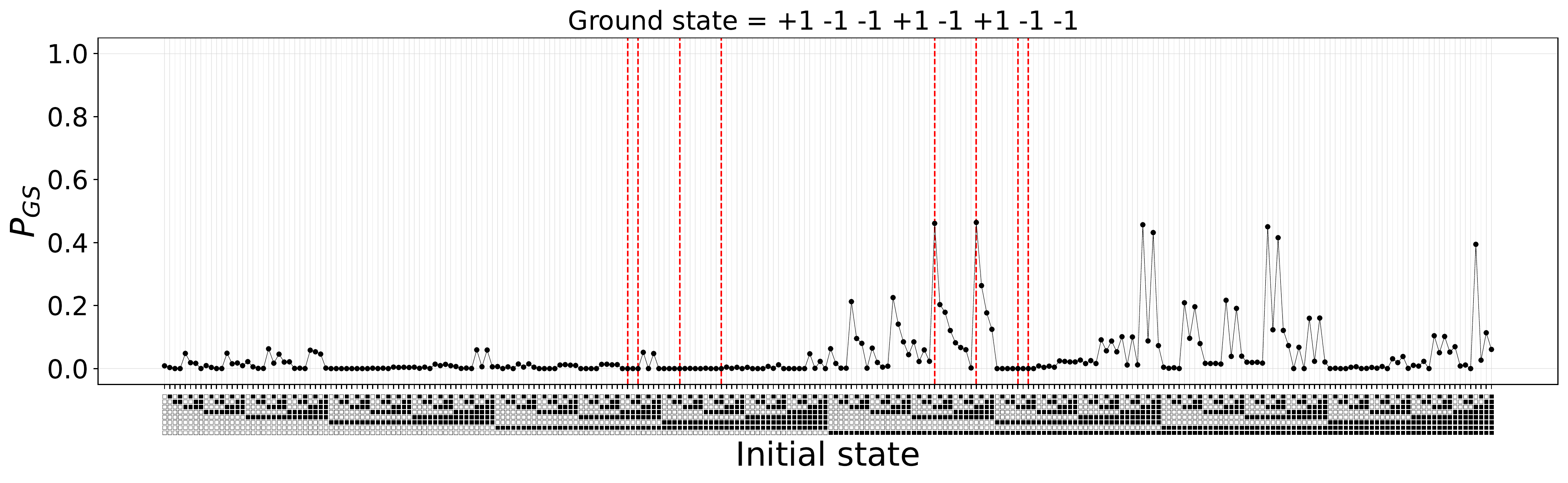}
    \includegraphics[width=0.7\textwidth]{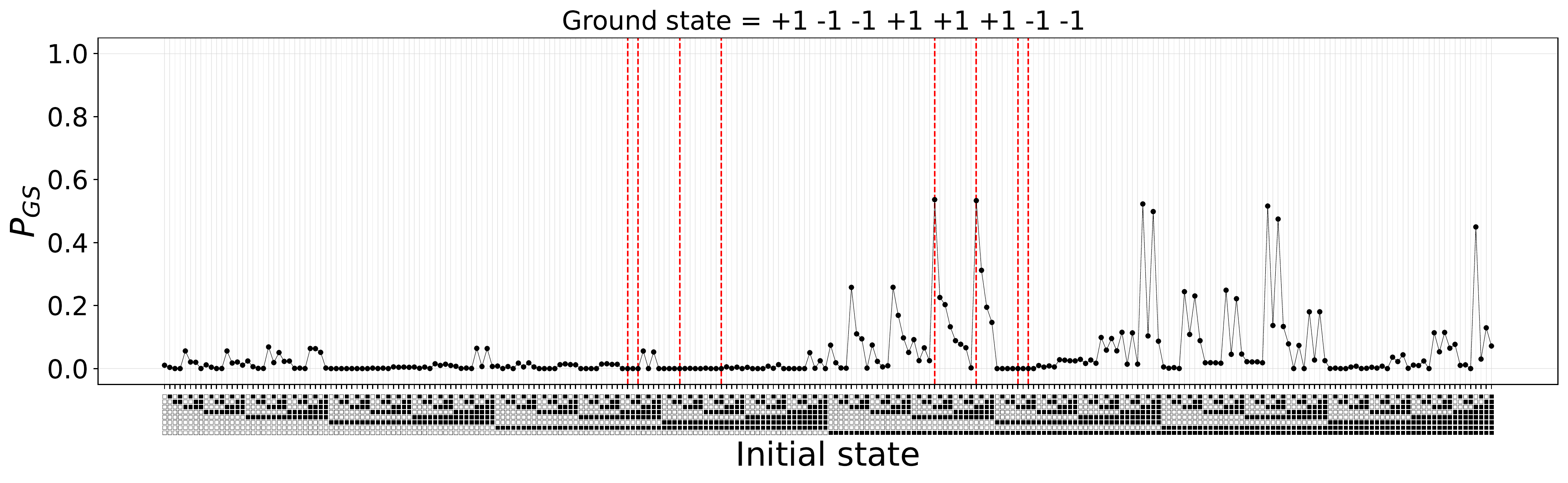}
    \includegraphics[width=0.7\textwidth]{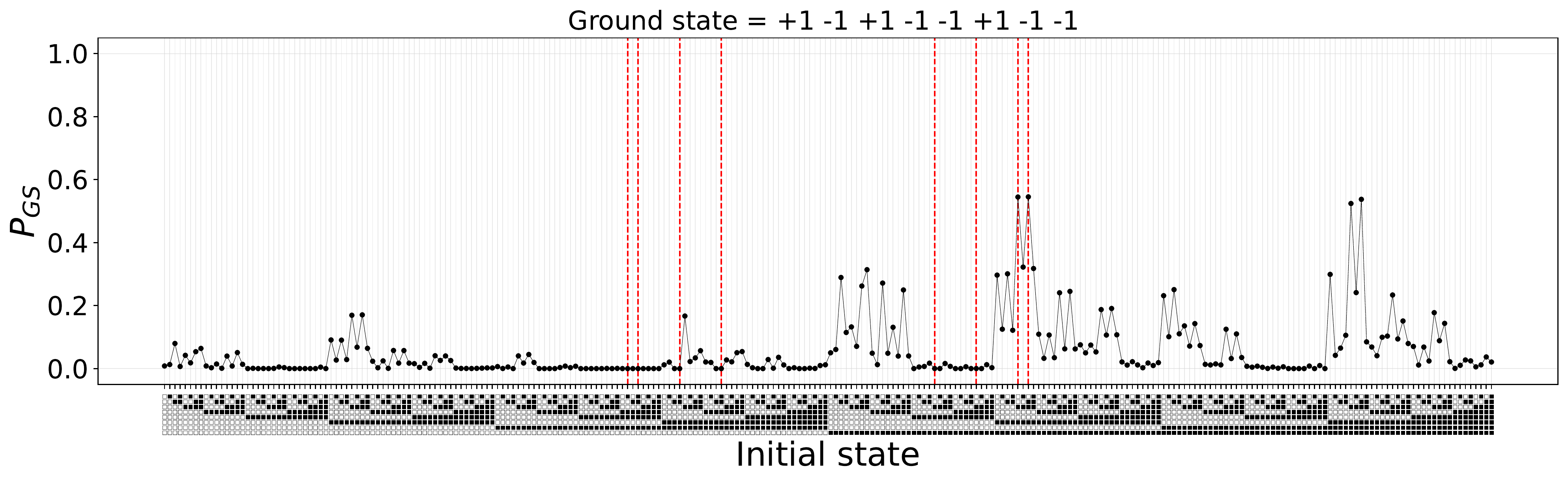}
    \includegraphics[width=0.7\textwidth]{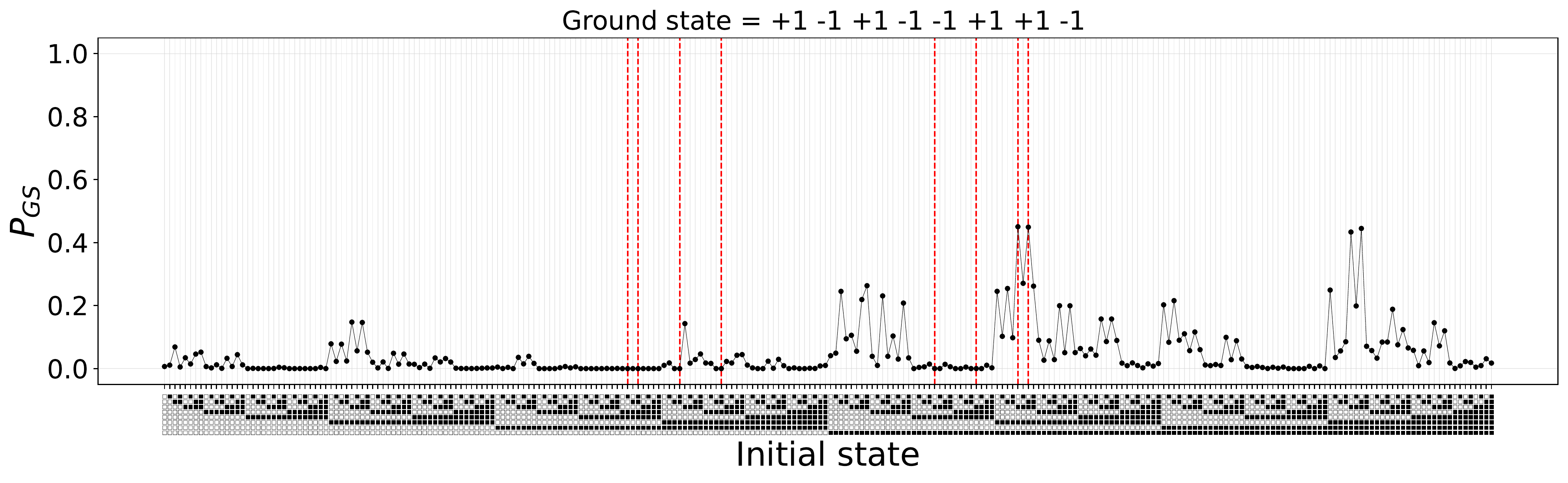}
    \caption{Continuation of Figure \ref{fig:RA_only_N8} showing the reverse annealing only ground state success proportions for the other six ground states of the $N_8$ Ising. In each plot all states on the x-axis which are groundstates are marked with vertical dashed red lines. }
    \label{fig:appendix_RA_only_N8}
\end{figure}

\end{document}